\documentclass[journal,a4paper,oneside,onecolumn]{IEEEtran} 

\usepackage{amsmath,amsfonts,amssymb,mathabx}

\newtheorem{thm}{Theorem}
\newtheorem{cor}[thm]{Corollary}
\newtheorem{lem}{Lemma}
\newtheorem{df}{Definition}
\newtheorem{rem}{Remark}

\newcommand{\lrB}[1]{\left[{#1}\right]}
\newcommand{\lrb}[1]{\left\{{#1}\right\}}
\newcommand{\lrsb}[1]{\left({#1}\right)}
\newcommand{\lrbar}[1]{\left|{#1}\right|}

\newcommand{\e}{\varepsilon}

\newcommand{\A}{\mathcal{A}}
\newcommand{\B}{\mathcal{B}}
\newcommand{\C}{\mathcal{C}}
\newcommand{\D}{\mathcal{D}}
\newcommand{\E}{\mathcal{E}}
\newcommand{\F}{\mathcal{F}}
\newcommand{\G}{\mathcal{G}}
\newcommand{\I}{\mathcal{I}}
\newcommand{\J}{\mathcal{J}}
\newcommand{\M}{\mathcal{M}}
\newcommand{\R}{\mathcal{R}}
\newcommand{\cS}{\mathcal{S}}
\newcommand{\T}{\mathcal{T}}
\newcommand{\U}{\mathcal{U}}
\newcommand{\V}{\mathcal{V}}
\newcommand{\W}{\mathcal{W}}
\newcommand{\X}{\mathcal{X}}
\newcommand{\Y}{\mathcal{Y}}
\newcommand{\Z}{\mathcal{Z}}

\newcommand{\mm}{\boldsymbol{m}}
\newcommand{\bp}{\boldsymbol{p}}
\newcommand{\uu}{\boldsymbol{u}}
\newcommand{\vv}{\boldsymbol{v}}
\newcommand{\ww}{\boldsymbol{w}}
\newcommand{\xx}{\boldsymbol{x}}
\newcommand{\yy}{\boldsymbol{y}}
\newcommand{\zz}{\boldsymbol{z}}
\newcommand{\cc}{\boldsymbol{c}}
\newcommand{\UU}{\boldsymbol{U}}
\newcommand{\VV}{\boldsymbol{V}}
\newcommand{\WW}{\boldsymbol{W}}
\newcommand{\YY}{\boldsymbol{Y}}
\newcommand{\ZZ}{\boldsymbol{Z}}
\newcommand{\bcF}{\boldsymbol{\F}}
\newcommand{\aalpha}{\boldsymbol{\alpha}}
\newcommand{\bbeta}{\boldsymbol{\beta}}

\newcommand{\NN}{\mathbb{N}}

\newcommand{\fC}{\mathfrak{C}}

\newcommand{\hr}{\widehat{r}}
\newcommand{\hu}{\widehat{u}}
\newcommand{\hM}{\widehat{M}}
\newcommand{\hU}{\widehat{U}}
\newcommand{\hZ}{\widehat{Z}}
\newcommand{\hzz}{\widehat{\zz}}

\newcommand{\tr}{\widetilde{r}}
\newcommand{\tw}{\widetilde{w}}
\newcommand{\tW}{\widetilde{W}}
\newcommand{\tZ}{\widetilde{Z}}
\newcommand{\tzz}{\widetilde{\zz}}

\newcommand{\tcD}{\widetilde{\D}}
\newcommand{\tcS}{\widetilde{\cS}}
\newcommand{\tcW}{\widetilde{\W}}
\newcommand{\tmu}{\widetilde{\mu}}

\newcommand{\hmm}{\widehat{\mm}}

\newcommand{\chu}{\widecheck{u}}
\newcommand{\chU}{\widecheck{U}}
\newcommand{\chZ}{\widecheck{Z}}
\newcommand{\chzz}{\widecheck{\zz}}

\newcommand{\oH}{\overline{H}}
\newcommand{\uH}{\underline{H}}
\newcommand{\uT}{\underline{\T}}
\newcommand{\oT}{\overline{\T}}
\newcommand{\orho}{\overline{\rho}}
\newcommand{\urho}{\underline{\rho}}

\newcommand{\omu}{\overline{\mu}}
\newcommand{\oO}{\overline{O}}
\newcommand{\oQ}{\overline{Q}}
\newcommand{\obigcup}{\overline{\bigcup}}

\newcommand{\limn}{\lim_{n\to\infty}}
\newcommand{\liminfn}{\liminf_{n\to\infty}}
\newcommand{\pliminfn}{\operatornamewithlimits{\text{p-liminf}}_{n\to\infty}}
\newcommand{\plimsupn}{\operatornamewithlimits{\text{p-limsup}}_{n\to\infty}}
\newcommand{\Prod}{\operatornamewithlimits{\text{\Large $\times$}}}
\newcommand{\im}{\mathrm{Im}}

\newcommand{\Error}{\mathrm{Error}}
\newcommand{\Prob}{\mathrm{Prob}}

\newcommand{\RJXG}{\R_{\mathrm{JXG}}}
\newcommand{\oRCRNG}{\overline{\R}_{\mathrm{CRNG}}}
\newcommand{\tRCRNG}{\widetilde{\R}_{\mathrm{CRNG}}}
\newcommand{\RCRNG}{\R_{\mathrm{CRNG}}}
\newcommand{\RHKCMG}{\R_{\mathrm{HKCMG}}}
\newcommand{\RHK}{\R_{\mathrm{HK}}}
\newcommand{\RCMG}{\R_{\mathrm{CMG}}}

\allowdisplaybreaks

\newcount\pictmp
\def\ave#1#2#3{
 \pictmp=#2
 \advance\pictmp by #3
 \ifodd\pictmp
 \divide\pictmp by 2
 \edef#1{\number\pictmp.5}
 \else
 \divide\pictmp by 2
 \edef#1{\number\pictmp}
 \fi
}
\def\qbline(#1,#2)(#3,#4){
 \ave{\picx}{#1}{#3}\ave{\picy}{#2}{#4}
 \qbezier(#1,#2)(\picx,\picy)(#3,#4)
}

\title{
 On the Achievability\\
 of Interference Channel Coding
}

\author{
 \IEEEauthorblockN{Jun~Muramatsu~\IEEEmembership{Senior Member,~IEEE}}
 \\
 \IEEEauthorblockA{
  NTT Communication Science Laboratories, NTT Corporation,\\
  2-4, Hikaridai, Seika-cho, Soraku-gun, Kyoto 619-0237, Japan\\
  E-mail: jun.muramatsu@ieee.org
 }
}

\usepackage{amsmath}
\begin{document}
\maketitle

\begin{abstract}
 This paper investigates the achievability
 of the interference channel coding.
 It is clarified that
 the rate-splitting technique is unnecessary
 to achieve Han-Kobayashi and Jian-Xin-Garg inner regions.
 Codes are constructed by using
 sparse matrices (with logarithmic column degree)
 and the constrained-random-number generators.
 By extending the problem,
 we can establish a possible extension of known inner regions.
\end{abstract}

\section{Introduction}

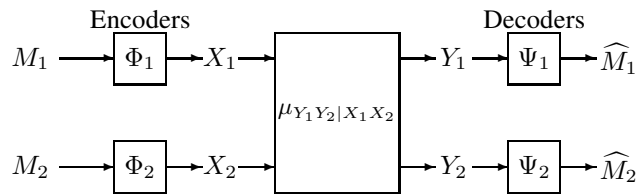
\begin{figure}[ht]
\begin{center}
 \unitlength 0.48mm
 \begin{picture}(178,55)(0,0)
  
  \put(33,48){\makebox(0,0){Encoders}}
  \put(141,48){\makebox(0,0){Decoders}}
  
  \put(3,37){\makebox(0,0){$M_1$}}
  \put(11,37){\vector(1,0){15}}
  \put(26,30){\framebox(14,14){$\Phi_1$}}
  \put(40,37){\vector(1,0){10}}
  \put(55,37){\makebox(0,0){$X_1$}}
  \put(60,37){\vector(1,0){10}}
  
  \put(3,7){\makebox(0,0){$M_2$}}
  \put(11,7){\vector(1,0){15}}
  \put(26,0){\framebox(14,14){$\Phi_2$}}
  \put(40,7){\vector(1,0){10}}
  \put(55,7){\makebox(0,0){$X_2$}}
  \put(60,7){\vector(1,0){10}}
  
  \put(70,0){\framebox(34,44){\small $\mu_{Y_1Y_2|X_1X_2}$}}
  
  \put(104,37){\vector(1,0){10}}
  \put(119,37){\makebox(0,0){$Y_1$}}
  \put(124,37){\vector(1,0){10}}
  \put(134,30){\framebox(14,14){$\Psi_1$}}
  \put(148,37){\vector(1,0){10}}
  \put(159,37){\makebox(0,0)[l]{$\hM_1$}}
  
  \put(104,7){\vector(1,0){10}}
  \put(119,7){\makebox(0,0){$Y_2$}}
  \put(124,7){\vector(1,0){10}}
  \put(134,0){\framebox(14,14){$\Psi_2$}}
  \put(148,7){\vector(1,0){10}}
  \put(159,7){\makebox(0,0)[l]{$\hM_2$}}
 \end{picture}
\end{center}
\caption{Interference Channel Coding}
\label{fig:ic}
\end{figure}

\begin{figure}[ht]
\begin{center}
 \unitlength 0.48mm
 \begin{picture}(178,55)(0,0)
  
  \put(33,48){\makebox(0,0){Encoders}}
  \put(141,48){\makebox(0,0){Decoders}}
  
  \put(3,22){\makebox(0,0){$M_0$}}
  \put(11,22){\line(2,3){10}}
  \put(11,22){\line(2,-3){10}}
  
  \put(3,37){\makebox(0,0){$M_1$}}
  \put(11,37){\vector(1,0){15}}
  \put(26,30){\framebox(14,14){$\Phi_1$}}
  \put(40,37){\vector(1,0){10}}
  \put(55,37){\makebox(0,0){$X_1$}}
  \put(60,37){\vector(1,0){10}}
  
  \put(3,7){\makebox(0,0){$M_2$}}
  \put(11,7){\vector(1,0){15}}
  \put(26,0){\framebox(14,14){$\Phi_2$}}
  \put(40,7){\vector(1,0){10}}
  \put(55,7){\makebox(0,0){$X_2$}}
  \put(60,7){\vector(1,0){10}}
  
  \put(70,0){\framebox(34,44){\small $\mu_{Y_1Y_2|X_1X_2}$}}
  
  \put(104,37){\vector(1,0){10}}
  \put(119,37){\makebox(0,0){$Y_1$}}
  \put(124,37){\vector(1,0){10}}
  \put(134,30){\framebox(14,14){$\Psi_1$}}
  \put(148,37){\vector(1,0){10}}
  \put(159,37){\makebox(0,0)[l]{$\hM_0,\hM_1$}}
  
  \put(104,7){\vector(1,0){10}}
  \put(119,7){\makebox(0,0){$Y_2$}}
  \put(124,7){\vector(1,0){10}}
  \put(134,0){\framebox(14,14){$\Psi_2$}}
  \put(148,7){\vector(1,0){10}}
  \put(159,7){\makebox(0,0)[l]{$\hM_0,\hM_2$}}
 \end{picture}
\end{center}
\caption{Interference Channel Coding with Common Message}
\label{fig:icc}
\end{figure}
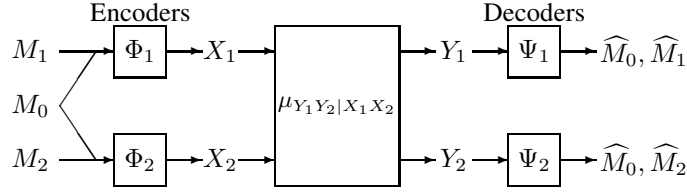

Clarifying the single-letterized capacity region
for general discrete memoryless interference channels (Fig. \ref{fig:ic})
has been one of the notable open problems for over forty years
from the first study by Ahlswede \cite{A74}.
The best known inner region was established by Han and Kobayashi \cite{HK81},
where they used the idea of rate splitting technique
introduced by Carleial \cite{C78}.
A simpler formula equivalent to the Han-Kobayashi region was established 
by Chong, Motani, Garg, and El Gamal \cite{CMGE08}.
It is shown numerically in \cite{NXY15} that
the Han-Kobayashi region is a proper subset of the capacity region.
Inner and outer regions of specific interference channels
are comprehensively summarized in \cite[Chapter 6]{EK11}.
The problem of interference channel coding with a common message
(Fig. \ref{fig:icc})
was introduced by Jiang, Xin and Garg \cite{JXG08},
where they established an inner region for this problem
by using the rate-splitting technique.
The multi-letterized capacity region
for general multiple-input-multiple-output channels
was established by Somekh-Baruch and Verd\'u \cite{SV06}
and Muramatsu and Miyake \cite{CRNG-MULTI}
by using information-spectrum methods \cite{HAN,CRNG,VH94}.
Although interference channels are included
in the class of general multiple-input-multiple-output channels,
the single-letterized formula for the capacity region remains unclear.

Contributions of this paper are summarized as follows.
\begin{itemize}
 \item
 It is clarified that
 the rate-splitting technique is unnecessary
 to achieve Han-Kobayashi and Jian-Xin-Garg inner regions.
 In fact, the rate-splitting technique is not used in the construction.
 Details are presented in Section \ref{sec:crng}.
 \item 
 Codes are constructed by using
 sparse matrices (with logarithmic column degree)
 and the constrained-random-number generators \cite{CRNG,CRNG-MULTI,SW2CC}.
 The proof of achievability is based on the hash property
 of the ensemble of functions.
 Details are presented in Section \ref{sec:proof-crng}.
 \item
 Construction of codes
 can be applied to channels with two or more inputs/outputs,
 where all encoders have access to all common messages.
 By extending the problem
 (hereafter we call the extension the {\em super-problem}),
 we can establish a possible extension of known inner regions.
 Details are presented in Section \ref{sec:crng-full}.
\end{itemize}

\section{Definitions and Notations}

Throughout this paper,
we use the notation $a_{\cS}\equiv\{a_s\}_{s\in\cS}$
to represent the set of all elements
(e.g.\ sets, symbols, sequences, random variables, functions)
$a_s$ with index $s$ belongs to set $\cS$.
We use the notation $|\cS|$ to represent the cardinality of set $\cS$.

Here, we introduce the definition of an achievable region
for a general multiple-input-multiple-output channel \cite{CRNG-MULTI}.
Let $\cS$ be the index set of multiple messages,
$\I$ be the index set of channel inputs,
and $\J$ be the index set of channel outputs.
A general channel is characterized 
by sequence $\{\mu_{Y^n_{\J}|X^n_{\I}}\}_{n=1}^{\infty}$
of conditional distributions,
where $X^n_{\I}\equiv\{X^n_i\}_{i\in\I}$
is a set of random variables of multiple channel inputs,
and $Y^n_{\J}\equiv\{Y^n_j\}_{j\in\J}$
is a set of random variables of multiple channel outputs.
For each $i\in\I$ and $n\in\NN$,
let $\X_i^n$ be the alphabet of random variable $X_i^n$.
For each $j\in\J$ and $n\in\NN$,
let $\Y_j^n$ be the alphabet of random variable $Y_j^n$.
For each $s\in\cS$ and $n\in\NN$,
let $M_{s,n}$ be a random variable of the $s$-th message
subject to the uniform distribution on alphabet $\M_{s,n}$.
We assume that
$\{M_{s,n}\}_{s\in\cS}$ are mutually independent and uniformly distributed.
For each $i\in\I$, 
let $\cS_i$ be the index set of sources available to Encoder $i$,
where $\cS_i\subset\cS$.
Encoder $i$ generates the channel input $X^n_i$
from the set of messages $M_{\cS_i,n}\equiv\{M_{s,n}\}_{s\in\cS_i}$.
For each $j\in\J$,
let $\D_j$ be the index set of messages reproduced by Decoder $j$,
where $\D_j\subset\cS$.
Decoder $j$ receives the channel output $Y^n_j$
and reproduces a set of messages
$\hM_{\D_j,n}\equiv\{\hM_{j,s,n}\}_{s\in\D_j}$,
where $\hM_{j,s,n}\in\M_{s,n}$
is the $s$-th message reproduced by Decoder $j$.
We expect that, with probability close to $1$, $\hM_{j,s,n}=M_{s,n}$
for all $j\in\J$ and $s\in\D_j$ by letting $n$ be sufficiently large.
We call rate vector $\{R_s\}_{s\in\cS}$ {\em achievable}
if there is a (possibly stochastic) code
$\{(\{\Phi_{i,n}\}_{i\in\I},\{\Psi_{j,n}\}_{j\in\J})\}_{n=1}^{\infty}$
consisting of encoders $\Phi_{i,n}:\Prod_{s\in\cS_i}\M_{s,n}\to\X^n_i$
and decoders $\Psi_{j,n}:\Y^n_j\to\Prod_{s\in\D_j}\M_{s,n}$ such that
\begin{gather*}
 \liminfn\frac{\log_2 |\M_{s,n}|}n\geq R_s
 \quad\text{for all}\ s\in\cS
 \\
 \limn
 \Prob\lrsb{
  \hM_{j,s,n}\neq M_{s,n}\ \text{for some}\ j\in\J\ \text{and}\ s\in\D_j
 }
 =0,
\end{gather*}
where $X_i^n\equiv\Phi_{i,n}(M_{\cS_i})$,
$\hM_{\D_j,n}\equiv\Psi_{j,n}(Y_j^n)$,
and the joint distribution of
$(M_{\cS,n},X_{\I}^n,Y_{\J}^n,\hM_{\D_{\J},n})$ is given as
\begin{align*}
 &
 \mu_{M_{\cS,n}X_{\I}^nY_{\J}^n\hM_{\D_{\J},n}}(
  \mm_{\cS},\xx_{\I},\yy_{\J},\hmm_{\D_{\J}}
 )
 \notag
 \\*
 &
 =
 \lrB{\prod_{j\in\J}\mu_{\hM_{\D_j,n}|Y_{j}^n}(\hmm_{\D_j}|\yy_{j})}
 \mu_{Y_{\J}^n|X_{\I}^n}(\yy_{\J}|\xx_{\I})
 \lrB{\prod_{i\in\I}\mu_{X_i^n|M_{\cS_i,n}}(\xx_i|\mm_{\cS_i})}
 \lrB{\prod_{s\in\cS}\frac 1{|\M_{s,n}|}}
\end{align*}
by letting $\hM_{\D_{\J},n}\equiv\{\hM_{j,s,n}\}_{j\in\J,s\in\D_j}$.
An achievable region is defined as the set of achievable rate vectors.

We use the following subscript rules.
When two digits $ij$ appear in the subscripts of message as $M_{ij}$,
the left digit $i\in\{1,2\}$ means that
only Encoder $i$ has access to this message,
and the right digit $j\in\{1,2\}$ means that
this message is aimed to be reproduced by Decoder $j$.
When $i=0$, both encoders have access to this message.
When $j=0$, this message is aimed to be reproduced by both decoders.
Then we have the following.
\begin{itemize}
 \item
 Encoders $1$ and $2$ have access to messages
 $M_{00}$, $M_{01}$, and $M_{02}$.
 \item
 Only Encoder $1$ has access to messages 
 $M_{10}$, $M_{11}$, and $M_{12}$.
 \item
 Only Encoder $2$ has access to messages
 $M_{20}$, $M_{21}$, and $M_{22}$.
 \item
 Messages $M_{00}$, $M_{10}$, and $M_{20}$
 are aimed to be reproduced by Decoders $1$ and $2$.
 \item 
 Messages $M_{01}$, $M_{11}$, and $M_{21}$
 are aimed to be reproduced by Decoder $1$.
 \item
 Messages $M_{02}$, $M_{12}$, and $M_{22}$
 are aimed to be reproduced by Decoder $2$.
\end{itemize}
The rate of message $M_{ij}$ is denoted by $R_{ij}\equiv\log_2(|\M_{ij,n}|)/n$
for $i,j\in\{0,1,2\}$.
It should be noted that
one-digit subscripts (e.g. $M_0,M_1,M_2,$ $R_0,R_1,R_2$)
do not follow this rule.

For a given $i\in\{1,2\}$, let $i'$ be defined as $1'\equiv2$ and $2'\equiv 1$
to identify another encoder/decoder of Encoder/Decoder $i$.

The entropy of random variable $A$ is defined as
\begin{equation*}
 H(A)\equiv\sum_{a\in\A}\mu_A(a)\log_2\frac1{\mu_A(a)},
\end{equation*}
where $\mu_A$ is the distribution of $A$.
The conditional entropy of random variable $A$
for a given random variable $B$ is defined as
\begin{equation*}
 H(A|B)\equiv\sum_{a\in\A,b\in\B}\mu_{AB}(a,b)\log_2\frac1{\mu_{A|B}(a|b)},
\end{equation*}
where $\mu_{AB}$ is the joint distribution of $(A,B)$ and
the conditional distribution $\mu_{A|B}$ is defined as
\begin{equation*}
 \mu_{A|B}(a|b)\equiv\frac{\mu_{AB}(a,b)}{\sum_{a\in\A}\mu_{AB}(a,b)}.
\end{equation*}
The mutual information and the conditional mutual information
are defined as
\begin{align*}
 I(A;B)
 &\equiv
 H(A) - H(A|B)
 \\
 I(A;B|C)
 &\equiv
 H(A|C) - H(A|B,C).
\end{align*}
Their basic properties are summarized in Appendix \ref{sec:entropy}.

\section{Han-Kobayashi-Chong-Motani-Garg Region}

This section reviews the inner regions introduced
by Han and Kobayashi \cite{HK81,KH07},
and Chong, Motani, and Garg \cite{CMG06,CMGE08}
for interference channel coding without common message (Fig.\ \ref{fig:ic}).

\begin{figure}[ht]
\begin{center}
 \unitlength 0.48mm
 \begin{picture}(225,55)(-20,0)
  
  \put(33,48){\makebox(0,0){Encoders}}
  \put(141,48){\makebox(0,0){Decoders}}
  
  \put(10,37){\makebox(0,0)[r]{$M_{10},M_{11}$}}
  \put(11,37){\vector(1,0){15}}
  \put(26,30){\framebox(14,14){$\Phi_1$}}
  \put(40,37){\vector(1,0){10}}
  \put(55,37){\makebox(0,0){$X_1$}}
  \put(60,37){\vector(1,0){10}}
  
  \put(10,7){\makebox(0,0)[r]{$M_{20},M_{22}$}}
  \put(11,7){\vector(1,0){15}}
  \put(26,0){\framebox(14,14){$\Phi_2$}}
  \put(40,7){\vector(1,0){10}}
  \put(55,7){\makebox(0,0){$X_2$}}
  \put(60,7){\vector(1,0){10}}
  
  \put(70,0){\framebox(34,44){\small $\mu_{Y_1Y_2|X_1X_2}$}}
  
  \put(104,37){\vector(1,0){10}}
  \put(119,37){\makebox(0,0){$Y_1$}}
  \put(124,37){\vector(1,0){10}}
  \put(134,30){\framebox(14,14){$\Psi_1$}}
  \put(148,37){\vector(1,0){10}}
  \put(159,37){\makebox(0,0)[l]{$\hM_{10},\hM_{20},\hM_{11}$}}
  
  \put(104,7){\vector(1,0){10}}
  \put(119,7){\makebox(0,0){$Y_2$}}
  \put(124,7){\vector(1,0){10}}
  \put(134,0){\framebox(14,14){$\Psi_2$}}
  \put(148,7){\vector(1,0){10}}
  \put(159,7){\makebox(0,0)[l]{$\hM_{10},\hM_{20},\hM_{22}$}}
 \end{picture}
\end{center}
\caption{Super-problem Introduced by Han-Kobayashi \cite{HK81}}
\label{fig:hk}
\end{figure}
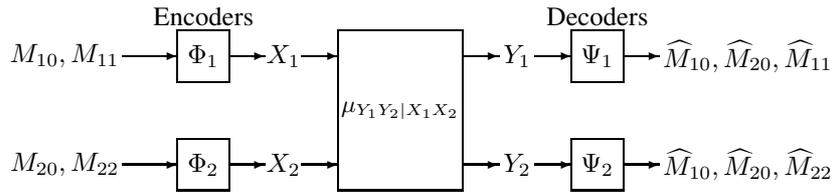

First, let us consider the following super-problem of
interference channel coding (Fig.\ \ref{fig:hk}).
Encoder $i$ has access to $(M_{i0},M_{ii})$ for each $i\in\{1,2\}$,
where $M_{ii}$ is a private message
which is reproduced only at Decoder $i$
and $M_{i0}$ is another private message which is reproduced at both decoders.
To summarize,
Encoder $1$ has access to $(M_{10},M_{11})$,
Encoder $2$ has access to $(M_{20},M_{21})$,
Decoder $1$ reproduces $(M_{10},M_{20},M_{11})$,
and Decoder $2$ reproduces $(M_{10},M_{20},M_{21})$
when messages are decoded successfully.

For the given joint distribution $p$ of
$(U_0,U_{10},U_{11},X_1,U_{20},U_{22},X_2)$ defined as
\begin{equation}
 p(u_0,u_{10},u_{11},x_1,u_{20},u_{22},x_2)
 \equiv
 p_{U_0}(u_0)
 \prod_{i\in\{1,2\}}
 p_{X_i|U_0U_{i0}U_{ii}}(x_i|u_0,u_{i0},u_{ii})
 p_{U_{ii}|U_0}(u_{ii}|u_0)
 p_{U_{i0}|U_0}(u_{i0}|u_0),
 \label{eq:joint-hk}
\end{equation}
where $U_0$ is a time-sharing random variable and
$p_{X_i|U_0U_{i0}U_{ii}}(x_i|u_0,u_{i0},u_{ii})\in\{0,1\}$,
let $\RHK(p)$ be the region defined as the set of all
non-negative rate quadruples $(R_{10},R_{11},R_{20},R_{22})$
satisfying
\begin{align*}
 R_{i0}
 &\leq
 I(U_{i0};Y_i|U_0,U_{i'0},U_{ii})
 \\
 R_{i'0}
 &\leq
 I(U_{i'0};Y_i|U_0,U_{i0},U_{ii})
 \\
 R_{ii}
 &\leq
 I(U_{ii};Y_i|U_0,U_{10},U_{20})
 \\
 R_{10}+R_{20}
 &\leq
 I(U_{10},U_{20};Y_i|U_0,U_{ii})
 \\
 R_{i0}+R_{ii}
 &\leq
 I(U_{i0},U_{ii};Y_i|U_0,U_{i'0})
 \\
 R_{i'0}+R_{ii}
 &\leq
 I(U_{i'0},U_{ii};Y_i|U_0,U_{i0})
 \\
 R_{10}+R_{20}+R_{ii}
 &\leq
 I(U_{10},U_{20},U_{ii};Y_i|U_0)
\end{align*}
for all $i\in\{1,2\}$.
By using the rate-splitting technique
\begin{equation*}
 R_i \equiv R_{i0} + R_{ii}\quad\text{for each}\ i\in\{0,1\},
\end{equation*}
and the Fourier-Motzkin method \cite[Appendix D]{EK11} to eliminate
$R_{10}$, $R_{11}$, $R_{20}$, $R_{22}$
and redundant inequalities,
we have the region of the interference channel coding
as the set of all non-negative rate pairs $(R_1,R_2)$ satisfying
\begin{align*}
 R_i
 &\leq
 I(U_{i0},U_{ii};Y_i|U_0,U_{i'0})
 \\
 R_i
 &\leq
 I(U_{ii};Y_i|U_0,U_{10},U_{20})
 +I(U_{i0};Y_{i'}|U_0,U_{i'0},U_{i'i'})
 \\
 R_1+R_2
 &\leq
 I(U_{ii};Y_i|U_0,U_{10},U_{20})
 +I(U_{10},U_{20},U_{i'i'};Y_{i'}|U_0)
 \\
 R_1+R_2
 &\leq
 I(U_{20},U_{11};Y_1|U_0,U_{10})
 +I(U_{10},U_{22};Y_2|U_0,U_{20})
 \\
 2R_i+R_{i'}
 &\leq
 I(U_{ii};Y_i|U_0,U_{10},U_{20})
 +I(U_{10},U_{20},U_{ii};Y_i|U_0)
 +I(U_{i0},U_{i'i'};Y_{i'}|U_0,U_{i'0})
 \\
 2R_i+R_{i'}
 &\leq
 2I(U_{ii};Y_i|U_0,U_{10},U_{20})
 +I(U_{i0},U_{i'i'};Y_{i'}|U_0,U_{i'0})
 +I(U_{10},U_{20};Y_{i'}|U_0,U_{i'i'})
\end{align*}
for all $i\in\{1,2\}$.
Then the region
\begin{equation*}
 \RHK\equiv\obigcup_{p}\RHK(p),
\end{equation*}
where $\obigcup_p$ is the convex closure of the union over all $p$
satisfying (\ref{eq:joint-hk}),
is shown to be achievable for the interference channel coding
in \cite[Theorem 3.1]{HK81}, \cite[Theorem B]{KH07}.

Next,
we introduce the modified Chong-Motani-Garg region \cite{CMG06,CMGE08,KH07}.
For the given joint distribution $p$ of $(U_0,U_1,X_1,U_2,X_2)$ defined as
\begin{align}
 p(u_0,u_1,x_1,u_2,x_2)
 &\equiv
 p_{U_0}(u_0)
 \prod_{i\in\{1,2\}}
 p_{X_iU_i|U_0}(x_i,u_i|u_0),
 \label{eq:joint-cmg}
\end{align}
let $\RCMG(p)$ be defined as the set of
all non-negative rate pairs $(R_1,R_2)$ satisfying
\begin{align*}
 R_i
 &\leq
 I(X_i;Y_i|U_0,U_{i'})
 \\
 R_1+R_2
 &\leq
 I(X_i;Y_i|U_0,U_1,U_2)
 + I(U_i,X_{i'};Y_{i'}|U_0)
 \\
 R_1+R_2
 &\leq
 I(U_2,X_1;Y_1|U_0,U_1)
 +I(U_1,X_2;Y_2|U_0,U_2)
 \\
 2R_i+R_{i'}
 &\leq
 I(U_{i'},X_i;Y_i|U_0)
 +I(X_i;Y_i|U_0,U_1,U_2)
 +I(U_i,X_{i'};Y_{i'}|U_0,U_{i'})
\end{align*}
for all $i\in\{1,2\}$.
We define $\RCMG$ as
\begin{equation*}
 \RCMG
 \equiv
 \obigcup_p\RCMG(p),
\end{equation*}
where $\obigcup_p$ is the convex closure of the union over all $p$
satisfying (\ref{eq:joint-cmg}).

It is shown in \cite[Theorem 2]{CMGE08}
that region $\RCMG$ is equivalent to $\RHK$.
In the following, we refer this region as
the {\it HKCMG (Han-Kobayashi-Chong-Motani-Garg) region}
denoted by $\RHKCMG$, that is, $\RHKCMG\equiv\RCMG=\RHK$.

\begin{rem}
It is also shown that the regions remain invariant with the 
constraint where the cardinalities of the support of random variables
have upper bounds.
Throughout the paper, we do not consider such constraints.
The evaluation of the upper bounds of the cardinalities
is a challenge we leave for the future.
\end{rem}

\section{Jiang-Xin-Garg Region}
\label{sec:jxg}

This section reviews the inner region introduced by
Jiang, Xin, and Garg \cite{JXG08}
for interference channel coding with a common message (Fig.\ \ref{fig:icc}).

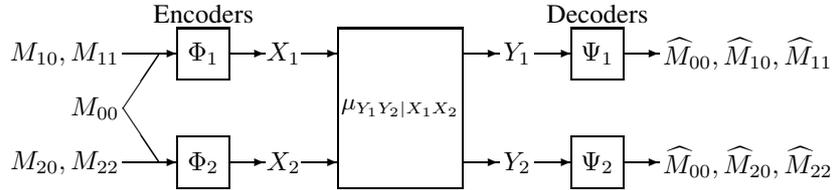
\begin{figure}[ht]
\begin{center}
 \unitlength 0.48mm
 \begin{picture}(225,55)(-20,0)
  
  \put(33,48){\makebox(0,0){Encoders}}
  \put(141,48){\makebox(0,0){Decoders}}
  
  \put(10,22){\makebox(0,0)[r]{$M_{00}$}}
  \put(11,22){\line(2,3){10}}
  \put(11,22){\line(2,-3){10}}
  
  \put(10,37){\makebox(0,0)[r]{$M_{10},M_{11}$}}
  \put(11,37){\vector(1,0){15}}
  \put(26,30){\framebox(14,14){$\Phi_1$}}
  \put(40,37){\vector(1,0){10}}
  \put(55,37){\makebox(0,0){$X_1$}}
  \put(60,37){\vector(1,0){10}}
  
  \put(10,7){\makebox(0,0)[r]{$M_{20},M_{22}$}}
  \put(11,7){\vector(1,0){15}}
  \put(26,0){\framebox(14,14){$\Phi_2$}}
  \put(40,7){\vector(1,0){10}}
  \put(55,7){\makebox(0,0){$X_2$}}
  \put(60,7){\vector(1,0){10}}
  
  \put(70,0){\framebox(34,44){\small $\mu_{Y_1Y_2|X_1X_2}$}}
  
  \put(104,37){\vector(1,0){10}}
  \put(119,37){\makebox(0,0){$Y_1$}}
  \put(124,37){\vector(1,0){10}}
  \put(134,30){\framebox(14,14){$\Psi_1$}}
  \put(148,37){\vector(1,0){10}}
  \put(159,37){\makebox(0,0)[l]{$\hM_{00},\hM_{10},\hM_{11}$}}
  
  \put(104,7){\vector(1,0){10}}
  \put(119,7){\makebox(0,0){$Y_2$}}
  \put(124,7){\vector(1,0){10}}
  \put(134,0){\framebox(14,14){$\Psi_2$}}
  \put(148,7){\vector(1,0){10}}
  \put(159,7){\makebox(0,0)[l]{$\hM_{00},\hM_{20},\hM_{22}$}}
 \end{picture}
\end{center}
\caption{Super-problem Introduced by Jiang-Xin-Garg \cite{JXG08}:
 Message $\M_{i0}$ is aimed to be reproduced by both decoders
 but not necessarily reproduced at Decoder $i'$.}
\label{fig:jxg}
\end{figure}

First, let us consider the following super-problem of
interference channel coding with a common message (Fig.\ \ref{fig:jxg}).
Encoder $i$ has access to $(M_{00},M_{i0},M_{ii})$ for each $i\in\{1,2\}$,
where $M_{00}$ is a common message reproduced at both decoders,
$M_{ii}$ is a private message which is reproduced only at Decoder $i$,
and $M_{i0}$ is another private message which is aimed to be reproduced
at both decoders but not necessarily reproduced at Decoder $i'$
in the sense that
the error of reproducing $M_{i0}$ at Decoder $i'$ is ignored.
To summarize,
Encoder $1$ has access to $(M_{00},M_{10},M_{11})$,
Encoder $2$ has access to $(M_{00},M_{20},M_{21})$,
Decoder $1$ reproduces $(M_{00},M_{10},M_{11})$,
and Decoder $2$ reproduces $(M_{00},M_{20},M_{21})$
when messages are decoded successfully,
where $M_{i0}$ may not be reproduced at Decoder $i'$.

For the given joint distribution $p$ of $(U_0,U_1,X_1,U_2,X_2)$ defined as
\begin{equation}
 p(u_0,u_1,x_1,u_2,x_2)
 \equiv
 p_{U_0}(u_0)
 \prod_{i\in\{1,2\}}
 p_{X_iU_i|U_0}(x_i,u_i|u_0),
 \label{eq:joint-jxg}
\end{equation}
let $\RJXG(p)$ be the region defined as the set of all
non-negative rate quintuples $(R_{00},R_{10},R_{11},R_{20},R_{22})$
satisfying
\begin{align*}
 R_{ii}
 &\leq
 I(X_i;Y_i|U_0,U_1,U_2)
 \\
 R_{i0}+R_{ii}
 &\leq
 I(X_i;Y_i|U_0,U_{i'})
 \\
 R_{ii}+R_{i'0}
 &\leq
 I(X_i,U_{i'};Y_i|U_0,U_i)
 \\
 R_{i0}+R_{ii}+R_{i'0}
 &\leq
 I(X_i,U_{i'};Y_i|U_0)
 \\
 R_{i0}+R_{ii}+R_{i'i}
 &\leq
 I(X_i,U_{i'};Y_i|U_0)
 \\
 R_{00}+R_{i0}+R_{ii}+R_{i'i}
 &\leq
 I(U_0,X_i,U_{i'};Y_i)
\end{align*}
for all $i\in\{1,2\}$.
By using the rate-splitting technique
\begin{align*}
 R_0
 &\equiv
 R_{00}
 \\
 R_i
 &\equiv
 R_{i0} + R_{ii}\quad\text{for each}\  i\in\{0,1\},
\end{align*}
and the Fourier-Motzkin method \cite[Appendix D]{EK11} to eliminate
$R_{10}$, $R_{11}$, $R_{20}$, $R_{22}$
and redundant inequalities,
we have the region
as the set of all non-negative rate triplets $(R_0,R_1,R_2)$ satisfying
\begin{align}
 R_i
 &\leq
 I(X_i;Y_i|U_0,U_{i'})
 \label{eq:jxg-Ri}
 \\
 R_0+R_i
 &\leq
 I(U_0,X_i,U_{i'};Y_i)
 \\
 R_1+R_2
 &\leq
 I(U_2,X_1;Y_1|U_0,U_1) 
 + I(U_1,X_2;Y_2|U_0,U_2)
 \\
 R_1+R_2
 &\leq
 I(U_{i'},X_i;Y_i|U_0)
 + I(X_{i'};Y_{i'}|U_0,U_1,U_2)
 \\
 R_0+R_1+R_2
 &\leq
 I(U_0,U_{i'},X_i;Y_i)
 + I(X_{i'};Y_{i'}|U_0,U_1,U_2)
 \\
 2R_i+R_{i'}
 &\leq
 I(U_{i'},X_i;Y_i|U_0)
 + I(X_i;Y_i|U_0,U_1,U_2)
 + I(U_i,X_{i'};Y_{i'}|U_0,U_{i'})
 \\
 R_0+2R_i+R_{i'}
 &\leq
 I(U_0,U_{i'},X_i;Y_i)
 + I(X_i;Y_i|U_0,U_1,U_2)
 + I(U_i,X_{i'};Y_{i'}|U_0,U_{i'})
 \label{eq:jxg-R0+2Ri+Ri'}
\end{align}
for all $i\in\{1,2\}$.
Then the region
\begin{equation*}
 \RJXG\equiv\obigcup_{p}\RJXG(p),
\end{equation*}
where $\obigcup_p$ is the convex closure of the union over all $p$
satisfying (\ref{eq:joint-jxg}),
is shown to be achievable for the interference channel coding
in \cite[Corollary 1]{JXG08}.
We refer this region as the {\it JXG (Jiang-Xin-Garg) region}.

Let $\RJXG^0(p)$ and $\RJXG^0$ be defined as
\begin{align*}
 \RJXG^0(p)
 &\equiv
 \lrb{
  (R_1,R_2): (0,R_1,R_2)\in\RJXG(p)
 }
 \\
 \RJXG^0
 &\equiv
 \obigcup_p
 \RJXG^0(p).
\end{align*}
Then we have the following lemma.
\begin{lem}[{\cite[Corollary 4 and Remark 11]{JXG08}}]
\label{lem:hk-jxg0}
$\RJXG^0=\RHKCMG$.
\end{lem}

\section{Inner Region Achievable with Codes
 Using Constrained-Random-Number Generator}
\label{sec:crng}

Regarding the JXG region, we consider the following question.
In the super-problem, we assumed that
message $M_{i0}$ may not be reproduced at decoder $i'$.
Then it seems unnecessary to split message $M_i$ into $M_{i0}$ and $M_{ii}$.
This is a motivation of this study.
In the following, it is shown that
the rate-splitting technique is unnecessary
for this inner region by constructing codes without rate splitting.

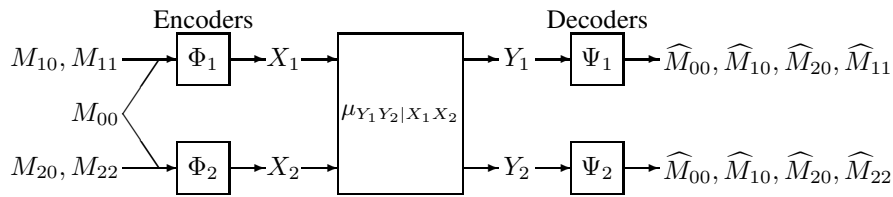
\begin{figure}[ht]
\begin{center}
 \unitlength 0.48mm
 \begin{picture}(242,55)(-20,0)
  
  \put(33,48){\makebox(0,0){Encoders}}
  \put(141,48){\makebox(0,0){Decoders}}
  
  \put(10,22){\makebox(0,0)[r]{$M_{00}$}}
  \put(11,22){\line(2,3){10}}
  \put(11,22){\line(2,-3){10}}
  
  \put(10,37){\makebox(0,0)[r]{$M_{10},M_{11}$}}
  \put(11,37){\vector(1,0){15}}
  \put(26,30){\framebox(14,14){$\Phi_1$}}
  \put(40,37){\vector(1,0){10}}
  \put(55,37){\makebox(0,0){$X_1$}}
  \put(60,37){\vector(1,0){10}}
  
  \put(10,7){\makebox(0,0)[r]{$M_{20},M_{22}$}}
  \put(11,7){\vector(1,0){15}}
  \put(26,0){\framebox(14,14){$\Phi_2$}}
  \put(40,7){\vector(1,0){10}}
  \put(55,7){\makebox(0,0){$X_2$}}
  \put(60,7){\vector(1,0){10}}
  
  \put(70,0){\framebox(34,44){\small $\mu_{Y_1Y_2|X_1X_2}$}}
  
  \put(104,37){\vector(1,0){10}}
  \put(119,37){\makebox(0,0){$Y_1$}}
  \put(124,37){\vector(1,0){10}}
  \put(134,30){\framebox(14,14){$\Psi_1$}}
  \put(148,37){\vector(1,0){10}}
  \put(159,37){\makebox(0,0)[l]{$\hM_{00},\hM_{10},\hM_{20},\hM_{11}$}}
  
  \put(104,7){\vector(1,0){10}}
  \put(119,7){\makebox(0,0){$Y_2$}}
  \put(124,7){\vector(1,0){10}}
  \put(134,0){\framebox(14,14){$\Psi_2$}}
  \put(148,7){\vector(1,0){10}}
  \put(159,7){\makebox(0,0)[l]{$\hM_{00},\hM_{10},\hM_{20},\hM_{22}$}}
 \end{picture}
\end{center}
\caption{Super-problem of Interference Channel Coding with Common Message:
 Message $M_{i0}$ is reproduced at Decoder $i'$.}
\label{fig:crng}
\end{figure}

First, let us consider the following super-problem of
interference channel coding with a common message (Fig.\ \ref{fig:crng})
which is a natural extension of the super-problem
introduced by Han and Kobayashi \cite{HK81}.
Encoder $i$ has access to $(M_{00},M_{i0},M_{ii})$ for each $i\in\{1,2\}$,
where $M_{00}$ is a common message reproduced at both decoders,
$M_{ii}$ is a private message which is reproduced only at Decoder $i$,
and $M_{i0}$ is another private message which is reproduced at both decoders.
To summarize,
Encoder $1$ has access to $(M_{00},M_{10},M_{11})$,
Encoder $2$ has access to $(M_{00},M_{20},M_{22})$,
Decoder $1$ reproduces $(M_{00},M_{10},M_{20},M_{11})$,
and Decoder $2$ reproduces $(M_{00},M_{10},M_{20},M_{22})$.
It should be noted that the error of reproducing $M_{i0}$
at Decoder $i'$ is {\it not} ignored.

For the given joint distribution $\mu$ of
$(Z_{00},Z_{10},Z_{11},X_1,Z_{20},Z_{22},X_2)$ defined as
\begin{align}
 \mu(z_{00},z_{10},z_{11},x_1,z_{20},z_{22},x_2)
 &\equiv
 \mu_{Z_{00}}(z_{00})
 \prod_{i\in\{1,2\}}
 \mu_{X_iZ_{i0}Z_{ii}|Z_{00}}(x_i,z_{i0},z_{ii}|z_{00}),
 \label{eq:joint-crng}
\end{align}
let $\RCRNG(\mu)$ be defined as the set of
all non-negative rate quintuples $(R_{00},R_{10},R_{11},R_{20},R_{22})$
ensuring that there is a non-negative rate quintuple
$(r_{00},r_{10},r_{11},r_{20},r_{22})$ such that
\begin{align}
 R_{00}+r_{00}
 &\leq
 H(Z_{00})
 \label{eq:crng-R0+r0}
 \\
 \sum_{s\in\tcS_i}
 \lrB{R_s+r_s}
 &\leq
 H(Z_{\tcS_i}|Z_{00})
 \label{eq:crng-sum-Rj+rj}
 \\
 \sum_{s\in\tcD_j} r_s
 &\geq H(Z_{\tcD_j}|Y_j,Z_{\D_j\setminus\tcD_j})
 \label{eq:crng-sum-rj}
\end{align}
for all $i,j\in\{1,2\}$,
$\tcS_i\subset\cS_i\equiv\{i0,ii\}$,
and $\tcD_j\subset\D_j\equiv\{00,10,20,jj\}$.
Let $\RCRNG$ be defined as
\begin{equation*}
 \RCRNG\equiv\obigcup_{\mu}\RCRNG(\mu),
\end{equation*}
where $\obigcup_{\mu}$ is the convex closure of the union over all $\mu$
satisfying (\ref{eq:joint-crng}).

Let $\RCRNG^0(\mu)$, $\RCRNG^0$,
$\RCRNG^{00}(\mu)$, and $\RCRNG^{00}$ be defined as
\begin{align*}
 \RCRNG^0(\mu)
 &\equiv
 \{(R_0,R_1,R_2): (R_0,0,R_1,0,R_2)\in\RCRNG(\mu)\}
 \\
 \RCRNG^0
 &\equiv
 \obigcup_{\mu}\RCRNG^0(\mu)
 \\
 \RCRNG^{00}(\mu)
 &\equiv
 \{(R_1,R_2): (0,0,R_1,0,R_2)\in\RCRNG(\mu)\}
 \\
 \RCRNG^{00}
 &\equiv
 \obigcup_{\mu}\RCRNG^{00}(\mu).
\end{align*}
Let $\tRCRNG(\tmu)$ be defined as the set of
all non-negative rate quintuples
$(R_{00},R_{10},R_{11},R_{20},R_{22})$ satisfying that
there is a non-negative rate quintuple
$(r_{00},r_{10},r_{11},r_{20},r_{22})$ such that
\begin{align}
 R_i
 &\leq
 I(V_{i0},V_{ii};Y_i|V_{00},V_{i'0})
 \label{eq:crng-Ri}
 \\
 R_0 + R_i
 &\leq
 I(V_{00},V_{10},V_{20},V_{ii};Y_i)
 \label{eq:crng-R0+Ri}
 \\
 R_1 + R_2
 &\leq
 I(V_{20},V_{11};Y_1|V_{00},V_{10})
 + I(V_{10},V_{22};Y_2|V_{00},V_{20})
 \label{eq:crng-R1+R2-1}
 \\
 R_1 + R_2
 &\leq
 I(V_{10},V_{20},V_{ii};Y_{i}|V_{00})
 + I(V_{i'i'};Y_{i'}|V_{00},V_{10},V_{20})
 \label{eq:crng-R1+R2-2}
 \\
 R_0 + R_1 + R_2
 &\leq
 I(V_{00},V_{10},V_{20},V_{ii};Y_i)
 + I(V_{i'i'};Y_{i'}|V_{00},V_{10},V_{20})
 \label{eq:crng-R0+R1+R2}
 \\
 2 R_i + R_{i'}
 &\leq
 I(V_{10},V_{20},V_{ii};Y_i|V_{00})
 + I(V_{ii};Y_i|V_{00},V_{10},V_{20})
 + I(V_{i0},V_{i'i'};Y_{i'}|V_{00},V_{i'0})
 \label{eq:crng-2Ri+Ri'}
 \\
 R_0 + 2 R_i + R_{i'}
 &\leq
 I(V_{00},V_{10},V_{20},V_{ii};Y_i)
 + I(V_{ii};Y_i|V_{00},V_{10},V_{20})
 + I(V_{i0},V_{i'i'};Y_{i'}|V_{00},V_{i'0}),
 \label{eq:crng-R0+2Ri+Ri'}
\end{align}
where the joint distribution $\tmu$ of
$(V_{00},V_{10},V_{11},X_1,V_{20},V_{22},X_2)$ is defined as
\begin{align}
 \tmu(v_{00},v_{10},v_{11},x_1,v_{20},v_{22},x_2)
 &\equiv
 \mu_{V_{00}}(v_{00})
 \prod_{i\in\{1,2\}}
 \mu_{X_iV_{i0}V_{ii}|V_{00}}(x_i,v_{i0},v_{ii}|v_{00}).
 \label{eq:joint-crng-simplified}
\end{align}
Let $\tRCRNG$, $\tRCRNG^0(\tmu)$, $\tRCRNG^0$,
$\tRCRNG^{00}(\tmu)$, and $\tRCRNG^{00}$ 
be defined similarly to $\RCRNG$, $\RCRNG^0(\mu)$, $\RCRNG^0$ 
$\RCRNG^{00}(\mu)$, and $\RCRNG^{00}$.
Then we have the following results, which are shown
in Sections \ref{sec:proof-crng}--\ref{sec:proof-jxg-crng}.
\begin{thm}
\label{thm:crng-achievable}
Region $\RCRNG$ is achievable for the above super-problem of
interference channel coding with a common message,
where the code is constructed by using the constrained-random-number-generators
\cite{CRNG,CRNG-MULTI,SW2CC}.
\end{thm}
\begin{thm}
\label{thm:crng-simplified}
$\tRCRNG^0=\RCRNG^0$ and $\tRCRNG^{00}=\RCRNG^{00}$.
\end{thm}

\begin{thm}
\label{thm:JXGsubsetCRNG}
$\RJXG=\RCRNG^0$.
\end{thm}

Immediately from Lemma \ref{lem:hk-jxg0} and Theorem \ref{thm:JXGsubsetCRNG},
we have the following corollary.
\begin{cor}
\label{thm:HKCMGsubsetCRNG00}
$\RHKCMG=\RJXG^0=\RCRNG^{00}$.
\end{cor}

It should be noted that
the definitions of $\RCRNG^0$ and $\RCRNG^{00}$ suggest that
the rate splitting technique is unnecessary
for the application of the super-problem
to interference channel coding with a common message
by letting $R_{10}=R_{20}=0$.
Thus, from Theorem \ref{thm:JXGsubsetCRNG}
and Corollary \ref{thm:HKCMGsubsetCRNG00},
we have the fact that regions $\RHKCMG$ and $\RJXG$
are achievable with codes without employing rate splitting.

\section{A Possible Extension of Inner Regions}
\label{sec:crng-full}

\begin{figure}[ht]
\begin{center}
 \unitlength 0.48mm
 \begin{picture}(292,55)(-37,0)
  
  \put(33,48){\makebox(0,0){Encoders}}
  \put(141,48){\makebox(0,0){Decoders}}
  
  \put(10,22){\makebox(0,0)[r]{$M_{00},M_{01},M_{02}$}}
  \put(11,22){\line(2,3){10}}
  \put(11,22){\line(2,-3){10}}
  
  \put(10,37){\makebox(0,0)[r]{$M_{10},M_{11},M_{12}$}}
  \put(11,37){\vector(1,0){15}}
  \put(26,30){\framebox(14,14){$\Phi_1$}}
  \put(40,37){\vector(1,0){10}}
  \put(55,37){\makebox(0,0){$X_1$}}
  \put(60,37){\vector(1,0){10}}
  
  \put(10,7){\makebox(0,0)[r]{$M_{20},M_{22},M_{21}$}}
  \put(11,7){\vector(1,0){15}}
  \put(26,0){\framebox(14,14){$\Phi_2$}}
  \put(40,7){\vector(1,0){10}}
  \put(55,7){\makebox(0,0){$X_2$}}
  \put(60,7){\vector(1,0){10}}
  
  \put(70,0){\framebox(34,44){\small $\mu_{Y_1Y_2|X_1X_2}$}}
  
  \put(104,37){\vector(1,0){10}}
  \put(119,37){\makebox(0,0){$Y_1$}}
  \put(124,37){\vector(1,0){10}}
  \put(134,30){\framebox(14,14){$\Psi_1$}}
  \put(148,37){\vector(1,0){10}}
  \put(159,37){\makebox(0,0)[l]{$\hM_{00},\hM_{01},\hM_{10},\hM_{20},\hM_{11},\hM_{21}$}}
  
  \put(104,7){\vector(1,0){10}}
  \put(119,7){\makebox(0,0){$Y_2$}}
  \put(124,7){\vector(1,0){10}}
  \put(134,0){\framebox(14,14){$\Psi_2$}}
  \put(148,7){\vector(1,0){10}}
  \put(159,7){\makebox(0,0)[l]{$\hM_{00},\hM_{02},\hM_{10},\hM_{20},\hM_{22},\hM_{12}$}}
 \end{picture}
\end{center}
\caption{Super-problem of Interference Channel Coding}
\label{fig:icc-full}
\end{figure}
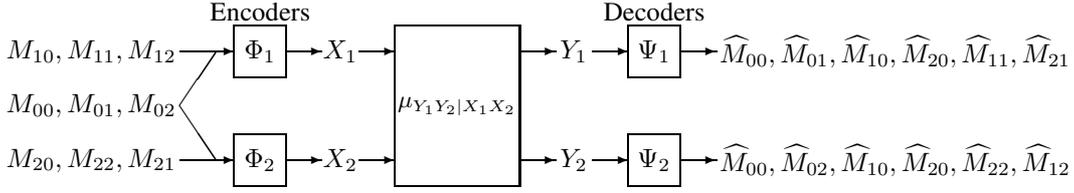

We could extend an inner region
by considering additional auxiliary messages
whose corresponding rates are equal to zero.
From this viewpoint,
we can consider the following super-problem
for the interference channel coding illustrated in Fig.~\ref{fig:icc-full}.

Regarding the interference channel coding with a common message
introduced in Section \ref{sec:jxg},
we consider the following additional messages.
In addition to message $M_{00}$ reproduced by both decoders,
both encoders have access to messages $M_{01}$ and $M_{02}$
which are reproduced only by Decoder $1$ and Decoder $2$, respectively.
For each $i\in\{1,2\}$,
in addition to private messages $M_{ii}$ and $M_{i0}$
reproduced by Decoder $i$ and both decoders, respectively,
Encoder $i$ has access to another private message $M_{ii'}$
which is reproduced only by Decoder $i'$.
To summarize,
Encoder $1$ has access to $(M_{00},M_{01},M_{02},M_{10},M_{11},M_{12})$,
Encoder $2$ has access to $(M_{00},M_{01},M_{02},M_{20},M_{22},M_{21})$,
Decoder $1$ reproduces $(M_{00},M_{01},M_{10},M_{20},M_{11},M_{21})$,
and Decoder $2$ reproduces $(M_{00},M_{02},M_{10},M_{20},M_{22},M_{12})$.

For the given joint distribution $\omu$ of
$(Z_{00},Z_{01},Z_{02},Z_{10},Z_{11},Z_{12},X_1,Z_{20},Z_{22},Z_{21},X_2)$
defined as
\begin{align}
 &
 \omu(z_{00},z_{01},z_{02},z_{10},z_{11},z_{12},x_1,z_{20},z_{22},z_{21},x_2)
 \notag
 \\*
 &\equiv
 \mu_{Z_{00}Z_{01}Z_{02}}(z_{00},z_{01},z_{02})
 \prod_{i\in\{1,2\}}
 \mu_{X_iZ_{i0}Z_{ii}Z_{ii'}|Z_{00}Z_{01}Z_{02}}(
  x_i,z_{i0},z_{ii},z_{ii'}|z_{00},z_{01},z_{02}
 ),
 \label{eq:joint-crng-full}
\end{align}
let $\oRCRNG(\omu)$ be defined as the set of
all non-negative rate nonuplets
$(R_{00},R_{01},R_{02},R_{10},R_{11},R_{12},R_{20},R_{22},R_{21})$
ensuring that there is a non-negative rate nonuplet
$(r_{00},r_{01},r_{02},r_{10},r_{11},r_{12},r_{20},r_{22},r_{21})$
such that
\begin{align}
 \sum_{s\in\tcS_0}
 \lrB{R_s+r_s}
 &\leq
 H(Z_{\tcS_0})
 \label{eq:crngfull-sum-S0}
 \\
 \sum_{s\in\tcS_i}
 \lrB{R_s+r_s}
 &\leq
 H(Z_{\tcS_i}|Z_{\cS_0})
 \label{eq:crngfull-sum-Si}
 \\
 \sum_{s\in\tcD_j} r_s
 &\geq H(Z_{\tcD_j}|Y_j,Z_{\D_j\setminus\tcD_j})
 \label{eq:crngfull-sum-Di}
\end{align}
for all $i,j\in\{1,2\}$,
$\tcS_0\subset\cS_0\equiv\{00,01,02\}$,
$\tcS_i\subset\cS_i\equiv\{i0,ii,ii'\}$,
and $\tcD_j\subset\D_j\equiv\{00,0j,10,20,jj,j'j\}$.
Let $\oRCRNG$ be defined as
\begin{equation*}
 \oRCRNG\equiv\obigcup_{\omu}\oRCRNG(\omu),
\end{equation*}
where $\obigcup_{\omu}$ is the convex closure of the union
over all $\omu$ satisfying (\ref{eq:joint-crng-full}).
Let $\oRCRNG^0(\omu)$, $\oRCRNG^0$,
$\oRCRNG^{00}(\omu)$, $\oRCRNG^{00}$,
$\oRCRNG^{000}(\omu)$, and $\oRCRNG^{000}$ be defined as
\begin{align*}
 \oRCRNG^0(\omu)
 &\equiv
 \{(R_{00},R_{10},R_{11},R_{20},R_{22}):
  (R_{00},0,0,R_{10},R_{11},0,R_{20},R_{22},0)\in\oRCRNG(\omu)\}
 \\
 \oRCRNG^0
 &\equiv
 \obigcup_{\omu}\oRCRNG'(\omu)
 \\
 \oRCRNG^{00}(\omu)
 &\equiv
 \{(R_0,R_1,R_2): (R_0,0,0,0,R_1,0,0,R_2,0)\in\oRCRNG(\omu)\}
 \\
 \oRCRNG^{00}
 &\equiv
 \obigcup_{\omu}\oRCRNG^{00}(\omu)
 \\
 \oRCRNG^{000}(\omu)
 &\equiv
 \{(R_1,R_2): (0,0,0,0,R_1,0,0,R_2,0)\in\oRCRNG(\omu)\}
 \\
 \oRCRNG^{000}
 &\equiv
 \obigcup_{\omu}\oRCRNG^{000}(\omu).
\end{align*}
The above yields the following results, which are shown
in Sections~\ref{sec:proof-crng} and \ref{sec:CRNGsubsetCRNGfull}.
\begin{thm}
\label{thm:crng-full-achievable}
Region $\oRCRNG$ is achievable for the above super-problem
of interference channel coding with common messages,
where the code is constructed by using constrained-random-number generators
\cite{CRNG,CRNG-MULTI,SW2CC}.
\end{thm}
\begin{thm}
\label{thm:CRNGsubsetCRNGfull}
\begin{gather*}
 \RCRNG\subset\oRCRNG^0
 \\
 \RJXG=\RCRNG^0\subset\oRCRNG^{00}
 \\
 \RHKCMG=\RJXG^0=\RCRNG^{00}\subset\oRCRNG^{000}.
\end{gather*}
\end{thm}

\section{Proof of Theorems \ref{thm:crng-achievable}
 and \ref{thm:crng-full-achievable}}
\label{sec:proof-crng}

In this section,
we prove Theorems \ref{thm:crng-achievable} and \ref{thm:crng-full-achievable}
by constructing a channel code for given joint distributions
$\mu$ and $\omu$, respectively.

The two proofs are given similarly
by defining index sets $\cS_0$, $\cS_1$, $\cS_2$, $\D_1$, and $\D_2$.
To summarize, we use the following definitions.
Let $\I\equiv\{1,2\}$ and $\J\equiv\{1,2\}$.
\begin{itemize}
 \item
 For the proof of Theorem \ref{thm:crng-achievable},
 let
 \begin{align*}
  \cS_0
  &\equiv\{00\}
  \\
  \cS_i
  &\equiv\{i0,ii\}
  \\
  \D_j
  &\equiv\{00,10,20,jj\}
 \end{align*}
 for each $i,j\in\{1,2\}$.
 \item
 For the proof of Theorem \ref{thm:crng-full-achievable},
 let
 \begin{align*}
  \cS_0
  &\equiv\{00,01,02\}
  \\
  \cS_i
  &\equiv\{i0,ii,ii'\}
  \\
  \D_j
  &\equiv\{00,10,20,0j,jj,j'j\}
 \end{align*}
 for each $i,j\in\{1,2\}$.
\end{itemize}
Let $\cS_{\I}$ and $\cS$ be defined as
$\cS_{\I}\equiv\bigcup_{i\in\I}\cS_i$
and $\cS\equiv\cS_0\cup\cS_{\I}$,
where $\cS$ is the disjoint union of $\{\cS_i\}_{i\in\{0\}\cup\I}$.
We have $\cS_{\I}=\cS\setminus\cS_0$.

It should be noted that
Lemma \ref{lem:source} and Theorem \ref{thm:channel-code},
which appear in the following sections,
are shown for arbitrary sets $\I$, $\J$, $\cS_0$, $\cS_i$, and $\D_j$.
Then the results can be extended to three or more encoders and decoders.
To prove the theorems,
we use the information-spectrum methods \cite{HAN,VH94},
which are reviewed in Appendix \ref{sec:ispec}.
It should be noted that
we can replace $\oH$ and $\uH$ by $H$ to represent
the entropy and conditional entropy for stationary memoryless sources
\cite[Remark 1.7.5 and Example 7.2.1]{HAN}.
We also use the hash property \cite{CRNG,HASH} of
ensembles of functions. This is reviewed in Appendix~\ref{sec:hash}.
For a set of functions $\F$,
let $\im\F\equiv \bigcup_{f\in\F}\{f(\zz): \zz\in\Z^n\}$.

For given $s\in\cS$ and $n\in\NN$,
let $\Z^n_s$ be the alphabet of
the sequence of random variables $Z^n_s\equiv(Z_{1,s},\ldots,Z_{n,s})$.
For given set $\cS$, let $\Z^n_{\cS}\equiv\Prod_{s\in\cS}\Z^n_s$.
For given functions $f_{s,n}:\Z^n_s\to\C_{s,n}$
and $g_{s,n}:\Z^n_s\to\M_{s,n}$,
where $\C_{s,n}$ and $\M_{s,n}$ are specified later,
let $(f_{s,n},g_{s,n})$ be a function defined as
\begin{equation*}
 (f_{s,n},g_{s_n})(\zz_s)\equiv(f_{s,n}(\zz_s),g_{s,n}(\zz_s))
 \quad\text{for each}\ \zz_s\in\Z^n_s.
\end{equation*}
For given $\cS$,
a set of functions $f_{\cS}\equiv\{f_{s,n}:\Z^n_s\to\C_{s,n}\}_{s\in\cS}$,
and $\cc_{\cS}\equiv\{\cc_s\}_{s\in\cS}\in\Prod_{s\in\cS}\C_{s,n}$,
let $\fC_{f_{\cS}}(\cc_{\cS})$ be defined as
\begin{equation*}
 \fC_{f_{\cS}}(\cc_{\cS})
 \equiv
 \lrb{
  \zz_s\in\Z^n_s: f_{s,n}(\zz_s)=\cc_s
  \ \text{for all}\ s\in\cS
 }.
\end{equation*}
For given $\cS$ and a set of functions
$(f,g)_{\cS}
\equiv
\{(f_{s,n},g_{s,n}):\Z^n_s\to\C_{s,n}\times\M_{s,n}\}_{s\in\cS}$,
where $f_{s,n}$ and $g_{s,n}$ are function on the same set $\Z^n_s$,
$\cc_{\cS}\equiv\{\cc_s\}_{s\in\cS}\in\Prod_{s\in\cS}\C_{s,n}$,
and $\mm_{\cS}\equiv\{\mm_s\}_{s\in\cS}\in\Prod_{s\in\cS}\M_{s,n}$,
let $\fC_{(f,g)_{\cS}}(\cc_{\cS},\mm_{\cS})$ be defined as
\begin{equation*}
 \fC_{(f,g)_{\cS}}(\cc_{\cS},\mm_{\cS})
 \equiv
 \fC_{f_{\cS}}(\cc_{\cS})\cap\fC_{g_{\cS}}(\mm_{\cS}).
\end{equation*}

Let $\chi$ be the support function of statement $\mathrm{S}$ defined as
\begin{equation*}
 \chi(\mathrm{S})
 \equiv
 \begin{cases}
  1
  &\text{$\mathrm{S}$ is true}
  \\
  0
  &\text{$\mathrm{S}$ is false.}
 \end{cases}
\end{equation*}

\subsection{Construction of Source Code}

\begin{figure}[ht]
\begin{center}
 \unitlength 0.55mm
 \begin{picture}(147,150)(-5,0)
  \put(27,146){\makebox(0,0){Encoders}}
  \put(4,135){\makebox(0,0){$\zz_{00}$}}
  \put(10,135){\vector(1,0){10}}
  \put(20,128){\framebox(14,14){$f_{00,n}$}}
  \put(34,135){\vector(1,0){10}}
  \put(50,135){\makebox(0,0){$\cc_{00}$}}
  \qbline(56,135)(76,137)
  \qbline(56,135)(76,65)
  
  \put(4,119){\makebox(0,0){$\zz_{01}$}}
  \put(10,119){\vector(1,0){10}}
  \put(20,112){\framebox(14,14){$f_{01,n}$}}
  \put(34,119){\vector(1,0){10}}
  \put(50,119){\makebox(0,0){$\cc_{01}$}}
  \qbline(56,119)(76,127)
  
  \put(4,103){\makebox(0,0){$\zz_{02}$}}
  \put(10,103){\vector(1,0){10}}
  \put(20,96){\framebox(14,14){$f_{02,n}$}}
  \put(34,103){\vector(1,0){10}}
  \put(50,103){\makebox(0,0){$\cc_{02}$}}
  \qbline(56,103)(76,55)
  
  \put(4,87){\makebox(0,0){$\zz_{10}$}}
  \put(10,87){\vector(1,0){10}}
  \put(20,80){\framebox(14,14){$f_{10,n}$}}
  \put(34,87){\vector(1,0){10}}
  \put(50,87){\makebox(0,0){$\cc_{10}$}}
  \qbline(56,87)(76,117)
  \qbline(56,87)(76,45)
  
  \put(4,71){\makebox(0,0){$\zz_{11}$}}
  \put(10,71){\vector(1,0){10}}
  \put(20,64){\framebox(14,14){$f_{11,n}$}}
  \put(34,71){\vector(1,0){10}}
  \put(50,71){\makebox(0,0){$\cc_{11}$}}
  \qbline(56,71)(76,107)
  
  \put(4,55){\makebox(0,0){$\zz_{12}$}}
  \put(10,55){\vector(1,0){10}}
  \put(20,48){\framebox(14,14){$f_{12,n}$}}
  \put(34,55){\vector(1,0){10}}
  \put(50,55){\makebox(0,0){$\cc_{12}$}}
  \qbline(56,55)(76,35)
  
  \put(4,39){\makebox(0,0){$\zz_{20}$}}
  \put(10,39){\vector(1,0){10}}
  \put(20,32){\framebox(14,14){$f_{20,n}$}}
  \put(34,39){\vector(1,0){10}}
  \put(50,39){\makebox(0,0){$\cc_{20}$}}
  \qbline(56,39)(76,97)
  \qbline(56,39)(76,25)
  
  \put(4,23){\makebox(0,0){$\zz_{22}$}}
  \put(10,23){\vector(1,0){10}}
  \put(20,16){\framebox(14,14){$f_{22,n}$}}
  \put(34,23){\vector(1,0){10}}
  \put(50,23){\makebox(0,0){$\cc_{22}$}}
  \qbline(56,23)(76,15)
  
  \put(4,8){\makebox(0,0){$\zz_{21}$}}
  \put(10,8){\vector(1,0){10}}
  \put(20,0){\framebox(14,14){$f_{21,n}$}}
  \put(34,8){\vector(1,0){10}}
  \put(50,8){\makebox(0,0){$\cc_{21}$}}
  \qbline(56,8)(76,87)
  
  \put(109,146){\makebox(0,0){Decoders}}
  
  \put(76,137){\vector(1,0){26}}
  \put(76,127){\vector(1,0){26}}
  \put(76,117){\vector(1,0){26}}
  \put(76,107){\vector(1,0){26}}
  \put(76,97){\vector(1,0){26}}
  \put(76,87){\vector(1,0){26}}
  \put(92,77){\vector(1,0){10}}
  \put(87,77){\makebox(0,0){$\yy_1$}}
  
  \put(102,72){\framebox(14,70){$\hZ^n_{\D_1}$}}
  
  \put(76,65){\vector(1,0){26}}
  
  \put(76,55){\vector(1,0){26}}
  \put(76,45){\vector(1,0){26}}
  \put(76,35){\vector(1,0){26}}
  \put(76,25){\vector(1,0){26}}
  \put(76,15){\vector(1,0){26}}
  \put(92,5){\vector(1,0){10}}
  \put(87,5){\makebox(0,0){$\yy_2$}}
  
  \put(102,0){\framebox(14,70){$\hZ^n_{\D_2}$}}
  
  \put(116,137){\vector(1,0){10}}
  \put(116,127){\vector(1,0){10}}
  \put(116,117){\vector(1,0){10}}
  \put(116,107){\vector(1,0){10}}
  \put(116,97){\vector(1,0){10}}
  \put(116,87){\vector(1,0){10}}
  \put(133,137){\makebox(0,0){$\hzz_{00}$}}
  \put(133,127){\makebox(0,0){$\hzz_{01}$}}
  \put(133,117){\makebox(0,0){$\hzz_{10}$}}
  \put(133,107){\makebox(0,0){$\hzz_{20}$}}
  \put(133,97){\makebox(0,0){$\hzz_{11}$}}
  \put(133,87){\makebox(0,0){$\hzz_{21}$}}
  
  \put(116,65){\vector(1,0){10}}
  \put(116,55){\vector(1,0){10}}
  \put(116,45){\vector(1,0){10}}
  \put(116,35){\vector(1,0){10}}
  \put(116,25){\vector(1,0){10}}
  \put(116,15){\vector(1,0){10}}
  \put(133,65){\makebox(0,0){$\hzz_{00}$}}
  \put(133,55){\makebox(0,0){$\hzz_{02}$}}
  \put(133,45){\makebox(0,0){$\hzz_{10}$}}
  \put(133,35){\makebox(0,0){$\hzz_{20}$}}
  \put(133,25){\makebox(0,0){$\hzz_{12}$}}
  \put(133,15){\makebox(0,0){$\hzz_{22}$}}
 \end{picture}
\end{center}
\caption{Construction of Source Code with Decoder Side Information
 for the Proof of Theorem \ref{thm:crng-full-achievable}:
 For the proof of Theorem \ref{thm:crng-achievable},
 the flows of vectors $\zz_s$, $f_{s,n}$, $\cc_s$, $s\in\{01,02,12,21\}$
 are omitted.}
\label{fig:source-code}
\end{figure}
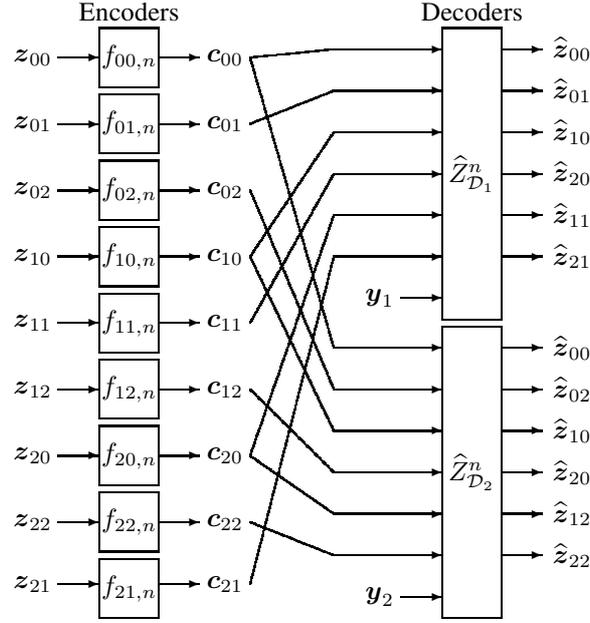

Before the construction of a channel code,
we introduce a single-hop multi-terminal source code
with decoder side information (Fig.~\ref{fig:source-code})
given in \cite{CRNG-MULTI}.

For index set $\cS$ of messages and index set $\J$ of decoders,
let
$(\ZZ_{\cS},\YY_{\J})
\equiv
\{(\{Z^n_s\}_{s\in\cS},\{Y^n_j\}_{j\in\J})\}_{n=1}^{\infty}$
be general correlated sources, which are characterized by
joint distributions $\mu_{Z_{\cS}^nY_{\J}^n}$ of
$Z_{\cS}^n,\equiv\{Z_s^n\}_{s\in\cS}$ and $Y_{\J}^n\equiv\{Y_j^n\}_{j\in\J}$.
For each $s\in\cS$ and $n\in\NN$,
let $\Z_s^n$ be the alphabet of message $Z_s^n$.
For each $j\in\J$ and $n\in\NN$,
let $\Y_j^n$ be the alphabet of side information $Y_j^n$
available for Decoder $j$.
It should be noted that we assume that
$\Z_s^n$ is a finite set,
while $\Y_j^n$ is allowed to be an arbitrary (infinite, continuous) set.

For each $s\in\cS$ and $n\in\NN$,
let $f_{s,n}:\Z_s^n\to\C_{s,n}$ be the $s$-th
encoding function, where $\C_{s,n}$ is the set of all codewords.
We can use a sparse matrix as function $f_{s,n}$ by assuming that
$\Z_s^n$ is an $n$-dimensional linear space on a finite field.
Let $C_{s,n}\equiv f_{s,n}(Z_s^n)$ be the codeword of Encoder $s$
and $r_s\equiv\log_2(|\C_{s,n}|)/n$ be the encoding rate of
the $s$-th message.
Let $\cc_s\in\C_{s,n}$ be the $s$-th codeword.

For each $j\in\J$,
let $\D_j$ be the index set of codewords available for Decoder $j$,
which is also the index set of messages reproduced by the decoder,
where $\D_j\subset\cS$.
Let $C_{\D_j,n}\equiv\{C_{s,n}\}_{s\in\D_j}$ be the set 
of codewords available for Decoder $j$.
For each $j\in\J$, the decoder generates $\hZ_{\D_j}^n(\cc_{\D_j},\yy_j)$
by using a constrained-random-number generator \cite{CRNG,CRNG-MULTI,SW2CC}
with a distribution given as
\begin{align}
 \mu_{\hZ_{\D_j}^n|Y_j^nC_{\D_j,n}}(\hzz_{\D_j}|\yy_j,\cc_{\D_j})
 &\equiv
 \frac{
  \mu_{Z_{\D_j}^n|Y_j^n}(\hzz_{\D_j}|\yy_j)
  \chi(f_{\D_j}(\hzz_{\D_j})=\cc_{\D_j})
 }{\mu_{Z_{\D_j}^n|Y_j^n}(\fC_{f_{\D_j}}(\cc_{\D_j})|\yy_j)}
 \label{eq:source-decoder}
\end{align}
for a given codeword $\cc_{\D_j}\equiv\{\cc_s\}_{s\in\D_j}$
and side information $\yy_j\in\Y_j^n$,
where $f_{\D_j}(\hzz_{\D_j})\equiv\lrb{f_{s,n}(\hzz_s)}_{s\in\D_j}$.
For $j\in\J$ and $s\in\D_j$,
let $\hZ^n_{j,s}$ be the random variable
of the reproduction of $Z^n_s$ at Decoder $j$.
Let $\hZ^n_{\D_j}\equiv\{\hZ^n_{j,s}\}_{s\in\D_j}$
be the random variable of all reproductions at Decoder $j$.
Let $\hZ^n_{\D_{\J}}\equiv\{\hZ^n_{j,s}\}_{j\in\J,s\in\D_j}$
be the random variable of all reproductions.
It follows the joint distribution of
$(Z_{\cS}^n,C_{\cS,n},Y_{\J}^n,\hZ^n_{\D_{\J}})$ is given as
\begin{equation}
 \mu_{Z_{\cS}^nC_{\cS,n}Y_{\J}^n\hZ^n_{\D_{\J}}}
 (\zz_{\cS},\cc_{\cS},\yy_{\J},\hzz_{\D_{\J}})
 \equiv
 \lrB{
  \prod_{j\in\J}
  \mu_{\hZ^n_{\D_j}|Y_j^nC_{\D_j,n}}(\hzz_{\D_j}|\yy_j,\cc_{\D_j})
 }
 \lrB{
  \prod_{s\in\cS}
  \mu_{C_{s,n}|Z_s^n}(\cc_s|\zz_s)
 }
 \mu_{Z_{\cS}^nY_{\J}^n}(\zz_{\cS},\yy_{\J}),
 \label{eq:source-joint}
\end{equation}
where $\mu_{C_{s,n}|Z_s^n}(\cc_s|\zz_s)\equiv\chi(f_{s,n}(\zz_s)=\cc_s)$.
The above yields the following lemma
which is shown in Section~\ref{sec:proof-source}.
\begin{lem}
\label{lem:source}
Let $(\ZZ_{\D_j},\YY_j)\equiv\{(\{Z^n_s\}_{s\in\D_j},Y^n_j)\}_{n=1}^{\infty}$
be a pair of general correlated sources.
Assume that $(\F_{s,n},p_{F_{s,n}})$ has the hash property
((\ref{eq:hash}) described in Appendix \ref{sec:hash})
for every $s\in\D_j$.
Then the expectation of the decoding error probability of Decoder~$j$
is evaluated as
\begin{align}
 &
 E_{F_{\D_j}}\lrB{
  \mu_{Z^n_{\D_j}\hZ^n_{\D_j}}\lrsb{
   \lrb{
    (\zz_{\D_j},\hzz_{\D_j}):
    \hzz_{\D_j}\neq\zz_{\D_j}
   }
  }
 }
 \notag
 \\*
 &\leq
 2
 \sum_{
  \tcD_j\subset\D_j:
  \tcD_j\neq\emptyset
 }
 \alpha_{F_{\tcD_j},n}\lrB{\beta_{F_{\D_j\setminus\tcD_j,n}}+1}
 2^{
  -n\lrB{
   \sum_{s\in\tcD_j}r_s-\oH(\ZZ_{\tcD_j}|\YY_j,\ZZ_{\D_j\setminus\tcD_j})
   -\e
  }
 }
 +2\beta_{F_{\D_j,n}}
 +2\mu_{Z^n_{\D_j}Y^n_j}(\oT_j^{\complement}),
 \label{eq:source-error}
\end{align}
where
$\C_{s,n}\equiv\im\F_{s,n}$,
\begin{align}
 \oT_j
 &\equiv
 \lrb{
  (\zz_{\D_j},\yy):
  \begin{aligned}
   &
   \frac1n\log_2
   \frac1{\mu_{Z^n_{\tcD_j}|Y^n_jZ^n_{\D_j\setminus\tcD_j}}
    (\zz_{\tcD_j}|\yy_j,\zz_{\D_j\setminus\tcD_j})}
   \leq
   \oH(\ZZ_{\tcD_j}|\YY_j,\ZZ_{\D_j\setminus\tcD_j})+\e
   \\
   &\text{for all}\ \tcD_j\ \text{satisfying}\ \emptyset\neq\tcD_j\subset\D_j
  \end{aligned}
 }
 \label{eq:oT}
\end{align}
for a given $\e>0$,
and the decoding error probability
$\mu_{Z^n_{\D_j}\hZ^n_{\D_j}}\lrsb{
 \lrb{
  (\zz_{\D_j},\hzz_{\D_j}):
  \hzz_{\D_j}\neq\zz_{\D_j}
 }
}$
depends on $f_{\D_j}$ through the joint distribution defined by
(\ref{eq:source-joint}).
\end{lem}

\begin{rem}
Let us assume that
$\alpha_{F_{\tcD_j},n}\to1$, $\beta_{F_{\tcD_j},n}\to0$,
and $\{r_s\}_{s\in\D_j}$ satisfies 
\begin{align*}
 \sum_{s\in\tcD_j}r_s
 &>
 \oH(\ZZ_{\tcD_j}|\YY_j,\ZZ_{\D_j\setminus\tcD_j})+\e
\end{align*}
for all $(j,\tcD_j)$ satisfying
$j\in\J$ and $\emptyset\neq\tcD_j\subset\D_j$,
where $\e$ is a sufficiently small positive number.
From this inequality and the fact that
$\mu_{Z_{\D_j}Y_j}(\oT_j^{\complement})\to0$,
we have the fact that
for all sufficiently large $n$ there are $\{f_{s,n}\}_{s\in\D_j}$
such that the right hand side of (\ref{eq:source-error}) is close to zero.
By taking the union bound and the random coding argument,
we have the fact that
there is a set of functions (sparse matrices)
$\{f_{s,n}\}_{s\in\cS}$ such that
the error probability is close to $0$ for all sufficiently large $n$.
However,
we break the proof of the source coding theorem
because additional conditions for $\{f_{s,n}\}_{s\in\cS}$
are required to prove the channel coding theorem.
\end{rem}

\subsection{Construction of Channel Code}

This section introduces a channel code for the proof of
Theorems~\ref{thm:crng-achievable} and~\ref{thm:crng-full-achievable}.
The idea for the construction is drawn from \cite{CRNG,SW2CC}.

Let $f_{\cS}\equiv\{f_{s,n}\}_{s\in\cS}$
and $\hZ^n_{\D_{\J}}\equiv\{\hZ^n_{j,s}\}_{j\in\J,s\in\D_j}$
be the encoders and the decoders of
the source code introduced in the last section.
For each $s\in\cS$ and $n\in\NN$,
let $g_{s,n}:\Z^n_s\to\M_{s,n}$ be a function
and $R_s\equiv\log_2(|\M_{s,n}|)/n$ be the rate of the $s$-th message.
We can use a sparse matrix as function $g_{s,n}$ by assuming that
$\Z_s^n$ is an $n$-dimensional linear space on a finite field.

We fix two sets of functions
$f_{\cS}\equiv\{f_{s,n}\}_{s\in\cS}$, $g_{\cS}\equiv\{g_{s,n}\}_{s\in\cS}$
and a set of vectors
$\cc_{\cS}\equiv\{\cc_s\}_{s\in\cS}$
such that they are available for constructing encoders and decoders.
For each $i\in\I$,
Encoder $i$ uses $f_{\cS_0}$, $g_{\cS_0}$, $\cc_{\cS_0}$,
$f_{\cS_i}$, $g_{\cS_i}$, and $\cc_{\cS_i}$.
For each $j\in\J$,
Decoder $j$ uses $\cc_{\D_j}$, $\hZ^n_{\D_{\J}}$, and $g_{\D_j}$,
where $f_{\D_j}$ is used implicitly in $\hZ^n_{\D_{\J}}$.
We fix the probability distributions
$\mu_{Z_{\cS_0}^n}$ and conditional probability distributions
$\{\mu_{Z_{\cS_i}^n|Z_{\cS_0}^n}\}_{i\in\I}$
and $\{\mu_{X_i^n|Z^n_{\cS_0}Z^n_{\cS_i}}\}_{i\in\I}$.

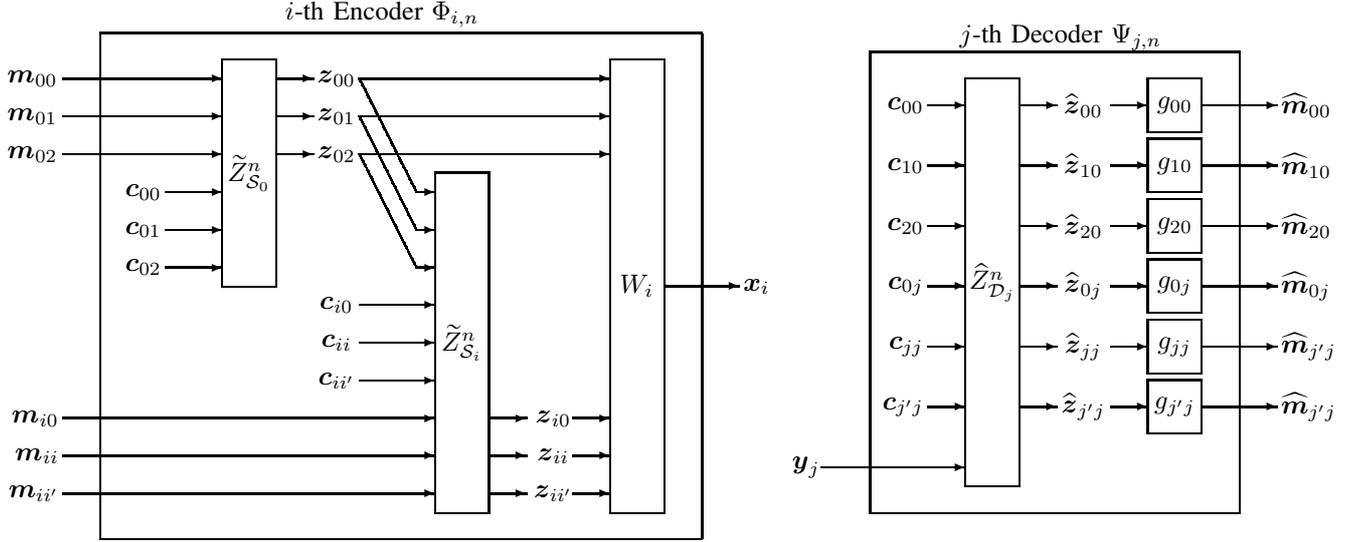
\begin{figure}[ht]
\begin{center}
 \unitlength 0.5mm
 \begin{picture}(200,146)(0,0)
  \put(94,139){\makebox(0,0){$i$-th Encoder $\Phi_{i,n}$}}
  
  \put(2,122){\makebox(0,0){$\mm_{00}$}}
  \put(2,112){\makebox(0,0){$\mm_{01}$}}
  \put(2,102){\makebox(0,0){$\mm_{02}$}}
  \put(10,122){\vector(1,0){42}}
  \put(10,112){\vector(1,0){42}}
  \put(10,102){\vector(1,0){42}}
  
  \put(36,92){\makebox(0,0)[r]{$\cc_{00}$}}
  \put(36,82){\makebox(0,0)[r]{$\cc_{01}$}}
  \put(36,72){\makebox(0,0)[r]{$\cc_{02}$}}
  \put(37,92){\vector(1,0){15}}
  \put(37,82){\vector(1,0){15}}
  \put(37,72){\vector(1,0){15}}
  
  \put(52,67){\framebox(14,60){$\tZ^n_{\cS_0}$}}
  
  \put(66,122){\vector(1,0){10}}
  \put(66,112){\vector(1,0){10}}
  \put(66,102){\vector(1,0){10}}
  \put(82,122){\makebox(0,0){$\zz_{00}$}}
  \put(82,112){\makebox(0,0){$\zz_{01}$}}
  \put(82,102){\makebox(0,0){$\zz_{02}$}}
  \put(88,122){\vector(1,0){66}}
  \put(88,112){\vector(1,0){66}}
  \put(88,102){\vector(1,0){66}}
  
  \qbline(88,122)(103,92)
  \qbline(88,112)(103,82)
  \qbline(88,102)(103,72)
  \put(103,92){\vector(1,0){5}}
  \put(103,82){\vector(1,0){5}}
  \put(103,72){\vector(1,0){5}}
  
  \put(82,62){\makebox(0,0){$\cc_{i0}$}}
  \put(82,52){\makebox(0,0){$\cc_{ii}$}}
  \put(82,42){\makebox(0,0){$\cc_{ii'}$}}
  \put(88,62){\vector(1,0){20}}
  \put(88,52){\vector(1,0){20}}
  \put(88,42){\vector(1,0){20}}

  \put(9,32){\makebox(0,0)[r]{$\mm_{i0}$}}
  \put(9,22){\makebox(0,0)[r]{$\mm_{ii}$}}
  \put(9,12){\makebox(0,0)[r]{$\mm_{ii'}$}}
  
  \put(108,7){\framebox(14,90){$\tZ^n_{\cS_i}$}}
  
  \put(10,32){\vector(1,0){98}}
  \put(10,22){\vector(1,0){98}}
  \put(10,12){\vector(1,0){98}}
  
  \put(144,32){\vector(1,0){10}}
  \put(144,22){\vector(1,0){10}}
  \put(144,12){\vector(1,0){10}}
  \put(139,32){\makebox(0,0){$\zz_{i0}$}}
  \put(139,22){\makebox(0,0){$\zz_{ii}$}}
  \put(139,12){\makebox(0,0){$\zz_{ii'}$}}
  \put(122,32){\vector(1,0){10}}
  \put(122,22){\vector(1,0){10}}
  \put(122,12){\vector(1,0){10}}
  
  \put(154,7){\framebox(14,120){$W_i$}}
  
  \put(189,67){\makebox(0,0)[l]{$\xx_i$}}
  \put(168,67){\vector(1,0){20}}
  
  \put(20,0){\framebox(158,134){}}
 \end{picture}
 \begin{picture}(134,146)(0,-7)
  \put(70,126){\makebox(0,0){$j$-th Decoder $\Psi_{j,n}$}}
  
  \put(34,108){\makebox(0,0)[r]{$\cc_{00}$}}
  \put(34,92){\makebox(0,0)[r]{$\cc_{10}$}}
  \put(34,76){\makebox(0,0)[r]{$\cc_{20}$}}
  \put(34,60){\makebox(0,0)[r]{$\cc_{0j}$}}
  \put(34,44){\makebox(0,0)[r]{$\cc_{jj}$}}
  \put(34,28){\makebox(0,0)[r]{$\cc_{j'j}$}}
  
  \put(76,108){\makebox(0,0){$\hzz_{00}$}}
  \put(76,92){\makebox(0,0){$\hzz_{10}$}}
  \put(76,76){\makebox(0,0){$\hzz_{20}$}}
  \put(76,60){\makebox(0,0){$\hzz_{0j}$}}
  \put(76,44){\makebox(0,0){$\hzz_{jj}$}}
  \put(76,28){\makebox(0,0){$\hzz_{j'j}$}}
  
  \put(35,108){\vector(1,0){10}}
  \put(35,92){\vector(1,0){10}}
  \put(35,76){\vector(1,0){10}}
  \put(35,60){\vector(1,0){10}}
  \put(35,44){\vector(1,0){10}}
  \put(35,28){\vector(1,0){10}}
  
  \put(59,108){\vector(1,0){10}}
  \put(59,92){\vector(1,0){10}}
  \put(59,76){\vector(1,0){10}}
  \put(59,60){\vector(1,0){10}}
  \put(59,44){\vector(1,0){10}}
  \put(59,28){\vector(1,0){10}}
  
  \put(83,108){\vector(1,0){10}}
  \put(83,92){\vector(1,0){10}}
  \put(83,76){\vector(1,0){10}}
  \put(83,60){\vector(1,0){10}}
  \put(83,44){\vector(1,0){10}}
  \put(83,28){\vector(1,0){10}}
  
  \put(107,108){\vector(1,0){20}}
  \put(107,92){\vector(1,0){20}}
  \put(107,76){\vector(1,0){20}}
  \put(107,60){\vector(1,0){20}}
  \put(107,44){\vector(1,0){20}}
  \put(107,28){\vector(1,0){20}}
  
  \put(93,101){\framebox(14,14){$g_{00}$}}
  \put(93,85){\framebox(14,14){$g_{10}$}}
  \put(93,69){\framebox(14,14){$g_{20}$}}
  \put(93,53){\framebox(14,14){$g_{0j}$}}
  \put(93,37){\framebox(14,14){$g_{jj}$}}
  \put(93,21){\framebox(14,14){$g_{j'j}$}}
  
  \put(128,108){\makebox(0,0)[l]{$\hmm_{00}$}}
  \put(128,92){\makebox(0,0)[l]{$\hmm_{10}$}}
  \put(128,76){\makebox(0,0)[l]{$\hmm_{20}$}}
  \put(128,60){\makebox(0,0)[l]{$\hmm_{0j}$}}
  \put(128,44){\makebox(0,0)[l]{$\hmm_{j'j}$}}
  \put(128,28){\makebox(0,0)[l]{$\hmm_{j'j}$}}
  
  \put(7,12){\vector(1,0){38}}
  \put(3,12){\makebox(0,0){$\yy_j$}}
  \put(45,7){\framebox(14,108){$\hZ^n_{\D_j}$}}
  \put(20,0){\framebox(97,122){}}
 \end{picture}
\end{center}
\caption{Construction of Code for 2-input-2-output Interference Channel
 for the Proof of Theorem \ref{thm:crng-full-achievable}:
 For the proof of Theorem \ref{thm:crng-achievable},
 the flows of vectors $\cc_s$, $\mm_s$, $\zz_s$, $s\in\{01,02,12,21\}$
 are omitted.}
\label{fig:channel-code}
\end{figure}

Here, we define the constrained-random-number generators
\cite{CRNG,CRNG-MULTI,SW2CC} used to construct encoders.
Let $\mu_{Z^n_{\cS}}$ be defined as
\begin{equation}
 \mu_{Z^n_{\cS}}(\zz_{\cS})
 \equiv
 \mu_{Z^n_{\cS_0}}(\zz_{\cS_0})
 \prod_{i\in\I}
 \mu_{Z^n_{\cS_i}|Z^n_{\cS_0}}(\zz_{\cS_i}|\zz_{\cS_0}).
 \label{eq:ZS}
\end{equation}
Let $\tZ_{\cS_0}^n$
be a random variable corresponding to the distribution
\begin{align}
 \mu_{\tZ^n_{\cS_0}|C_{\cS_0,n}M_{\cS_0,n}}(\zz_{\cS_0}|\cc_{\cS_0},\mm_{\cS_0})
 &\equiv
 \frac{
  \mu_{Z_{\cS_0}^n}(\zz_{\cS_0})
  \chi(f_{\cS_0}(\zz_{\cS_0})=\cc_{\cS_0},g_{\cS_0}(\zz_{\cS_0})=\mm_{\cS_0})
 }{
  \mu_{Z_{\cS_0}^n}(\fC_{(f,g)_{\cS_0}}(\cc_{\cS_0},\mm_{\cS_0}))
 }.
 \label{eq:crng-S0}
\end{align}
The constrained-random-number generator of $\tZ_{\cS_0}^n$
is used by all encoders
and it is assumed that all encoders have the same output $\zz_{\cS_0}$.
Let $\tZ_{\cS_i}^n$ be a random variable corresponding to the distribution
\begin{align}
 \mu_{\tZ_{\cS_i}^n|Z^n_{\cS_0}C_{\cS_i,n}M_{\cS_i,n}}
 (\zz_{\cS_i}|\zz_{\cS_0},\cc_{\cS_i},\mm_{\cS_i})
 &\equiv
 \frac{
  \mu_{Z_{\cS_i}^n|Z_{\cS_0}^n}(\zz_{\cS_i}|\zz_{\cS_0})
  \chi(f_{\cS_i}(\zz_{\cS_i})=\cc_{\cS_i},g_{\cS_i}(\zz_{\cS_i})=\mm_{\cS_i})
 }{
  \mu_{Z_{\cS_i}^n|Z_{\cS_0}^n}(
   \fC_{(f,g)_{\cS_i}}(\cc_{\cS_i},\mm_{\cS_i})
   |\zz_{\cS_0}
  )
 }.
 \label{eq:crng-Si}
\end{align}
The constrained-random-number generatorof $\tZ_{\cS_i}^n$
is used by Encoder $i$.
Let $W_i:\Z^n_{\cS_0}\times\Z^n_{\cS_i}\to\X_i^n$
be a channel (non-deterministic function)
with inputs $(\zz_{\cS_0},\zz_{\cS_i})$ and an output $\xx_i$
subject to the conditional probability distribution
$\mu_{X_i|Z^n_{\cS_0}Z^n_{\cS_i}}$.
We define the encoding function $\Phi_{i,n}:\M_{\cS_i,n}\to\X_i^n$
of Encoder $i$ as
\begin{equation*}
 \Phi_{i,n}(\mm_{\cS_0},\mm_{\cS_i})
 \equiv
 W_i(\tZ^n_{\cS_0},\tZ^n_{\cS_i}),
\end{equation*}
where the encoder claims an error
when at least one of the denominators of (\ref{eq:crng-S0})
and (\ref{eq:crng-Si}) is zero.
By using the decoder of the source code $\hZ^n_{\D_j}$,
we define the decoding function $\Psi_{j,n}:\Y_j^n\to\M_{\D_j,n}$
of Decoder $j$ as
\begin{equation*}
 \Psi_{j,n}(\yy_j)
 \equiv
 \{g_{s,n}(\hZ^n_{j,s})\}_{s\in\D_j}.
\end{equation*}
The flow of vectors is illustrated in Fig.\ \ref{fig:channel-code}.

Let $\hM_{\D_j,n}\equiv \Psi_j(Y^n_j)$
and $\Error(f_{\cS},g_{\cS},\cc_{\cS})$ 
be the decoding error probability.
We have the following theorem, where the proof is given in
Section \ref{sec:proof-channel}.

\begin{thm}
\label{thm:channel-code}
Let us assume that $\cS$ is the disjoint union of $\{\cS_i\}_{i\in\{0\}\cup\I}$,
that is, $\cS=\bigcup_{i\in\{0\}\cup\I}\cS_i$ and $\cS_i\cap\cS_{i'}=\emptyset$
for all $i\neq i'$.
Let us assume that $\{(r_s,R_s)\}_{s\in\cS}$ satisfies
\begin{align}
 \sum_{s\in\tcS_0}[R_s+r_s]
 &<
 \uH(\ZZ_{\tcS_0})
 \label{eq:channel-Rs+rs}
 \\
 \sum_{s\in\tcS_i}[R_s+r_s]
 &<
 \uH(\ZZ_{\tcS_i}|\ZZ_{\cS_0})
 \\
 \sum_{s\in\tcD_j}r_s
 &>
 \oH(\ZZ_{\tcD_j}|\YY_j,\ZZ_{\D_j\setminus\tcD_j})
 \label{eq:channel-rs}
\end{align}
for all $\tcS_0$ satisfying $\emptyset\neq\tcS_0\subset\cS_0$,
all $(i,\tcS_i)$ satisfying $i\in\I$ and $\emptyset\neq\tcS_i\subset\cS_i$,
and for all $(j,\tcD_j)$
satisfying $j\in\J$ and $\emptyset\neq\tcD_j\subset\D_j$,
where the joint distribution of $(Z^n_{\cS},X^n_{\I})$ is given by
(\ref{eq:ZS}).
Then for any $\delta>0$ and all sufficiently large $n$
there are functions (sparse matrices)
$f_{\cS}$, $g_{\cS}$ and vectors $\cc_{\cS}$ such that
$\Error(f_{\cS},g_{\cS},\cc_{\cS})\leq\delta$.
\end{thm}

\begin{rem}
In the construction of encoders,
it is assumed that 
the output $\zz_{\cS_0}$
of the constrained random number generator $\tZ^n_{\cS_0}$
is shared by all encoders.
This means that all encoders have shared randomness
except the common messages.
However, 
we can show the fact by using the random coding argument
that it is possible to fix $\zz_{\cS_0}$
to obtain the same error probability at worst \cite[Remark 20]{CRNG}.
\end{rem}

\subsection{Proof of Theorems \ref{thm:crng-achievable}
 and \ref{thm:crng-full-achievable}}

By applying Theorem~\ref{thm:channel-code}
to stationary memoryless sources/channels,
we have the single-letterized conditions
\begin{align*}
 \sum_{s\in\tcS_0}[R_s+r_s]
 &<
 H(Z_{\tcS_0})
 \\
 \sum_{s\in\tcS_i}[R_s+r_s]
 &<
 H(Z_{\tcS_i}|Z_{\cS_0})
 \\
 \sum_{s\in\tcD_j}r_s
 &>
 H(Z_{\tcD_j}|Y_j,Z_{\D_j\setminus\tcD_j})
\end{align*}
for the existence of a channel code.
By letting $\{R_s\}_{s\in\cS}$ approach arbitrary close
to the boundary of the region,
we have the conditions (\ref{eq:crng-R0+r0})--(\ref{eq:crng-sum-rj})
and (\ref{eq:crngfull-sum-S0})--(\ref{eq:crngfull-sum-Di}).
By taking the union over $\mu$ and $\omu$
satisfying (\ref{eq:joint-crng}) and (\ref{eq:joint-crng-full})
and using the time-sharing arguments,
we have Theorems \ref{thm:crng-achievable} and \ref{thm:crng-full-achievable},
respectively.
\hfill\IEEEQED

\subsection{Proof of Lemma~\ref{lem:source}}
\label{sec:proof-source}

First, we show the following lemma.
We omit the dependence of $\F$, $F$, $f$, $\C$, $C$,
$Y$, and $Z$ on $n$.

\begin{lem}[{\cite[Eq. (58)]{CRNG-MULTI}}]
\label{lem:source-map}
Let $(\ZZ_{\D_j},\YY_j)\equiv\{(\{Z^n_s\}_{s\in\D_j},Y^n_j)\}_{n=1}^{\infty}$
be a pair of general correlated sources.
Assume that $(\F_{s,n},p_{F_{s,n}})$ has the hash property
((\ref{eq:hash}) in Appendix \ref{sec:hash})
for every $s\in\D_j$.
Then we have
\begin{align*}
 &
 E_{F_{\D_j}}\lrB{
  \mu_{Z_{\D_j}Y_j}\lrsb{
   \lrb{
    (\zz_{\D_j},\yy_j): \chzz_{\D_j}(F_{\D_j}(\zz_{\D_j})|\yy_j)\neq\zz_{\D_j}
   }
  }
 }
 \notag
 \\*
 &\leq
 \sum_{
  \tcD_j\subset\D_j:
  \tcD_j\neq\emptyset
 }
 \alpha_{F_{\tcD_j}}\lrB{\beta_{F_{\D_j\setminus\tcD_j}}+1}
 \frac{
  2^{
   n\lrB{\oH(\ZZ_{\tcD_j}|\YY_j,\ZZ_{\D_j\setminus\tcD_j})+\e}
  }
 }{
  \prod_{s\in\tcD_j}|\im\F_s|
 }
 +\beta_{F_{\D_j}}
 +\mu_{Z_{\D_j}Y_j}(\oT_j^{\complement}),
\end{align*}
where
$\im\F_s\equiv\bigcup_{f_s\in\F_s}\{f_s(\zz_s):\zz_s\in\Z^n_s\}$,
$\oT_j$ is defined by (\ref{eq:oT}),
and $\chzz_{\D_j}(\cc_{\D_j}|\yy_j)$ outputs one of the elements
in $\oT_j\cap\fC_{f_{\D_j}}(\cc_{\D_j})$
and declares an error
when $\oT_j\cap\fC_{f_{\D_j}}(\cc_{\D_j})=\emptyset$.
\end{lem}
\begin{IEEEproof}
 Let
 $\oT_j(\yy_j)
 \equiv
 \{\zz_{\D_j}: (\zz_{\D_j},\yy_j)\in\oT_j\}$
 and assume that $(\zz_{\D_j},\yy)\in\oT_j$
 and $\chzz_{\D_j}(f_{\D_j}(\zz_{\D_j})|\yy_j)\neq\zz_{\D_j}$.
 Then we have
 $\lrB{\oT_j(\yy_j)\setminus\{\zz_{\D_j}\}}
 \cap\fC_{f_{\D_j}}(f_{\D_j}(\zz_{\D_j}))\neq\emptyset$.
 We have
 \begin{align}
  E_{F_{\D_j}}\lrB{
   \chi(\chzz_{\D_j}(F_{\D_j}(\zz_{\D_j})|\yy_j)\neq\zz_{\D_j})
  }
  &\leq
  p_{F_{\D_j}}\lrsb{\lrb{
    f_{\D_j}:
    \lrB{\oT_{\Z_{\D_j}}(\yy_j)\setminus\{\zz_{\D_j}\}}
    \cap\C_{f_{\D_j}}(f_{\D_j}(\zz_{\D_j}))\neq\emptyset
  }}
  \notag
  \\
  &\leq
  \sum_{
   \tcD_j\subset\D_j:
   \tcD_j\neq\emptyset
  }
  \frac{
   \alpha_{F_{\tcD_j}}\lrB{\beta_{F_{\D_j\setminus\tcD_j}}+1}
   \oO_{\tcD_j}
  }{
   \prod_{s\in\tcD_j}\lrbar{\im\F_s}
  }
  +\beta_{F_{\D_j}}
  \notag
  \\
  &\leq
  \sum_{
   \tcD_j\subset\D_j:
   \tcD_j\neq\emptyset
  }
  \alpha_{F_{\tcD_j}}\lrB{\beta_{F_{\D_j\setminus\tcD_j}}+1}
  \frac{
   2^{
    n\lrB{\oH(\ZZ_{\tcD_j}|\YY_j,\ZZ_{\D_j\setminus\tcD_j})+\e}
   }
  }{
   \prod_{s\in\tcD_j}\lrbar{\im\F_s}
  }
  +\beta_{F_{\D_j}},
 \end{align}
 where
 the second inequality comes from Lemma~\ref{lem:mCRP} 
 in Appendix \ref{sec:hash}
 by letting
 \begin{equation*}
  \T\equiv\oT_j
 \end{equation*}
 and the third inequality comes from the fact that
 \begin{equation*}
  \oO_{\tcD_j}
  \leq
  2^{n\lrB{\oH(\ZZ_{\tcD_j}|\YY_j,\ZZ_{\D_j\setminus\tcD_j})+\e}}.
 \end{equation*}
 We have
 \begin{align}
  &
  E_{F_{\D_j}}\lrB{
   \mu_{Z_{\D_j}Y_j}\lrsb{
    \lrb{
     (\zz_{\D_j},\yy_j): \chzz_{\D_j}(F_{\D_j}(\zz_{\D_j})|\yy_j)\neq\zz_{\D_j}
    }
   }
  }
  \notag
  \\*
  &=
  E_{F_{\D_j}}\lrB{
   \sum_{
    \zz_{\D_j}\in\Z^n_{\D_j},
    \yy_j\in\Y^n_j
   }
   \mu_{Z_{\D_j}Y_j}(\zz_{\D_j},\yy_j)
   \chi(\chzz_{\D_j}(F_{\D_j}(\zz_{\D_j})|\yy_j)\neq \zz_{\D_j})
  }
  \notag
  \\*
  &=
  \sum_{(\zz_{\D_j},\yy_j)\in\oT_{Z_{\D_j}}}
  \mu_{Z_{\D_j}Y_j}(\zz_{\D_j},\yy_j)
  E_{F_{\D_j}}\lrB{
   \chi(\chzz_{\D_j}(F_{\D_j}(\zz_{\D_j})|\yy_j)\neq\zz_{\D_j})
  }
  \notag
  \\*
  &\quad
  +
  \sum_{(\zz_{\D_j},\yy_j)\in\oT_{Z_{\D_j}}^{\complement}}
  \mu_{Z_{\D_j}Y_j}(\zz_{\D_j},\yy_j)
  E_{F_{\D_j}}\lrB{
   \chi(\chzz_{\D_j}(F_{\D_j}(\zz_{\D_j})|\yy_j)\neq \zz_{\D_j})
  }
  \notag
  \\
  &\leq
  \sum_{
   \tcD_j\subset\D_j:
   \tcD_j\neq\emptyset
  }
  \alpha_{F_{\tcD_j}}\lrB{\beta_{F_{\D_j\setminus\tcD_j}}+1}
  \frac{
   2^{
    n\lrB{\oH(\ZZ_{\tcD_j}|\YY_j,\ZZ_{\D_j\setminus\tcD_j})+\e}
   }
  }{
   \prod_{s\in\tcD_j}\lrbar{\im\F_s}
  }
  +\beta_{F_{\D_j}}
  +\mu_{Z_{\D_j}Y_j}(\oT_j^{\complement}).
 \end{align}
\end{IEEEproof}

Next, we introduce the following lemma without the proof.
\begin{lem}[{\cite[Corollary 2]{SDECODING}}]
\label{lem:sdecoding}
Let $(U,V)$ be a pair consisting of state $U\in\U$
and observation $V\in\V$,
where $\mu_{UV}$ is the joint distribution of $(U,V)$.
We make a stochastic decision with $\mu_{U|V}$,
that is, the joint distribution of $(U,V)$ and a guess $\hU\in\U$ of the
state $U$ is given as
\begin{equation*}
 \mu_{UV\hU}(u,v,\hu)\equiv\mu_{UV}(u,v)\mu_{U|V}(\hu|v).
\end{equation*}
Then the decision error probability of this rule
is at most twice the decision error probability
of {\it any} (possibly stochastic) decision, 
that is,
\begin{align*}
 \sum_{\substack{
   u\in\U,v\in\V,\hu\in\U:
   \\
   \hu\neq u
 }}
 \mu_{UV}(u,v)
 \mu_{U|V}(\hu|v)
 &\leq
 2
 \sum_{\substack{
   u\in\U,v\in\V,\chu\in\U:
   \\
   \chu\neq u
 }}
 \mu_{UV}(u,v)
 \mu_{\chU|V}(\chu|v)
\end{align*}
for an arbitrary probability distribution $\mu_{\chU|V}$.
\end{lem}

Finally, we prove Lemma~\ref{lem:source}.
For a given $f_{\D_j}$,
the joint distribution of $(Z_{\D_j}^n,C_{\D_j},Y_j^n)$ is defined as
\begin{equation*}
 \mu_{Z_{\D_j}C_{\D_j}Y_j}(\zz_{\D_j},\cc_{\D_j},\yy_j)
 =
 \mu_{Z_{\D_j}Y_j}(\zz_{\D_j},\yy_j)
 \chi(f_{\D_j}(\zz_{\D_j})=\cc_{\D_j}).
\end{equation*}
Then we have
\begin{align}
 \mu_{Z_{\D_j}|Y_jC_{\D_j}}(\zz_{\D_j}|\yy_j,\cc_{\D_j})
 &\equiv
 \frac{
  \mu_{Z_{\D_j}C_{\D_j}Y_j}(\zz_{\D_j},\cc_{\D_j},\yy_j)
 }{
  \sum_{\zz_{\D_j}\in\Z^n_{\D_j}}
  \mu_{Z_{\D_j}C_{\D_j}Y_j}(\zz_{\D_j},\cc_{\D_j},\yy_j)
 }
 \notag
 \\
 &=
 \frac{
  \mu_{Z_{\D_j},Y_j}(\zz_{\D_j},\yy_j)\chi(f_{\D_j}(\zz_{\D_j})=\cc_{\D_j})
 }{
  \sum_{\zz_{\D_j}\in\Z^n_{\D_j}}
  \mu_{Z_{\D_j},Y_j}(\zz_{\D_j},\yy_j)
  \chi(f_{\D_j}(\zz_{\D_j})=\cc_{\D_j})
 }
 \notag
 \\
 &=
 \frac{
  \mu_{Z_{\D_j}|Y_j}(\zz_{\D_j}|\yy_j)\chi(f_{\D_j}(\zz_{\D_j})=\cc_{\D_j})
 }{
  \sum_{\zz_{\D_j}\in\Z^n_{\D_j}}
  \mu_{Z_{\D_j}|Y_j}(\zz_{\D_j}|\yy_j)
  \chi(f_{\D_j}(\zz_{\D_j})=\cc_{\D_j})
 }
 \notag
 \\
 &=
 \mu_{\hZ_{\D_j}|Y_jC_{\D_j}}(\zz_{\D_j}|\yy_j,\cc_{\D_j}),
\end{align}
that is, the constrained-random-number generator defined by
(\ref{eq:source-decoder}) is a stochastic decision with
$\mu_{Z_{\D_j}|Y_jC_{\D_j}}$.
By applying Lemma~\ref{lem:sdecoding}, we have the fact that
\begin{align}
 \mu_{Z_{\D_j}\hZ_{\D_j}}\lrsb{
  \lrb{
   (\zz_{\D_j},\hzz_{\D_j}):
   \hzz_{\D_j}\neq\zz_{\D_j}
  }
 }
 &=
 \sum_{\substack{
   \zz_{\D_j}\in\Z^n_{\D_j},
   \cc_{\D_j}\in\C_{\D_j},
   \\
   \yy_j\in\Y^n_j,
   \hzz_{\D_j}\in\Z^n_{\D_j}:
   \\
   \hzz_{\D_j}\neq\zz_{D_j}
 }}
 \mu_{Z_{\D_j}C_{\D_j}Y_j\hZ_{\D_j}}(\zz_{\D_j},\cc_{\D_j},\yy_j,\hzz_{\D_j})
 \notag
 \\
 &=
 \sum_{\substack{
   \zz_{\D_j}\in\Z^n_{\D_j},
   \cc_{\D_j}\in\C_{\D_j},
   \\
   \yy_j\in\Y^n_j,
   \hzz_{\D_j}\in\Z^n_{\D_j}:
   \\
   \hzz_{\D_j}\neq\zz_{D_j}
 }}
 \mu_{Z_{\D_j}C_{\D_j}Y_j}(\zz_{\D_j},\cc_{\D_j},\yy_j)
 \mu_{Z_{\D_j}|Y_jC_{\D_j}}(\hzz_{\D_j}|\yy_j,\cc_{\D_j})
 \notag
 \\
 &\leq
 2
 \sum_{\substack{
   \zz_{\D_j}\in\Z^n_{\D_j},
   \cc_{\D_j}\in\C_{\D_j},
   \\
   \yy_j\in\Y^n_j,
   \chzz_{\D_j}\in\Z^n_{\D_j}:
   \\
   \chzz_{\D_j}\neq\zz_{D_j}
 }}
 \mu_{Z_{\D_j}C_{\D_j}Y_j}(\zz_{\D_j},\cc_{\D_j},\yy_j)
 \mu_{\chZ_{\D_j}|Y_jC_{\D_j}}(\chzz_{\D_j}|\yy_j,\cc_{\D_j})
 \notag
 \\
 &=
 2
 \sum_{\substack{
   \zz_{\D_j}\in\Z^n_{\D_j},
   \cc_{\D_j}\in\C_{\D_j},
   \\
   \yy_j\in\Y^n_j,
   \chzz_{\D_j}\in\Z^n_{\D_j}:
   \\
   \chzz_{\D_j}\neq\zz_{D_j}
 }}
 \mu_{Z_{\D_j}Y_j}(\zz_{\D_j},\yy_j)
 \chi(f_{\D_j}(\zz_{\D_j})=\cc_{\D_j})
 \chi(\chzz_{\D_j}(\cc_{\D_j}|\yy_j)=\chzz_{\D_j})
 \notag
 \\
 &=
 2
 \sum_{\substack{
   \zz_{\D_j}\in\Z^n_{\D_j},
   \yy_j\in\Y^n_j:
   \\
   \chzz_{\D_j}(f_{\D_j}(\zz_{\D_j})|\yy_j)\neq\zz_{\D_j}
 }}
 \mu_{Z_{\D_j}Y_j}(\zz_{\D_j},\yy_j)
 \notag
 \\
 &=
 2
 \mu_{Z_{\D_j}Y_j}
 \lrsb{
  \lrb{
   (\zz_{\D_j},\yy_j):
   \chzz_{\D_j}(f_{\D_j}(\zz_{\D_j})|\yy_j)\neq\zz_{\D_j}
  }
 },
\end{align}
where
$\mu_{\chZ_{\D_j}|Y_jC_{\D_j}}(\chzz_{\D_j}|\cc_{\D_j},\yy_j)
\equiv
\chi(\chzz_{\D_j}(\cc_{\D_j}|\yy_j)=\chzz_{\D_j})$
and the second equality comes from (\ref{eq:source-joint}).
By applying Lemma~\ref{lem:source-map}, we have the fact that
\begin{align}
 &
 E_{F_{\D_j}}\lrB{
  \mu_{Z_{\D_j}\hZ_{\D_j}}\lrsb{
   \lrb{
    (\zz_{\D_j},\hzz_{\D_j}):
    \hzz_{\D_j}\neq\zz_{\D_j}
   }
  }
 }
 \notag
 \\*
 &\leq
 2
 E_{F_{\D_j}}\lrB{
  \mu_{Z_{\D_j}Y_j}
  \lrsb{
   \lrb{
    (\zz_{\D_j},\yy_j):
    \chzz_{\D_j}(f_{\D_j}(\zz_{\D_j})|\yy_j)\neq\zz_{\D_j}
   }
  }
 }
 \notag
 \\
 &\leq
 2 \sum_{
  \tcD_j\subset\D_j:
  \tcD_j\neq\emptyset
 }
 \alpha_{F_{\tcD_j}}\lrB{\beta_{F_{\D_j\setminus\tcD_j}}+1}
 2^{
  -n\lrB{
   \sum_{s\in\tcD_j}r_s-\oH(\ZZ_{\tcD_j}|\YY_j,\ZZ_{\D_j\setminus\tcD_j})-\e
  }
 }
 +2 \beta_{F_{\D_j}}
 +2 \mu_{Z_{\D_j}Y_j}(\oT_j^{\complement}),
\end{align}
where we use the relation $r_s=\log_2(|\C_s|)/n=\log_2(|\im\F_s|)/n$
in the last inequality.
\hfill\IEEEQED

\subsection{Proof of Theorem~\ref{thm:channel-code}}
\label{sec:proof-channel}

Let us assume that ensembles $(\F_s,p_{F_s})$ and $(\G_s,p_{G_s})$
have the hash property ((\ref{eq:hash}) in Appendix \ref{sec:hash})
for every $s\in\cS$.
From Lemma \ref{lem:hash-FG} in Appendix \ref{sec:hash},
we have the fact that the joint 
ensemble $(\F_{s,n}\times\G_{s,n},p_{(F,G)_{s,n}})$
also satisfies the hash property.
Let $\C_{s,n}\equiv\im\F_{n,s}$ and $\M_{s,n}\equiv\im\G_{n,s}$,
where they are used without notice.
We use the fact without notice that
$\{\fC_{(f,g)_{\tcS}}(\cc_{\tcS},\mm_{\tcS}))\}_{
\cc_{\tcS}\in\C_{\tcS},\mm_{\tcS}\in\M_{\tcS}
}$
is a partition of $\Z_{\tcS}^n$ for a given $\tcS\subset\cS$,
and $\{\cS_i\}_{i\in\{0\}\cup\I}$ is a partition of $\cS$.
Let
\begin{align*}
\M_{\cS_0,n}
&\equiv
\Prod_{s\in\cS_0}\M_{s,n}
\\
\M_{\cS_{\I},n}
&\equiv
\Prod_{i\in\I}\Prod_{s\in\cS_i}\M_{s,n}
\\
\C_{\cS_0,n}
&\equiv
\Prod_{s\in\cS_0}\C_{s,n}
\\
\C_{\cS_{\I},n}
&\equiv
\Prod_{i\in\I}\Prod_{s\in\cS_i}\C_{s,n}
\\
\X_{\I}^n
&\equiv
\Prod_{i\in\I}\X_i^n
\\
\Y_{\J}^n
&\equiv
\Prod_{j\in\J}\Y_j^n
\\
\Z_{\cS}^n
&\equiv
\Prod_{s\in\cS}\Z_s^n.
\end{align*}
In the following, we omit the dependence of
$\F$, $F$, $f$, $\G$, $G$, $g$, $\C$, $C$, $\M$, $M$,
$X$, $Y$, and $Z$ on $n$.

Let 
\begin{align*}
&\E((f,g)_{\cS_0},\cc_{\cS_0})
\equiv
\lrb{
 \mm_{\cS_0}:
 \mu_{Z_{\cS_0}}
 (\fC_{(f,g)_{\cS_0}}(\cc_{\cS_0},\mm_{\cS_0}))
 =0
}
\\
&\E((f,g)_{\cS_{\I}},\cc_{\cS_{\I}})
\equiv
\lrb{
 \mm_{\cS_{\I}}:
 \mu_{Z_{\cS_i}|Z_{\cS_0}}
 (\fC_{(f,g)_{\cS_i}}(\cc_{\cS_i},\mm_{\cS_i})|\zz_{\cS_0})=0
 \ \text{for some}\ i\in\I\ \text{and}\ \zz_{\cS_0}\in\Z^n_{\cS}
}
\\
&\E(g_{\D_{\J}},\mm_{\D_{\J}})
\equiv
\lrb{
 \hzz_{\D_{\J}}:
 g_s(\hzz_{j,s})\neq \mm_s
 \ \text{for some}\ j\in\J, s\in\D_j
}.
\end{align*}

Then the error probability 
$\Error(f_{\cS},g_{\cS},\cc_{\cS})$ is evaluated as
\begin{align}
&
\Error(f_{\cS},g_{\cS},\cc_{\cS})
\notag
\\*
&\leq
\sum_{
 \mm_{\cS_0}\in\E((f,g)_{\cS_0},\cc_{\cS_0})
}
\frac1{\prod_{s\in\cS_0}|\M_s|}
+
\sum_{
 \mm_{\cS_{\I}}\in\E((f,g)_{\cS_{\I}},\cc_{\cS_{\I}})
}
\frac1{\prod_{s\in\cS_{\I}}|\M_s|}
\notag
\\*
&\quad
+
\sum_{\substack{
  \mm_{\cS_0}\notin\E((f,g)_{\cS_0},\cc_{\cS_0}),
  \zz_{\cS_0}\in\fC_{(f,g)_{\cS_0}}(\cc_{\cS_0},\mm_{\cS_0}),
  \\
  \mm_{\cS_{\I}}\notin\E((f,g)_{\cS_{\I}},\cc_{\cS_{\I}}),
  \zz_{\cS_{\I}}\in\fC_{(f,g)_{\cS_{\I}}}(\cc_{\cS_{\I}},\mm_{\cS_{\I}}),
  \\
  \xx_{\I}\in\X^n_{\I},
  \yy_{\J}\in\Y^n_{\J},
  \hzz_{\D_{\J}}\in\E(g_{\D_{\J}},\mm_{\D_{\J}})
}}
\lrB{
 \prod_{j\in\J}
 \mu_{\hZ_{\D_j}|Y_jC_{\D_j}}(\hzz_{\D_j}|\yy_j,\cc_{\D_j})
}
\mu_{Y_{\J}|X_{\I}}(\yy_{\J}|\xx_{\I})
\notag
\\*
&\qquad\cdot
\lrB{
 \prod_{i\in\I}
 \mu_{X_i|Z_{\cS_i}Z_{\cS_0}}(\xx_i|\zz_{\cS_i},\zz_{\cS_0})
 \mu_{\tZ_{\cS_i}|Z_{\cS_0}C_{\cS_i}M_{\cS_i}}
 (\zz_{\cS_i}|\zz_{\cS_0},\cc_{\cS_i},\mm_{\cS_i})
}
\mu_{\tZ_{\cS_0}|C_{\cS_0}M_{\cS_0}}(\zz_{\cS_0}|\cc_{\cS_0},\mm_{\cS_0})
\prod_{s\in\cS}\frac1{|\M_s|}
\notag
\\
&\leq
\sum_{
 \mm_{\cS_0}\in\E((f,g)_{\cS_0},\cc_{\cS_0})
}
\frac1{|\M_{\cS_0}|}
+
\sum_{
 \mm_{\cS_{\I}}\in\E((f,g)_{\cS_{\I}},\cc_{\cS_{\I}})
}
\frac1{|\M_{\cS_{\I}}|}
\notag
\\*
&\quad
+
\sum_{\substack{
  \mm_{\cS_0}\notin\E((f,g)_{\cS_0},\cc_{\cS_0}),
  \\
  \zz_{\cS_0}\in\fC_{(f,g)_{\cS_0}}(\cc_{\cS_0},\mm_{\cS_0}),
  \\
  \mm_{\cS_{\I}}\notin\E((f,g)_{\cS_{\I}},\cc_{\cS_{\I}}),
  \\
  \zz_{\cS_{\I}}\in\fC_{(f,g)_{\cS_{\I}}}(\cc_{\cS_{\I}},\mm_{\cS_{\I}}),
  \\
  \yy_{\J}\in\Y^n_{\J},
  \hzz_{\D_{\J}}\in\E(g_{\D_{\J}},\mm_{\D_{\J}})
}}
\frac{
 \mu_{Z_{\cS}Y_{\J}}(\zz_{\cS},\yy_{\J})
 \mu_{\hZ_{\D_{\J}}|Y_{\J}C_{\D_{\J}}}
 (\hzz_{\D_{\J}}|\yy_{\J},\cc_{\D_{\J}})
}{
 \mu_{Z_{\cS_{\I}}|Z_{\cS_0}}(
  \fC_{(f,g)_{\cS_{\I}}}(\cc_{\cS_{\I}},\mm_{\cS_{\I}})
  |\zz_{\cS_0})
 \mu_{Z_{\cS_0}}(\fC_{(f,g)_{\cS_0}}(\cc_{\cS_0},\mm_{\cS_0}))
 |\M_{\cS_0}|
 |\M_{\cS_{\I}}|
},
\label{eq:channel-error}
\end{align}
where
$\mu_{\hZ_{\D_j}|C_{\D_j}Y_j}$, $\mu_{\tZ_{\cS_0}|C_{\cS_0}M_{\cS_0}}$,
and $\mu_{\tZ_{\cS_i}|Z_{\cS_0}C_{\cS_i}M_{\cS_i}}$ are defined by
(\ref{eq:source-decoder}), (\ref{eq:crng-S0}), and (\ref{eq:crng-Si}),
respectively,
the first and the second terms
on the right hand side correspond to the encoding error probability,
the third term on the right hand side
corresponds to the decoding error probability,
the union bound is used
in the first inequality,
and the relations (\ref{eq:crng-S0}), (\ref{eq:crng-Si}), and 
\begin{align*}
\mu_{Z_{\cS_{\I}}|Z_{\cS_0}}(\zz_{\cS_{\I}}|\zz_{\cS_0})
&=
\prod_{i\in\I}
\mu_{Z_{\cS_i}|Z_{\cS_0}}(\zz_{\cS_i}|\zz_{\cS_0})
\notag
\\
\mu_{Z_{\cS}Y_{\J}}(\zz_{\cS},\yy_{\J})
&=
\sum_{\xx_{\I}\in\X^n_{\I}}
\mu_{Y_{\J}|X_{\I}}(\yy_{\J}|\xx_{\I})
\lrB{
 \prod_{i\in\I}
 \mu_{X_i|Z_{\cS_i}Z_{\cS_0}}(\xx_i|\zz_{\cS_i},\zz_{\cS_0})
 \mu_{Z_{\cS_i}|Z_{\cS_0}}(\zz_{\cS_i}|\zz_{\cS_0})
}
\mu_{Z_{\cS_0}}(\zz_{\cS_0})
\end{align*}
are used in the second inequality.
The third term on the right hand side of the inequality
of (\ref{eq:channel-error}) is evaluated as
\begin{align}
&
\text{[the third term of (\ref{eq:channel-error})]}
\notag
\\*
&\leq
\sum_{
 \mm_{\cS_0}\notin\E((f,g)_{\cS_0},\cc_{\cS_0}),
 \zz_{\cS_0}\in\fC_{(f,g)_{\cS_0}}(\cc_{\cS_0},\mm_{\cS_0}),
 \mm_{\cS_{\I}}\notin\E((f,g)_{\cS_{\I}},\cc_{\cS_{\I}}),
 \zz_{\cS_{\I}}\in\fC_{(f,g)_{\cS_{\I}}}(\cc_{\cS_{\I}},\mm_{\cS_{\I}}),
 \yy_{\J}\in\Y^n_{\J},
 \hzz_{\D_{\J}}\in\E(g_{\D_{\J}},\mm_{\D_{\J}})
}
\Biggl[
 \notag
 \\*
 &\qquad
 \frac{
  \mu_{Z_{\cS}Y_{\J}}(\zz_{\cS},\yy_{\J})
  \mu_{\hZ_{\D_{\J}}|Y_{\J}C_{\D_{\J}}}
  (\hzz_{\D_{\J}}|\yy_{\J},\cc_{\D_{\J}})
  |\C_{\cS_0}|
 }{
  \mu_{Z_{\cS_{\I}}|Z_{\cS_0}}(
   \fC_{(f,g)_{\cS_{\I}}}(\cc_{\cS_{\I}},\mm_{\cS_{\I}})
   |\zz_{\cS_0})
  |\M_{\cS_{\I}}|
 }
 \lrbar{
  \frac1{
   \mu_{Z_{\cS_0}}(\fC_{(f,g)_{\cS_{0}}}(\cc_{\cS_0},\mm_{\cS_0}))
   |\C_{\cS_0}|
   |\M_{\cS_0}|
  }
  -1
 }
 \notag
 \\
 &\qquad
 +
 \mu_{Z_{\cS}Y_{\J}}(\zz_{\cS},\yy_{\J})
 \mu_{\hZ_{\D_{\J}}|Y_{\J}C_{\D_{\J}}}
 (\hzz_{\D_{\J}}|\yy_{\J},\cc_{\D_{\J}})
 |\C_{\cS_0}|
 |\C_{\cS_{\I}}|
 \lrbar{
  \frac{
   1
  }{
   \mu_{Z_{\cS_{\I}}|Z_{\cS_0}}(
    \fC_{(f,g)_{\cS_{\I}}}(\cc_{\cS_{\I}},\mm_{\cS_{\I}})
    |\zz_{\cS_0})
   |\C_{\cS_{\I}}|
   |\M_{\cS_{\I}}|
  }
  -
  1
 }
 \notag
 \\*
 &\qquad
 +
 \mu_{Z_{\cS}Y_{\J}}(\zz_{\cS},\yy_{\J})
 \mu_{\hZ_{\D_{\J}}|Y_{\J}C_{\D_{\J}}}
 (\hzz_{\D_{\J}}|\yy_{\J},\cc_{\D_{\J}})
 |\C_{\cS_0}|
 |\C_{\cS_{\I}}|
 \Biggr]
\notag
\\
&\leq
\sum_{\substack{
  \mm_{\cS_0}\notin\E((f,g)_{\cS_0},\cc_{\cS_0}),
  \mm_{\cS_{\I}}\notin\E((f,g)_{\cS_{\I}},\cc_{\cS_{\I}})
}}
\frac{
 |\C_{\cS_0}|
}{
 |\M_{\cS_{\I}}|
}
\lrbar{
 \frac1{
  |\C_{\cS_0}|
  |\M_{\cS_0}|
 }
 -
 \mu_{Z_{\cS_0}}(\fC_{(f,g)_{\cS_{0}}}(\cc_{\cS_0},\mm_{\cS_0}))
}
\notag
\\*
&\quad
+
\sum_{\substack{
  \mm_{\cS_0}\notin\E((f,g)_{\cS_0},\cc_{\cS_0}),
  \zz_{\cS_0}\in\fC_{(f,g)_{\cS_0}}(\cc_{\cS_0},\mm_{\cS_0}),
  \\
  \mm_{\cS_{\I}}\notin\E((f,g)_{\cS_{\I}},\cc_{\cS_{\I}})
}}
\mu_{Z_{\cS_0}}(\zz_{\cS_0})
|\C_{\cS_0}|
|\C_{\cS_{\I}}|
\lrbar{
 \frac{
  1
 }{
  |\C_{\cS_{\I}}|
  |\M_{\cS_{\I}}|
 }
 -
 \mu_{Z_{\cS_{\I}}|Z_{\cS_0}}(
  \fC_{(f,g)_{\cS_{\I}}}(\cc_{\cS_{\I}},\mm_{\cS_{\I}})
  |\zz_{\cS_0})
}
\notag
\\*
&\quad
+
\sum_{
 \mm_{\cS}\in\M_{\cS},
 \zz_{\cS}\in\fC_{(f,g)_{\cS}}(\cc_{\cS},\mm_{\cS}),
 \yy_{\J}\in\Y^n_{\J},
 \hzz_{\D_{\J}}\in\E(g_{\D_{\J}},\mm_{\D_{\J}})
}
\mu_{Z_{\cS}Y_{\J}}(\zz_{\cS},\yy_{\J})
\mu_{\hZ_{\D_{\J}}|Y_{\J}C_{\D_{\J}}}
(\hzz_{\D_{\J}}|\yy_{\J},\cc_{\D_{\J}})
|\C_{\cS_0}|
|\C_{\cS_{\I}}|,
\label{eq:channel-error-3}
\end{align}
where the first inequality comes from the triangular inequality,
and the second inequality comes from the fact that
\begin{gather*}
\sum_{\hzz_{\D_{\J}}\in\E(g_{\D_{\J}},\mm_{\D_{\J}})}
\mu_{\hZ_{\D_{\J}}|Y_{\J}C_{\D_{\J}}}
(\hzz_{\D_{\J}}|\yy_{\J},\cc_{\D_{\J}})
\leq
1
\\
\sum_{
 \zz_{\cS_{\I}}\in\fC_{(f,g)_{\cS_{\I}}}(\cc_{\cS_{\I}},\mm_{\cS_{\I}}),
 \yy_{\J}\in\Y^n_{\J}
}
\mu_{Z_{\cS}Y_{\J}}(\zz_{\cS},\yy_{\J})
=
\mu_{Z_{\cS_{\I}}|Z_{\cS_0}}(
 \fC_{(f,g)_{\cS_{\I}}}(\cc_{\cS_{\I}},\mm_{\cS_{\I}})
 |\zz_{\cS_0})
\mu_{Z_{\cS_0}}(\zz_{\cS_0})
\\
\sum_{
 \zz_{\cS_0}\in\fC_{(f,g)_{\cS_0}}(\cc_{\cS_0},\mm_{\cS_0})
}
\mu_{Z_{\cS_0}}(\zz_{\cS_0})
=
\mu_{Z_{\cS_0}}(
 \fC_{(f,g)_{\cS_0}}(\cc_{\cS_0},\mm_{\cS_0})
),
\end{gather*}
$\mu_{Z_{\cS_0}}(\fC_{(f,g)_{\cS_0}}(\cc_{\cS_0},\mm_{\cS_0}))>0$
for all $\mm_{\cS_0}\notin\E((f,g)_{\cS_0},\cc_{\cS_0})$,
and
$\mu_{Z_{\cS_{\I}}|Z_{\cS_0}}(
\fC_{(f,g)_{\cS_{\I}}}(\cc_{\cS_{\I}},\mm_{\cS_{\I}})
|\zz_{\cS_0})>0$
for all $\mm_{\cS_{\I}}\notin\E((f,g)_{\cS_{\I}},\cc_{\cS_{\I}})$.
The first term on the right hand side of
the last inequality of (\ref{eq:channel-error-3})
is evaluated as
\begin{align}
[\text{the first term of (\ref{eq:channel-error-3})}]
&\leq
\sum_{
 \mm_{\cS_0}\notin\E((f,g)_{\cS_0},\cc_{\cS_0})
}
|\C_{\cS_0}|
\lrbar{
 \frac1{
  |\C_{\cS_0}|
  |\M_{\cS_0}|
 }
 -
 \mu_{Z_{\cS_0}}(\fC_{(f,g)_{\cS_{0}}}(\cc_{\cS_0},\mm_{\cS_0}))
}
\notag
\\
&=
\sum_{\substack{
  \mm_{\cS_0}\in\M_{\cS_0}
}}
|\C_{\cS_0}|
\lrbar{
 \frac{
  1
 }{
  |\C_{\cS_0}|
  |\M_{\cS_0}|
 }
 -
 \mu_{Z_{\cS_0}}(\fC_{(f,g)_{\cS_{0}}}(\cc_{\cS_0},\mm_{\cS_0}))
}
-
\sum_{\substack{
  \mm_{\cS_0}\in\E((f,g)_{\cS_0},\cc_{\cS_0})
}}
\frac1{
 |\M_{\cS_0}|
},
\label{eq:channel-error-3-1}
\end{align}
where
the equality comes from the fact that
$\mu_{Z_{\cS_0}}(\fC_{(f,g)_{\cS_0}}(\cc_{\cS_0},\mm_{\cS_0}))=0$
when $\mm_{\cS_0}\in\E((f,g)_{\cS_0},\cc_{\cS_0})$.
The second term
on the right hand side of
the last inequality of (\ref{eq:channel-error-3})
is evaluated as
\begin{align}
&
\text{[the second term of (\ref{eq:channel-error-3})]}
\notag
\\
&\leq
\sum_{\substack{
  \zz_{\cS_0}\in\fC_{f_{\cS_0}}(\cc_{\cS_0}),
  \mm_{\cS_{\I}}\notin\E((f,g)_{\cS_{\I}},\cc_{\cS_{\I}})
}}
\mu_{Z_{\cS_0}}(\zz_{\cS_0})
|\C_{\cS_0}|
|\C_{\cS_{\I}}|
\lrbar{
 \frac{
  1
 }{
  |\C_{\cS_{\I}}|
  |\M_{\cS_{\I}}|
 }
 -
 \mu_{Z_{\cS_{\I}}|Z_{\cS_0}}(
  \fC_{(f,g)_{\cS_{\I}}}(\cc_{\cS_{\I}},\mm_{\cS_{\I}})
  |\zz_{\cS_0})
}
\notag
\\
&=
\sum_{\substack{
  \zz_{\cS_0}\in\fC_{f_{\cS_0}}(\cc_{\cS_0}),
  \mm_{\cS_{\I}}\in\M_{\cS_{\I}}
}}
\mu_{Z_{\cS_0}}(\zz_{\cS_0})
|\C_{\cS_0}|
|\C_{\cS_{\I}}|
\lrbar{
 \frac{
  1
 }{
  |\C_{\cS_{\I}}|
  |\M_{\cS_{\I}}|
 }
 -
 \mu_{Z_{\cS_{\I}}|Z_{\cS_0}}(
  \fC_{(f,g)_{\cS_{\I}}}(\cc_{\cS_{\I}},\mm_{\cS_{\I}})
  |\zz_{\cS_0})
}
\notag
\\*
&\quad
-
\sum_{\substack{
  \zz_{\cS_0}\in\fC_{f_{\cS_0}}(\cc_{\cS_0}),
  \mm_{\cS_{\I}}\in\E((f,g)_{\cS_{\I}},\cc_{\cS_{\I}})
}}
\frac{
 \mu_{Z_{\cS_0}}(\zz_{\cS_0})
 |\C_{\cS_0}|
}{
 |\M_{\cS_{\I}}|
}
\label{eq:channel-error-3-2}
\end{align}
where the equality comes from the fact that
$\mu_{Z_{\cS_{\I}}|Z_{\cS_0}}
(\fC_{(f,g)_{\cS_{\I}}}(\cc_{\cS_{\I}},\mm_{\cS_{\I}})|\zz_{\cS_0})
=0$
when $\mm_{\cS_{\I}}\in\E((f,g)_{\cS_{\I}},\cc_{\cS_{\I}})$.

From here, we use without notice the fact that $C_{\cS}$ is generated at random
subject to the uniform distribution on $\C_{\cS}$.
From (\ref{eq:channel-error})--(\ref{eq:channel-error-3-2})
and the fact that
\begin{align}
E_{(F,G)_{\cS}C_{\cS}}\lrB{
 \sum_{\substack{
   \zz_{\cS_0}\in\fC_{F_{\cS_0}}(C_{\cS_0}),
   \mm_{\cS_{\I}}\in\E((F,G)_{\cS_{\I}},C_{\cS_{\I}})
 }}
 \frac{
  \mu_{Z_{\cS_0}}(\zz_{\cS_0})
  |\C_{\cS_0}|
 }{
  |\M_{\cS_{\I}}|
 }
}
&=
E_{(F,G)_{\cS}C_{\cS_{\I}}}\lrB{
 \sum_{\substack{
   \cc_{\cS_0}\in\C_{\cS_0},
   \zz_{\cS_0}\in\fC_{F_{\cS_0}}(C_{\cS_0}),
   \\
   \mm_{\cS_{\I}}\in\E((F,G)_{\cS_{\I}},C_{\cS_{\I}})
 }}
 \frac{
  \mu_{Z_{\cS_0}}(\zz_{\cS_0})
 }{
  |\M_{\cS_{\I}}|
 }
}
\notag
\\
&=
E_{(F,G)_{\cS}C_{\cS_{\I}}}\lrB{
 \sum_{
  \mm_{\cS_{\I}}\in\E((F,G)_{\cS_{\I}},C_{\cS_{\I}})
 }
 \frac1{
  |\M_{\cS_{\I}}|
 }
},
\end{align}
we have
\begin{align}
&
E_{(F,G)_{\cS}C_{\cS}}\lrB{
 \Error(F_{\cS},G_{\cS},C_{\cS})
}
\notag
\\*
&\leq
E_{(F,G)_{\cS}C_{\cS}}\lrB{
 \sum_{
  \mm_{\cS_0}\in\M_{\cS_0}
 }
 |\C_{\cS_0}|
 \lrbar{
  \frac1{
   |\C_{\cS_0}|
   |\M_{\cS_0}|
  }
  -
  \mu_{Z_{\cS_0}}(\fC_{(F,G)_{\cS_{0}}}(C_{\cS_0},\mm_{\cS_0}))
 }
}
\notag
\\*
&\quad
+
E_{(F,G)_{\cS}C_{\cS}}\lrB{
 \sum_{
  \zz_{\cS_0}\in\fC_{F_{\cS_0}}(C_{\cS_0}),
  \mm_{\cS_{\I}}\in\M_{\cS_{\I}}
 }
 \mu_{Z_{\cS_0}}(\zz_{\cS_0})
 |\C_{\cS_0}|
 |\C_{\cS_{\I}}|
 \lrbar{
  \frac1{
   |\C_{\cS_{\I}}|
   |\M_{\cS_{\I}}|
  }
  -
  \mu_{Z_{\cS_{\I}}|Z_{\cS_0}}(
   \fC_{(F,G)_{\cS_{\I}}}(C_{\cS_{\I}},\mm_{\cS_{\I}})
   |\zz_{\cS_0})
 }
}
\notag
\\*
&\quad
+
E_{(F,G)_{\cS}C_{\cS}}\lrB{
 \sum_{\substack{
   \mm_{\cS}\in\M_{\cS},
   \zz_{\cS}\in\fC_{(F,G)_{\cS}}(C_{\cS},\mm_{\cS}),
   \\
   \yy_{\J}\in\Y^n_{\J},
   \hzz_{\D_{\J}}\in\E(G_{\D_{\J}},\mm_{\D_{\J}})
 }}
 \mu_{Z_{\cS}Y_{\J}}(\zz_{\cS},\yy_{\J})
 \mu_{\hZ_{\D_{\J}}|C_{\D_{\J}}Y_{\J}}
 (\hzz_{\D_{\J}}|C_{\D_{\J}},\yy_{\J})
 |\C_{\cS_0}|
 |\C_{\cS_{\I}}|
}.
\label{eq:channel-ave-error}
\end{align}

Let $\uT_0$  be defined as
\begin{align*}
\uT_0
&\equiv
\lrb{
 \zz_{\cS_0}:
 \frac 1n\log_2\frac1{\mu_{Z_{\tcS_0}}(\zz_{\tcS_0})}\geq \uH(\ZZ_{\tcS_0})-\e
 \ \text{for all}\ \tcS_0\subset\cS_0
}.
\end{align*}
Then the first term on the right hand side of 
(\ref{eq:channel-ave-error})
is evaluated as
\begin{align}
&
\text{[the first term of (\ref{eq:channel-ave-error})]}
\notag
\\*
&=
E_{(F,G)_{\cS}}\lrB{
 \sum_{
  \cc_{\cS_0}\in\C_{\cS_0},
  \cc_{\cS_{\I}}\in\C_{\cS_{\I}},
  \mm_{\cS_0}\in\M_{\cS_0}
 }
 \frac1{
  |\C_{\cS_{\I}}|
 }
 \lrbar{
  \frac1{
   |\C_{\cS_0}|
   |\M_{\cS_0}|
  }
  -
  \mu_{Z_{\cS_0}}(\fC_{(F,G)_{\cS_{0}}}(\cc_{\cS_0},\mm_{\cS_0}))
 }
}
\notag
\\
&=
E_{(F,G)_{\cS_0}}\lrB{
 \sum_{
  \cc_{\cS_0}\in\C_{\cS_0},
  \mm_{\cS_0}\in\M_{\cS_0}
 }
 \lrbar{
  \frac1{
   |\C_{\cS_0}|
   |\M_{\cS_0}|
  }
  -
  \mu_{Z_{\cS_0}}(\fC_{(F,G)_{\cS_{0}}}(\cc_{\cS_0},\mm_{\cS_0}))
 }
}
\notag
\\
&\leq
E_{(F,G)_{\cS_0}}\lrB{
 \sum_{
  \cc_{\cS_0}\in\C_{\cS_0},
  \mm_{\cS_0}\in\M_{\cS_0}
 }
 \lrbar{
  \mu_{Z_{\cS_0}}(
   \uT_0\cap\fC_{(F,G)_{\cS_0}}(\cc_{\cS_0},\mm_{\cS_0})
  )
  -
  \frac{
   \mu_{Z_{\cS_0}}(\uT_0)
  }{
   |\C_{\cS_0}||\M_{\cS_0}|
  }
 }
}
\notag
\\*
&\quad
+
E_{(F,G)_{\cS_0}}\lrB{
 \sum_{
  \cc_{\cS_0}\in\C_{\cS_0},
  \mm_{\cS_0}\in\M_{\cS_0}
 }
 \mu_{Z_{\cS_0}}(
  \uT_0^{\complement}
  \cap\fC_{(F,G)_{\cS_0}}(\cc_{\cS_0},\mm_{\cS_0})
 )
}
+
E_{(F,G)_{\cS_0}}\lrB{
 \sum_{
  \cc_{\cS_0}\in\C_{\cS_0},
  \mm_{\cS_0}\in\M_{\cS_0}
 }
 \frac{
  \mu_{Z_{\cS_0}}(\uT_0^{\complement})
 }{
  |\C_{\cS_0}||\M_{\cS_0}|
 }
}
\notag
\\
&=
\mu_{Z_{\cS_0}}(\uT_0)
E_{(F,G)_{\cS_0}}\lrB{
 \sum_{
  \cc_{\cS_0}\in\C_{\cS_0},
  \mm_{\cS_0}\in\M_{\cS_0}
 }
 \lrbar{
  \frac{
   \mu_{Z_{\cS_0}}(
    \uT_0\cap\fC_{(F,G)_{\cS_0}}(\cc_{\cS_0},\mm_{\cS_0})
   )
  }{
   \mu_{Z_{\cS_0}}(\uT_0)
  }
  -
  \frac1{
   \prod_{s\in\cS_0}|\im\F_s||\im\G_s|
  }
 }
}
+
2\mu_{Z_{\cS_0}}(\uT_0^{\complement})
\notag
\\
&\leq
\mu_{Z_{\cS_0}}(
 \uT_0
)
\sqrt{
 \alpha_{(F,G)_{\cS_0}}-1
 +
 \sum_{
  \tcS_0\subset\cS_0:
  \tcS_0\neq\emptyset
 }
 \alpha_{(F,G)_{\cS_0\setminus\tcS_0}}
 [\beta_{(F,G)_{\tcS_0}}+1]
 \lrB{
  \prod_{s\in\tcS_0}|\im\F_s||\im\G_s|
 }
 \frac{
  2^{-n[\uH(\ZZ_{\tcS_0})-\e]}
 }{
  \mu_{Z_{\cS_0}}(\uT_0)
 }
}
+
2\mu_{Z_{\cS_0}}(\uT_0^{\complement})
\notag
\\
&\leq
\sqrt{
 \textstyle
 \alpha_{(F,G)_{\cS_0}}-1
 +\sum_{
  \tcS_0\subset\cS_0:
  \tcS_0\neq\emptyset
 }
 \alpha_{(F,G)_{\cS_0\setminus\tcS_0}}
 [\beta_{(F,G)_{\tcS_0}}+1]
 \lrB{
  \prod_{s\in\tcS_0}|\im\F_s||\im\G_s|
 }
 2^{-n[\uH(\ZZ_{\tcS_0})-\e]}
}
+2\mu_{Z_{\cS_0}}(\uT_0^{\complement}),
\label{eq:channel-ave-error-1}
\end{align}
where the first inequality comes from the triangular inequality,
and the second inequality comes from
Lemma~\ref{lem:mBCP} in Appendix~\ref{sec:hash}
for joint ensembles $\{(\F_s\times\G_s,p_{(F_s,G_s)})\}_{s\in\cS_0}$
by letting
\begin{align*}
\T
&\equiv
\uT_0
\\
Q
&\equiv
\mu_{Z_{\cS_0}}
\end{align*}
and using the relations
\begin{align}
\T_{\tcS_0}
&\subset
\lrb{
 \zz_{\tcS_0}:
 \frac 1n\log_2\frac1{\mu_{Z_{\tcS_0}}(\zz_{\tcS_0})}
 \geq \uH(\ZZ_{\tcS_0})-\e
}
\notag
\\
\oQ_{\tcS_0^{\complement}}
&=
\max_{\zz_{\tcS_0}\in\T_{\tcS_0}}
\sum_{
 \zz_{\tcS_0^{\complement}}\in\T_{\tcS_0^{\complement}}
}
\mu_{Z_{\tcS_0}Z_{\tcS_0^{\complement}}}
(\zz_{\tcS_0},\zz_{\tcS_0^{\complement}})
\notag
\\
&\leq
\max_{\zz_{\tcS_0}\in\T_{\tcS_0}}\mu_{Z_{\tcS_0}}(\zz_{\tcS_0})
\notag
\\
&\leq
2^{-n[\uH(\ZZ_{\tcS_0})-\e]}.
\end{align}

Let $\uT_{\I}$ be defined as
\begin{align*}
\uT_{\I}
&\equiv
\lrb{
 (\zz_{\cS_0},\zz_{\cS_{\I}}):
 \frac 1n
 \log_2\frac1{\mu_{Z_{\tcS_{\I}}|Z_{\cS_0}}(\zz_{\tcS_{\I}}|\zz_{\cS_0})}
 \geq
 \uH(\ZZ_{\tcS_{\I}}|\ZZ_{\cS_0})-\e
 \ \text{for all}\ \tcS_{\I}\subset\cS_{\I}
}.
\end{align*}
Let $\uT_{\I}(\zz_{\cS_0})$ be defined as
\begin{equation*}
\uT_{\I}(\zz_{\cS_0})
\equiv
\lrb{
 \zz_{\cS_{\I}}:
 (\zz_{\cS_0},\zz_{\cS_{\I}})\in\uT_{\I}
}.
\end{equation*}
Then the second term
on the right hand side of (\ref{eq:channel-ave-error})
is evaluated as
\begin{align}
&
\text{[the second term of (\ref{eq:channel-ave-error})]}
\notag
\\*
&=
E_{(F,G)_{\cS}}\lrB{
 \sum_{
  \cc_{\cS_0}\in\C_{\cS_0},
  \cc_{\cS_{\I}}\in\C_{\cS_{\I}},
  \zz_{\cS_0}\in\fC_{F_{\cS_0}}(\cc_{\cS_0}),
  \mm_{\cS_{\I}}\in\M_{\cS_{\I}}
 }
 \mu_{Z_{\cS_0}}(\zz_{\cS_0})
 \lrbar{
  \frac1{
   |\C_{\cS_{\I}}|
   |\M_{\cS_{\I}}|
  }
  -
  \mu_{Z_{\cS_{\I}}|Z_{\cS_0}}(
   \fC_{(F,G)_{\cS_{\I}}}(\cc_{\cS_{\I}},\mm_{\cS_{\I}})
   |\zz_{\cS_0})
 }
}
\notag
\\
&=
E_{(F,G)_{\cS_{\I}}}\lrB{
 \sum_{
  \zz_{\cS_0}\in\Z^n_{\cS_0},
  \cc_{\cS_{\I}}\in\C_{\cS_{\I}},
  \mm_{\cS_{\I}}\in\M_{\cS_{\I}}
 }
 \mu_{Z_{\cS_0}}(\zz_{\cS_0})
 \lrbar{
  \frac1{
   |\C_{\cS_{\I}}|
   |\M_{\cS_{\I}}|
  }
  -
  \mu_{Z_{\cS_{\I}}|Z_{\cS_0}}(
   \fC_{(F,G)_{\cS_{\I}}}(\cc_{\cS_{\I}},\mm_{\cS_{\I}})
   |\zz_{\cS_0})
 }
}
\notag
\\
&\leq
E_{(F,G)_{\cS_{\I}}}
\lrB{
 \sum_{\substack{
   \zz_{\cS_0}\in\Z^n_{\cS_0},
   \\
   \cc_{\cS_{\I}}\in\C_{\cS_{\I}},
   \mm_{\cS_{\I}}\in\M_{\cS_{\I}}
 }}
 \mu_{Z_{\cS_0}}(\zz_{\cS_0})
 \lrbar{
  \mu_{Z_{\cS_{\I}}|Z_{\cS_0}}(
   \uT_{\I}(\zz_{\cS_0})
   \cap\fC_{(F,G)_{\cS_{\I}}}(\cc_{\cS_{\I}},\mm_{\cS_{\I}})
   |\zz_{\cS_0}
  )
  -
  \frac{
   \mu_{Z_{\cS_{\I}}|Z_{\cS_0}}(
    \uT_{\I}(Z_{\cS_0})
    |\zz_{\cS_0}
   )
  }{
   |\C_{\cS_{\I}}|
   |\M_{\cS_{\I}}|
  }
 }
}
\notag
\\*
&\quad
+
E_{(F,G)_{\cS_{\I}}}
\lrB{
 \sum_{
  \zz_{\cS_0}\in\Z^n_{\cS_0},
  \cc_{\cS_{\I}}\in\C_{\cS_{\I}},
  \mm_{\cS_{\I}}\in\M_{\cS_{\I}}
 }
 \mu_{Z_{\cS_{\I}}|Z_{\cS_0}}(
  \uT_{\I}(\zz_{\cS_0})^{\complement}
  \cap\fC_{(F,G)_{\cS_{\I}}}(\cc_{\cS_{\I}},\mm_{\cS_{\I}})
  |\zz_{\cS_0}
 )
 \mu_{Z_{\cS_0}}(\zz_{\cS_0})
}
\notag
\\*
&\quad
+
E_{(F,G)_{\cS_{\I}}}
\lrB{
 \sum_{
  \zz_{\cS_0}\in\Z^n_{\cS_0},
  \cc_{\cS_{\I}}\in\C_{\cS_{\I}},
  \mm_{\cS_{\I}}\in\M_{\cS_{\I}}
 }
 \frac{
  \mu_{Z_{\cS_{\I}}|Z_{\cS_0}}(
   \uT_{\I}(\zz_{\cS_0})^{\complement}
   |\zz_{\cS_0}
  )
  \mu_{Z_{\cS_0}}(\zz_{\cS_0})
 }{
  |\C_{\cS_{\I}}|
  |\M_{\cS_{\I}}|
 }
}
\notag
\\
&=
\sum_{
 \zz_{\cS_0}\in\Z_{\cS_0}
}
\mu_{Z_{\cS_{\I}}|Z_{\cS_0}}(
 \uT_{\I}(\zz_{\cS_0})
 |\zz_{\cS_0}
)
\mu_{Z_{\cS_0}}(\zz_{\cS_0})
\notag
\\*
&\qquad\cdot
E_{(F,G)_{\cS_{\I}}}\lrB{
 \sum_{
  \cc_{\cS_{\I}}\in\C_{\cS_{\I}},
  \mm_{\cS_{\I}}\in\M_{\cS_{\I}}
 }
 \lrbar{
  \frac{
   \mu_{Z_{\cS_{\I}}|Z_{\cS_0}}(
    \uT_{\I}(\zz_{\cS_0})
    \cap\fC_{(F,G)_{\cS_{\I}}}(\cc_{\cS_{\I}},\mm_{\cS_{\I}})
    |\zz_{\cS_0}
   )
  }{
   \mu_{Z_{\cS_{\I}}|Z_{\cS_0}}(
    \uT_{\I}(\zz_{\cS_0})
    |\zz_{\cS_0}
   )
  }
  -
  \frac1
  {\prod_{s\in\cS_{\I}}|\im\F_s||\im\G_s|}
 }
}
\notag
\\*
&\quad
+
2\sum_{
 \zz_{\cS_0}\in\Z_{\cS_0}
}
\mu_{Z_{\cS_{\I}}|Z_{\cS_0}}(
 \uT_{\I}(\zz_{\cS_0})^{\complement}
 |\zz_{\cS_0}
)
\mu_{Z_{\cS_0}}(\zz_{\cS_0})
\notag
\\
&\leq
\sum_{
 \zz_{\cS_0}\in\Z_{\cS_0}
}
\mu_{Z_{\cS_{\I}}|Z_{\cS_0}}(
 \uT_{\I}(\zz_{\cS_0})
 |\zz_{\cS_0}
)
\mu_{Z_{\cS_0}}(\zz_{\cS_0})
\notag
\\*
&\qquad\cdot
\sqrt{
 \alpha_{(F,G)_{\cS_{\I}}}-1
 +
 \sum_{
  \tcS_{\I}\subset\cS_{\I}:
  \tcS_{\I}\neq\emptyset
 }
 \alpha_{(F,G)_{\tcS_{\I}^{\complement}}}
 [\beta_{(F,G)_{\tcS_{\I}}}+1]
 \lrB{
  \prod_{s\in\tcS_{\I}}|\im\F_s||\im\G_s|
 }
 \frac{
  2^{-n[\uH(\ZZ_{\tcS_{\I}}|\ZZ_{\cS_0})-\e]}
 }{
  \mu_{Z_{\cS_{\I}}|Z_{\cS_0}}(
   \uT_{\I}(\zz_{\cS_0})
   |\zz_{\cS_0}
  )
 }
}
\notag
\\*
&\quad
+
2\mu_{Z_{\cS}}(\uT_{\I}^{\complement})
\notag
\\
&\leq
\sum_{
 \zz_{\cS_0}\in\Z_{\cS_0}
}
\mu_{Z_{\cS_0}}(\zz_{\cS_0})
\sqrt{
 \alpha_{(F,G)_{\cS_{\I}}}-1
 +
 \sum_{
  \tcS_{\I}\subset\cS_{\I}:
  \tcS_{\I}\neq\emptyset
 }
 \alpha_{(F,G)_{\cS_{\I}\setminus\tcS_{\I}}}
 [\beta_{(F,G)_{\tcS_{\I}}}+1]
 \lrB{
  \prod_{s\in\tcS_{\I}}|\im\F_s||\im\G_s|
 }
 2^{-n[\uH(\ZZ_{\tcS_{\I}}|\ZZ_{\cS_0})-\e]}
}
\notag
\\*
&\quad
+
2\mu_{Z_{\cS}}(\uT_{\I}^{\complement})
\notag
\\
&=
\sqrt{
 \textstyle
 \alpha_{(F,G)_{\cS_{\I}}}-1
 +\sum_{
  \tcS_{\I}\subset\cS_{\I}:
  \tcS_{\I}\neq\emptyset
 }
 \alpha_{(F,G)_{\cS_{\I}\setminus\tcS_{\I}}}
 [\beta_{(F,G)_{\tcS_{\I}}}+1]
 \lrB{
  \prod_{s\in\tcS_{\I}}|\im\F_s||\im\G_s|
 }
}
+
2\mu_{Z_{\cS}}(\uT_{\I}^{\complement}),
\label{eq:channel-ave-error-2}
\end{align}
where the first inequality comes from the triangular inequality,
and the second inequality comes from
Lemma~\ref{lem:mBCP} in Appendix~\ref{sec:hash}
for joint ensembles $\{(\F_s\times\G_s,p_{(F_s,G_s)})\}_{s\in\cS_{\I}}$
by letting
\begin{align*}
\T
&\equiv
\uT_{\I}(\zz_{\cS_0})
\\
Q
&\equiv
\mu_{Z_{\cS_{\I}}|Z_{\cS_0}}(\cdot|\zz_{\cS_0})
\end{align*}
and using the relations
\begin{align}
\T_{\tcS_{\I}}
&\subset
\lrb{
 \zz_{\tcS_{\I}}:
 \frac 1n
 \log_2\frac1{\mu_{Z_{\tcS_{\I}}|Z_{\cS_0}}(\zz_{\tcS_{\I}}|\zz_{\cS_0})}
 \geq
 \uH(\ZZ_{\tcS_{\I}}|\ZZ_{\cS_0})-\e
}
\notag
\\
\oQ_{\tcS_{\I}^{\complement}}
&=
\max_{\zz_{\tcS_{\I}}\in\T_{\tcS_{\I}}}
\sum_{
 \zz_{\tcS_{\I}^{\complement}}
 \in\T_{\tcS_{\I}^{\complement}|\tcS_{\I}}(\zz_{\tcS_{\I}})
}
\mu_{Z_{\tcS_{\I}}Z_{\tcS_{\I}^{\complement}}|Z_{\cS_0}}
(\zz_{\tcS_{\I}},\zz_{\tcS_{\I}^{\complement}}|\zz_{\cS_0})
\notag
\\
&\leq
\max_{\zz_{\tcS_{\I}}\in\T_{\tcS_{\I}}}
\mu_{Z_{\tcS_{\I}}|Z_{\cS_0}}(\zz_{\tcS_{\I}}|\zz_{\cS_0})
\notag
\\
&\leq
2^{-n[\uH(\ZZ_{\tcS_{\I}}|\ZZ_{\cS_0})-\e]}.
\end{align}

The third term
on the right hand side of (\ref{eq:channel-ave-error})
is evaluated as
\begin{align}
\lrB{\text{the third term of (\ref{eq:channel-ave-error})}}
&=
E_{(F,G)_{\cS}}\lrB{
 \sum_{\substack{
   \cc_{\cS}\in\C_{\cS},
   \mm_{\cS}\in\M_{\cS},
   \zz_{\cS}\in\fC_{(F,G)_{\cS}}(\cc_{\cS},\mm_{\cS}),
   \\
   \yy_{\J}\in\Y^n_{\J},
   \hzz_{\D_{\J}}\in\E(G_{\D_{\J}},\mm_{\D_{\J}})
 }}
 \mu_{Z_{\cS}Y_{\J}}(\zz_{\cS},\yy_{\J})
 \mu_{\hZ_{\D_{\J}}|\cc_{\D_{\J}}Y_{\J}}
 (\hzz_{\D_{\J}}|\cc_{\D_{\J}},\yy_{\J})
}
\notag
\\
&\leq
\sum_{j\in\J}
E_{(F,G)_{\cS}}\lrB{
 \sum_{\substack{
   \cc_{\cS}\in\C_{\cS},
   \mm_{\cS}\in\M_{\cS},
   \zz_{\cS}\in\fC_{(F,G)_{\cS}}(\cc_{\cS},\mm_{\cS}),
   \\
   \yy_{\J}\in\Y^n_{\J},
   \hzz_{\D_{\J}}\in\Z^n_{\D_{\J}}:
   \\
   \hzz_{\D_j}\neq \zz_{\D_j}
 }}
 \mu_{Z_{\cS}Y_{\J}}(\zz_{\cS},\yy_{\J})
 \prod_{j\in\J}
 \mu_{\hZ_{\D_j}|Y_jC_{\D_j}}
 (\hzz_{\D_j}|\yy_j,\cc_{\D_j})
}
\notag
\\
&=
\sum_{j\in\J}
E_{F_{\D_j}}\lrB{
 \sum_{\substack{
   \cc_{\D_j}\in\C_{\D_j},
   \zz_{\D_j}\in\fC_{F_{\D_j}}(\cc_{\D_j}),
   \\
   \yy_j\in\Y^n_j,
   \hzz_{\D_j}\in\Z^n_{\D_j}:
   \\
   \hzz_{\D_j}\neq \zz_{\D_j}
 }}
 \mu_{Z_{\D_j}Y_j}(\zz_{\D_j},\yy_j)
 \mu_{\hZ_{\D_j}|Y_jC_{\D_j}}
 (\hzz_{\D_j}|\yy_j,\cc_{\D_j})
}
\notag
\\
&=
\sum_{j\in\J}
E_{F_{\D_j}}\lrB{
 \mu_{Z^n_{\D_j}\hZ^n_{\D_j}}\lrsb{\lrb{
   (\zz_{\D_j},\hzz_{\D_j}):
   \hzz_{\D_j}\neq\zz_{\D_j}
  }
 }
},
\label{eq:channel-ave-error-3}
\end{align}
where
$\hzz_{\D_j}\equiv\{\hzz_{j,s}\}_{s\in\D_j}$,
and the inequality 
comes from the union bound and the fact that
$g(\hzz_{j,s})\neq\mm_s$ and $s\in\D_j$ implies $\hzz_{\D_j}\neq\zz_{\D_j}$,
and the probability
$\mu_{Z^n_{\D_j}\hZ^n_{\D_j}}\lrsb{\lrb{
 (\zz_{\D_j},\hzz_{\D_j}):
 \hzz_{\D_j}\neq\zz_{\D_j}
}}$
depends on $F_{\D_j}$ through the joint distribution
defined by (\ref{eq:source-joint})\footnote{It should be noted that
random variable $C_{\cS}$ that appears in (\ref{eq:source-joint})
should be denoted by a different symbol from that subject to the uniform
distribution on $\C_{\cS}$.}
in the last inequality.

From
(\ref{eq:channel-ave-error}),
(\ref{eq:channel-ave-error-1}),
(\ref{eq:channel-ave-error-2}),
(\ref{eq:channel-ave-error-3}),
Lemma \ref{lem:source},
and the relations
$r_s=\log_2(|\C_s|)=\log_2(|\im\F_s|)/n$
and $R_s=\log_2(|\M_s|)/n=\log_2(|\im\G_s|)/n$,
we have
\begin{align}
&
E_{(F,G)_{\cS}C_{\cS}}\lrB{\Error(F_{\cS},G_{\cS},C_{\cS})}
\notag
\\*
&\leq
\sqrt{
 \textstyle
 \alpha_{(F,G)_{\cS_0}}-1
 +\sum_{
  \tcS_0\subset\cS_0:
  \tcS_0\neq\emptyset
 }
 \alpha_{(F,G)_{\cS_0\setminus\tcS_0}}
 [\beta_{(F,G)_{\tcS_0}}+1]
 2^{-n\gamma(\tcS_0)}
}
+2\mu_{Z_{\cS_0}}(\uT_0^{\complement})
\notag
\\*
&\quad
+
\sqrt{
 \textstyle
 \alpha_{(F,G)_{\cS_{\I}}}-1
 +\sum_{
  \tcS_{\I}\subset\cS_{\I}:
  \tcS_{\I}\neq\emptyset
 }
 \alpha_{(F,G)_{\cS_{\I}\setminus\tcS_{\I}}}
 [\beta_{(F,G)_{\tcS_{\I}}}+1]
 2^{-n\gamma(\tcS_{\I})}
}
+2\mu_{Z_{\cS}}(\uT_{\I}^{\complement})
\notag
\\*
&\quad
+2\sum_{j\in\J}
\sum_{
 \tcD_j\subset\D_j:
 \tcD_j\neq\emptyset
}
\alpha_{F_{\tcD_j}}\lrB{\beta_{F_{\D_j\setminus\tcD_j}}+1}
2^{
 -n\gamma(\tcD_j)
}
+2 \sum_{j\in\J}
\beta_{F_{\D_j}}
+2 \sum_{j\in\J}
\mu_{Z_{\D_j}Y_j}(\oT_j^{\complement}),
\end{align}
where
\begin{align*}
\gamma(\tcS_0)
&\equiv
\uH(\ZZ_{\tcS_0})-\sum_{s\in\tcS_0}\lrB{r_s+R_s}-\e
\\
\gamma(\tcS_{\I})
&\equiv
\uH(\ZZ_{\tcS_{\I}}|\ZZ_{\cS_0})-\sum_{s\in\tcS_{\I}}[r_s+R_s]-\e
\\
\gamma(\tcD_j)
&\equiv
\sum_{s\in\tcD_j}r_s-\oH(\ZZ_{\tcD_j}|\YY_j,\ZZ_{\D_j\setminus\tcD_j})-\e.
\end{align*}

Finally, let us assume that
$\{(r_s,R_s)\}_{s\in\cS}$ satisfies
(\ref{eq:channel-Rs+rs})--(\ref{eq:channel-rs})
for all $\tcS_0$ satisfying $\emptyset\neq\tcS_0\subset\cS_0$,
all $(i,\tcS_i)$ satisfying $i\in\I$ and $\emptyset\neq\tcS_i\subset\cS_i$,
and for all $(j,\tcD_j)$ satisfying $j\in\J$ $\emptyset\neq\tcD_j\subset\D_j$.
We have
\begin{align}
\sum_{s\in\tcS_0}\lrB{r_s+R_s}
&<
\uH(\ZZ_{\tcS_0})
\\
\sum_{s\in\tcS_{\I}}\lrB{r_s+R_s}
&=
\sum_{i\in\I}\sum_{s\in\tcS_{\I}\cap\cS_i}\lrB{r_s+R_s}
\notag
\\
&<
\sum_{i\in\I}\uH(\ZZ_{\tcS_{\I}\cap\cS_i}|\ZZ_{\cS_0})
\notag
\\
&\leq
\uH(\ZZ_{\tcS_{\I}}|\ZZ_{\cS_0})
\label{eq:channel-condition-bcp}
\\
\sum_{s\in\tcD_j}r_s
&>
\oH(\ZZ_{\tcD_j}|\YY_j,\ZZ_{\D_j\setminus\tcD_j}).
\end{align} 
where the last inequality of (\ref{eq:channel-condition-bcp})
comes from Lemma \ref{lem:ispec-markov} in Appendix~\ref{sec:ispec}.
Then, by letting
$\alpha_{F_s}\to1$, $\beta_{F_s}\to0$,
$\alpha_{G_s}\to1$, $\beta_{G_s}\to0$,
$\mu_{Z_{\cS_0}}(\uT_0^{\complement})\to0$,
$\mu_{Z_{\cS}}(\uT_{\I}^{\complement})\to0$,
$\mu_{Z_{\D_j}Y_j}(\oT_j^{\complement})\to0$,
$\e\to0$, and using the random coding argument,
we have the fact that
for all $\delta>0$ and sufficiently large $n$
there are $f_{\cS}=\{f_s\}_{s\in\cS}$, $g_{\cS}=\{g_s\}_{s\in\cS}$,
and $\cc_{\cS}=\{\cc_s\}_{s\in\cS}$
such that $\Error(f_{\cS},g_{\cS},\cc_{\cS})\leq\delta$.
\hfill\IEEEQED

\section{Proof of Theorem \ref{thm:crng-simplified}}

This section outlines the proof of
Theorem \ref{thm:crng-simplified},
where the details are given in
Appendices \ref{sec:crng-fme}--\ref{sec:proof-crng-crng-case1-appendix}.

From here, we use the rule of reference number.
When the reference number has an ornament `\ ${}'$\ ',
it means that corresponding relation
is applied after exchanging $i$ and $i'$.
For example, `(\ref{eq:crng-case3})$'$' means
\begin{equation*}
R_{i'}
\geq
I(V_{i'i'};Y_{i'}|V_{00},V_{10},V_{20})
+ I(V_{i'0};Y_i|V_{00},V_{i0},V_{ii}).
\end{equation*}
This rule is applied to the following sections and the appendices.

By letting
\begin{align*}
R_{00}
&\equiv R_0
\\
R_{i0}
&\equiv 0
\\
R_{ii}
&\equiv R_i,
\end{align*}
for each $i\in\{1,2\}$,
we have
\begin{align}
R_0 + r_{00}
&\leq 
H(Z_{00})
\label{eq:crng-region0-first}
\\
r_{i0}
&\leq
H(Z_{i0}|Z_{00})
\\
R_i + r_{ii}
&\leq
H(Z_{ii}|Z_{00})
\\
r_{i0} + R_i + r_{ii}
&\leq
H(Z_{i0},Z_{ii}|Z_{00})
\\
r_{00}
&\geq
H(Z_{00}|Y_i,Z_{10},Z_{20},Z_{ii})
\\
r_{i0}
&\geq
H(Z_{i0}|Y_i,Z_{00},Z_{i'0},Z_{ii})
\\
r_{i'0}
&\geq
H(Z_{i'0}|Y_i,Z_{00},Z_{i0},Z_{ii})
\\
r_{ii}
&\geq
H(Z_{ii}|Y_i,Z_{00},Z_{10},Z_{20})
\\
r_{00} + r_{i0}
&\geq
H(Z_{00},Z_{i0}|Y_i,Z_{i'0},Z_{ii})
\\
r_{00} + r_{i'0}
&\geq
H(Z_{00},Z_{i'0}|Y_i,Z_{i0},Z_{ii})
\\
r_{00} + r_{ii}
&\geq
H(Z_{00},Z_{ii}|Y_i,Z_{10},Z_{20})
\\
r_{10} + r_{20}
&\geq
H(Z_{10},Z_{20}|Y_i,Z_{00},Z_{ii})
\\
r_{i0} + r_{ii}
&\geq
H(Z_{i0},Z_{ii}|Y_i,Z_{00},Z_{i'0})
\\
r_{i'0} + r_{ii}
&\geq
H(Z_{i'0},Z_{ii}|Y_i,Z_{00},Z_{i0})
\\
r_{00} + r_{10} + r_{20}
&\geq
H(Z_{00},Z_{10},Z_{20}|Y_i,Z_{ii})
\\
r_{00} + r_{i0} + r_{ii}
&\geq
H(Z_{00},Z_{i0},Z_{ii}|Y_i,Z_{i'0})
\\
r_{00} + r_{i'0} + r_{ii}
&\geq
H(Z_{00},Z_{i'0},Z_{ii}|Y_i,Z_{i0})
\\
r_{10} + r_{20} + r_{ii}
&\geq
H(Z_{10},Z_{20},Z_{ii}|Y_i,Z_{00})
\\
r_{00} + r_{10} + r_{20} + r_{ii}
&\geq
H(Z_{00},Z_{10},Z_{20},Z_{ii}|Y_i)
\label{eq:crng-region0-last}
\end{align}
for all $i\in\{0,1\}$
by expanding (\ref{eq:crng-R0+r0})--(\ref{eq:crng-sum-rj}).
Then we show the following lemma.

\begin{lem}
\label{lem:crng-fme}
By applying the Fourier-Motzkin method \cite[Appendix D]{EK11} to eliminate
variables $r_{00}$, $r_{11}$, $r_{22}$, $r_{10}$, $r_{20}$
from (\ref{eq:crng-region0-first})--(\ref{eq:crng-region0-last})
for all $i\in\{0,1\}$,
we have
\begin{align}
R_0
&\leq  
I(Z_{00};Y_i,Z_{10},Z_{20},Z_{ii})
\label{eq:crng-fme0}
\\
R_0
&\leq
I(Z_{00},Z_{i0};Y_i,Z_{i'0},Z_{ii})
\label{eq:crng-fme1}
\\
R_0
&\leq
I(Z_{00},Z_{i'0};Y_i,Z_{i0},Z_{ii})
\label{eq:crng-fme2}
\\
R_0
&\leq
I(Z_{00},Z_{10},Z_{20};Y_i,Z_{ii})
\label{eq:crng-fme3}
\\
R_i
&\leq
I(Z_{ii};Y_i,Z_{10},Z_{20}|Z_{00})
\label{eq:crng-fme4}
\\
R_i
&\leq
I(Z_{i0},Z_{ii};Y_i,Z_{i'0}|Z_{00})
\label{eq:crng-fme5}
\\
R_i
&\leq
I(Z_{ii};Y_i,Z_{10},Z_{20}|Z_{00})
+ I(Z_{i0};Y_{i'},Z_{i'0},Z_{i'i'}|Z_{00})
- I(Z_{i0};Z_{ii}|Z_{00})
\label{eq:crng-fme6}
\\
R_0 + R_i
&\leq
I(Z_{00},Z_{ii};Y_i,Z_{10},Z_{20})
\label{eq:crng-fme7}
\\
R_0 + R_i
&\leq
I(Z_{00},Z_{i0},Z_{ii};Y_i,Z_{i'0})
\label{eq:crng-fme8}
\\
R_0 + R_i
&\leq
I(Z_{00},Z_{i'0},Z_{ii};Y_i,Z_{i0})
\label{eq:crng-fme9}
\\
R_0 + R_i
&\leq
I(Z_{00},Z_{10},Z_{20},Z_{ii};Y_i)
\label{eq:crng-fme10}
\\
R_0 + R_i
&\leq
I(Z_{ii};Y_i,Z_{10},Z_{20}|Z_{00})
+ I(Z_{00},Z_{i0};Y_{i'},Z_{i'0},Z_{i'i'})
- I(Z_{i0};Z_{ii}|Z_{00})
\label{eq:crng-fme11}
\\
R_0 + R_i
&\leq
I(Z_{ii};Y_i,Z_{10},Z_{20}|Z_{00})
+ I(Z_{00},Z_{10},Z_{20};Y_{i'},Z_{i'i'})
- I(Z_{i0};Z_{ii}|Z_{00})
\label{eq:crng-fme12}
\\
R_0 + R_i
&\leq
I(Z_{00},Z_{ii};Y_i,Z_{10},Z_{20})
+ I(Z_{i0};Y_{i'},Z_{i'0},Z_{i'i'}|Z_{00})
- I(Z_{i0};Z_{ii}|Z_{00})
\label{eq:crng-fme13}
\\
R_0 + R_i
&\leq
I(Z_{00},Z_{i'0},Z_{ii};Y_i,Z_{i0})
+ I(Z_{i0};Y_{i'},Z_{i'0},Z_{i'i'}|Z_{00})
- I(Z_{i0};Z_{ii}|Z_{00})
\label{eq:crng-fme14}
\\
R_1 + R_2
&\leq
I(Z_{ii};Y_i,Z_{10},Z_{20}|Z_{00})
+ I(Z_{i0},Z_{i'i'};Y_{i'},Z_{i'0}|Z_{00})
- I(Z_{i0};Z_{ii}|Z_{00})
\label{eq:crng-fme15}
\\
R_1 + R_2
&\leq
I(Z_{ii};Y_i,Z_{10},Z_{20}|Z_{00})
+ I(Z_{10},Z_{20},Z_{i'i'};Y_{i'}|Z_{00})
- I(Z_{i0};Z_{ii}|Z_{00})
\label{eq:crng-fme16}
\\
R_1 + R_2
&\leq
I(Z_{20},Z_{11};Y_1,Z_{10}|Z_{00})
+ I(Z_{10},Z_{22};Y_2,Z_{20}|Z_{00})
- I(Z_{10};Z_{11}|Z_{00}) - I(Z_{20};Z_{22}|Z_{00})
\label{eq:crng-fme17}
\\
2R_0 + R_i
&\leq
I(Z_{00},Z_{ii};Y_i,Z_{10},Z_{20})
+ I(Z_{00},Z_{i0};Y_{i'},Z_{i'0},Z_{i'i'})
-I(Z_{i0};Z_{ii}|Z_{00})
\label{eq:crng-fme18}
\\
2R_0 + R_i
&\leq
I(Z_{00},Z_{ii};Y_i,Z_{10},Z_{20})
+ I(Z_{00},Z_{10},Z_{20};Y_{i'},Z_{i'i'})
- I(Z_{i0};Z_{ii}|Z_{00})
\label{eq:crng-fme19}
\\
2R_0 + R_i
&\leq
I(Z_{00},Z_{i'0},Z_{ii};Y_i,Z_{i0})
+ I(Z_{00},Z_{i0};Y_{i'},Z_{i'0},Z_{i'i'})
- I(Z_{i0};Z_{ii}|Z_{00})
\label{eq:crng-fme20}
\\
2R_0 + R_i
&\leq
I(Z_{00},Z_{i'0},Z_{ii};Y_i,Z_{i0})
+ I(Z_{00},Z_{10},Z_{20};Y_{i'},Z_{i'i'})
- I(Z_{i0};Z_{ii}|Z_{00})
\label{eq:crng-fme21}
\\
2 R_i + R_{i'}
&\leq
I(Z_{10},Z_{20},Z_{ii};Y_i|Z_{00})
+ I(Z_{ii};Y_i,Z_{10},Z_{20}|Z_{00})
+ I(Z_{i0},Z_{i'i'};Y_{i'},Z_{i'0}|Z_{00}) 
\notag
\\*
&\quad
- I(Z_{10};Z_{11}|Z_{00})
- I(Z_{20};Z_{22}|Z_{00})
\label{eq:crng-fme22}
\\
R_0 + R_1 + R_2
&\leq
I(Z_{ii};Y_i,Z_{10},Z_{20}|Z_{00})
+ I(Z_{00},Z_{i0},Z_{i'i'};Y_{i'},Z_{i'0})
- I(Z_{i0};Z_{ii}|Z_{00})
\label{eq:crng-fme23}
\\
R_0 + R_1 + R_2
&\leq
I(Z_{ii};Y_i,Z_{10},Z_{20}|Z_{00})
+ I(Z_{00},Z_{10},Z_{20},Z_{i'i'};Y_{i'})
- I(Z_{i0};Z_{ii}|Z_{00})
\label{eq:crng-fme24}
\\
R_0 + R_1 + R_2
&\leq
I(Z_{00},Z_{ii};Y_i,Z_{10},Z_{20})
+ I(Z_{i0},Z_{i'i'};Y_{i'},Z_{i'0}|Z_{00})
- I(Z_{i0};Z_{ii}|Z_{00})
\label{eq:crng-fme25}
\\
R_0 + R_1 + R_2
&\leq
I(Z_{00},Z_{ii};Y_i,Z_{10},Z_{20})
+ I(Z_{10},Z_{20},Z_{i'i'};Y_{i'}|Z_{00})
- I(Z_{i0};Z_{ii}|Z_{00})
\label{eq:crng-fme26}
\\
R_0 + R_1 + R_2
&\leq
I(Z_{00},Z_{i'0},Z_{ii};Y_i,Z_{i0})
+ I(Z_{i0},Z_{i'i'};Y_{i'},Z_{i'0}|Z_{00})
- I(Z_{10};Z_{11}|Z_{00})
- I(Z_{20};Z_{22}|Z_{00})
\label{eq:crng-fme27}
\\
2R_0 + R_1 + R_2
&\leq
I(Z_{00},Z_{ii};Y_i,Z_{10},Z_{20})
+ I(Z_{00},Z_{i0},Z_{i'i'};Y_{i'},Z_{i'0})
- I(Z_{i0};Z_{ii}|Z_{00})
\label{eq:crng-fme28}
\\
2R_0 + R_1 + R_2
&\leq
I(Z_{00},Z_{ii};Y_i,Z_{10},Z_{20})
+ I(Z_{00},Z_{10},Z_{20},Z_{i'i'};Y_{i'})
- I(Z_{i0};Z_{ii}|Z_{00})
\label{eq:crng-fme29}
\\
2R_0 + R_1 + R_2
&\leq
I(Z_{00},Z_{i'0},Z_{ii};Y_i,Z_{i0})
+ I(Z_{00},Z_{10},Z_{20},Z_{i'i'};Y_{i'})
- I(Z_{i0};Z_{ii}|Z_{00})
\label{eq:crng-fme30}
\\
2R_0 + R_1 + R_2
&\leq
I(Z_{00},Z_{20},Z_{11};Y_1,Z_{10})
+ I(Z_{00},Z_{10},Z_{22};Y_2,Z_{20})
- I(Z_{10};Z_{11}|Z_{00}) - I(Z_{20};Z_{22}|Z_{00})
\label{eq:crng-fme31}
\\
R_0 + 2R_i + R_{i'}
&\leq
I(Z_{10},Z_{20},Z_{ii};Y_i|Z_{00})
+ I(Z_{ii};Y_i,Z_{10},Z_{20}|Z_{00})
+ I(Z_{00},Z_{i0},Z_{i'i'};Y_{i'},Z_{i'0})
\notag
\\*
&\quad
- I(Z_{10};Z_{11}|Z_{00})
- I(Z_{20};Z_{22}|Z_{00})
\label{eq:crng-fme32}
\\
R_0 + 2R_i + R_{i'}
&\leq
I(Z_{00},Z_{10},Z_{20},Z_{ii};Y_i)
+ I(Z_{ii};Y_i,Z_{10},Z_{20}|Z_{00})
+ I(Z_{i0},Z_{i'i'};Y_{i'},Z_{i'0}|Z_{00}) 
\notag
\\*
&\quad
- I(Z_{10};Z_{11}|Z_{00})
- I(Z_{20};Z_{22}|Z_{00})
\label{eq:crng-fme33}
\\
2R_0 + 2R_i + R_{i'}
&\leq
I(Z_{00},Z_{10},Z_{20},Z_{ii};Y_i)
+ I(Z_{ii};Y_i,Z_{10},Z_{20}|Z_{00})
+ I(Z_{00},Z_{i0},Z_{i'i'};Y_{i'},Z_{i'0})
\notag
\\*
&\quad
- I(Z_{10};Z_{11}|Z_{00})
- I(Z_{20};Z_{22}|Z_{00})
\label{eq:crng-fme34}
\\
2R_0 + 2R_i + R_{i'}
&\leq
I(Z_{00},Z_{10},Z_{20},Z_{ii};Y_i)
+ I(Z_{00},Z_{ii};Y_i,Z_{10},Z_{20})
+ I(Z_{i0},Z_{i'i'};Y_{i'},Z_{i'0}|Z_{00})
\notag
\\*
&\quad
- I(Z_{10};Z_{11}|Z_{00})
- I(Z_{20};Z_{22}|Z_{00})
\label{eq:crng-fme35}
\\
3R_0 + 2R_i + R_{i'}
&\leq
I(Z_{00},Z_{10},Z_{20},Z_{ii};Y_i)
+ I(Z_{00},Z_{ii};Y_i,Z_{10},Z_{20})
+ I(Z_{00},Z_{i0},Z_{i'i'};Y_{i'},Z_{i'0})
\notag
\\*
&\quad
- I(Z_{10};Z_{11}|Z_{00})
- I(Z_{20};Z_{22}|Z_{00}).
\label{eq:crng-fme36}
\end{align}
\end{lem}

\begin{IEEEproof}
Here, we simply outline the proof, where
details are given in Appendix~\ref{sec:crng-fme}.
First, we apply Fourier-Motzkin method \cite[Appendix D]{EK11}
to eliminate $r_{00}$.
Then, we apply Fourier-Motzkin method again to eliminate $r_{11}$ and $r_{22}$.
Then, we apply Lemma \ref{lem:fme} in Appendix \ref{sec:crng-fme}
to eliminate $r_{10}$
and $r_{20}$, where the Fourier-Motzkin method is
applied twice to obtain the lemma.
Finally,
by using Shannon-type (in)equalities \cite[Chap.\ 14]{Y08}
to eliminate redundant inequalities,
we have (\ref{eq:crng-fme0})--(\ref{eq:crng-fme36}).
\end{IEEEproof}

By letting $V_s\equiv Z_s$ for each $s\in\{00,10,11,20,22\}$,
we have the fact that (\ref{eq:crng-fme0})--(\ref{eq:crng-fme36})
implies (\ref{eq:crng-Ri})--(\ref{eq:crng-R0+2Ri+Ri'}).
This implies the relation $\RCRNG^0(\mu)\subset\tRCRNG^0(\tmu)$.
where the joint distribution $\tmu$ of
$(Z_{00},Z_{10},Z_{11},X_1,Z_{20},Z_{22},X_2)$
satisfies (\ref{eq:joint-crng-full}) from (\ref{eq:joint-crng}).
Hence, we have the fact that $\RCRNG^0\subset\tRCRNG^0$.

To complete the proof of the theorem,
we show the relation $\RCRNG^0(\mu)\supset\tRCRNG^0(\tmu)$,
which implies $\RCRNG^0\supset\tRCRNG^0$.
We can shown relations $\tRCRNG^{00}(\tmu)=\RCRNG^{00}(\mu)$
and $\tRCRNG^{00}=\RCRNG^{00}$
immediately from the fact that
$(R_1,R_2)\in\tRCRNG^{00}(\tmu)$,
$(0,R_1,R_2)\in\tRCRNG^0(\tmu)$,
and $(0,0,R_1,0,R_2)\in\tRCRNG(\tmu)$
are equivalent,
and $(R_1,R_2)\in\RCRNG^{00}(\mu)$, $(0,R_1,R_2)\in\RCRNG^0(\mu)$,
and $(0,0,R_1,0,R_2)\in\RCRNG^0(\mu)$
are equivalent.
For given random variables
$(V_{00},V_{10},V_{11},X_1,$ $V_{20},V_{22},X_2,Y_1,Y_2)$
and $(R_0,R_1,R_2)$
satisfying (\ref{eq:crng-Ri})--(\ref{eq:joint-crng-simplified}),
let $Q$ be a random variable satisfying
\begin{gather}
I(Q;V_{00},V_{10},V_{11},X_1,V_{20},V_{22},X_2,Y_1,Y_2)
=
0
\label{eq:Qindependence}
\\
\min_{i\in\{1,2\}}I(V_{00},V_{10},V_{20},V_{ii};Y_i)
\leq
H(Q).
\label{eq:crng-H(Q)'}
\end{gather}
From (\ref{eq:crng-R0+Ri}) and (\ref{eq:crng-H(Q)'}), we have
\begin{align}
R_0
&\leq
R_0 + R_j
\notag
\\
&\leq
\min_{i\in\{1,2\}}I(V_{00},V_{10},V_{20},V_{ii};Y_i)
\notag
\\
&\leq
H(Q).
\label{eq:crng-H(Q)}
\end{align}
Let $(Z_{00},Z_{10},Z_{20},Z_{11},Z_{22})$
be defined as
\begin{align}
Z_{00}
&\equiv
(Q,V_{00})
\label{eq:def-Z00}
\\
Z_{i0}
&\equiv
\begin{cases}
 (Q,V_{i0})
 &\text{if}
 \ R_i
 \leq
 I(V_{ii};Y_i|V_{00},V_{10},V_{20}) + I(V_{i0};Y_{i'}|V_{00},V_{i'0},V_{i'i'})
 \\
 Q
 &\text{if}
 \ R_i
 \geq
 I(V_{ii};Y_i|V_{00},V_{10},V_{20}) + I(V_{i0};Y_{i'}|V_{00},V_{i'0},V_{i'i'})
\end{cases}
\label{eq:def-Zi0}
\\
Z_{ii}
&\equiv
(Q,V_{i0},V_{ii})
\label{eq:def-Zii}
\end{align}
for each $i\in\{1,2\}$.
From (\ref{eq:joint-crng-simplified}), we have
\begin{equation}
I(V_{10},V_{11},X_1;V_{20},V_{22},X_2|V_{00})
=
0.
\label{eq:I(VX1;VX2|V0)=0}
\end{equation}
Then we have the fact that
\begin{align}
I(Z_{10},Z_{11},X_1;Z_{20},Z_{22},X_2|Z_{00})
&=
I(Q,V_{10},V_{11},X_1;Q,V_{20},V_{22},X_2|Q,V_{00})
\notag
\\
&=
I(V_{10},V_{11},X_1;V_{20},V_{22},X_2|V_{00})
\notag
\\
&=
0,
\label{eq:I(ZX1;ZX2|Z0)=0}
\end{align}
where
the second equality coems from
(\ref{eq:Qindependence}) and Lemma \ref{lem:Q} in Appendix \ref{sec:entropy},
and the last inequality comes from (\ref{eq:I(VX1;VX2|V0)=0}).
Then the joint distribution $\tmu$ of
$(Z_{00},Z_{10},Z_{11},X_1,Z_{20},Z_{22},X_2)$
satisfies (\ref{eq:joint-crng-simplified}).

Here, we show the fact that
(\ref{eq:crng-Ri})--(\ref{eq:crng-R0+2Ri+Ri'}),
and (\ref{eq:crng-H(Q)})
imply (\ref{eq:crng-fme0})--(\ref{eq:crng-fme36}),
where this fact implies the relation $\tRCRNG^0(\tmu)\subset\RCRNG^0(\mu)$.
We have (\ref{eq:crng-fme0})--(\ref{eq:crng-fme3})
from (\ref{eq:crng-H(Q)}), (\ref{eq:def-Z00}),
and (\ref{eq:def-Zii}),
where details are presented in Appendix \ref{sec:proof-crng-crng-0-3}.
It is sufficient to show (\ref{eq:crng-fme4})--(\ref{eq:crng-fme36})
by separating the following four cases:
\begin{description}
\item[{\bf Case $0$:}]
\begin{align}
 R_i 
 &\leq
 I(V_{ii};Y_i|V_{00},V_{10},V_{20}) + I(V_{i0};Y_{i'}|V_{00},V_{i'0},V_{i'i'})
 \quad\text{for all}\ i\in\{1,2\}
 \label{eq:crng-case0}
\end{align}
\item[{\bf Case $i\in\{1,2\}$:}]
\begin{align}
 R_i 
 &\leq
 I(V_{ii};Y_i|V_{00},V_{10},V_{20}) + I(V_{i0};Y_{i'}|V_{00},V_{i'0},V_{i'i'})
 \label{eq:crng-casei-i}
 \\
 R_{i'}
 &\geq
 I(V_{i'i'};Y_{i'}|V_{00},V_{10},V_{20}) + I(V_{i'0};Y_i|V_{00},V_{i0},V_{ii})
 \label{eq:crng-casei-io}
\end{align}
\item[{\bf Case $3$:}]
\begin{align}
 R_i
 &\geq
 I(V_{ii};Y_i|V_{00},V_{10},V_{20}) + I(V_{i0};Y_{i'}|V_{00},V_{i'0},V_{i'i'})
 \quad\text{for all}\ i\in\{1,2\}.
 \label{eq:crng-case3}
\end{align}
\end{description}

We have the fact that 
(\ref{eq:crng-R1+R2-1}) and (\ref{eq:crng-case3})$'$
implies (\ref{eq:crng-case0}) as in
\begin{align}
R_i
&=
R_1 + R_2 - R_{i'}
\notag
\\
&\leq
I(V_{i'0},V_{ii};Y_i|V_{00},V_{i0})
+ I(V_{i0},V_{i'i'};Y_{i'}|V_{00},V_{i'0})
- I(V_{i'i'};Y_{i'}|V_{00},V_{10},V_{20})
- I(V_{i'0};Y_i|V_{00},V_{i0},V_{ii})
\notag
\\
&=
I(V_{ii};Y_i|V_{00},V_{i0})
+ I(V_{i0};Y_{i'}|V_{00},V_{i'0})
\notag
\\
&=
H(V_{ii}|V_{00},V_{i0})
- H(V_{ii}|Y_i,V_{00},V_{i0})
+ H(V_{i0}|V_{00},V_{i'0})
- H(V_{i0}|Y_{i'},V_{00},V_{i'0})
\notag
\\
&\leq
H(V_{ii}|V_{00},V_{10},V_{20})
- H(V_{ii}|Y_i,V_{00},V_{10},V_{20})
+ H(V_{i0}|V_{00},V_{i'0},V_{i'i'})
- H(V_{i0}|Y_{i'},V_{00},V_{i'0},V_{i'i'})
\notag
\\
&=
I(V_{ii};Y_i|V_{00},V_{10},V_{20})
+ I(V_{i0};Y_{i'}|V_{00},V_{i'0},V_{i'i'}),
\end{align}
where the second inequality comes from
Lemma \ref{lem:markov} in Appendix \ref{sec:entropy}.
This fact implies that Case 3 is reduced to Case 0.

In the following sections, we focus on Cases 0 and 1,
where Case 2 follows similarly to the proof for Case 1.
Then we have the fact that (\ref{eq:crng-Ri})--(\ref{eq:crng-R0+2Ri+Ri'}),
and (\ref{eq:crng-H(Q)})
imply (\ref{eq:crng-fme0})--(\ref{eq:crng-fme36}).
Hence, we have the relations $\tRCRNG^0(\tmu)\subset\RCRNG^0(\mu)$
and $\tRCRNG^0\subset\RCRNG^0$, which completes the proof of the theorem.
\hfill\IEEEQED

\subsection{Case 0}

Here we present an outline of the proof for Case 0,
where the details are given 
in Appendix \ref{sec:proof-crng-crng-case0-appendix}.
From (\ref{eq:def-Zi0}), we have
\begin{align}
Z_{i0}
&=
(Q,V_{i0})
\label{eq:def-Zi0-case0}
\end{align}
for all $i\in\{1,2\}$.

We have the following (in)equalities
from (\ref{eq:Qindependence}), (\ref{eq:def-Z00}),
(\ref{eq:def-Zii})--(\ref{eq:I(ZX1;ZX2|Z0)=0}),
(\ref{eq:def-Zi0-case0}), 
and Shannon-type (in)equalities:
\begin{align}
I(V_{ii};Y_i|V_{00},V_{10},V_{20})
&=
I(Z_{ii};Y_i,Z_{10},Z_{20}|Z_{00})
- I(Z_{i0};Z_{ii}|Z_{00})
\label{eq:crng-case0-I(V11;Y1|V00,V10,V20)}
\\
H(Q) + I(V_{ii};Y_i|V_{00},V_{10},V_{20})
&\leq
\min\lrb{
 \begin{aligned}
  &I(Z_{00},Z_{ii};Y_i,Z_{10},Z_{20}),
  \\
  &I(Z_{00},Z_{i'0},Z_{ii};Y_i,Z_{i0})
 \end{aligned}
}
- I(Z_{i0};Z_{ii}|Z_{00})
\label{eq:crng-case0-H(Q)+I(V11;Y1|V00,V10,V20)}
\\
I(V_{i0},V_{ii};Y_i|V_{00},V_{i'0})
&\leq
\min\lrb{
 \begin{aligned}
  &I(Z_{ii};Y_i,Z_{10},Z_{20}|Z_{00}),
  \\
  &I(Z_{i0},Z_{ii};Y_i,Z_{i'0}|Z_{00})
 \end{aligned}
}
\label{eq:crng-case0-I(V10,V11;Y1|V00,V20)}
\\
H(Q) + I(V_{i0},V_{ii};Y_i|V_{00},V_{i'0})
&\leq
\min\lrb{
 \begin{aligned}
  &I(Z_{00},Z_{ii};Y_i,Z_{10},Z_{20}),
  \\
  &I(Z_{00},Z_{i0},Z_{ii};Y_i,Z_{i'0}),
  \\
  &I(Z_{00},Z_{i'0},Z_{ii};Y_i,Z_{i0})
 \end{aligned}
}
\label{eq:crng-case0-H(Q)+I(V10,V11;Y1|V00,V20)}
\\
I(V_{i'0},V_{ii};Y_i|V_{00},V_{i0})
&=
I(Z_{i'0},Z_{ii};Y_i,Z_{i0}|Z_{00})
- I(Z_{i0};Z_{ii}|Z_{00})
\label{eq:crng-case0-I(V20,V11;Y1|V00,V10)}
\\
H(Q) + I(V_{i'0},V_{ii};Y_i|V_{00},V_{i0})
&\leq
I(Z_{00},Z_{i'0},Z_{ii};Y_i,Z_{i0})
- I(Z_{i0};Z_{ii}|Z_{00})
\label{eq:crng-case0-H(Q)+I(V20,V11;Y1|V00,V10)}
\\
I(V_{10},V_{20},V_{ii};Y_i|V_{00})
&\leq
\min\lrb{
 \begin{aligned}
  &I(Z_{i'0},Z_{ii};Y_i,Z_{i0}|Z_{00}),
  \\
  &I(Z_{10},Z_{20},Z_{ii};Y_i|Z_{00})
 \end{aligned}
}
\label{eq:crng-case0-I(V10,V20,V11;Y1|V00)}
\\
I(V_{00},V_{10},V_{20},V_{ii};Y_i)
&\leq
\min\lrb{
 \begin{aligned}
  &I(Z_{00},Z_{i'0},Z_{ii};Y_i,Z_{i0}),
  \\
  &I(Z_{00},Z_{10},Z_{20},Z_{ii};Y_i)
 \end{aligned}
}
\label{eq:crng-case0-I(V00,V10,V20,V11;Y1)}
\\
I(V_{i0};Y_{i'}|V_{00},V_{i'0},V_{i'i'})
&\leq
I(Z_{i0};Y_{i'},Z_{i'0},Z_{i'i'}|Z_{00})
\label{eq:crng-case0-I(V10;Y2|V00,V20,V22)}
\\
H(Q) + I(V_{i0};Y_{i'}|V_{00},V_{i'0},V_{i'i'})
&\leq
\min\lrb{
 \begin{aligned}
  &I(Z_{00},Z_{i0};Y_{i'},Z_{i'0},Z_{i'i'}),
  \\
  &I(Z_{00},Z_{10},Z_{20};Y_{i'},Z_{i'i'})
 \end{aligned}
}.
\label{eq:crng-case0-H(Q)+I(V10;Y2|V00,V20,V22)}
\end{align}
From the above (in)equalities,
we have the fact that
(\ref{eq:crng-Ri})--(\ref{eq:crng-R0+2Ri+Ri'}),
and (\ref{eq:crng-H(Q)})
imply (\ref{eq:crng-fme4})--(\ref{eq:crng-fme36}).
Derivations are summarized as follows.

\begin{itemize}
\item
Inequalities (\ref{eq:crng-fme4}) and (\ref{eq:crng-fme5})
come from (\ref{eq:crng-Ri}) and
(\ref{eq:crng-case0-I(V10,V11;Y1|V00,V20)}).
\item
Inequality (\ref{eq:crng-fme6}) comes from 
(\ref{eq:crng-case0}),
(\ref{eq:crng-case0-I(V11;Y1|V00,V10,V20)}),
and (\ref{eq:crng-case0-I(V10;Y2|V00,V20,V22)}).
\item
Inequalities (\ref{eq:crng-fme7})--(\ref{eq:crng-fme9})
come from (\ref{eq:crng-Ri}), (\ref{eq:crng-H(Q)}),
and (\ref{eq:crng-case0-H(Q)+I(V10,V11;Y1|V00,V20)}).
\item
Inequality (\ref{eq:crng-fme10}) comes from
(\ref{eq:crng-R0+Ri})
and (\ref{eq:crng-case0-I(V00,V10,V20,V11;Y1)}).
\item
Inequalities (\ref{eq:crng-fme11}) and (\ref{eq:crng-fme12}) come from
(\ref{eq:crng-H(Q)}),
(\ref{eq:crng-case0}),
(\ref{eq:crng-case0-I(V11;Y1|V00,V10,V20)}),
and (\ref{eq:crng-case0-H(Q)+I(V10;Y2|V00,V20,V22)}).
\item
Inequalities (\ref{eq:crng-fme13}) and (\ref{eq:crng-fme14}) come from
(\ref{eq:crng-H(Q)}),
(\ref{eq:crng-case0}),
(\ref{eq:crng-case0-I(V10;Y2|V00,V20,V22)}),
and (\ref{eq:crng-case0-H(Q)+I(V11;Y1|V00,V10,V20)}).
\item
Inequalities (\ref{eq:crng-fme15}) and (\ref{eq:crng-fme16})
come from (\ref{eq:crng-R1+R2-2})$'$,
(\ref{eq:crng-case0-I(V11;Y1|V00,V10,V20)}),
and (\ref{eq:crng-case0-I(V10,V20,V11;Y1|V00)})$'$.
\item
Inequality (\ref{eq:crng-fme17})
comes from (\ref{eq:crng-R1+R2-1})
and (\ref{eq:crng-case0-I(V20,V11;Y1|V00,V10)}).
\item
Inequalities (\ref{eq:crng-fme18})--(\ref{eq:crng-fme21}) come from
(\ref{eq:crng-H(Q)}),
(\ref{eq:crng-case0}),
(\ref{eq:crng-case0-H(Q)+I(V11;Y1|V00,V10,V20)}),
and (\ref{eq:crng-case0-H(Q)+I(V10;Y2|V00,V20,V22)}).
\item
Inequality (\ref{eq:crng-fme22})
comes from (\ref{eq:crng-2Ri+Ri'}),
(\ref{eq:crng-case0-I(V11;Y1|V00,V10,V20)}),
(\ref{eq:crng-case0-I(V20,V11;Y1|V00,V10)})$'$,
and (\ref{eq:crng-case0-I(V10,V20,V11;Y1|V00)}).
\item
Inequalities (\ref{eq:crng-fme23}) and (\ref{eq:crng-fme24})
come from (\ref{eq:crng-R0+R1+R2})$'$,
(\ref{eq:crng-case0-I(V11;Y1|V00,V10,V20)}),
and (\ref{eq:crng-case0-I(V00,V10,V20,V11;Y1)})$'$.
\item
Inequalities (\ref{eq:crng-fme25}) and (\ref{eq:crng-fme26})
come from (\ref{eq:crng-R1+R2-2})$'$,
(\ref{eq:crng-H(Q)}),
and (\ref{eq:crng-case0-I(V10,V20,V11;Y1|V00)})$'$.
\item
Inequality (\ref{eq:crng-fme27})
comes from (\ref{eq:crng-R1+R2-1}), (\ref{eq:crng-H(Q)})
(\ref{eq:crng-case0-I(V20,V11;Y1|V00,V10)})$'$,
and (\ref{eq:crng-case0-H(Q)+I(V20,V11;Y1|V00,V10)}).
\item
Inequality (\ref{eq:crng-fme28})
comes from (\ref{eq:crng-R0+R1+R2})$'$,
(\ref{eq:crng-H(Q)}),
(\ref{eq:crng-case0-H(Q)+I(V11;Y1|V00,V10,V20)}),
and (\ref{eq:crng-case0-I(V00,V10,V20,V11;Y1)})$'$,
\item
Inequalities (\ref{eq:crng-fme29}) and (\ref{eq:crng-fme30})
come from (\ref{eq:crng-R0+R1+R2})$'$,
(\ref{eq:crng-H(Q)}),
(\ref{eq:crng-case0-H(Q)+I(V11;Y1|V00,V10,V20)}),
and (\ref{eq:crng-case0-I(V00,V10,V20,V11;Y1)})$'$.
\item
Inequality (\ref{eq:crng-fme31})
comes from (\ref{eq:crng-R1+R2-1}), (\ref{eq:crng-H(Q)}),
and (\ref{eq:crng-case0-H(Q)+I(V20,V11;Y1|V00,V10)}).
\item
Inequality (\ref{eq:crng-fme32})
comes from (\ref{eq:crng-2Ri+Ri'}), (\ref{eq:crng-H(Q)}),
(\ref{eq:crng-case0-I(V11;Y1|V00,V10,V20)}),
(\ref{eq:crng-case0-H(Q)+I(V20,V11;Y1|V00,V10)})$'$,
and (\ref{eq:crng-case0-I(V10,V20,V11;Y1|V00)}).
\item 
Inequality (\ref{eq:crng-fme33})
comes from (\ref{eq:crng-R0+2Ri+Ri'}),
(\ref{eq:crng-case0-I(V11;Y1|V00,V10,V20)}),
(\ref{eq:crng-case0-I(V20,V11;Y1|V00,V10)})$'$,
and (\ref{eq:crng-case0-I(V00,V10,V20,V11;Y1)}).
\item 
Inequality (\ref{eq:crng-fme34})
comes from (\ref{eq:crng-R0+2Ri+Ri'}),
(\ref{eq:crng-H(Q)}),
(\ref{eq:crng-case0-I(V11;Y1|V00,V10,V20)}),
(\ref{eq:crng-case0-H(Q)+I(V20,V11;Y1|V00,V10)})$'$,
and (\ref{eq:crng-case0-I(V00,V10,V20,V11;Y1)}).
\item
Inequality (\ref{eq:crng-fme35})
comes from (\ref{eq:crng-R0+2Ri+Ri'}),
(\ref{eq:crng-H(Q)}),
(\ref{eq:crng-case0-H(Q)+I(V11;Y1|V00,V10,V20)}),
(\ref{eq:crng-case0-I(V20,V11;Y1|V00,V10)}),
and (\ref{eq:crng-case0-I(V00,V10,V20,V11;Y1)}).
\item
Inequality (\ref{eq:crng-fme36})
comes from (\ref{eq:crng-R0+2Ri+Ri'}),
(\ref{eq:crng-H(Q)}),
(\ref{eq:crng-case0-H(Q)+I(V11;Y1|V00,V10,V20)}),
(\ref{eq:crng-case0-H(Q)+I(V20,V11;Y1|V00,V10)})$'$,
and (\ref{eq:crng-case0-I(V00,V10,V20,V11;Y1)}).
\end{itemize}

\subsection{Case 1}

Here we outline the proof for Case 1,
where the details are given
in Appendix \ref{sec:proof-crng-crng-case1-appendix}.
From (\ref{eq:def-Z00})-(\ref{eq:def-Zii}), we have
\begin{align}
Z_{10}
&=
(Q,V_{10})
\label{eq:def-Z10-case1}
\\
Z_{20}
&=
Q
\label{eq:def-Z20-case1}
\end{align}
and
\begin{align}
I(Z_{20};Z_{22}|Z_{00})
&=
0.
\label{eq:crng-case1-I(Z20;Z22|Z00)}
\end{align}

From (\ref{eq:crng-casei-io}), we have
\begin{equation}
R_2
\geq
I(V_{22};Y_2|V_{00},V_{10},V_{20}) + I(V_{20};Y_1|V_{00},V_{10},V_{11}).
\label{eq:crng-case1-R2lower}
\end{equation}
Then we have the following facts.
\begin{itemize}
\item
From (\ref{eq:crng-R1+R2-1}) and (\ref{eq:crng-case1-R2lower}),
we have
\begin{equation}
 R_1
 \leq
 I(V_{11};Y_1|V_{00},V_{10})
 + I(V_{10};Y_2|V_{00},V_{20}).
 \label{eq:crng-case1-R1-1}
\end{equation}
\item
From (\ref{eq:crng-R1+R2-2}) and (\ref{eq:crng-case1-R2lower}),
we have
\begin{align}
 R_1
 &\leq
 I(V_{10},V_{11};Y_1|V_{00}).
 \label{eq:crng-case1-R1-2}
\end{align}
\item
From (\ref{eq:crng-R0+R1+R2}) and (\ref{eq:crng-case1-R2lower}),
we have
\begin{align}
 R_0 + R_1
 &\leq
 I(V_{00},V_{10},V_{11};Y_1).
 \label{eq:crng-case1-R0+R1}
\end{align}
\item
From (\ref{eq:crng-2Ri+Ri'}) and (\ref{eq:crng-case1-R2lower}),
we have
\begin{align}
 R_1 + R_2
 &\leq
 I(V_{11};Y_1|V_{00},V_{10})
 + I(V_{10},V_{20},V_{22};Y_2|V_{00}).
 \label{eq:crng-case1-R1+R2}
\end{align}
\item
From (\ref{eq:crng-R0+2Ri+Ri'}) and (\ref{eq:crng-case1-R2lower}),
we have
\begin{align}
 R_0 + R_1 + R_2
 &\leq
 I(V_{11};Y_1|V_{00},V_{10})
 + I(V_{00},V_{10},V_{20},V_{22};Y_2).
 \label{eq:crng-case1-R0+R1+R2}
\end{align}
\end{itemize}

We have the following (in)equalities from
(\ref{eq:Qindependence}), (\ref{eq:def-Z00}),
(\ref{eq:def-Zii})--(\ref{eq:I(ZX1;ZX2|Z0)=0}),
(\ref{eq:def-Z10-case1}), (\ref{eq:def-Z20-case1}),
and Shannon-type (in)equalities:
\begin{align}
H(Q)
&\leq
\min\lrb{
 \begin{aligned}
  &I(Z_{00},Z_{20};Y_1,Z_{10},Z_{11}),
  \\
  &I(Z_{00},Z_{10},Z_{20};Y_1,Z_{11})
 \end{aligned}
}
\label{eq:crng-case1-H(Q)}
\\
I(V_{11};Y_1|V_{00},V_{10})
&\leq
\min\lrb{
 \begin{aligned}
  &I(Z_{11};Y_1,Z_{10},Z_{20}|Z_{00}),
  \\
  &I(Z_{20},Z_{11};Y_1,Z_{10}|Z_{00})
 \end{aligned}
}
- I(Z_{10};Z_{11}|Z_{00})
\label{eq:crng-case1-I(V11;Y1|V00,V10)}
\\
H(Q) + I(V_{11};Y_1|V_{00},V_{10})
&\leq
\min\lrb{
 \begin{aligned}
  &I(Z_{00},Z_{11};Y_1,Z_{10},Z_{20}),
  \\
  &I(Z_{00},Z_{20},Z_{11};Y_1,Z_{10})
 \end{aligned}
}
- I(Z_{10};Z_{11}|Z_{00})
\label{eq:crng-case1-H(Q)+I(V11;Y1|V00,V10)}
\\
I(V_{10},V_{11};Y_1|V_{00})
&\leq
\min\lrb{
 \begin{aligned}
  &I(Z_{11};Y_1,Z_{10},Z_{20}|Z_{00}),
  \\
  &I(Z_{10},Z_{11};Y_1,Z_{20}|Z_{00}),
  \\
  &I(Z_{20},Z_{11};Y_1,Z_{10}|Z_{00}),
  \\
  &I(Z_{10},Z_{20},Z_{11};Y_1|Z_{00})
 \end{aligned}
}
\label{eq:crng-case1-I(V10,V11;Y1|V00)}
\\
H(Q) + I(V_{10},V_{11};Y_1|V_{00})
&\leq
I(Z_{00},Z_{20},Z_{11};Y_1,Z_{10})
\label{eq:crng-case1-H(Q)+I(V10,V11;Y1|V00)}
\\
I(V_{00},V_{10},V_{11};Y_1)
&\leq
\min\lrb{
 \begin{aligned}
  &I(Z_{00},Z_{11};Y_1,Z_{10},Z_{20}),
  \\
  &I(Z_{00},Z_{10},Z_{11};Y_1,Z_{20}),
  \\
  &I(Z_{00},Z_{20},Z_{11};Y_1,Z_{10}),
  \\
  &I(Z_{00},Z_{10},Z_{20},Z_{11};Y_1)
 \end{aligned}
}
\label{eq:crng-case1-I(V00,V10,V11;Y1)}
\\
I(V_{00},V_{10},V_{20},V_{11};Y_1)
&=
I(Z_{00},Z_{10},Z_{20},Z_{11};Y_1)
\label{eq:crng-case1-I(V00,V10,V20,V11;Y1)}
\\
I(V_{10};Y_2|V_{00},V_{20})
&\leq
I(Z_{10};Y_2,Z_{20},Z_{22}|Z_{00})
\label{eq:crng-case1-I(V10;Y2|V00,V20)}
\\
H(Q)
+ I(V_{10};Y_2|V_{00},V_{20})
&\leq
\min\lrb{
 \begin{aligned}
  &I(Z_{00},Z_{10};Y_2,Z_{20},Z_{22}),
  \\
  &I(Z_{00},Z_{10},Z_{20},;Y_2,Z_{22})
 \end{aligned}
}
\label{eq:crng-case1-H(Q)+I(V10;Y2|V00,V20)}
\\
I(V_{20},V_{22};Y_2|V_{00},V_{10})
&\leq
\min\lrb{
 \begin{aligned}
  &I(Z_{22};Y_2,Z_{10},Z_{20}|Z_{00}),
  \\
  &I(Z_{20},Z_{22};Y_2,Z_{10}|Z_{00})
 \end{aligned}
}
\label{eq:crng-case1-I(V20,V22;Y2|V00,V10)}
\\
H(Q) + I(V_{20},V_{22};Y_2|V_{00},V_{10})
&\leq
\min\lrb{
 \begin{aligned}
  &I(Z_{00},Z_{22};Y_2,Z_{10},Z_{20}),
  \\
  &I(Z_{00},Z_{20},Z_{22};Y_2,Z_{10}),
  \\
  &I(Z_{00},Z_{10},Z_{22};Y_2,Z_{20})
 \end{aligned}
}
\label{eq:crng-case1-H(Q)+I(V20,V22;Y2|V00,V10)}
\\
I(V_{10},V_{20},V_{22};Y_2|V_{00})
&\leq
\min\lrb{
 \begin{aligned}
  &I(Z_{10},Z_{22};Y_2,Z_{20}|Z_{00}),
  \\
  &I(Z_{10},Z_{20},Z_{22};Y_2|Z_{00})
 \end{aligned}
}
\label{eq:crng-case1-I(V10,V20,V22;Y2|V00)}
\\
H(Q)+ I(V_{10},V_{20},V_{22};Y_2|V_{00})
&\leq
I(Z_{00},Z_{10},Z_{22};Y_2,Z_{20})
\label{eq:crng-case1-H(Q)+I(V10,V20,V22;Y2|V00)}
\\
I(V_{00},V_{10},V_{20},V_{22};Y_2)
&\leq
\min\lrb{
 \begin{aligned}
  &I(Z_{00},Z_{10},Z_{22};Y_2,Z_{20})
  \\
  &I(Z_{00},Z_{10},Z_{20},Z_{22};Y_2)
 \end{aligned}
}.
\label{eq:crng-case1-I(V00,V10,V20,V22;Y2)}
\end{align}
From the above (in)equalities
we have the fact that
(\ref{eq:crng-Ri})--(\ref{eq:crng-R0+2Ri+Ri'}),
and (\ref{eq:crng-H(Q)})
imply (\ref{eq:crng-fme4})--(\ref{eq:crng-fme36}).
Derivations are summarized as follows.
\begin{itemize}
\item
Inequalities (\ref{eq:crng-fme4}) and (\ref{eq:crng-fme5}) come from
(\ref{eq:crng-case1-R1-2}) and
(\ref{eq:crng-case1-I(V10,V11;Y1|V00)})
when $i=1$,
and they come from
(\ref{eq:crng-Ri})$'$ and
(\ref{eq:crng-case1-I(V20,V22;Y2|V00,V10)})
when $i=2$.
\item
Inequality (\ref{eq:crng-fme6}) comes from 
(\ref{eq:crng-case1-R0+R1}),
(\ref{eq:crng-case1-I(V11;Y1|V00,V10)}),
and (\ref{eq:crng-case1-I(V10;Y2|V00,V20)})
when $i=1$,
and it comes from 
(\ref{eq:crng-Ri})$'$,
(\ref{eq:crng-case1-I(Z20;Z22|Z00)})
and (\ref{eq:crng-case1-I(V20,V22;Y2|V00,V10)})
when $i=2$.
\item
Inequalities (\ref{eq:crng-fme7})--(\ref{eq:crng-fme9}) come from
(\ref{eq:crng-case1-R0+R1}),
and (\ref{eq:crng-case1-I(V00,V10,V11;Y1)})
when $i=1$,
and they come from
(\ref{eq:crng-Ri})$'$, (\ref{eq:crng-H(Q)}),
and (\ref{eq:crng-case1-H(Q)+I(V20,V22;Y2|V00,V10)})
when $i=2$.
\item
Inequality (\ref{eq:crng-fme10}) comes from
(\ref{eq:crng-R0+Ri}),
(\ref{eq:crng-case1-I(V00,V10,V20,V11;Y1)}),
and (\ref{eq:crng-case1-I(V00,V10,V20,V22;Y2)}).
\item
Inequalities (\ref{eq:crng-fme11}) and (\ref{eq:crng-fme12}) come from
(\ref{eq:crng-H(Q)}),
(\ref{eq:crng-case1-R1-1}),
(\ref{eq:crng-case1-I(V11;Y1|V00,V10)}),
and (\ref{eq:crng-case1-H(Q)+I(V10;Y2|V00,V20)})
when $i=1$,
and they come from
(\ref{eq:crng-Ri})$'$, (\ref{eq:crng-H(Q)}),
(\ref{eq:crng-case1-I(Z20;Z22|Z00)}),
(\ref{eq:crng-case1-H(Q)}),
and (\ref{eq:crng-case1-I(V20,V22;Y2|V00,V10)})
when $i=2$.
\item
Inequalities (\ref{eq:crng-fme13}) and (\ref{eq:crng-fme14}) come from
(\ref{eq:crng-H(Q)}),
(\ref{eq:crng-case1-R1-1}),
(\ref{eq:crng-case1-H(Q)+I(V11;Y1|V00,V10)}),
and (\ref{eq:crng-case1-I(V10;Y2|V00,V20)})
when $i=1$,
and they come from
(\ref{eq:crng-Ri})$'$, (\ref{eq:crng-H(Q)}),
(\ref{eq:crng-case1-I(Z20;Z22|Z00)}),
(\ref{eq:crng-case1-H(Q)+I(V20,V22;Y2|V00,V10)})
when $i=2$.
\item
Inequalities (\ref{eq:crng-fme15}) and (\ref{eq:crng-fme16}) come from
(\ref{eq:crng-case1-R1+R2}),
(\ref{eq:crng-case1-I(V11;Y1|V00,V10)}),
and (\ref{eq:crng-case1-I(V10,V20,V22;Y2|V00)})
when $i=1$,
and they come from (\ref{eq:crng-Ri})$'$,
(\ref{eq:crng-case1-I(Z20;Z22|Z00)}),
(\ref{eq:crng-case1-R1-2}),
(\ref{eq:crng-case1-I(V10,V11;Y1|V00)}),
and (\ref{eq:crng-case1-I(V20,V22;Y2|V00,V10)})
when $i=2$.
\item
Inequality (\ref{eq:crng-fme17}) comes from
(\ref{eq:crng-case1-I(Z20;Z22|Z00)}),
(\ref{eq:crng-case1-R1+R2}),
(\ref{eq:crng-case1-I(V11;Y1|V00,V10)}),
and (\ref{eq:crng-case1-I(V10,V20,V22;Y2|V00)}).
\item
Inequalities (\ref{eq:crng-fme18})--(\ref{eq:crng-fme21}) come from
(\ref{eq:crng-H(Q)}),
(\ref{eq:crng-case1-R0+R1}),
(\ref{eq:crng-case1-H(Q)+I(V11;Y1|V00,V10)}),
and (\ref{eq:crng-case1-H(Q)+I(V10;Y2|V00,V20)})
when $i=1$,
and they come from
(\ref{eq:crng-Ri})$'$, (\ref{eq:crng-H(Q)}),
(\ref{eq:crng-case1-I(Z20;Z22|Z00)}),
(\ref{eq:crng-case1-H(Q)}),
and (\ref{eq:crng-case1-H(Q)+I(V20,V22;Y2|V00,V10)})
when $i=2$.
\item
Inequality (\ref{eq:crng-fme22}) comes from
(\ref{eq:crng-case1-I(Z20;Z22|Z00)}),
(\ref{eq:crng-case1-R1-2}),
(\ref{eq:crng-case1-R1+R2}),
(\ref{eq:crng-case1-I(V11;Y1|V00,V10)}),
(\ref{eq:crng-case1-I(V10,V11;Y1|V00)}),
and (\ref{eq:crng-case1-I(V10,V20,V22;Y2|V00)})
when $i=1$,
and it comes from
(\ref{eq:crng-Ri})$'$,
(\ref{eq:crng-case1-I(Z20;Z22|Z00)}),
(\ref{eq:crng-case1-R1+R2}),
(\ref{eq:crng-case1-I(V11;Y1|V00,V10)}),
(\ref{eq:crng-case1-I(V20,V22;Y2|V00,V10)}),
and (\ref{eq:crng-case1-I(V10,V20,V22;Y2|V00)})
when $i=2$.
\item
Inequalities (\ref{eq:crng-fme23}) and (\ref{eq:crng-fme24}) come from
(\ref{eq:crng-case1-R0+R1+R2}),
(\ref{eq:crng-case1-I(V11;Y1|V00,V10)}),
and (\ref{eq:crng-case1-I(V00,V10,V20,V22;Y2)})
when $i=1$,
and they come from
(\ref{eq:crng-Ri})$'$,
(\ref{eq:crng-case1-I(Z20;Z22|Z00)}),
(\ref{eq:crng-case1-R0+R1}),
(\ref{eq:crng-case1-I(V00,V10,V11;Y1)}),
and (\ref{eq:crng-case1-I(V20,V22;Y2|V00,V10)})
when $i=2$.
\item
Inequalities (\ref{eq:crng-fme25}) and (\ref{eq:crng-fme26}) come from
(\ref{eq:crng-H(Q)}),
(\ref{eq:crng-case1-R1+R2}),
(\ref{eq:crng-case1-H(Q)+I(V11;Y1|V00,V10)}),
and (\ref{eq:crng-case1-I(V10,V20,V22;Y2|V00)}),
when $i=1$,
and they come from
(\ref{eq:crng-Ri})$'$, (\ref{eq:crng-H(Q)}),
(\ref{eq:crng-case1-I(Z20;Z22|Z00)}),
(\ref{eq:crng-case1-R1-2}),
(\ref{eq:crng-case1-I(V10,V11;Y1|V00)}),
and (\ref{eq:crng-case1-H(Q)+I(V20,V22;Y2|V00,V10)})
when $i=2$.
\item
Inequality (\ref{eq:crng-fme27}) comes from
(\ref{eq:crng-H(Q)}),
(\ref{eq:crng-case1-I(Z20;Z22|Z00)}),
(\ref{eq:crng-case1-R1+R2}),
(\ref{eq:crng-case1-H(Q)+I(V11;Y1|V00,V10)}),
and (\ref{eq:crng-case1-I(V10,V20,V22;Y2|V00)})
when $i=1$,
and it comes from
(\ref{eq:crng-H(Q)}),
(\ref{eq:crng-case1-I(Z20;Z22|Z00)}),
(\ref{eq:crng-case1-R1+R2}),
(\ref{eq:crng-case1-I(V11;Y1|V00,V10)}),
and (\ref{eq:crng-case1-H(Q)+I(V10,V20,V22;Y2|V00)})
when $i=2$.
\item
Inequality (\ref{eq:crng-fme28}) comes from
(\ref{eq:crng-H(Q)}), (\ref{eq:crng-case1-R1+R2}),
(\ref{eq:crng-case1-H(Q)+I(V11;Y1|V00,V10)}),
and (\ref{eq:crng-case1-H(Q)+I(V10,V20,V22;Y2|V00)})
when $i=1$,
and it comes from
(\ref{eq:crng-Ri})$'$, (\ref{eq:crng-H(Q)}),
(\ref{eq:crng-case1-I(Z20;Z22|Z00)}),
(\ref{eq:crng-case1-R1-2}),
(\ref{eq:crng-case1-H(Q)+I(V10,V11;Y1|V00)}),
and (\ref{eq:crng-case1-H(Q)+I(V20,V22;Y2|V00,V10)})
when $i=2$
\item
Inequalities (\ref{eq:crng-fme29}) and (\ref{eq:crng-fme30}) come from
(\ref{eq:crng-H(Q)}),
(\ref{eq:crng-case1-R0+R1+R2}),
(\ref{eq:crng-case1-H(Q)+I(V11;Y1|V00,V10)}),
and (\ref{eq:crng-case1-I(V00,V10,V20,V22;Y2)})
when $i=1$,
and they come from (\ref{eq:crng-Ri})$'$,
(\ref{eq:crng-H(Q)}),
(\ref{eq:crng-case1-I(Z20;Z22|Z00)}),
(\ref{eq:crng-case1-R0+R1}),
(\ref{eq:crng-case1-I(V00,V10,V11;Y1)}),
and (\ref{eq:crng-case1-H(Q)+I(V20,V22;Y2|V00,V10)})
when $i=2$.
\item
Inequality (\ref{eq:crng-fme31}) comes from
(\ref{eq:crng-H(Q)}),
(\ref{eq:crng-case1-I(Z20;Z22|Z00)}),
(\ref{eq:crng-case1-R0+R1+R2}),
(\ref{eq:crng-case1-H(Q)+I(V11;Y1|V00,V10)}),
and (\ref{eq:crng-case1-H(Q)+I(V10,V20,V22;Y2|V00)}).
\item
Inequality (\ref{eq:crng-fme32}) comes from
(\ref{eq:crng-H(Q)}),
(\ref{eq:crng-case1-I(Z20;Z22|Z00)}),
(\ref{eq:crng-case1-R1-2}), 
(\ref{eq:crng-case1-R1+R2}),
(\ref{eq:crng-case1-I(V11;Y1|V00,V10)}),
(\ref{eq:crng-case1-I(V10,V11;Y1|V00)}),
and (\ref{eq:crng-case1-H(Q)+I(V10,V20,V22;Y2|V00)})
when $i=1$,
and it comes from (\ref{eq:crng-Ri})$'$,
(\ref{eq:crng-H(Q)}),
(\ref{eq:crng-case1-I(Z20;Z22|Z00)}),
(\ref{eq:crng-case1-R1+R2}),
(\ref{eq:crng-case1-H(Q)+I(V11;Y1|V00,V10)}),
(\ref{eq:crng-case1-I(V20,V22;Y2|V00,V10)}),
and (\ref{eq:crng-case1-I(V10,V20,V22;Y2|V00)})
when $i=2$.
\item
Inequality (\ref{eq:crng-fme33}) comes from
(\ref{eq:crng-R0+Ri}),
(\ref{eq:crng-case1-I(Z20;Z22|Z00)}),
(\ref{eq:crng-case1-R1+R2}),
(\ref{eq:crng-case1-I(V11;Y1|V00,V10)}),
(\ref{eq:crng-case1-I(V00,V10,V20,V11;Y1)}),
and (\ref{eq:crng-case1-I(V10,V20,V22;Y2|V00)})
when $i=1$,
and it comes from
(\ref{eq:crng-Ri})$'$,
(\ref{eq:crng-case1-I(Z20;Z22|Z00)}),
(\ref{eq:crng-case1-R0+R1+R2}),
(\ref{eq:crng-case1-I(V11;Y1|V00,V10)}),
(\ref{eq:crng-case1-I(V20,V22;Y2|V00,V10)}),
and (\ref{eq:crng-case1-I(V00,V10,V20,V22;Y2)})
when $i=2$.
\item
Inequality (\ref{eq:crng-fme34}) comes from
(\ref{eq:crng-H(Q)}),
(\ref{eq:crng-case1-I(Z20;Z22|Z00)}),
(\ref{eq:crng-case1-R0+R1}),
(\ref{eq:crng-case1-R1+R2}),
(\ref{eq:crng-case1-I(V11;Y1|V00,V10)}),
(\ref{eq:crng-case1-I(V00,V10,V11;Y1)}),
and (\ref{eq:crng-case1-H(Q)+I(V10,V20,V22;Y2|V00)}),
when $i=1$,
and it comes from
(\ref{eq:crng-Ri})$'$,
(\ref{eq:crng-H(Q)}),
(\ref{eq:crng-case1-I(Z20;Z22|Z00)}),
(\ref{eq:crng-case1-R0+R1+R2}),
(\ref{eq:crng-case1-H(Q)+I(V11;Y1|V00,V10)}),
(\ref{eq:crng-case1-I(V20,V22;Y2|V00,V10)}),
and (\ref{eq:crng-case1-I(V00,V10,V20,V22;Y2)})
when $i=2$.
\item
Inequality (\ref{eq:crng-fme35}) comes from
(\ref{eq:crng-H(Q)}),
(\ref{eq:crng-case1-I(Z20;Z22|Z00)}),
(\ref{eq:crng-case1-R0+R1}),
(\ref{eq:crng-case1-R1+R2}),
(\ref{eq:crng-case1-H(Q)+I(V11;Y1|V00,V10)}),
(\ref{eq:crng-case1-I(V00,V10,V11;Y1)}),
and (\ref{eq:crng-case1-I(V10,V20,V22;Y2|V00)})
when $i=1$,
and it comes from
(\ref{eq:crng-Ri})$'$,
(\ref{eq:crng-H(Q)}),
(\ref{eq:crng-case1-I(Z20;Z22|Z00)}),
(\ref{eq:crng-case1-R0+R1+R2}),
(\ref{eq:crng-case1-I(V11;Y1|V00,V10)}),
(\ref{eq:crng-case1-H(Q)+I(V20,V22;Y2|V00,V10)}),
and (\ref{eq:crng-case1-I(V00,V10,V20,V22;Y2)})
when $i=2$.
\item
Inequality (\ref{eq:crng-fme36}) comes from
(\ref{eq:crng-H(Q)}),
(\ref{eq:crng-case1-R0+R1}),
(\ref{eq:crng-case1-R1+R2}),
(\ref{eq:crng-case1-I(Z20;Z22|Z00)}),
(\ref{eq:crng-case1-H(Q)+I(V11;Y1|V00,V10)}),
(\ref{eq:crng-case1-I(V00,V10,V11;Y1)}),
and (\ref{eq:crng-case1-H(Q)+I(V10,V20,V22;Y2|V00)})
when $i=1$,
and it comes from
(\ref{eq:crng-Ri})$'$,
(\ref{eq:crng-H(Q)}),
(\ref{eq:crng-case1-I(Z20;Z22|Z00)}),
(\ref{eq:crng-case1-R0+R1+R2}),
(\ref{eq:crng-case1-H(Q)+I(V11;Y1|V00,V10)}),
(\ref{eq:crng-case1-H(Q)+I(V20,V22;Y2|V00,V10)}),
and (\ref{eq:crng-case1-I(V00,V10,V20,V22;Y2)})
when $i=2$.
\end{itemize}

\section{Proof of Theorem \ref{thm:JXGsubsetCRNG}}
\label{sec:proof-jxg-crng}

Since we have the relation $\tRCRNG^0=\RCRNG^0$
from Theorem \ref{thm:crng-simplified},
it is sufficient to show that $\RJXG=\tRCRNG^0$.

First, we show the fact that $\RJXG\subset\tRCRNG^0$.
By letting
\begin{align*}
V_{00}
&\equiv
U_0
\\
V_{i0}
&\equiv
U_i
\\
V_{ii}
&\equiv
X_i
\end{align*}
for each $i\in\{0,1\}$,
we have the fact that (\ref{eq:jxg-Ri})--(\ref{eq:jxg-R0+2Ri+Ri'})
imply (\ref{eq:crng-Ri})--(\ref{eq:crng-R0+2Ri+Ri'}).
Then we have the fact that $\RJXG\subset\tRCRNG^0$.

Next, we show the fact that $\RJXG\supset\tRCRNG^0$.
Let $\tmu$ satisfies (\ref{eq:joint-crng-simplified}).
From the functional representation lemma \cite[Page 626]{EK11},
we have the fact that
for a given $(V_{00},V_{i0},X_i)$
there is random variable $\tW_{i0}\in\tcW_{i0}$
and function $\lambda_{i0}$ such that
the joint distribution of $\tW_{i0}$
is independent from all other random variables
and the joint distribution of 
$(V_{00},V_{i0},X_i)$ can be represented
as that of $(V_{00},\lambda_{i0}(V_{00},X_i,\tW_{i0}),X_i)$.
By applying the functional representation lemma again,
we have the fact that
for a given $(V_{00},V_{i0},V_{ii},X_i)$
there is random variable $\tW_{ii}\in\tcW_{ii}$
and function $\lambda_{ii}$ such that
$\tW_{ii}$ is independent from all other random variables
and the joint distribution of 
$(V_{00},V_{i0},V_{10},X_i)$ can be represented
as that of $(V_{00},V_{i0},\lambda_{ii}(V_{00},V_{i0},X_i,\tW_{ii}),X_i)$.
Then we have the fact that
\begin{align*}
&
\mu_{X_iV_{i0}V_{ii}|V_{00}}(x_i,v_{i0},v_{ii}|v_{00})
\notag
\\
&=
\sum_{\tw_{ii}\in\tcW_{ii}}
\chi(\lambda_{ii}(v_{00},v_{i0},x_i,\tw_{ii})=v_{ii})\mu_{\tW_{ii}}(\tw_{ii})
\sum_{\tw_{i0}\in\tcW_{i0}}
\chi(\lambda_{i0}(v_{00},x_i,\tw_{i0})=v_{i0})\mu_{\tW_{i0}}(\tw_{i0})
\mu_{X_i|V_{00}}(x_i|v_{00})
\end{align*}
for all $(v_{00},v_{i0},v_{ii},x_i)$,
where
\begin{equation*}
\mu_{X_i|V_{00}}(x_i|u_0)
\equiv
\sum_{v_{i0},\in\V_{i0},v_{ii}\in\V_{ii}}
\mu_{X_iV_{i0}V_{ii}|V_{00}}(v_{i0},v_{ii},x_i|u_0).
\end{equation*}
By letting
\begin{align*}
U_0 &\equiv V_{00}
\\
U_i &\equiv V_{i0}\quad\text{for each}\ i\in\{1,2\},
\end{align*}
we have the fact that the joint distribution of $(U_0,U_1,X_1,U_2,X_2)$
satisfies (\ref{eq:joint-jxg}),
where $p_{U_0}$ and $p_{U_iX_i|U_0}$ are given as
\begin{align*}
p_{U_0}(u_0)
&=
\mu_{V_{00}}(u_0)
\\
p_{X_i|U_0}(x_i|u_0)
&=
\mu_{X_i|V_{00}}(x_i|u_0)
\\
p_{U_i|X_iU_0}(u_i|x_i,u_0)
&=
\sum_{\tw_{i0}\in\tcW_{i0}}
\chi(\lambda_{i0}(u_0,x_i,\tw_{i0})=u_i)
\mu_{\tW_{i0}}(\tw_{i0}).
\end{align*}
We have
\begin{align}
I(V_{00},V_{10},V_{20},V_{ii};Y_i)
&\leq
I(V_{00},V_{10},V_{20},V_{ii},X_i,\tW_{i0},\tW_{ii};Y_i)
\notag
\\
&=
I(V_{00},V_{i'0},X_i,\tW_{i0},\tW_{ii};Y_i)
\notag
\\
&=
I(V_{00},V_{i'0},X_i;Y_i)
\notag
\\
&=
I(U_0,U_{i'},X_i;Y_i)
\label{eq:crng-jxg-I(V00,V10,V20,Vii;Yi)}
\\
I(V_{10},V_{20},V_{ii};Y_i|V_{00})
&\leq
I(V_{10},V_{20},V_{ii},X_i,\tW_{i0},\tW_{ii};Y_i|V_{00})
\notag
\\
&=
I(V_{i'0},X_i,\tW_{i0},\tW_{ii};Y_i|V_{00})
\notag
\\
&=
I(V_{i'0},X_i;Y_i|V_{00})
\notag
\\
&=
I(U_{i'},X_i;Y_i|U_0)
\\
I(V_{i0},V_{ii};Y_i|V_{00},V_{i'0})
&\leq
I(V_{i0},V_{ii},X_i,\tW_{i0},\tW_{ii};Y_i|V_{00},V_{i'0})
\notag
\\
&=
I(X_i,\tW_{i0},\tW_{ii};Y_i|V_{00},V_{i'0})
\notag
\\
&=
I(X_i;Y_i|V_{00},V_{i'0})
\notag
\\
&=
I(X_i;Y_i|U_0,U_{i'})
\\
I(V_{ii};Y_i|V_{00},V_{10},V_{20})
&\leq
I(V_{i0},V_{ii},X_i,\tW_{i0},\tW_{ii};Y_i|V_{00},V_{10},V_{20})
\notag
\\
&=
I(X_i,\tW_{i0},\tW_{ii};Y_i|V_{00},V_{10},V_{20})
\notag
\\
&=
I(X_i;Y_i|V_{00},V_{10},V_{20})
\notag
\\
&=
I(X_i;Y_i|U_0,U_1,U_2)
\label{eq:crng-jxg-I(Vii;Yi|V00,V10,V20)}
\end{align}
where the first equalities
come from
Lemma \ref{lem:function} in Appendix \ref{sec:entropy} and
the fact that
$(V_{i_0},V_{ii})$ is a function of $(V_{00},X_i,\tW_{i0},\tW_{ii})$
as
\begin{equation*}
(V_{i_0},V_{ii})
\equiv
(\lambda_{i0}(V_{00},X_i,\tW_{i0}),
 \lambda_{ii}(V_{00},\lambda_{i0}(V_{00},X_i,\tW_{i0}),X_i,\tW_{ii})),
\end{equation*}
and the second equalities
come from Lemma \ref{lem:Q} in Appendix \ref{sec:entropy}.
From (\ref{eq:crng-jxg-I(V00,V10,V20,Vii;Yi)})--(\ref{eq:crng-jxg-I(Vii;Yi|V00,V10,V20)}),
we have the fact that
(\ref{eq:crng-Ri})--(\ref{eq:crng-R0+2Ri+Ri'})
imply (\ref{eq:jxg-Ri})--(\ref{eq:jxg-R0+2Ri+Ri'}).
Then we have the fact that
$\RJXG(p)\supset\tRCRNG^0(\mu)$
and $\RJXG\supset\tRCRNG^0=\RCRNG^0$.

\section{Proof of Theorem \ref{thm:CRNGsubsetCRNGfull}}
\label{sec:CRNGsubsetCRNGfull}

First, we show $\RCRNG\subset\oRCRNG^0$
by showing that
for a given $\mu$ defined by (\ref{eq:joint-crng}),
there is a joint distribution $\omu$ of
$(Z_{00},Z_{01},Z_{02},Z_{10},Z_{11},Z_{12},X_1,Z_{20},Z_{22},Z_{21},X_2)$
satisfying (\ref{eq:joint-crng-full}) and $\RCRNG(\mu)\subset\oRCRNG^0(\omu)$.
For a given $\mu$ defined by (\ref{eq:joint-crng}),
let $(R_{00},R_{10},R_{11},R_{20},R_{22})\in\RCRNG(\mu)$ and assume that
$R_s=r_s=0$ and $Z_s$ is a constant random variable
for all $s\in\{01,02,12,21\}$.
Then we have the joint distribution $\omu$ of
$(Z_{00},Z_{01},Z_{02},Z_{10},Z_{11},Z_{12},X_1,Z_{20},Z_{22},Z_{21},X_2)$
such that the marginal distribution of $\omu$
with respect to $(Z_{00},Z_{10},Z_{11},X_1,Z_{20},Z_{22},X_2)$
is $\mu$.
For all $\tcS_0\subset\cS_0\equiv\{00,01,02\}$, 
we have (\ref{eq:crngfull-sum-S0}) as
\begin{align}
\sum_{s\in\tcS_0}
\lrB{R_s+r_s}
&=
\sum_{s\in\tcS_0\cap\{00\}}
\lrB{R_s+r_s}
\notag
\\
&\leq
\begin{cases}
 0
 &\text{if}\ \tcS_0\cap\{00\}=\emptyset
 \\
 H(Z_{00})
 &\text{if}\ \tcS_0\cap\{00\}=\{00\}
\end{cases}
\notag
\\
&=
H(Z_{\tcS_0\cap\{00\}})
\notag
\\
&=
H(Z_{\tcS_0}),
\end{align}
where the first equality comes from the fact that
$R_s=r_s=0$ for all $s\in\{01,02\}$,
the inequality comes from (\ref{eq:crng-R0+r0}),
and the last equality comes from
the fact that $Z_s$ is constant for all $s\in\{01,02\}$.
For all $i\in\{1,2\}$ and $\tcS_i\subset\cS_i\equiv\{i0,ii,ii'\}$, 
we have (\ref{eq:crngfull-sum-Si}) as
\begin{align}
\sum_{s\in\tcS_i}
\lrB{R_s+r_s}
&=
\sum_{s\in\tcS_i\cap\{i0,ii\}}
\lrB{R_s+r_s}
\notag
\\
&\leq
H(Z_{\tcS_i\cap\{i0,ii\}}|Z_{\cS_0})
\notag
\\
&=
H(Z_{\tcS_i}|Z_{\cS_0}),
\end{align}
where the first equality comes from the fact that
$R_s=r_s=0$ for all $s\in\{12,21\}$,
the inequality comes from (\ref{eq:crng-sum-Rj+rj}),
and the last equality comes from
the fact that $Z_s$ is constant for all $s\in\{01,02,12,21\}$.
For all $j\in\{1,2\}$ and $\tcD_j\subset\D_j\equiv\{00,0j,10,20,jj,j'j\}$,
we have (\ref{eq:crngfull-sum-Di}) as
\begin{align}
\sum_{s\in\tcD_j} r_s
&=
\sum_{s\in\tcD_j\cap\{00,10,20,jj\}} r_s
\notag
\\
&\geq
H(Z_{\tcD_j\cap\{00,10,20,jj\}}|Y_i,Z_{\{00,10,20,jj\}\setminus\tcD_j})
\notag
\\
&=
H(Z_{\tcD_j}|Y_j,Z_{\D_j\setminus\tcD_j}),
\end{align}
where the first equality comes from the fact that
$r_s=0$ for all $s\in\{0j,j'j\}$,
the inequality comes from (\ref{eq:crng-sum-rj}),
and the last equality comes from
the fact that $Z_s$ is constant for all $s\in\{0j,j'j\}$.
Hence, we have the fact that
$(R_{00},R_{10},R_{11},R_{20},R_{22})\in\oRCRNG(\omu)$,
which implies $\RCRNG(\mu)\subset\oRCRNG^0(\omu)$
and $\RCRNG\subset\oRCRNG^0$.

Next, we show $\RCRNG^0\subset\oRCRNG^{00}$.
For a given $\mu$ satisfying (\ref{eq:joint-crng}),
assume that $(R_0,R_1,R_2)\in\RCRNG^0(\mu)$
and let $\omu$ be defined as above.
Then, we have $(R_0,0,R_1,0,R_2)\in\RCRNG(\mu)\subset\oRCRNG^0(\omu)$
from the definition of $\RCRNG^0(\mu)$.
In addition, we have $(R_0,0,0,R_1,0,0,R_2,0)\in\oRCRNG(\omu)$
from the definition of $\oRCRNG^0(\omu)$.
This implies that $(R_0,R_1,R_2)\in\oRCRNG^{00}(\omu)$ 
and $(R_0,R_1,R_2)\in\oRCRNG^{00}$ from the definition of $\oRCRNG^{00}(\omu)$
and $\oRCRNG^{00}$.
The relation $\RJXG=\RCRNG^0$ comes from Theorem~\ref{thm:JXGsubsetCRNG}.

Similarly to the proof of $\RCRNG^0\subset\oRCRNG^{00}$,
we can show 
$\RHKCMG=\RCRNG^{00}\subset\oRCRNG^{000}$,
where the relation $\RHKCMG=\RCRNG^{00}$
comes from Corollary \ref{thm:HKCMGsubsetCRNG00}.
\hfill\IEEEQED

\section{Concluding Remarks}

Finally, let us close this paper with some remarks.

To derive the standard formula of achievable region
of $\oRCRNG(\omu)$,
we can apply the Fourier-Motzkin method
to eliminate variables $\{r_s: s\in\cS_0\cup\cS_1\cup\cS_2\}$,
which appear in the definition of $\oRCRNG(\omu)$.
However, it seems hard to specify the redundant inequalities
from the set of inequalities obtained by using the method.
Information Theoretic Inequality Prover (ITIP)
\cite{GGP15,GP17,HLTY20,L21,PPD,Y08} might solve the problem.

The following questions are challenges for the future.
\begin{enumerate}
\item Clarify the smallest set of inequalities
without variables $\{r_s: s\in\cS_0\cup\cS_1\cup\cS_2\}$
that can represent $\oRCRNG(\omu)$.
\item Is 
$\RCRNG$ the proper subset of $\oRCRNG^0$?
\item Can we extend regions $\RCRNG^{0}$, $\RCRNG^{00}$,
$\oRCRNG^{00}$, and $\oRCRNG^{000}$
by applying the rate-splitting technique to $\RCRNG$ and $\oRCRNG$,
respectively?
\item Clarify the single-letterized capacity region
of interference channel with common message(s).
\end{enumerate}

It should be noted here that
for given specific $\mu$ satisfying (\ref{eq:ZS}) and
specific real vector $\{R_s\}_{s\in\cS}$,
we can obtain $\{r_s\}_{s\in\cS}$
satisfying (\ref{eq:channel-Rs+rs})--(\ref{eq:channel-rs})
by using linear programming whenever it exists.
From this viewpoint, the expression
(\ref{eq:channel-Rs+rs})--(\ref{eq:channel-rs})
with variables $\{r_s\}_{s\in\cS}$
could be computable
(by assuming the finite support size of random variables),
sufficiently simple,
and intuitively understandable
formula of the regions.

\appendix

\subsection{Basic Information-Theoretic Equalities/Inequalities}
\label{sec:entropy}

In this section, we review some information-theoretic (in)equalities.
In the proofs, we use the following trivial facts without notice,
where they are sometimes applied many times in a single
(in)equality.

\begin{lem}
Let $A,B,C,D$ be random variables.
Then we have
\begin{align*}
H(A|B)&\geq 0
\\
I(A;B|C)&\geq 0
\\
H(A,B|C)&= H(B,A|C)
\\
I(A;B|C)&= I(B;A|C)
\\
H(A,A|B)&= H(A|B)
\\
I(A,A;B|C)&= I(A;B|C)
\\
H(A|B,B)&= H(A|B)
\\
I(A;B|C,C)&= I(A;B|C)
\\
H(A,B|B,C)&= H(A|B,C)
\\
I(A;B,C|C,D)
&=
I(A;B|C,D)
\\
I(A;A|B)&= H(A|B)
\\
H(A,B|C)
&=
H(A|C) + H(B|C) - I(A;B|C)
\\
H(A,B|C)
&=
H(A|B,C) + H(B|C)
\\
I(A;B,C|D)
&=
I(A;B|D)
+
I(A;C|B,D)
\\
H(A|B)
&\geq
H(A|B,C)
\\
H(A,B|C)
&\geq
H(A|C)
\\
I(A,B;C|D)
&\geq
I(A;C|D)
\\
H(A,B|C)
&\geq
H(A|B,C)
\\
I(A;B,C|D)
&\geq
I(A;B|C,D)
\\
H(A|C)
+
H(B|C)
&\geq
H(A,B|C)
\\
H(\lambda(A)|A)
&=
0,
\end{align*}
where $\lambda$ is an arbitrary deterministic function,
and we can use notations
$H(A|)\equiv H(A)$, $I(A;B|)\equiv I(A;B)$,
$H(|A)\equiv 0$, $I(;A|B)\equiv 0$
to apply the above relations.
\end{lem}

In the proof, we use the following lemmas.

\begin{lem}
\label{lem:xor}
Let $A,B,C$ be random variables. Then we have
\begin{equation*}
H(A,B|C) \geq H(A|B,C) + H(B|A,C).
\end{equation*}
\end{lem}
\begin{IEEEproof}
We have
\begin{align}
 H(A,B|C)
 &=
 H(A|B,C) + H(B|C)
 \notag
 \\
 &\geq
 H(A|B,C) + H(B|A,C).
\end{align}
\end{IEEEproof}

\begin{lem}
\label{lem:promotion}
Let $A,B,C,D$ be random variables. Then we have
\begin{equation*}
H(A|B,C,D) \geq H(A,B|C,D) - H(B|C).
\end{equation*}
\end{lem}
\begin{IEEEproof}
We have
\begin{align}
 H(A|B,C,D)
 &=
 H(A,B|C,D) - H(B|C,D)
 \notag
 \\
 &\geq
 H(A,B|C,D) - H(B|C).
\end{align}
\end{IEEEproof}

\begin{lem}
\label{lem:relocation}
Let $A,B,C,D$ be random variables. Then we have
\begin{equation*}
H(A,B,C|D) + H(A|B,C,D)
\geq
H(A,B|C,D) + H(A,C|B,D).
\end{equation*}
\end{lem}
\begin{IEEEproof}
We have
\begin{align}
 H(A,B,C|D) + H(A|B,C,D)
 &=
 H(A,B|C,D) + H(C|D) + H(A|B,C,D)
 \notag
 \\
 &\geq
 H(A,B|C,D) + H(C|B,D) + H(A|B,C,D)
 \notag
 \\
 &=
 H(A,B|C,D) + H(A,C|B,D).
\end{align}
\end{IEEEproof}

\begin{lem}
\label{lem:markov}
Let $A,B,C$ be random variables satisfying $I(A;B|C)=0$.
Then we have
\begin{equation*}
H(A|B,C)
=
H(A|C).
\end{equation*}
\end{lem}
\begin{IEEEproof}
We have
\begin{align}
 H(A|B,C)
 &=
 H(A|B,C)
 + I(A;B|C)
 \notag
 \\
 &=
 H(A|B,C)
 + H(A|C)
 - H(A|B,C)
 \notag
 \\
 &=
 H(A|C),
\end{align}
where the equality comes from the assumption.
\end{IEEEproof}

\begin{lem}
\label{lem:c-independence-H(A,B|C)}
Let $A,B,C$ be random variables
satisfying $I(A;B|C) = 0$.
Then we have
\begin{equation*}
H(A,B|C)
=
H(A|C) + H(B|C).
\end{equation*}
\end{lem}
\begin{IEEEproof}
We have
\begin{align}
 H(A,B|C)
 &=
 H(A|C) + H(B|C) - I(A;B|C)
 \notag
 \\
 &=
 H(A|C) + H(B|C),
\end{align}
where the equality comes from the assumption.
\end{IEEEproof}

\begin{lem}
\label{lem:c-independence-I(A;C|D)}
Let $A,B,C,D$ be random variables satisfying $I(A,B;C|D)=0$.
Then we have
\begin{equation*}
I(A;C|D)=0.
\end{equation*}
\end{lem}
\begin{IEEEproof}
We have $I(A;C|D)=0$
the fact that $I(A;C|D)\geq 0$ and
\begin{align}
 I(A;C|D)
 &\leq
 I(A,B;C|D)
 \notag
 \\*
 &= 
 0,
\end{align}
where the equality comes from the assumption.
\end{IEEEproof}

\begin{lem}
\label{lem:c-independence-I(A;B,C|D)}
Let $A,B,C,D$ be random variables.
satisfying $I(A,B;C|D)=0$.
Then we have
\begin{align*}
I(A;B,C|D)
&= 
I(A;B|D).
\end{align*}
\end{lem}
\begin{IEEEproof}
We have
\begin{align}
 I(A;B,C|D)
 &= 
 I(A;B|D)+I(A;C|B,D)
 \notag
 \\*
 &= 
 I(A;B|D)+I(A,B;C|D) - I(B;C|D)
 \notag
 \\*
 &=
 I(A;B|D),
\end{align}
where the last equality comes from the assumption
and Lemma \ref{lem:c-independence-I(A;C|D)}.
\end{IEEEproof}

\begin{lem}
\label{lem:c-independence-I(A;B|C,D)}
Let $A,B,C,D$ be random variables satisfying $I(A,B;C|D)=0$.
Then we have
\begin{equation*}
I(A;B|C,D)
=
I(A;B|D).
\end{equation*}
\end{lem}
\begin{IEEEproof}
We have
\begin{align}
 I(A;B|C,D)
 &=
 I(A;B,C|D) - I(A;C|D)
 \notag
 \\
 &=
 I(A;B|D),
\end{align}
where the last equality comes from the assumption
and Lemma \ref{lem:c-independence-I(A;C|D)}.
\end{IEEEproof}
\begin{lem}
\label{lem:Q}
Let $A,B,C,D$ be random variables satisfying $I(Q;A,B,C)=0$.
Then we have
\begin{align*}
I(Q;A)
&= 0
\\
I(Q,A;B)
&=
I(A;B)
\\
I(Q,A;Q,B)
&=
H(Q) + I(A;B)
\\
I(Q,A;B|C)
&=
I(A;B|C)
\\
I(A;B|Q,C)
&=
I(A;B|C)
\\
I(Q,A;B|Q,C)
&=
I(A;B|C).
\end{align*}
\end{lem}
\begin{IEEEproof}
We have $I(Q;A)=0$ from the fact that $I(Q;A)\geq 0$ and
\begin{align}
 I(Q;A)
 &\leq
 I(Q;A,B,C)
 \notag
 \\
 &= 
 0,
\end{align}
where the equality comes from the assumption.
Other inequalities are shown as
\begin{align}
 I(Q,A;B)
 &=
 H(Q,A) + H(B) - H(Q,A,B)
 \notag
 \\
 &=
 H(Q) + H(A) - I(Q;A)
 + H(B) - H(Q) - H(A,B)
 + I(Q;A,B)
 \notag
 \\
 &=
 H(A) + H(B) - H(A,B)
 \notag
 \\
 &=
 I(A;B)
 \\
 I(Q,A;Q,B)
 &=
 H(Q,A) + H(Q,B) - H(Q,A,B)
 \notag
 \\
 &=
 H(Q) + H(A)
 - I(Q;A)
 + H(Q) + H(B) - I(Q;B)
 - H(Q) - H(A,B) + I(Q;A,B)
 \notag
 \\
 &=
 H(Q) + H(A)
 + H(B) - H(A,B)
 \notag
 \\
 &=
 H(Q) + I(A;B)
 \\
 I(Q,A;B|C)
 &=
 I(Q,A;B,C) - I(Q,A;C)
 \notag
 \\
 &=
 I(A;B,C) - I(A;C)
 \notag
 \\
 &=
 I(A;B|C)
 \\
 I(A;B|Q,C)
 &=
 I(Q,A,C;B) - I(Q,C;B)
 \notag
 \\
 &=
 I(A,C;B) - I(C;B)
 \notag
 \\
 &=
 I(A;B|C)
 \\
 I(Q,A;B|Q,C)
 &=
 I(A;B|Q,C)
 \notag
 \\
 &=
 I(A;B|C).
\end{align}
\end{IEEEproof}

\begin{lem}
\label{lem:function}
For random variables $A$ and $B$ and a deterministic function $\lambda$,
we have
\begin{equation*}
I(A,\lambda(A,C);B|C)
=
I(A;B|C).
\end{equation*}
\end{lem}
\begin{IEEEproof}
Since
\begin{align}
 H(A,\lambda(A,C)|C)
 &=
 H(A,C,\lambda(A,C))
 - H(C)
 \notag
 \\
 &=
 H(\lambda(A,C)|A,C)
 + H(A,C)
 - H(C)
 \notag
 \\
 &=
 H(A|C),
\end{align}
we have
\begin{align}
 I(A,\lambda(A,C);B|C)
 &=
 H(A,\lambda(A,C)|C)
 -
 H(A,\lambda(A,C)|B,C)
 \notag
 \\
 &=
 H(A|C)
 -
 H(A|B,C)
 \notag
 \\
 &=
 I(A;B|C).
\end{align}
\end{IEEEproof}

\subsection{Entropy and Mutual Information for General Sources}
\label{sec:ispec}

First, we review the definition of
the limit superior/inferior in probability
introduced in \cite{HAN}.
For sequence $\{U_n\}_{n=1}^{\infty}$ of random variables,
the {\em limit superior in probability} $\plimsupn U_n$
and the {\em limit inferior in probability} $\pliminfn U_n$ are defined
as
\begin{align*}
\plimsupn U_n
&\equiv 
\inf\lrb{\theta: \limn \Prob\lrsb{U_n>\theta}=0}
\\
\pliminfn U_n
&\equiv
\sup\lrb{\theta: \limn \Prob\lrsb{U_n<\theta}=0}.
\end{align*}
We have the following relations~\cite[Section 1.3]{HAN}:
\begin{align}
\plimsupn\lrB{-U_n}
&=\pliminfn U_n.
\label{eq:plimsup-pliminf}
\\
\pliminfn\lrB{U_n+V_n}
&\leq \plimsupn U_n + \pliminfn V_n
\label{eq:pliminf-upper}
\\
\pliminfn\lrB{U_n+V_n}
&\geq \pliminfn U_n + \pliminfn V_n.
\label{eq:pliminf-lower}
\end{align}

For sequence $\{\mu_{U_n}\}_{n=1}^{\infty}$
of probability distributions corresponding to $\UU\equiv\{U_n\}_{n=1}^{\infty}$,
we define the spectral inf-entropy rate $\uH(\UU)$ as
\begin{align*}
\uH(\UU)
&\equiv
\pliminfn\frac 1n\log_2\frac1{\mu_{U_n}(U^n)}.
\end{align*}

For general sequence $\{\mu_{U_nV_n}\}_{n=1}^{\infty}$ of
joint probability distributions
corresponding to $(\UU,\VV)=\{(U_n,V_n)\}_{n=1}^{\infty}$,
we define the spectral conditional sup-entropy rate $\oH(\UU|\VV)$,
and the spectral conditional inf-entropy rate $\uH(\UU|\VV)$,
as
\begin{align*}
\oH(\UU|\VV)
&\equiv
\plimsupn\frac 1n\log_2\frac1{\mu_{U_n|V_n}(U_n|V_n)}
\\
\uH(\UU|\VV)
&\equiv
\pliminfn\frac 1n\log_2\frac1{\mu_{U_n|V_n}(U_n|V_n)}.
\end{align*}

We show the following lemmas.
\begin{lem}
$\uH(\UU|\VV)\geq \uH(\UU|\VV,\WW)$.
\end{lem}
\begin{IEEEproof}
From \cite[Lemma 3.2.1]{HAN},
we have
\begin{equation}
 \pliminfn\frac 1n\log_2\frac{\mu_{U_n}(U_n)}{\nu_n(U_n)}\geq 0,
 \label{eq:pliminf-div}
\end{equation}
where $\nu_n$ is an arbitrary distribution on
the support of $U_n$.
Then we have
\begin{align}
 \uH(\UU|\VV)-\uH(\UU|\VV,\WW)
 &=
 \pliminfn\frac 1n\log_2
 \frac1{\mu_{U_n|V_n}(U_n|V_n)}
 -
 \pliminfn\frac 1n\log_2\frac1{\mu_{U_n|V_nW_n}(U_n|V_n,W_n)}
 \notag
 \\*
 &=
 \pliminfn\frac 1n\log_2
 \frac1{\mu_{U_n|V_n}(U_n|V_n)}
 +
 \plimsupn\frac 1n\log_2\mu_{U_n|V_nW_n}(U_n|V_n,W_n)
 \notag
 \\
 &\geq
 \pliminfn\frac 1n\log_2
 \frac{\mu_{U_n|V_nW_n}(U_n|V_n,W_n)}{\mu_{U_n|V_n}(U_n|V_n)}
 \notag
 \\
 &=
 \pliminfn\frac 1n\log_2
 \frac{\mu_{U_nV_nW_n}(U_n,V_n,W_n)}
 {\mu_{U_n|V_n}(U_n|V_n)\mu_{V_nW_n}(V_n,W_n)}
 \notag
 \\
 &\geq
 0,
\end{align}
where the second equality comes from (\ref{eq:plimsup-pliminf}),
the first inequality comes from (\ref{eq:pliminf-upper}),
and the second inequality comes from
(\ref{eq:pliminf-div}).
\end{IEEEproof}

\begin{lem}
\label{lem:ispec-markov}
When
$\mu_{U_nW_n|V_n}(\uu,\ww|\vv)=\mu_{U_n|V_n}(\uu|\vv)\mu_{W_n|V_n}(\ww|\vv)$
for all $n\in\NN$ and $(\uu,\vv,\ww)\in\U_n\times\V_n\times W_n$, we have
\begin{equation*}
\uH(\UU,\WW|\VV)
\geq
\uH(\UU|\VV) + \uH(\WW|\VV).
\end{equation*}
\end{lem}
\begin{IEEEproof}
We have
\begin{align}
 \uH(\UU,\WW|\VV)
 &=
 \pliminfn\frac 1n\log_2
 \frac1{\mu_{U_n,W_n|V_n}(U_n,W_n|V_n)}
 \notag
 \\
 &=
 \pliminfn\frac 1n\log_2
 \frac1{\mu_{U_n|V_n}(U_n|V_n)\mu_{W_n|V_n}(W_n|V_n)}
 \notag
 \\
 &\geq
 \pliminfn\frac 1n\log_2
 \frac1{\mu_{U_n|V_n}(U_n|V_n)}
 + \pliminfn\frac 1n\log_2
 \frac1{\mu_{W_n|V_n}(W_n|V_n)}
 \notag
 \\
 &\geq
 \uH(\UU|\VV) + \uH(\WW|\VV),
\end{align}
where the first inequality comes from (\ref{eq:pliminf-lower}).
\end{IEEEproof}

\subsection{$(\aalpha,\bbeta)$-hash property}
\label{sec:hash}

In this section we review the hash property
introduced in \cite{CRNG,ISIT2010,SW2CC,CRNG-MULTI} and basic lemmas.

\begin{df}[{\cite[Definition~3]{CRNG}}]
Let $\F_n$ be a set of functions on $\U^n$.
For probability distribution $p_{F_n}$ on $\F_n$, we
call the pair $(\F_n,p_{F_n})$ an {\em ensemble}.
Then, $(\F_n,p_{F_n})$ has an
$(\alpha_{F_n},\beta_{F_n})$-{\em hash property}
if there is a pair $(\alpha_{F_n},\beta_{F_n})$
depending on $p_{F_n}$ such that
\begin{align}
\sum_{\substack{
  \tzz\in\U^n\setminus\{\zz\}:
  \\
  p_{F_n}(\{f: f(\zz) = f(\tzz)\})>\frac{\alpha_{F_n}}{|\im\F_n|}
}}
p_{F_n}\lrsb{\lrb{f: f(\zz) = f(\tzz)}}
\leq
\beta_{F_n}
\label{eq:hash}
\end{align}
for any $\zz\in\Z^n$,
where
\begin{equation*}
\im\F_n\equiv\bigcup_{f\in\F_n}\{f(\zz):\zz\in\Z^n\}.
\end{equation*}
Then, we say that 
$(\bcF,\bp_F)$ has an $(\aalpha_F,\bbeta_F)$-{\em hash property}
if
$\aalpha_F\equiv\{\alpha_{F_n}\}_{n=1}^{\infty}$ and
$\bbeta_F\equiv\{\beta_{F_n}\}_{n=1}^{\infty}$
satisfy
\begin{align}
\limn \alpha_{F_n}
&=1
\label{eq:alpha-bcp}
\\
\limn \beta_{F_n}
&=0.
\label{eq:beta-crp}
\end{align}
Throughout this paper,
we omit the dependence of $\F$ and $F$ on $n$.
\end{df}

It should be noted that
when $\F$ is a two-universal class of hash functions \cite{CW}
and $p_F$ is the uniform distribution on $\F$,
then $(\F,p_F)$ has a $(1,0)$-hash property.
Random binning \cite{C75}
and the set of all linear functions \cite{CSI82} are
typical examples of two-universal class of hash functions.
It is proved in \cite[Section III-B of the extended version]{ISIT2010} that
an ensemble of sparse matrices (with logarithmic column degree)
has a hash property.

First, we introduce the lemma for a joint ensemble.
\begin{lem}[
{\cite[Lemma 4 of the extended version]{ISIT2010}\cite[Lemma 3]{CRNG}}
]
\label{lem:hash-FG}
Let $(\F,p_F)$ and $(\G,p_G)$ be ensembles of functions
on the same set $\Z^n$.
Assume that $(\F,p_F)$ (resp. $(\G,p_G)$) has an $(\alpha_F,\beta_F)$-hash
(resp. $(\alpha_G,\beta_G)$-hash) property.
Let $(f,g)\in\F\times\G$ be a function defined as
\begin{equation*}
(f,g)(\zz)\equiv(f(\uu),g(\uu))\quad\text{for each}\ \zz\in\Z^n.
\end{equation*}
Let $p_{(F,G)}$  be a joint distribution on $\F\times\G$ defined as
\begin{equation*}
p_{(F,G)}(f,g)\equiv p_F(f)p_G(g)\quad\text{for each}\ (f,g)\in\F\times\G.
\end{equation*}
Then the ensemble $(\F\times\G, p_{(F,G)})$ has an
$(\alpha_{(F,G)},\beta_{(F,G)})$-hash property,
where $(\alpha_{(F,G)},\beta_{(F,G)})$ is defined as
\begin{align*}
\alpha_{(F,G)}
&\equiv
\alpha_F\alpha_G
\\
\beta_{(F,G)}
&\equiv
\beta_F+\beta_G.
\end{align*}
\end{lem}

Next, we introduce lemmas that are multiple extensions of
the {\it collision-resistant property} and
the {\it balanced-coloring property}.
We use the following notations.
For each $s\in\cS$, let $\F_s$ be a set of functions on $\Z_s^n$.
For given set $\cS$, let $\im\F_{\cS}\equiv\Prod_{s\in\cS}\im\F_s$.
For given $f_{\cS}\equiv\{f_s\}_{s\in\cS}\in\F_{\cS}$
and $\cc_{\cS}\equiv\{\cc_s\}_{s\in\cS}\in\im\F_{\cS}$,
let
\begin{equation*}
\fC_{f_{\cS}}(\cc_{\cS})
\equiv
\lrb{
 \zz_s\in\Z^n_s: f_{s,n}(\zz_s)=\cc_s
 \ \text{for all}\ s\in\cS
}.
\end{equation*}

Let $\Z_{\tcS}^n\equiv\Prod_{s\in\tcS}\Z_s^n$ and 
\begin{align*}
\alpha_{F_{\tcS}}
&\equiv
\prod_{s\in\tcS}\alpha_{F_s}
\\
\beta_{F_{\tcS}}
&\equiv
\prod_{s\in\tcS}\lrB{\beta_{F_s}+1}-1,
\end{align*}
where $\prod_{s\in\emptyset}\theta_s\equiv1$.
It should be noted that
\begin{align*}
\limn \alpha_{F_{\tcS}}=1
\\
\limn \beta_{F_{\tcS}}=0
\end{align*}
for every $\tcS\subset\cS$
when $(\aalpha_{F_s},\bbeta_{F_s})$ satisfies 
(\ref{eq:alpha-bcp}) and (\ref{eq:beta-crp}) for all $s\in\cS$.
For $\T\subset\Z_{\cS}^n$ and $\zz_{\tcS}\in\Z^n_{\tcS}$,
let $\T_{\tcS}$ and $\T_{\tcS^{\complement}|\tcS}(\zz_{\tcS})$ be defined as
\begin{align*}
&
\T_{\tcS}
\equiv\{\zz_{\tcS}:
 (\zz_{\tcS},\zz_{\tcS^{\complement}})\in\T
 \ \text{for some}\ \zz_{\tcS^{\complement}}\in\Z_{\tcS^{\complement}}
 \}
\\
&
\T_{\tcS^{\complement}|\tcS}(\zz_{\tcS})
\equiv
\{\zz_{\tcS^{\complement}}: (\zz_{\tcS},\zz_{\tcS^{\complement}})\in\T\},
\end{align*}
where $\tcS^{\complement}\equiv\cS\setminus\tcS$.

The following lemma is a multiple extension of
the {\it collision-resistant property}.
This lemma implies that
there is an assignment such that every bin contains at most one item.
\begin{lem}[{\cite[Lemma 7 of the extended version]{ISIT2010}}]
\label{lem:mCRP}
For each $s\in\cS$, let $\F_s$ be a set
of functions on $\Z_s^n$
and $p_{F_s}$ be the probability distribution on $\F_s$,
where $(\F_s,p_{F_s})$ satisfies (\ref{eq:hash}).
We assume that random variables in $F_{\cS}\equiv\{F_s\}_{s\in\cS}$
are mutually independent.
Then
\begin{align*}
&
p_{F_{\cS}}\lrsb{\lrb{
  f_{\cS}:
  \lrB{\T\setminus\{\zz_{\cS}\}}\cap\fC_{f_{\cS}}(f_{\cS}(\zz_{\cS}))
  \neq
  \emptyset
}}
\leq
\sum_{
 \tcS\subset\cS:
 \tcS\neq\emptyset
}
\frac{
 \alpha_{F_{\tcS}}\lrB{\beta_{F_{\tcS^{\complement}}}+1}
 \oO_{\tcS}
}
{
 \prod_{s\in\tcS}\lrbar{\im\F_s}
}
+\beta_{F_{\cS}}
\end{align*}
for all $\T\subset\Z_{\cS}^n$ and $\zz_{\cS}\in\Z_{\cS}^n$,
where
\begin{align*}
\oO_{\tcS}
&\equiv
\begin{cases}
 |\T|
 &\text{if}\ \tcS=\cS,
 \\
 \displaystyle\max_{\zz_{\tcS^{\complement}}\in\T_{\cS^{\complement}}}
 \lrbar{\T_{\tcS|\tcS^{\complement}}\lrsb{\zz_{\tcS^{\complement}}}},
 &\text{if}\ \emptyset\neq\tcS\subsetneq\cS.
\end{cases}
\end{align*}
\end{lem}

The following lemma is related to the {\em balanced-coloring property},
which is an extension of the leftover hash lemma~\cite{IZ89}
and the balanced-coloring lemma~\cite[Lemma 3.1]{AC98}\cite[Lemma 17.3]{CK11}.
This lemma implies that there is an assignment that divides
a set equally.
\begin{lem}[{\cite[Lemma 4 of the extended version]{CRNG-MULTI}}]
\label{lem:mBCP}
For each $s\in\cS$, let $\F_s$ be a set of functions on $\Z_s^n$
and $p_{F_s}$ be the probability distribution on $\F_s$,
where $(\F_s,p_{F_s})$ satisfies (\ref{eq:hash}).
We assume that random variables in $F_{\cS}\equiv\{F_s\}_{s\in\cS}$
are mutually independent.
Then
\begin{align*}
E_{F_{\cS}}\lrB{
 \sum_{\cc_{\cS}\in\im\F_{\cS}}
 \lrbar{
  \frac{Q(\T\cap\fC_{F_{\cS}}(\cc_{\cS}))}
  {Q(\T)}
  -
  \frac1
  {\prod_{s\in\cS}|\im\F_s|}
 }
}
&\leq
\sqrt{
 \alpha_{F_{\cS}}-1
 +
 \sum_{
  \tcS\subset\cS:
  \tcS\neq\emptyset
 }
 \alpha_{F_{\tcS^{\complement}}}
 \lrB{\beta_{F_{\cS}}+1}
 \lrB{\prod_{s\in\tcS}|\im\F_s|}
 \frac{
  \oQ_{\tcS^{\complement}}
 }
 {Q(\T)}
}
\end{align*}
for any function $Q:\Z_{\cS}\to[0,\infty)$ and $\T\subset\Z_{\cS}^n$,
where
\begin{equation*}
\oQ_{\tcS^{\complement}}
\equiv
\begin{cases}
 \displaystyle
 \max_{\zz_{\cS}\in\T}Q(\zz_{\cS})
 &\text{if}\ \tcS^{\complement}=\cS
 \\
 \displaystyle
 \max_{\zz_{\tcS}\in\T_{\tcS}}
 \sum_{\zz_{\tcS^{\complement}}\in\T_{\tcS^{\complement}|\tcS}(\zz_{\tcS})}
 Q(\zz_{\tcS},\zz_{\tcS^{\complement}})
 &\text{if}\ \emptyset\neq\tcS^{\complement}\subsetneq\cS.
\end{cases}
\end{equation*}
\end{lem}

\subsection{Proof of Lemma \ref{lem:crng-fme}}
\label{sec:crng-fme}

First, we show the following lemma.
\begin{lem}
\label{lem:fme}
By using the Fourier-Motzkin method to eliminate $\rho_1$ and $\rho_2$ from
\begin{gather}
\urho_i\leq \rho_i \leq \orho_i
\label{eq:fme-rho}
\\
\rho_i-\rho_{i'}\leq \delta_i
\\
\rho_1 + \rho_2\geq \sigma_i
\label{eq:fme-sigma}
\end{gather}
for all $i\in\{1,2\}$,
we have the set of conditions
\begin{align}
0
&\leq
\delta_1 + \delta_2
\label{eq:fme-delta1+delta2}
\\
\urho_i
&\leq
\min\{\orho_i,\orho_{i'} + \delta_i\}
\\
\sigma_i
&\leq
\min\{
 \orho_1 + \orho_2,
 2\orho_{i'} + \delta_i,
 2\orho_i + \delta_{i'}
 \}
\label{eq:fme-sigma<min}
\end{align}
for all $i\in\{1,2\}$.
\end{lem}
\begin{IEEEproof}
By using the Fourier-Motzkin method to eliminate $\rho_1$ from
\begin{align*}
 \rho_1
 &\leq
 \orho_1
 \\
 \rho_1
 &\leq
 \rho_2 + \delta_1
 \\
 \rho_1
 &\geq 
 \urho_1
 \\
 \rho_1
 &\geq
 \rho_2 - \delta_2
 \\
 \rho_1
 &\geq
 \sigma_1 - \rho_2
 \\
 \rho_1
 &\geq
 \sigma_2 - \rho_2
 \\
 \rho_2
 &\leq
 \orho_2
 \\
 \rho_2
 &\geq
 \urho_2,
\end{align*}
which are equivalent to
the set of conditions (\ref{eq:fme-rho})--(\ref{eq:fme-sigma}),
we have
\begin{align}
 \urho_1
 &\leq
 \orho_1
 \label{eq:fme-urho1<orho1}
 \\
 \rho_2 - \delta_2
 &\leq
 \orho_1
 \\
 \sigma_1 - \rho_2
 &\leq
 \orho_1
 \\
 \sigma_2 - \rho_2
 &\leq
 \orho_1
 \\
 \urho_1
 &\leq
 \rho_2 + \delta_1
 \\
 \rho_2 - \delta_2
 &\leq
 \rho_2 + \delta_1
 \\
 \sigma_1 - \rho_2
 &\leq
 \rho_2 + \delta_1
 \\
 \sigma_2 - \rho_2
 &\leq
 \rho_2 + \delta_1
 \\
 \rho_2
 &\leq
 \orho_2
 \\
 \rho_2
 &\geq
 \urho_2.
 \label{eq:fme-rho2>urho2}
\end{align}
By using the Fourier-Motzkin method to eliminate $\rho_2$ from
\begin{align*}
 0
 &\leq
 \delta_1 + \delta_2
 \\
 \urho_1
 &\leq
 \orho_1
 \\
 \rho_2
 &\leq
 \orho_2
 \\
 \rho_2
 &\leq
 \orho_1 + \delta_2
 \\
 \rho_2
 &\geq
 \urho_2
 \\
 \rho_2
 &\geq
 \sigma_1-\orho_1
 \\
 \rho_2
 &\geq
 \sigma_2 - \orho_1
 \\
 \rho_2
 &\geq
 \urho_1 - \delta_1
 \\
 2\rho_2
 &\geq
 \sigma_1 - \delta_1
 \\
 2\rho_2
 &\geq
 \sigma_2 - \delta_1,
\end{align*}
which are equivalent to the set of conditions
(\ref{eq:fme-urho1<orho1})--(\ref{eq:fme-rho2>urho2}),
we have
\begin{align}
 0
 &\leq
 \delta_1 + \delta_2
 \label{eq:fme-0<delta1+delta2}
 \\
 \urho_1
 &\leq
 \orho_1
 \\
 \urho_2
 &\leq
 \orho_2
 \\
 \sigma_1 - \orho_1
 &\leq
 \orho_2
 \\
 \sigma_2 - \orho_1
 &\leq
 \orho_2
 \\
 \urho_1 - \delta_1
 &\leq
 \orho_2
 \\
 \sigma_1 - \delta_1
 &\leq
 2\orho_2
 \\
 \sigma_2 - \delta_1
 &\leq
 2\orho_2
 \\
 \urho_2
 &\leq
 \orho_1 + \delta_2
 \\
 \sigma_1 - \orho_1
 &\leq
 \orho_1 + \delta_2
 \label{eq:fme-sigma1-orho1<orho1+delta2}
 \\
 \sigma_2 - \orho_1
 &\leq
 \orho_1 + \delta_2
 \label{eq:fme-sigma2-orho1<orho1+delta2}
 \\
 \urho_1 - \delta_1
 &\leq
 \orho_1 + \delta_2
 \label{eq:fme-urho1-delta1<orho1+delta2}
 \\
 \sigma_1 - \delta_1
 &\leq
 2[\orho_1 + \delta_2]
 \label{eq:fme-sigma1-delta1<2[orho1+delta2]}
 \\
 \sigma_2 - \delta_1
 &\leq
 2[\orho_1 + \delta_2],
 \label{eq:fme-sigma2-delta1<2[orho1+delta2]}
\end{align}
where
(\ref{eq:fme-sigma1-delta1<2[orho1+delta2]})
and 
(\ref{eq:fme-sigma2-delta1<2[orho1+delta2]})
are redundant because it comes from
(\ref{eq:fme-0<delta1+delta2}),
(\ref{eq:fme-sigma1-orho1<orho1+delta2})
and (\ref{eq:fme-sigma2-orho1<orho1+delta2}),
as
\begin{align}
 \sigma_i - \delta_1
 &=
 [\sigma_i - \orho_1] + \orho_1
 - [\delta_1 + \delta_2] + \delta_2
 \notag
 \\
 &\leq
 \orho_1 + \delta_2 + \orho_1 + \delta_2
 \notag
 \\
 &=
 2[\orho_1 + \delta_2]
\end{align}
for all $i\in\{1,2\}$.
From (\ref{eq:fme-0<delta1+delta2})--(\ref{eq:fme-urho1-delta1<orho1+delta2}),
we have
\begin{align*}
 0
 &\leq
 \delta_1 + \delta_2
 \\
 \urho_1
 &\leq
 \orho_1
 \\
 \urho_2
 &\leq
 \orho_2
 \\
 \sigma_1
 &\leq
 \orho_1 + \orho_2
 \\
 \sigma_2
 &\leq
 \orho_1 + \orho_2
 \\
 \urho_1
 &\leq
 \orho_2 + \delta_1
 \\
 \urho_2
 &\leq
 \orho_1 + \delta_2
 \\
 \sigma_1
 &\leq
 2\orho_1 + \delta_2
 \\
 \sigma_2
 &\leq
 2\orho_1 + \delta_2
 \\
 \sigma_1
 &\leq
 2\orho_2 + \delta_1
 \\
 \sigma_2
 &\leq
 2\orho_2 + \delta_1,
\end{align*}
which are equivalent to
(\ref{eq:fme-delta1+delta2})--(\ref{eq:fme-sigma<min})
for all $i\in\{1,2\}$.
\end{IEEEproof}

From here, we show Lemma \ref{lem:crng-fme}.
Let
\begin{align}
\hr_0
&\equiv
H(Z_{00})-R_0
\label{eq:def-hr0}
\\
\hr_i
&\equiv
H(Z_{ii}|Z_{00})-R_i
\\
\tr_i
&\equiv
H(Z_{i0},Z_{ii}|Z_{00})-R_i
\label{eq:def-tri}
\end{align}
for each $i\in\{1,2\}$,
where we have the relation
\begin{equation}
\hr_i = \tr_i - H(Z_{i0}|Z_{00},Z_{ii}).
\label{eq:hr1=tr1-10|0011}
\end{equation}
Then we have
\begin{align*}
r_{00}
&\leq
\hr_0
\\
r_{i0}
&\leq
H(Z_{i0}|Z_{00})
\\
r_{ii}
&\leq
\hr_i
\\
r_{i0} + r_{ii}
&\leq
\tr_i
\\
r_{00}
&\geq
H(Z_{00}|Y_i,Z_{10},Z_{20},Z_{ii})
\\
r_{i0}
&\geq
H(Z_{i0}|Y_i,Z_{00},Z_{i'0},Z_{ii})
\\
r_{i'0}
&\geq
H(Z_{i'0}|Y_i,Z_{00},Z_{i0},Z_{ii})
\\
r_{ii}
&\geq
H(Z_{ii}|Y_i,Z_{00},Z_{10},Z_{20})
\\
r_{00} + r_{i0}
&\geq
H(Z_{00},Z_{i0}|Y_i,Z_{i'0},Z_{ii})
\\
r_{00} + r_{i'0}
&\geq
H(Z_{00},Z_{i'0}|Y_i,Z_{i0},Z_{ii})
\\
r_{00} + r_{ii}
&\geq
H(Z_{00},Z_{ii}|Y_i,Z_{10},Z_{20})
\\
r_{10} + r_{20}
&\geq
H(Z_{10},Z_{20}|Y_i,Z_{00},Z_{ii})
\\
r_{i0} + r_{ii}
&\geq
H(Z_{i0},Z_{ii}|Y_i,Z_{00},Z_{i'0})
\\
r_{i'0} + r_{ii}
&\geq
H(Z_{i'0},Z_{ii}|Y_i,Z_{00},Z_{i0})
\\
r_{00} + r_{10} + r_{20}
&\geq
H(Z_{00},Z_{10},Z_{20}|Y_i,Z_{ii})
\\
r_{00} + r_{i0} + r_{ii}
&\geq
H(Z_{00},Z_{i0},Z_{ii}|Y_i,Z_{i'0})
\\
r_{00} + r_{i'0} + r_{ii}
&\geq
H(Z_{00},Z_{i'0},Z_{ii}|Y_i,Z_{i0})
\\
r_{10} + r_{20} + r_{ii}
&\geq
H(Z_{10},Z_{20},Z_{ii}|Y_i,Z_{00})
\\
r_{00} + r_{10} + r_{20} + r_{ii}
&\geq
H(Z_{00},Z_{10},Z_{20},Z_{ii}|Y_i)
\end{align*}
from (\ref{eq:crng-region0-first})--(\ref{eq:crng-region0-last}).

By using the Fourier-Motzkin method to eliminate $r_{00}$, we have
\begin{align*}
r_{i0}
&\leq
H(Z_{i0}|Z_{00})
\\
r_{ii}
&\leq
\hr_i
\\
r_{i0} + r_{ii}
&\leq
\tr_i 
\\
\hr_0
&\geq
H(Z_{00}|Y_i,Z_{10},Z_{20},Z_{ii})
\\
r_{i0}
&\geq
H(Z_{i0}|Y_i,Z_{00},Z_{i'0},Z_{ii})
\\
r_{i'0}
&\geq
H(Z_{i'0}|Y_i,Z_{00},Z_{i0},Z_{ii})
\\
r_{ii}
&\geq
H(Z_{ii}|Y_i,Z_{00},Z_{10},Z_{20})
\\
\hr_0 + r_{i0}
&\geq
H(Z_{00},Z_{i0}|Y_i,Z_{i'0},Z_{ii})
\\
\hr_0 + r_{i'0}
&\geq
H(Z_{00},Z_{i'0}|Y_i,Z_{i0},Z_{ii})
\\
\hr_0 + r_{ii}
&\geq
H(Z_{00},Z_{ii}|Y_i,Z_{10},Z_{20})
\\
r_{10} + r_{20}
&\geq
H(Z_{10},Z_{20}|Y_i,Z_{00},Z_{ii})
\\
r_{i0} + r_{ii}
&\geq
H(Z_{i0},Z_{ii}|Y_i,Z_{00},Z_{i'0})
\\
r_{i'0} + r_{ii}
&\geq
H(Z_{i'0},Z_{ii}|Y_i,Z_{00},Z_{i0})
\\
\hr_0 + r_{10} + r_{20}
&\geq
H(Z_{00},Z_{10},Z_{20}|Y_i,Z_{ii})
\\
\hr_0 + r_{i0} + r_{ii}
&\geq
H(Z_{00},Z_{i0},Z_{ii}|Y_i,Z_{i'0})
\\
\hr_0 + r_{i'0} + r_{ii}
&\geq
H(Z_{00},Z_{i'0},Z_{ii}|Y_i,Z_{i0})
\\
r_{10} + r_{20} + r_{ii}
&\geq
H(Z_{10},Z_{20},Z_{ii}|Y_i,Z_{00})
\\
\hr_0 + r_{10} + r_{20} + r_{ii}
&\geq
H(Z_{00},Z_{10},Z_{20},Z_{ii}|Y_i).
\end{align*}
By using the Fourier-Motzkin method to eliminate $r_{ii}$, we have
\begin{align*}
r_{i0}
&\leq
H(Z_{i0}|Z_{00})
\\
\hr_0
&\geq
H(Z_{00}|Y_i,Z_{10},Z_{20},Z_{ii})
\\
r_{i0}
&\geq
H(Z_{i0}|Y_i,Z_{00},Z_{i'0},Z_{ii})
\\
r_{i'0}
&\geq
H(Z_{i'0}|Y_i,Z_{00},Z_{i0},Z_{ii})
\\
\hr_i
&\geq
H(Z_{ii}|Y_i,Z_{00},Z_{10},Z_{20})
\\
\tr_i - r_{i0}
&\geq 
H(Z_{ii}|Y_i,Z_{00},Z_{10},Z_{20})
\\
\hr_0 + r_{i0}
&\geq
H(Z_{00},Z_{i0}|Y_i,Z_{i'0},Z_{ii})
\\
\hr_0 + r_{i'0}
&\geq
H(Z_{00},Z_{i'0}|Y_i,Z_{i0},Z_{ii})
\\
\hr_0 + \hr_i
&\geq
H(Z_{00},Z_{ii}|Y_i,Z_{10},Z_{20})
\\
\hr_0 + \tr_i - r_{i0}
&\geq
H(Z_{00},Z_{ii}|Y_i,Z_{10},Z_{20})
\\
r_{10} + r_{20}
&\geq
H(Z_{10},Z_{20}|Y_i,Z_{00},Z_{ii})
\\
r_{i0} 
+ \hr_i
&\geq
H(Z_{i0},Z_{ii}|Y_i,Z_{00},Z_{i'0})
\\
r_{i0} 
+ \tr_i - r_{i0}
&\geq
H(Z_{i0},Z_{ii}|Y_i,Z_{00},Z_{i'0})
\\
r_{i'0} 
+ \hr_i
&\geq
H(Z_{i'0},Z_{ii}|Y_i,Z_{00},Z_{i0})
\\
r_{i'0} 
+ \tr_i - r_{i0}
&\geq
H(Z_{i'0},Z_{ii}|Y_i,Z_{00},Z_{i0})
\\
\hr_0
+ r_{10} + r_{20}
&\geq 
H(Z_{00},Z_{10},Z_{20}|Y_i,Z_{ii})
\\
\hr_0
+ r_{i0}
+ \hr_i
&\geq
H(Z_{00},Z_{i0},Z_{ii}|Y_i,Z_{i'0})
\\
\hr_0
+ r_{i0}
+ \tr_i - r_{i0}
&\geq
H(Z_{00},Z_{i0},Z_{ii}|Y_i,Z_{i'0})
\\
\hr_0
+ r_{i'0}
+ \hr_i
&\geq
H(Z_{00},Z_{i'0},Z_{ii}|Y_i,Z_{i0})
\\
\hr_0
+ r_{i'0}
+ \tr_i - r_{i0}
&\geq
H(Z_{00},Z_{i'0},Z_{ii}|Y_i,Z_{i0})
\\
r_{10} + r_{20}
+ \hr_i
&\geq
H(Z_{10},Z_{20},Z_{ii}|Y_i,Z_{00})
\\
r_{10} + r_{20}
+ \tr_i - r_{i0}
&\geq
H(Z_{10},Z_{20},Z_{ii}|Y_i,Z_{00})
\\
\hr_0
+ r_{10} + r_{20}
+ \hr_i
&\geq
H(Z_{00},Z_{10},Z_{20},Z_{ii}|Y_i)
\\
\hr_0
+ r_{10} + r_{20}
+ \tr_i - r_{i0}
&\geq
H(Z_{00},Z_{10},Z_{20},Z_{ii}|Y_i).
\end{align*}
Then we have
\begin{align}
r_{i0}
&\leq
H(Z_{i0}|Z_{00})
\label{eq:r10>10|00}
\\
\hr_0
&\geq 
H(Z_{00}|Y_i,Z_{10},Z_{20},Z_{ii})
\label{eq:hr0>00|1}
\\
r_{i0}
&\geq
H(Z_{i0}|Y_i,Z_{00},Z_{i'0},Z_{ii})
\label{eq:r10>10|1}
\\
r_{i'0}
&\geq
H(Z_{i'0}|Y_i,Z_{00},Z_{i0},Z_{ii})
\label{eq:r20>20|1}
\\
\hr_i
&\geq
H(Z_{ii}|Y_i,Z_{00},Z_{10},Z_{20})
\label{eq:hr1>11|1}
\\
r_{i0}
&\leq
\tr_i
- H(Z_{ii}|Y_i,Z_{00},Z_{10},Z_{20})
\label{eq:r10<tr1-11|1}
\\
r_{i0}
&\geq
H(Z_{00},Z_{i0}|Y_i,Z_{i'0},Z_{ii})
- \hr_0
\label{eq:r10>0010|1}
\\
r_{i'0}
&\geq
H(Z_{00},Z_{i'0}|Y_i,Z_{i0},Z_{ii})
- \hr_0
\label{eq:r20>0020|1}
\\
\hr_0 + \hr_i
&\geq
H(Z_{00},Z_{ii}|Y_i,Z_{10},Z_{20})
\label{eq:hr0+hr1>0011|1}
\\
r_{i0}
&\leq
\hr_0 + \tr_i
- H(Z_{00},Z_{ii}|Y_i,Z_{10},Z_{20})
\label{eq:r10<hr0+tr1-0011|1}
\\
r_{10} + r_{20}
&\geq
H(Z_{10},Z_{20}|Y_i,Z_{00},Z_{ii})
\label{eq:r10+r20>1020|1}
\\
r_{i0} 
&\geq
H(Z_{i0},Z_{ii}|Y_i,Z_{00},Z_{i'0})
- \hr_i
\label{eq:r10>1011|1}
\\
\tr_i
&\geq
H(Z_{i0},Z_{ii}|Y_i,Z_{00},Z_{i'0})
\label{eq:tr1>1011|1}
\\
r_{i'0} 
&\geq
H(Z_{i'0},Z_{ii}|Y_i,Z_{00},Z_{i0})
- \hr_i
\label{eq:r20>2011|1}
\\
r_{i'0} - r_{i0} 
&\geq
H(Z_{i'0},Z_{ii}|Y_i,Z_{00},Z_{i0})
- \tr_i
\label{eq:r20-r10>2011|1-tr1}
\\
r_{10} + r_{20}
&\geq
H(Z_{00},Z_{10},Z_{20}|Y_i,Z_{ii})
- \hr_0
\label{eq:r10+r20>001020|1}
\\
r_{i0}
&\geq
H(Z_{00},Z_{i0},Z_{ii}|Y_i,Z_{i'0})
- \hr_0
- \hr_i
\label{eq:r10>001011|1}
\\
\hr_0 + \tr_i
&\geq
H(Z_{00},Z_{i0},Z_{ii}|Y_i,Z_{i'0})
\label{eq:hr0+tr1>001011|1}
\\
r_{i'0}
&\geq
H(Z_{00},Z_{i'0},Z_{ii}|Y_i,Z_{i0})
- \hr_0
- \hr_i
\label{eq:r20>002011|1}
\\
r_{i'0} - r_{i0}
&\geq 
H(Z_{00},Z_{i'0},Z_{ii}|Y_i,Z_{i0})
- \hr_0 - \tr_i
\label{eq:r20-r10>002011|1-hr0-tr1}
\\
r_{10} + r_{20}
&\geq
H(Z_{10},Z_{20},Z_{ii}|Y_i,Z_{00})
- \hr_i
\label{eq:r10+r20>102011|1-hr1}
\\
r_{i'0}
&\geq
H(Z_{10},Z_{20},Z_{ii}|Y_i,Z_{00})
- \tr_i
\label{eq:r20>102011|1}
\\
r_{10} + r_{20}
&\geq
H(Z_{00},Z_{10},Z_{20},Z_{ii}|Y_i)
- \hr_0 - \hr_i
\label{eq:r10+r20>00102011|1}
\\
r_{i'0}
&\geq
H(Z_{00},Z_{10},Z_{20},Z_{ii}|Y_i)
- \hr_0
- \tr_i.
\label{eq:r20>00102011|1}
\end{align}

Let
\begin{align*}
\rho_i
&\equiv r_{i0}
\\
\orho_i
&\equiv
\min\lrb{
 \begin{aligned}
  &H(Z_{i0}|Z_{00}),
  \\
  &\tr_i - H(Z_{ii}|Y_i,Z_{00},Z_{10},Z_{20}),
  \\
  &\hr_0 + \tr_i - H(Z_{00},Z_{ii}|Y_i,Z_{10},Z_{20})
 \end{aligned}
}
\\
\urho_i
&\equiv
\max\lrb{
 \begin{aligned}
  &
  H(Z_{i0}|Y_i,Z_{00},Z_{i'0},Z_{ii}),
  \\
  &
  H(Z_{i0}|Y_{i'},Z_{00},Z_{i'0},Z_{i'i'}),
  \\
  &
  H(Z_{00},Z_{i0}|Y_i,Z_{i'0},Z_{ii})
  - \hr_0,
  \\
  &
  H(Z_{00},Z_{i0}|Y_{i'},Z_{i'0},Z_{i'i'})
  - \hr_0,
  \\
  &
  H(Z_{i0},Z_{ii}|Y_i,Z_{00},Z_{i'0})
  - \hr_i,
  \\
  &
  H(Z_{i0},Z_{i'i'}|Y_{i'},Z_{00},Z_{i'0})
  - \hr_{i'},
  \\
  &
  H(Z_{00},Z_{i0},Z_{ii}|Y_i,Z_{i'0})
  - \hr_0
  - \hr_i,
  \\
  &
  H(Z_{00},Z_{i0},Z_{i'i'}|Y_{i'},Z_{i'0})
  - \hr_0
  - \hr_{i'},
  \\
  &H(Z_{10},Z_{20},Z_{i'i'}|Y_{i'},Z_{00})
  - \tr_{i'},
  \\
  &H(Z_{00},Z_{10},Z_{20},Z_{i'i'}|Y_{i'})
  - \hr_0 - \tr_{i'}
 \end{aligned}
}
\\
\delta_i
&\equiv
\min\lrb{
 \begin{aligned}
  &\tr_i
  - H(Z_{i'0},Z_{ii}|Y_{i},Z_{00},Z_{i0}),
  \\
  &\hr_0 + \tr_i
  - H(Z_{00},Z_{i'0},Z_{ii}|Y_i,Z_{i0})
 \end{aligned}
}
\\
\sigma_i
&\equiv
\max\lrb{
 \begin{aligned}
  &H(Z_{10},Z_{20}|Y_i,Z_{00},Z_{ii}),
  \\
  &H(Z_{00},Z_{10},Z_{20}|Y_i,Z_{ii}) - \hr_0,
  \\
  &H(Z_{10},Z_{20},Z_{ii}|Y_i,Z_{00}) - \hr_i,
  \\
  &H(Z_{00},Z_{10},Z_{20},Z_{ii}|Y_i) - \hr_0 - \hr_i
 \end{aligned}
}.
\end{align*}
Then we have the fact that
\begin{itemize}
\item the set of conditions
(\ref{eq:r10>10|00}), (\ref{eq:r10<tr1-11|1}),
and (\ref{eq:r10<hr0+tr1-0011|1})
is equivalent to the condition $\rho_i\leq\orho_i$,
\item the set of conditions
(\ref{eq:r10>10|1}), (\ref{eq:r20>20|1})$'$, (\ref{eq:r10>0010|1}),
(\ref{eq:r20>0020|1})$'$, (\ref{eq:r10>1011|1}), (\ref{eq:r20>2011|1})$'$,
(\ref{eq:r10>001011|1}), (\ref{eq:r20>002011|1})$'$, (\ref{eq:r20>102011|1}),
and (\ref{eq:r20>00102011|1})$'$
is equivalent to the condition $\rho_i\geq\urho_i$,
\item the set of conditions
(\ref{eq:r20-r10>2011|1-tr1}) and (\ref{eq:r20-r10>002011|1-hr0-tr1})
is equivalent to the condition $\rho_i-\rho_{i'}\leq\delta_i$,
\item the set of conditions
(\ref{eq:r10+r20>1020|1}), (\ref{eq:r10+r20>001020|1}),
(\ref{eq:r10+r20>102011|1-hr1}),  and (\ref{eq:r10+r20>00102011|1})
is equivalent to the condition $\rho_1+\rho_2\geq\sigma_i$.
\end{itemize}
Since $r_{i0}$ does not appear in
the conditions
(\ref{eq:hr0>00|1}), 
(\ref{eq:hr1>11|1}),
(\ref{eq:hr0+hr1>0011|1}),
(\ref{eq:tr1>1011|1}),
and (\ref{eq:hr0+tr1>001011|1}),
we ignore them temporarily
and focus on the elimination of $r_{10}$ and $r_{20}$.
By applying Lemma~\ref{lem:fme} in Appendix \ref{sec:crng-fme},
we have
\begin{align}
0
&\leq
\delta_1 + \delta_2.
\label{eq:delta1+delta2}
\\
\urho_i
&\leq
\min\lrb{\orho_i, \orho_{i'} + \delta_i}
\\
\sigma_i
&\leq
\min\lrb{
 \orho_1 + \orho_2,
 2\orho_{i} + \delta_{i'},
 2\orho_{i'} + \delta_i
},
\label{eq:sigma}
\end{align}
which are equivalent to the set of conditions
(\ref{eq:r10>10|00}), (\ref{eq:r10>10|1}), (\ref{eq:r20>20|1}),
(\ref{eq:r10<tr1-11|1})--(\ref{eq:r20>0020|1}),
(\ref{eq:r10<hr0+tr1-0011|1})--(\ref{eq:r10>1011|1}),
(\ref{eq:r20>2011|1})--(\ref{eq:r10>001011|1}),
(\ref{eq:r20>002011|1})--(\ref{eq:r20>00102011|1}).
Since
\begin{align}
\orho_{i'}+\delta_i
&=
\min\lrb{
 \begin{aligned}
  &H(Z_{i'0}|Z_{00})
  +\tr_i
  - H(Z_{i'0},Z_{ii}|Y_i,Z_{00},Z_{i0}),
  \\
  &H(Z_{i'0}|Z_{00})
  + \hr_0 + \tr_i
  - H(Z_{00},Z_{i'0},Z_{ii}|Y_i,Z_{i0})
  \\
  &\tr_{i'}
  - H(Z_{i'i'}|Y_{i'},Z_{00},Z_{10},Z_{20})
  + \tr_i
  - H(Z_{i'0},Z_{ii}|Y_i,Z_{00},Z_{i0}),
  \\
  &\tr_{i'}
  - H(Z_{i'i'}|Y_{i'},Z_{00},Z_{10},Z_{20})
  + \hr_0 + \tr_i
  - H(Z_{00},Z_{i'0},Z_{ii}|Y_i,Z_{i0}),
  \\
  &\hr_0 + \tr_{i'} - H(Z_{00},Z_{i'i'}|Y_{i'},Z_{10},Z_{20})
  + \tr_i
  - H(Z_{i'0},Z_{ii}|Y_i,Z_{00},Z_{i0}),
  \\
  &\hr_0 + \tr_{i'} - H(Z_{00},Z_{i'i'}|Y_{i'},Z_{10},Z_{20})
  + \hr_0 + \tr_i
  - H(Z_{00},Z_{i'0},Z_{ii}|Y_i,Z_{i0}),
 \end{aligned}
}
\notag
\\
&=
\min\lrb{
 \begin{aligned}
  &\tr_i + H(Z_{i'0}|Z_{00})
  - H(Z_{i'0},Z_{ii}|Y_i,Z_{00},Z_{i0})
  \\
  &\hr_0 + \tr_i
  + H(Z_{i'0}|Z_{00})
  - H(Z_{00},Z_{i'0},Z_{ii}|Y_i,Z_{i0}),
  \\
  &\tr_1 + \tr_2
  - H(Z_{i'0},Z_{ii}|Y_i,Z_{00},Z_{i0}) 
  - H(Z_{i'i'}|Y_{i'},Z_{00},Z_{10},Z_{20}),
  \\
  &\hr_0 + \tr_1 + \tr_2
  - H(Z_{00},Z_{i'0},Z_{ii}|Y_i,Z_{i0})
  - H(Z_{i'i'}|Y_{i'},Z_{00},Z_{10},Z_{20}),
  \\
  &\hr_0 + \tr_1 + \tr_2
  - H(Z_{i'0},Z_{ii}|Y_i,Z_{00},Z_{i0})
  - H(Z_{00},Z_{i'i'}|Y_{i'},Z_{10},Z_{20}),
  \\
  &2\hr_0 + \tr_1 + \tr_2
  - H(Z_{00},Z_{i'0},Z_{ii}|Y_i,Z_{i0})
  - H(Z_{00},Z_{i'i'}|Y_{i'},Z_{10},Z_{20})
 \end{aligned}
}
\end{align}
and
\begin{align}
\tr_i - H(Z_{ii}|Y_i,Z_{00},Z_{10},Z_{20})
&\leq
\tr_i
+ H(Z_{i'0}|Z_{00})
- H(Z_{i'0}|Y_i,Z_{00},Z_{i0})
- H(Z_{ii}|Y_i,Z_{00},Z_{10},Z_{20})
\notag
\\
&\leq
\tr_i + H(Z_{i'0}|Z_{00}) - H(Z_{i'0},Z_{ii}|Y_i,Z_{00},Z_{i0}),
\end{align}
we have
\begin{equation}
\min\{\orho_i,\orho_{i'}+\delta_i\}
=
\min\{\orho_i,\eta_i\},
\label{eq:eta}
\end{equation}
where
\begin{align}
\eta_i
&\equiv
\min\lrb{
 \begin{aligned}
  &\hr_0 + \tr_i
  + H(Z_{i'0}|Z_{00})
  - H(Z_{00},Z_{i'0},Z_{ii}|Y_i,Z_{i0}),
  \\
  &\tr_1 + \tr_2
  - H(Z_{i'0},Z_{ii}|Y_i,Z_{00},Z_{i0}) 
  - H(Z_{i'i'}|Y_{i'},Z_{00},Z_{10},Z_{20}),
  \\
  &\hr_0 + \tr_1 + \tr_2
  - H(Z_{00},Z_{i'0},Z_{ii}|Y_i,Z_{i0})
  - H(Z_{i'i'}|Y_{i'},Z_{00},Z_{10},Z_{20}),
  \\
  &\hr_0 + \tr_1 + \tr_2
  - H(Z_{i'0},Z_{ii}|Y_i,Z_{00},Z_{i0})
  - H(Z_{00},Z_{i'i'}|Y_{i'},Z_{10},Z_{20}),
  \\
  &2\hr_0 + \tr_1 + \tr_2
  - H(Z_{00},Z_{i'0},Z_{ii}|Y_i,Z_{i0})
  - H(Z_{00},Z_{i'i'}|Y_{i'},Z_{10},Z_{20})
 \end{aligned}
}.
\end{align}
Since
\begin{align}
\orho_1 + \orho_2
&=
\min\lrb{
 \begin{aligned}
  &H(Z_{10}|Z_{00}) + H(Z_{20}|Z_{00}),
  \\
  &H(Z_{10}|Z_{00})
  +\tr_2 - H(Z_{22}|Y_2,Z_{00},Z_{10},Z_{20}),
  \\
  &H(Z_{10}|Z_{00})
  +\hr_0 + \tr_2 - H(Z_{00},Z_{22}|Y_2,Z_{10},Z_{20}),
  \\
  &\tr_1 - H(Z_{11}|Y_1,Z_{00},Z_{10},Z_{20})
  +H(Z_{20}|Z_{00}),
  \\
  &\tr_1 - H(Z_{11}|Y_1,Z_{00},Z_{10},Z_{20})
  + \tr_2 - H(Z_{22}|Y_2,Z_{00},Z_{10},Z_{20}),
  \\
  &\tr_1 - H(Z_{11}|Y_1,Z_{00},Z_{10},Z_{20})
  + \hr_0 + \tr_2 - H(Z_{00},Z_{22}|Y_2,Z_{10},Z_{20}),
  \\
  &\hr_0 + \tr_1 - H(Z_{00},Z_{11}|Y_1,Z_{10},Z_{20})
  +H(Z_{20}|Z_{00}),
  \\
  &\hr_0 + \tr_1 - H(Z_{00},Z_{11}|Y_1,Z_{10},Z_{20})
  +\tr_2 - H(Z_{22}|Y_2,Z_{00},Z_{10},Z_{20}),
  \\
  &\hr_0 + \tr_1 - H(Z_{00},Z_{11}|Y_1,Z_{10},Z_{20})
  +\hr_0 + \tr_2 - H(Z_{00},Z_{22}|Y_2,Z_{10},Z_{20})
 \end{aligned}
}
\notag
\\
&=
\min\lrb{
 \begin{aligned}
  &H(Z_{10}|Z_{00}) + H(Z_{20}|Z_{00}),
  \\
  &\tr_1
  + H(Z_{20}|Z_{00}) - H(Z_{11}|Y_1,Z_{00},Z_{10},Z_{20}),
  \\
  &\tr_2
  + H(Z_{10}|Z_{00}) - H(Z_{22}|Y_2,Z_{00},Z_{10},Z_{20}),
  \\
  &\hr_0 + \tr_1
  +H(Z_{20}|Z_{00})
  - H(Z_{00},Z_{11}|Y_1,Z_{10},Z_{20}),
  \\
  &\hr_0 + \tr_2
  + H(Z_{10}|Z_{00}) - H(Z_{00},Z_{22}|Y_2,Z_{10},Z_{20}),
  \\
  &\tr_1 +\tr_2
  - H(Z_{11}|Y_1,Z_{00},Z_{10},Z_{20})
  - H(Z_{22}|Y_2,Z_{00},Z_{10},Z_{20}),
  \\
  &\hr_0 + \tr_1 + \tr_2
  - H(Z_{11}|Y_1,Z_{00},Z_{10},Z_{20})
  - H(Z_{00},Z_{22}|Y_2,Z_{10},Z_{20}),
  \\
  &\hr_0 + \tr_1 + \tr_2
  - H(Z_{00},Z_{11}|Y_1,Z_{10},Z_{20})
  - H(Z_{22}|Y_2,Z_{00},Z_{10},Z_{20}),
  \\
  &2\hr_0 + \tr_1 + \tr_2
  - H(Z_{00},Z_{11}|Y_1,Z_{10},Z_{20})
  - H(Z_{00},Z_{22}|Y_2,Z_{10},Z_{20})
 \end{aligned}
}
\end{align}
\begin{align}
2\orho_{i'}+\delta_i
&=
\min\lrb{
 \begin{aligned}
  &2H(Z_{i'0}|Z_{00})
  + \tr_i
  - H(Z_{i'0},Z_{ii}|Y_i,Z_{00},Z_{i0}),
  \\
  &2H(Z_{i'0}|Z_{00})
  + \hr_0 + \tr_i
  - H(Z_{00},Z_{i'0},Z_{ii}|Y_i,Z_{i0}),
  \\
  &2\lrB{\tr_{i'} - H(Z_{i'i'}|Y_{i'},Z_{00},Z_{10},Z_{20})}
  + \tr_i - H(Z_{i'0},Z_{ii}|Y_i,Z_{00},Z_{i0}),
  \\
  &2\lrB{\tr_{i'} - H(Z_{i'i'}|Y_{i'},Z_{00},Z_{10},Z_{20})}
  + \hr_0 + \tr_i
  - H(Z_{00},Z_{i'0},Z_{ii}|Y_i,Z_{i0}),
  \\
  &2\lrB{\hr_0 + \tr_{i'} - H(Z_{00},Z_{i'i'}|Y_{i'},Z_{10},Z_{20})}
  + \tr_i - H(Z_{i'0},Z_{ii}|Y_i,Z_{00},Z_{i0}),
  \\
  &2\lrB{\hr_0 + \tr_{i'} - H(Z_{00},Z_{i'i'}|Y_{i'},Z_{10},Z_{20})}
  + \hr_0 + \tr_i
  - H(Z_{00},Z_{i'0},Z_{ii}|Y_i,Z_{i0})
 \end{aligned}
}
\notag
\\
&=
\min\lrb{
 \begin{aligned}
  &\tr_i + 2H(Z_{i'0}|Z_{00})
  - H(Z_{i'0},Z_{ii}|Y_i,Z_{00},Z_{i0}),
  \\
  &\hr_0 + \tr_i
  + 2H(Z_{i'0}|Z_{00})
  - H(Z_{00},Z_{i'0},Z_{ii}|Y_i,Z_{i0}),
  \\
  &\tr_i + 2\tr_{i'}
  - H(Z_{i'0},Z_{ii}|Y_i,Z_{00},Z_{i0})
  - 2H(Z_{i'i'}|Y_{i'},Z_{00},Z_{10},Z_{20}),
  \\
  &\hr_0 + \tr_i + 2\tr_{i'}
  - H(Z_{00},Z_{i'0},Z_{ii}|Y_i,Z_{i0})
  - 2 H(Z_{i'i'}|Y_{i'},Z_{00},Z_{10},Z_{20}),
  \\
  &2\hr_0 + \tr_i + 2\tr_{i'}
  - H(Z_{i'0},Z_{ii}|Y_i,Z_{00},Z_{i0})
  - 2 H(Z_{00},Z_{i'i'}|Y_{i'},Z_{10},Z_{20}),
  \\
  &3\hr_0 + \tr_i + 2\tr_{i'}
  - H(Z_{00},Z_{i'0},Z_{ii}|Y_i,Z_{i0})
  - 2 H(Z_{00},Z_{i'i'}|Y_{i'},Z_{10},Z_{20})
 \end{aligned}
}
\end{align}
and
\begin{align}
&
\tr_i
+ 2H(Z_{i'0}|Z_{00})
- H(Z_{i'0},Z_{ii}|Y_i,Z_{00},Z_{i0})
\notag
\\*
&=
\tr_i
+ H(Z_{i'0}|Z_{00})
- H(Z_{ii}|Y_i,Z_{00},Z_{10},Z_{20})
+ H(Z_{i'0}|Z_{00})
- H(Z_{i'0}|Y_i,Z_{00},Z_{i0})
\notag
\\
&\geq
\tr_i + H(Z_{i'0}|Z_{00})
- H(Z_{ii}|Y_i,Z_{00},Z_{10},Z_{20}),
\end{align}
we have
\begin{equation}
\min\lrb{\orho_1+\orho_2,2\orho_{i'}+\delta_i,2\orho_i+\delta_{i'}}
=
\min\lrb{\orho_1+\orho_2,\zeta_i},
\label{eq:zeta}
\end{equation}
where
\begin{align}
\zeta_i
&\equiv
\min\lrb{
 \begin{aligned}
  &\hr_0 + \tr_i
  + 2H(Z_{i'0}|Z_{00})
  - H(Z_{00},Z_{i'0},Z_{ii}|Y_i,Z_{i0}),
  \\
  &\hr_0 + \tr_{i'}
  + 2H(Z_{i0}|Z_{00})
  - H(Z_{00},Z_{i0},Z_{i'i'}|Y_{i'},Z_{i'0}),
  \\
  &\tr_i
  + 2\tr_{i'}
  - H(Z_{i'0},Z_{ii}|Y_i,Z_{00},Z_{i0}) 
  - 2H(Z_{i'i'}|Y_{i'},Z_{00},Z_{10},Z_{20}),
  \\
  &2\tr_{i}
  + \tr_{i'}
  - 2H(Z_{ii}|Y_{i},Z_{00},Z_{10},Z_{20})
  - H(Z_{i0},Z_{i'i'}|Y_{i'},Z_{00},Z_{i'0}) ,
  \\
  &\hr_0 + \tr_i + 2\tr_{i'}
  - H(Z_{00},Z_{i'0},Z_{ii}|Y_i,Z_{i0})
  - 2H(Z_{i'i'}|Y_{i'},Z_{00},Z_{10},Z_{20}),
  \\
  &\hr_0 + 2\tr_i + \tr_{i'}
  - 2H(Z_{ii}|Y_i,Z_{00},Z_{10},Z_{20})
  - H(Z_{00},Z_{i0},Z_{i'i'}|Y_{i'},Z_{i'0}),
  \\
  &2\hr_0 + \tr_i + 2\tr_{i'}
  - H(Z_{i'0},Z_{ii}|Y_i,Z_{00},Z_{i0})
  - 2H(Z_{00},Z_{i'i'}|Y_{i'},Z_{10},Z_{20}),
  \\
  &2\hr_0 + 2\tr_i + \tr_{i'}
  - 2H(Z_{00},Z_{ii}|Y_{i},Z_{10},Z_{20})
  - H(Z_{i0},Z_{i'i'}|Y_{i'},Z_{00},Z_{i'0}),
  \\
  &3\hr_0
  + \tr_i
  + 2\tr_{i'}
  - H(Z_{00},Z_{i'0},Z_{ii}|Y_i,Z_{i0})
  - 2H(Z_{00},Z_{i'i'}|Y_{i'},Z_{10},Z_{20}),
  \\
  &3\hr_0
  + 2\tr_i
  + \tr_{i'}
  - 2H(Z_{00},Z_{ii}|Y_i,Z_{10},Z_{20})
  - H(Z_{00},Z_{i0},Z_{i'i'}|Y_{i'},Z_{i'0})
 \end{aligned}
}.
\end{align}
Then, from
(\ref{eq:eta}) and (\ref{eq:zeta}),
we have equivalent conditions to
the set of conditions (\ref{eq:delta1+delta2})--(\ref{eq:sigma}) as
\begin{align*}
\delta_1 + \delta_2&\geq 0
\\
\urho_i
&\leq
\orho_i
\\
\urho_i
&\leq
\eta_i
\\
\sigma_i
&\leq
\orho_1 + \orho_2
\\
\sigma_i
&\leq
\zeta_i.
\end{align*}

Since
\begin{align}
\delta_1+\delta_2
&=
\min\lrb{
 \begin{aligned}
  &\tr_1
  - H(Z_{20},Z_{11}|Y_1,Z_{00},Z_{10})
  +\tr_2
  - H(Z_{10},Z_{22}|Y_2,Z_{00},Z_{20}),
  \\
  &\tr_1
  - H(Z_{20},Z_{11}|Y_1,Z_{00},Z_{10})
  +\hr_0 + \tr_2
  - H(Z_{00},Z_{10},Z_{22}|Y_2,Z_{20}),
  \\
  &\hr_0 + \tr_1
  - H(Z_{00},Z_{20},Z_{11}|Y_1,Z_{10})
  +\tr_2
  - H(Z_{10},Z_{22}|Y_2,Z_{00},Z_{20}),
  \\
  &\hr_0 + \tr_1
  - H(Z_{00},Z_{20},Z_{11}|Y_1,Z_{10})
  +\hr_0 + \tr_2
  - H(Z_{00},Z_{10},Z_{22}|Y_2,Z_{20})
 \end{aligned}
}
\notag
\\
&=
\min\lrb{
 \begin{aligned}
  &\tr_1
  +\tr_2
  - H(Z_{20},Z_{11}|Y_1,Z_{00},Z_{10})
  - H(Z_{10},Z_{22}|Y_2,Z_{00},Z_{20}),
  \\
  &\hr_0 + \tr_1 + \tr_2
  - H(Z_{20},Z_{11}|Y_1,Z_{00},Z_{10})
  - H(Z_{00},Z_{10},Z_{22}|Y_2,Z_{20}),
  \\
  &\hr_0 + \tr_1
  +\tr_2
  - H(Z_{00},Z_{20},Z_{11}|Y_1,Z_{10})
  - H(Z_{10},Z_{22}|Y_2,Z_{00},Z_{20}),
  \\
  &2\hr_0 + \tr_1 + \tr_2
  - H(Z_{00},Z_{20},Z_{11}|Y_1,Z_{10})
  - H(Z_{00},Z_{10},Z_{22}|Y_2,Z_{20})
 \end{aligned}
},
\end{align}
we have equivalent conditions to
$\delta_1 + \delta_2\geq 0$
as
\begin{align}
\tr_1
+\tr_2
&\geq
H(Z_{20},Z_{11}|Y_1,Z_{00},Z_{10})
+H(Z_{10},Z_{22}|Y_2,Z_{00},Z_{20})
\label{eq:tr1+tr2>2011|1+1022|2}
\\
\hr_0 + \tr_1 +\tr_2
&\geq
H(Z_{00},Z_{i'0},Z_{ii}|Y_i,Z_{i0})
+ H(Z_{i0},Z_{i'i'}|Y_{i'},Z_{00},Z_{i'0})
\quad\text{for all}\ i\in\{1,2\}
\label{eq:hr0+tr1+tr2>002011|1+1022|2}
\\
2\hr_0 + \tr_1 + \tr_2
&\geq
H(Z_{00},Z_{20},Z_{11}|Y_1,Z_{10})
+H(Z_{00},Z_{10},Z_{22}|Y_2,Z_{20}).
\label{eq:2hr0+tr1+tr2>002011|1+001022|2}
\end{align}

We have equivalent conditions to $\urho_i\leq\orho_i$ as
\begin{align}
H(Z_{i0}|Y_i,Z_{00},Z_{i'0},Z_{ii})
&\leq
H(Z_{i0}|Z_{00})
\label{eq:urho-orho-00}
\\
H(Z_{i0}|Y_{i'},Z_{00},Z_{i'0},Z_{i'i'})
&\leq
H(Z_{i0}|Z_{00})
\label{eq:urho-orho-01}
\\
H(Z_{00},Z_{i0}|Y_i,Z_{i'0},Z_{ii})
- \hr_0
&\leq
H(Z_{i0}|Z_{00})
\label{eq:urho-orho-02}
\\
H(Z_{00},Z_{i0}|Y_{i'},Z_{i'0},Z_{i'i'})
- \hr_0
&\leq
H(Z_{i0}|Z_{00})
\label{eq:urho-orho-03}
\\
H(Z_{i0},Z_{ii}|Y_i,Z_{00},Z_{i'0})
- \hr_i
&\leq
H(Z_{i0}|Z_{00})
\label{eq:urho-orho-04}
\\
H(Z_{i0},Z_{i'i'}|Y_{i'},Z_{00},Z_{i'0})
- \hr_{i'}
&\leq
H(Z_{i0}|Z_{00})
\label{eq:urho-orho-05}
\\
H(Z_{00},Z_{i0},Z_{ii}|Y_i,Z_{i'0})
- \hr_0
- \hr_i
&\leq
H(Z_{i0}|Z_{00})
\label{eq:urho-orho-06}
\\
H(Z_{00},Z_{i0},Z_{i'i'}|Y_{i'},Z_{i'0})
- \hr_0
- \hr_{i'}
&\leq
H(Z_{i0}|Z_{00})
\label{eq:urho-orho-07}
\\
H(Z_{10},Z_{20},Z_{i'i'}|Y_{i'},Z_{00})
- \tr_{i'}
&\leq
H(Z_{i0}|Z_{00})
\label{eq:urho-orho-08}
\\
H(Z_{00},Z_{10},Z_{20},Z_{i'i'}|Y_{i'})
- \hr_0 - \tr_{i'}
&\leq
H(Z_{i0}|Z_{00})
\label{eq:urho-orho-09}
\\
H(Z_{i0}|Y_i,Z_{00},Z_{i'0},Z_{ii})
&\leq
\tr_i - H(Z_{ii}|Y_i,Z_{00},Z_{10},Z_{20})
\label{eq:urho-orho-10}
\\
H(Z_{i0}|Y_{i'},Z_{00},Z_{i'0},Z_{i'i'})
&\leq
\tr_i - H(Z_{ii}|Y_i,Z_{00},Z_{10},Z_{20})
\label{eq:urho-orho-11}
\\
H(Z_{00},Z_{i0}|Y_i,Z_{i'0},Z_{ii})
- \hr_0
&\leq
\tr_i - H(Z_{ii}|Y_i,Z_{00},Z_{10},Z_{20})
\label{eq:urho-orho-12}
\\
H(Z_{00},Z_{i0}|Y_{i'},Z_{i'0},Z_{i'i'})
- \hr_0
&\leq
\tr_i - H(Z_{ii}|Y_i,Z_{00},Z_{10},Z_{20})
\label{eq:urho-orho-13}
\\
H(Z_{i0},Z_{ii}|Y_i,Z_{00},Z_{i'0})
- \hr_i
&\leq
\tr_i - H(Z_{ii}|Y_i,Z_{00},Z_{10},Z_{20})
\label{eq:urho-orho-14}
\\
H(Z_{i0},Z_{i'i'}|Y_{i'},Z_{00},Z_{i'0})
- \hr_{i'}
&\leq
\tr_i - H(Z_{ii}|Y_i,Z_{00},Z_{10},Z_{20})
\label{eq:urho-orho-15}
\\
H(Z_{00},Z_{i0},Z_{ii}|Y_i,Z_{i'0})
- \hr_0
- \hr_i
&\leq
\tr_i - H(Z_{ii}|Y_i,Z_{00},Z_{10},Z_{20})
\label{eq:urho-orho-16}
\\
H(Z_{00},Z_{i0},Z_{i'i'}|Y_{i'},Z_{i'0})
- \hr_0
- \hr_{i'}
&\leq
\tr_i - H(Z_{ii}|Y_i,Z_{00},Z_{10},Z_{20})
\label{eq:urho-orho-17}
\\
H(Z_{10},Z_{20},Z_{i'i'}|Y_{i'},Z_{00})
- \tr_{i'}
&\leq
\tr_i - H(Z_{ii}|Y_i,Z_{00},Z_{10},Z_{20})
\label{eq:urho-orho-18}
\\
H(Z_{00},Z_{10},Z_{20},Z_{i'i'}|Y_{i'})
- \hr_0 - \tr_{i'}
&\leq
\tr_i - H(Z_{ii}|Y_i,Z_{00},Z_{10},Z_{20})
\label{eq:urho-orho-19}
\\
H(Z_{i0}|Y_i,Z_{00},Z_{i'0},Z_{ii})
&\leq
\hr_0 + \tr_i - H(Z_{00},Z_{ii}|Y_i,Z_{10},Z_{20})
\label{eq:urho-orho-20}
\\
H(Z_{i0}|Y_{i'},Z_{00},Z_{i'0},Z_{i'i'})
&\leq
\hr_0 + \tr_i - H(Z_{00},Z_{ii}|Y_i,Z_{10},Z_{20})
\label{eq:urho-orho-21}
\\
H(Z_{00},Z_{i0}|Y_i,Z_{i'0},Z_{ii})
- \hr_0
&\leq
\hr_0 + \tr_i - H(Z_{00},Z_{ii}|Y_i,Z_{10},Z_{20})
\label{eq:urho-orho-22}
\\
H(Z_{00},Z_{i0}|Y_{i'},Z_{i'0},Z_{i'i'})
- \hr_0
&\leq
\hr_0 + \tr_i - H(Z_{00},Z_{ii}|Y_i,Z_{10},Z_{20})
\label{eq:urho-orho-23}
\\
H(Z_{i0},Z_{ii}|Y_i,Z_{00},Z_{i'0})
- \hr_i
&\leq
\hr_0 + \tr_i - H(Z_{00},Z_{ii}|Y_i,Z_{10},Z_{20})
\label{eq:urho-orho-24}
\\
H(Z_{i0},Z_{i'i'}|Y_{i'},Z_{00},Z_{i'0})
- \hr_{i'}
&\leq
\hr_0 + \tr_i - H(Z_{00},Z_{ii}|Y_i,Z_{10},Z_{20})
\label{eq:urho-orho-25}
\\
H(Z_{00},Z_{i0},Z_{ii}|Y_i,Z_{i'0})
- \hr_0
- \hr_i
&\leq
\hr_0 + \tr_i - H(Z_{00},Z_{ii}|Y_i,Z_{10},Z_{20})
\label{eq:urho-orho-26}
\\
H(Z_{00},Z_{i0},Z_{i'i'}|Y_{i'},Z_{i'0})
- \hr_0
- \hr_{i'}
&\leq
\hr_0 + \tr_i - H(Z_{00},Z_{ii}|Y_i,Z_{10},Z_{20})
\label{eq:urho-orho-27}
\\
H(Z_{10},Z_{20},Z_{i'i'}|Y_{i'},Z_{00})
- \tr_{i'}
&\leq
\hr_0 + \tr_i - H(Z_{00},Z_{ii}|Y_i,Z_{10},Z_{20})
\label{eq:urho-orho-28}
\\
H(Z_{00},Z_{10},Z_{20},Z_{i'i'}|Y_{i'})
- \hr_0 - \tr_{i'}
&\leq
\hr_0 + \tr_i - H(Z_{00},Z_{ii}|Y_i,Z_{10},Z_{20}),
\label{eq:urho-orho-29}
\end{align}
where
\begin{itemize}
\item
(\ref{eq:urho-orho-00}) and (\ref{eq:urho-orho-01}) are trivially redundant,
\item 
(\ref{eq:urho-orho-02}) is equivalent to
\begin{align}
 \hr_0
 &\geq
 H(Z_{00},Z_{i0}|Y_i,Z_{i'0},Z_{ii})
 -H(Z_{i0}|Z_{00}),
 \label{eq:hr0>0010|1-10|00}
\end{align}
\item
(\ref{eq:urho-orho-03}) is equivalent to
\begin{align}
 \hr_0
 &\geq
 H(Z_{00},Z_{i0}|Y_{i'},Z_{i'0},Z_{i'i'})
 -H(Z_{i0}|Z_{00}),
 \label{eq:hr0>0010|2-10|00}
\end{align}
\item
(\ref{eq:urho-orho-04}) is redundant because it comes from
(\ref{eq:hr1>11|1}) and Lemma \ref{lem:promotion} as
\begin{align}
 \hr_i
 &\geq
 H(Z_{ii}|Y_i,Z_{00},Z_{10},Z_{20})
 \notag
 \\
 &\geq
 H(Z_{i0},Z_{ii}|Y_i,Z_{00},Z_{i'0}) - H(Z_{i0}|Z_{00}),
\end{align}
\item
(\ref{eq:urho-orho-05}) is redundant because it comes from
(\ref{eq:hr1>11|1})$'$ and Lemma \ref{lem:promotion} as
\begin{align}
 \hr_{i'}
 &\geq
 H(Z_{i'i'}|Y_{i'},Z_{00},Z_{10},Z_{20})
 \notag
 \\
 &\geq
 H(Z_{i0},Z_{i'i'}|Y_{i'},Z_{00},Z_{i'0}) - H(Z_{i0}|Z_{00}),
\end{align}
\item
(\ref{eq:urho-orho-06}) is redundant because it comes from
(\ref{eq:hr1=tr1-10|0011})
and
(\ref{eq:hr0+tr1>001011|1})
as
\begin{align}
 \hr_0 + \hr_i
 &=
 [\hr_0 + \tr_i] - H(Z_{i0}|Z_{00},Z_{ii})
 \notag
 \\
 &\geq
 H(Z_{00},Z_{i0},Z_{ii}|Y_i,Z_{i'0}) - H(Z_{i0}|Z_{00},Z_{ii}) 
 \notag
 \\
 &\geq
 H(Z_{00},Z_{i0},Z_{ii}|Y_i,Z_{i'0}) - H(Z_{i0}|Z_{00}),
\end{align}
\item
(\ref{eq:urho-orho-07}) is equivalent to
\begin{align}
 \hr_0
 + \hr_{i'}
 &\geq
 H(Z_{00},Z_{i0},Z_{i'i'}|Y_{i'},Z_{i'0})
 - H(Z_{i0}|Z_{00}),
 \label{eq:hr0+hr2>001022|2-10|00}
\end{align}
\item
(\ref{eq:urho-orho-08}) is redundant because it comes from
(\ref{eq:tr1>1011|1})$'$ and Lemma \ref{lem:promotion} as
\begin{align}
 \tr_{i'}
 &\geq
 H(Z_{i'0},Z_{i'i'}|Y_{i'},Z_{00},Z_{i0})
 \notag
 \\
 &\geq
 H(Z_{10},Z_{20},Z_{i'i'}|Y_{i'},Z_{00}) - H(Z_{i0}|Z_{00}),
\end{align}
\item
(\ref{eq:urho-orho-09}) is equivalent to
\begin{align}
 \hr_0 + \tr_{i'}
 &\geq
 H(Z_{00},Z_{10},Z_{20},Z_{i'i'}|Y_{i'})
 - H(Z_{i0}|Z_{00}),
 \label{eq:hr0+tr2>00102022|2-10|00}
\end{align}
\item
(\ref{eq:urho-orho-10}) is redundant because it comes from
(\ref{eq:tr1>1011|1}) and Lemma \ref{lem:xor} as
\begin{align}
 \tr_i
 &\geq
 H(Z_{i0},Z_{ii}|Y_i,Z_{00},Z_{i'0})
 \notag
 \\
 &\geq
 H(Z_{i0}|Y_i,Z_{00},Z_{i'0},Z_{ii})
 + H(Z_{ii}|Y_i,Z_{00},Z_{10},Z_{20}),
\end{align}
\item
(\ref{eq:urho-orho-11}) is equivalent to
\begin{align}
 \tr_i
 &\geq
 H(Z_{ii}|Y_i,Z_{00},Z_{10},Z_{20})
 + H(Z_{i0}|Y_{i'},Z_{00},Z_{i'0},Z_{i'i'}),
 \label{eq:tr1>10|2+11|1}
\end{align}
\item
(\ref{eq:urho-orho-12}) is redundant because it comes from
(\ref{eq:hr0+tr1>001011|1}) and Lemma \ref{lem:xor} as
\begin{align}
 \hr_0 + \tr_i
 &\geq
 H(Z_{00},Z_{i0},Z_{ii}|Y_i,Z_{i'0})
 \notag
 \\
 &\geq
 H(Z_{00},Z_{i0}|Y_i,Z_{i'0},Z_{ii})
 + H(Z_{ii}|Y_i,Z_{00},Z_{10},Z_{20}),
\end{align}
\item
(\ref{eq:urho-orho-13}) is equivalent to
\begin{align}
 \hr_0
 + \tr_i
 &\geq
 H(Z_{ii}|Y_i,Z_{00},Z_{10},Z_{20})
 + H(Z_{00},Z_{i0}|Y_{i'},Z_{i'0},Z_{i'i'}),
 \label{eq:hr0+tr1>0010|2+11|1}
\end{align}
\item
(\ref{eq:urho-orho-14}) is redundant because it comes from
(\ref{eq:hr1>11|1}) and
(\ref{eq:tr1>1011|1}) as
\begin{align}
 \hr_i + \tr_i
 &\geq
 H(Z_{ii}|Y_i,Z_{00},Z_{10},Z_{20})
 + H(Z_{i0},Z_{ii}|Y_i,Z_{00},Z_{i'0}),
\end{align}
\item
(\ref{eq:urho-orho-15}) is equivalent to
\begin{align}
 \hr_{i'}
 + \tr_i
 &\geq
 H(Z_{ii}|Y_i,Z_{00},Z_{10},Z_{20})
 + H(Z_{i0},Z_{i'i'}|Y_{i'},Z_{00},Z_{i'0}),
 \label{eq:hr2+tr1>1022|2+11|1}
\end{align}
\item
(\ref{eq:urho-orho-16}) is redundant because it comes from
(\ref{eq:hr1>11|1}) and
(\ref{eq:hr0+tr1>001011|1}) as
\begin{align}
 \hr_i + [ \hr_0 + \tr_i ]
 &\geq
 H(Z_{ii}|Y_i,Z_{00},Z_{10},Z_{20})
 + H(Z_{00},Z_{i0},Z_{ii}|Y_i,Z_{i'0}),
\end{align}
\item
(\ref{eq:urho-orho-17}) is equivalent to
\begin{align}
 \hr_0
 + \hr_{i'}
 + \tr_i
 &\geq
 H(Z_{ii}|Y_i,Z_{00},Z_{10},Z_{20})
 + H(Z_{00},Z_{i0},Z_{i'i'}|Y_{i'},Z_{i'0}),
 \label{eq:hr0+hr2+tr1>001022|2+11|1}
\end{align}
\item
(\ref{eq:urho-orho-18}) is equivalent to
\begin{align}
 \tr_1
 + \tr_2
 &\geq
 H(Z_{ii}|Y_i,Z_{00},Z_{10},Z_{20})
 + H(Z_{10},Z_{20},Z_{i'i'}|Y_{i'},Z_{00}),
 \label{eq:tr1+tr2>102022|2+11|1}
\end{align}
\item
(\ref{eq:urho-orho-19}) is equivalent to
\begin{align}
 \hr_0
 + \tr_1
 + \tr_2
 &\geq
 H(Z_{ii}|Y_i,Z_{00},Z_{10},Z_{20})
 + H(Z_{00},Z_{10},Z_{20},Z_{i'i'}|Y_{i'}),
 \label{eq:hr0+tr1+tr2>00102022|2+11|1}
\end{align}
\item
(\ref{eq:urho-orho-20}) is redundant because it comes from
(\ref{eq:hr0+tr1>001011|1}) and Lemma \ref{lem:xor} as
\begin{align}
 \hr_0 + \tr_i
 &\geq
 H(Z_{00},Z_{i0},Z_{ii}|Y_i,Z_{i'0})
 \notag
 \\
 &\geq
 H(Z_{00},Z_{ii}|Y_i,Z_{10},Z_{20})
 + H(Z_{i0}|Y_i,Z_{00},Z_{i'0},Z_{ii}),
\end{align}
\item
(\ref{eq:urho-orho-21}) is equivalent to
\begin{align}
 \hr_0 + \tr_i
 &\geq
 H(Z_{00},Z_{ii}|Y_i,Z_{10},Z_{20})
 + H(Z_{i0}|Y_{i'},Z_{00},Z_{i'0},Z_{i'i'}),
 \label{eq:hr0+tr1>10|2+0011|1}
\end{align}
\item
(\ref{eq:urho-orho-22}) is redundant because it comes from
(\ref{eq:hr0>00|1}), (\ref{eq:hr0+tr1>001011|1}),
and Lemma \ref{lem:relocation} as
\begin{align}
 \hr_0 + [ \hr_0 + \tr_i ]
 &\geq
 H(Z_{00}|Y_i,Z_{10},Z_{20},Z_{ii}) + H(Z_{00},Z_{i0},Z_{ii}|Y_i,Z_{i'0})
 \notag
 \\
 &\geq
 H(Z_{00},Z_{ii}|Y_i,Z_{10},Z_{20})
 + H(Z_{00},Z_{i0}|Y_i,Z_{i'0},Z_{ii}),
\end{align}
\item
(\ref{eq:urho-orho-23}) is equivalent to
\begin{align}
 2\hr_0 + \tr_i
 &\geq
 H(Z_{00},Z_{ii}|Y_i,Z_{10},Z_{20})
 + H(Z_{00},Z_{i0}|Y_{i'},Z_{i'0},Z_{i'i'}),
 \label{eq:2hr0+tr1>0010|2+0011|1}
\end{align}
\item
(\ref{eq:urho-orho-24}) is redundant because it comes from
(\ref{eq:hr0+hr1>0011|1}) and
(\ref{eq:tr1>1011|1}) as
\begin{align}
 [ \hr_0 + \hr_i ] + \tr_i
 &\geq
 H(Z_{00},Z_{ii}|Y_i,Z_{10},Z_{20})
 + H(Z_{i0},Z_{ii}|Y_i,Z_{00},Z_{i'0}),
\end{align}
\item
(\ref{eq:urho-orho-25}) is equivalent to
\begin{align}
 \hr_0 + \hr_{i'} + \tr_i
 &\geq
 H(Z_{00},Z_{ii}|Y_i,Z_{10},Z_{20})
 + H(Z_{i0},Z_{i'i'}|Y_{i'},Z_{00},Z_{i'0}),
 \label{eq:hr0+hr2+tr1>0011|1+1022|2}
\end{align}
\item
(\ref{eq:urho-orho-26}) is redundant because it comes from
(\ref{eq:hr0+hr1>0011|1}) and
(\ref{eq:hr0+tr1>001011|1}) as
\begin{align}
 [ \hr_0 + \hr_i ] + [ \hr_0 + \tr_i ]
 &\geq
 H(Z_{00},Z_{ii}|Y_i,Z_{10},Z_{20})
 + H(Z_{00},Z_{i0},Z_{ii}|Y_i,Z_{i'0}),
\end{align}
\item
(\ref{eq:urho-orho-27}) is equivalent to
\begin{align}
 2\hr_0
 + \hr_{i'}
 + \tr_i
 &\geq
 H(Z_{00},Z_{ii}|Y_i,Z_{10},Z_{20})
 + H(Z_{00},Z_{i0},Z_{i'i'}|Y_{i'},Z_{i'0}),
 \label{eq:2hr0+hr2+tr1>001022|2+0011|1}
\end{align}
\item
(\ref{eq:urho-orho-28}) is equivalent to
\begin{align}
 \hr_0 + \tr_1 + \tr_2
 &\geq
 H(Z_{00},Z_{ii}|Y_i,Z_{10},Z_{20})
 + H(Z_{10},Z_{20},Z_{i'i'}|Y_{i'},Z_{00}),
 \label{eq:hr0+tr1+tr2>102022|2+0011|1}
\end{align}
\item
(\ref{eq:urho-orho-29}) is equivalent to
\begin{align}
 2\hr_0 + \tr_1 + \tr_2
 &\geq
 H(Z_{00},Z_{ii}|Y_i,Z_{10},Z_{20})
 + H(Z_{00},Z_{10},Z_{20},Z_{i'i'}|Y_{i'}).
 \label{eq:2hr0+tr1+tr2>00102022|2+0011|1}
\end{align}
\end{itemize}

We have equivalent conditions to $\urho_i\leq\eta_i$ as
\begin{align}
H(Z_{i0}|Y_i,Z_{00},Z_{i'0},Z_{ii})
&\leq
\hr_0 + \tr_i
+ H(Z_{i'0}|Z_{00})
- H(Z_{00},Z_{i'0},Z_{ii}|Y_i,Z_{i0})
\label{eq:urho<eta-00}
\\
H(Z_{i0}|Y_{i'},Z_{00},Z_{i'0},Z_{i'i'})
&\leq
\hr_0 + \tr_i
+ H(Z_{i'0}|Z_{00})
- H(Z_{00},Z_{i'0},Z_{ii}|Y_i,Z_{i0})
\label{eq:urho<eta-01}
\\
H(Z_{00},Z_{i0}|Y_i,Z_{i'0},Z_{ii})
- \hr_0
&\leq
\hr_0 + \tr_i
+ H(Z_{i'0}|Z_{00})
- H(Z_{00},Z_{i'0},Z_{ii}|Y_i,Z_{i0})
\label{eq:urho<eta-02}
\\
H(Z_{00},Z_{i0}|Y_{i'},Z_{i'0},Z_{i'i'})
- \hr_0
&\leq
\hr_0 + \tr_i
+ H(Z_{i'0}|Z_{00})
- H(Z_{00},Z_{i'0},Z_{ii}|Y_i,Z_{i0})
\label{eq:urho<eta-03}
\\
H(Z_{i0},Z_{ii}|Y_i,Z_{00},Z_{i'0})
- \hr_i
&\leq
\hr_0 + \tr_i
+ H(Z_{i'0}|Z_{00})
- H(Z_{00},Z_{i'0},Z_{ii}|Y_i,Z_{i0})
\label{eq:urho<eta-04}
\\
H(Z_{i0},Z_{i'i'}|Y_{i'},Z_{00},Z_{i'0})
- \hr_{i'}
&\leq
\hr_0 + \tr_i
+ H(Z_{i'0}|Z_{00})
- H(Z_{00},Z_{i'0},Z_{ii}|Y_i,Z_{i0})
\label{eq:urho<eta-05}
\\
H(Z_{00},Z_{i0},Z_{ii}|Y_i,Z_{i'0})
- \hr_0
- \hr_i
&\leq
\hr_0 + \tr_i
+ H(Z_{i'0}|Z_{00})
- H(Z_{00},Z_{i'0},Z_{ii}|Y_i,Z_{i0})
\label{eq:urho<eta-06}
\\
H(Z_{00},Z_{i0},Z_{i'i'}|Y_{i'},Z_{i'0})
- \hr_0
- \hr_{i'}
&\leq
\hr_0 + \tr_i
+ H(Z_{i'0}|Z_{00})
- H(Z_{00},Z_{i'0},Z_{ii}|Y_i,Z_{i0})
\label{eq:urho<eta-07}
\\
H(Z_{10},Z_{20},Z_{i'i'}|Y_{i'},Z_{00})
- \tr_{i'}
&\leq
\hr_0 + \tr_i
+ H(Z_{i'0}|Z_{00})
- H(Z_{00},Z_{i'0},Z_{ii}|Y_i,Z_{i0})
\label{eq:urho<eta-08}
\\
H(Z_{00},Z_{10},Z_{20},Z_{i'i'}|Y_{i'})
- \hr_0 - \tr_{i'}
&\leq
\hr_0 + \tr_i
+ H(Z_{i'0}|Z_{00})
- H(Z_{00},Z_{i'0},Z_{ii}|Y_i,Z_{i0})
\label{eq:urho<eta-09}
\\
H(Z_{i0}|Y_i,Z_{00},Z_{i'0},Z_{ii})
&\leq
\tr_1 + \tr_2
- H(Z_{i'0},Z_{ii}|Y_i,Z_{00},Z_{i0}) 
- H(Z_{i'i'}|Y_{i'},Z_{00},Z_{10},Z_{20})
\label{eq:urho<eta-10}
\\
H(Z_{i0}|Y_{i'},Z_{00},Z_{i'0},Z_{i'i'})
&\leq
\tr_1 + \tr_2
- H(Z_{i'0},Z_{ii}|Y_i,Z_{00},Z_{i0})
- H(Z_{i'i'}|Y_{i'},Z_{00},Z_{10},Z_{20})
\label{eq:urho<eta-11}
\\
H(Z_{00},Z_{i0}|Y_i,Z_{i'0},Z_{ii})
- \hr_0
&\leq
\tr_1 + \tr_2
- H(Z_{i'0},Z_{ii}|Y_i,Z_{00},Z_{i0}) 
- H(Z_{i'i'}|Y_{i'},Z_{00},Z_{10},Z_{20})
\label{eq:urho<eta-12}
\\
H(Z_{00},Z_{i0}|Y_{i'},Z_{i'0},Z_{i'i'})
- \hr_0
&\leq
\tr_1 + \tr_2
- H(Z_{i'0},Z_{ii}|Y_i,Z_{00},Z_{i0}) 
- H(Z_{i'i'}|Y_{i'},Z_{00},Z_{10},Z_{20})
\label{eq:urho<eta-13}
\\
H(Z_{i0},Z_{ii}|Y_i,Z_{00},Z_{i'0})
- \hr_i
&\leq
\tr_1 + \tr_2
- H(Z_{i'0},Z_{ii}|Y_i,Z_{00},Z_{i0}) 
- H(Z_{i'i'}|Y_{i'},Z_{00},Z_{10},Z_{20})
\label{eq:urho<eta-14}
\\
H(Z_{i0},Z_{i'i'}|Y_{i'},Z_{00},Z_{i'0})
- \hr_{i'}
&\leq
\tr_1 + \tr_2
- H(Z_{i'0},Z_{ii}|Y_i,Z_{00},Z_{i0}) 
- H(Z_{i'i'}|Y_{i'},Z_{00},Z_{10},Z_{20})
\label{eq:urho<eta-15}
\\
H(Z_{00},Z_{i0},Z_{ii}|Y_i,Z_{i'0})
- \hr_0
- \hr_i
&\leq
\tr_1 + \tr_2
- H(Z_{i'0},Z_{ii}|Y_i,Z_{00},Z_{i0}) 
- H(Z_{i'i'}|Y_{i'},Z_{00},Z_{10},Z_{20})
\label{eq:urho<eta-16}
\\
H(Z_{00},Z_{i0},Z_{i'i'}|Y_{i'},Z_{i'0})
- \hr_0
- \hr_{i'}
&\leq
\tr_1 + \tr_2
- H(Z_{i'0},Z_{ii}|Y_i,Z_{00},Z_{i0}) 
- H(Z_{i'i'}|Y_{i'},Z_{00},Z_{10},Z_{20})
\label{eq:urho<eta-17}
\\
H(Z_{10},Z_{20},Z_{i'i'}|Y_{i'},Z_{00})
- \tr_{i'}
&\leq
\tr_1 + \tr_2
- H(Z_{i'0},Z_{ii}|Y_i,Z_{00},Z_{i0}) 
- H(Z_{i'i'}|Y_{i'},Z_{00},Z_{10},Z_{20})
\label{eq:urho<eta-18}
\\
H(Z_{00},Z_{10},Z_{20},Z_{i'i'}|Y_{i'})
- \hr_0 - \tr_{i'}
&\leq
\tr_1 + \tr_2
- H(Z_{i'0},Z_{ii}|Y_i,Z_{00},Z_{i0}) 
- H(Z_{i'i'}|Y_{i'},Z_{00},Z_{10},Z_{20})
\label{eq:urho<eta-19}
\\
H(Z_{i0}|Y_i,Z_{00},Z_{i'0},Z_{ii})
&\leq
\hr_0 + \tr_1 + \tr_2
- H(Z_{00},Z_{i'0},Z_{ii}|Y_i,Z_{i0})
- H(Z_{i'i'}|Y_{i'},Z_{00},Z_{10},Z_{20})
\label{eq:urho<eta-20}
\\
H(Z_{i0}|Y_{i'},Z_{00},Z_{i'0},Z_{i'i'})
&\leq
\hr_0 + \tr_1 + \tr_2
- H(Z_{00},Z_{i'0},Z_{ii}|Y_i,Z_{i0})
- H(Z_{i'i'}|Y_{i'},Z_{00},Z_{10},Z_{20})
\label{eq:urho<eta-21}
\\
H(Z_{00},Z_{i0}|Y_i,Z_{i'0},Z_{ii})
- \hr_0
&\leq
\hr_0 + \tr_1 + \tr_2
- H(Z_{00},Z_{i'0},Z_{ii}|Y_i,Z_{i0})
- H(Z_{i'i'}|Y_{i'},Z_{00},Z_{10},Z_{20})
\label{eq:urho<eta-22}
\\
H(Z_{00},Z_{i0}|Y_{i'},Z_{i'0},Z_{i'i'})
- \hr_0
&\leq
\hr_0 + \tr_1 + \tr_2
- H(Z_{00},Z_{i'0},Z_{ii}|Y_i,Z_{i0})
- H(Z_{i'i'}|Y_{i'},Z_{00},Z_{10},Z_{20})
\label{eq:urho<eta-23}
\\
H(Z_{i0},Z_{ii}|Y_i,Z_{00},Z_{i'0})
- \hr_i
&\leq
\hr_0 + \tr_1 + \tr_2
- H(Z_{00},Z_{i'0},Z_{ii}|Y_i,Z_{i0})
- H(Z_{i'i'}|Y_{i'},Z_{00},Z_{10},Z_{20})
\label{eq:urho<eta-24}
\\
H(Z_{i0},Z_{i'i'}|Y_{i'},Z_{00},Z_{i'0})
- \hr_{i'}
&\leq
\hr_0 + \tr_1 + \tr_2
- H(Z_{00},Z_{i'0},Z_{ii}|Y_i,Z_{i0})
- H(Z_{i'i'}|Y_{i'},Z_{00},Z_{10},Z_{20})
\label{eq:urho<eta-25}
\\
H(Z_{00},Z_{i0},Z_{ii}|Y_i,Z_{i'0})
- \hr_0
- \hr_i
&\leq
\hr_0 + \tr_1 + \tr_2
- H(Z_{00},Z_{i'0},Z_{ii}|Y_i,Z_{i0})
- H(Z_{i'i'}|Y_{i'},Z_{00},Z_{10},Z_{20})
\label{eq:urho<eta-26}
\\
H(Z_{00},Z_{i0},Z_{i'i'}|Y_{i'},Z_{i'0})
- \hr_0
- \hr_{i'}
&\leq
\hr_0 + \tr_1 + \tr_2
- H(Z_{00},Z_{i'0},Z_{ii}|Y_i,Z_{i0})
- H(Z_{i'i'}|Y_{i'},Z_{00},Z_{10},Z_{20})
\label{eq:urho<eta-27}
\\
H(Z_{10},Z_{20},Z_{i'i'}|Y_{i'},Z_{00})
- \tr_{i'}
&\leq
\hr_0 + \tr_1 + \tr_2
- H(Z_{00},Z_{i'0},Z_{ii}|Y_i,Z_{i0})
- H(Z_{i'i'}|Y_{i'},Z_{00},Z_{10},Z_{20})
\label{eq:urho<eta-28}
\\
H(Z_{00},Z_{10},Z_{20},Z_{i'i'}|Y_{i'})
- \hr_0 - \tr_{i'}
&\leq
\hr_0 + \tr_1 + \tr_2
- H(Z_{00},Z_{i'0},Z_{ii}|Y_i,Z_{i0})
- H(Z_{i'i'}|Y_{i'},Z_{00},Z_{10},Z_{20})
\label{eq:urho<eta-29}
\\
H(Z_{i0}|Y_i,Z_{00},Z_{i'0},Z_{ii})
&\leq
\hr_0 + \tr_1 + \tr_2
- H(Z_{i'0},Z_{ii}|Y_i,Z_{00},Z_{i0})
- H(Z_{00},Z_{i'i'}|Y_{i'},Z_{10},Z_{20})
\label{eq:urho<eta-30}
\\
H(Z_{i0}|Y_{i'},Z_{00},Z_{i'0},Z_{i'i'})
&\leq
\hr_0 + \tr_1 + \tr_2
- H(Z_{i'0},Z_{ii}|Y_i,Z_{00},Z_{i0})
- H(Z_{00},Z_{i'i'}|Y_{i'},Z_{10},Z_{20})
\label{eq:urho<eta-31}
\\
H(Z_{00},Z_{i0}|Y_i,Z_{i'0},Z_{ii})
- \hr_0
&\leq
\hr_0 + \tr_1 + \tr_2
- H(Z_{i'0},Z_{ii}|Y_i,Z_{00},Z_{i0})
- H(Z_{00},Z_{i'i'}|Y_{i'},Z_{10},Z_{20})
\label{eq:urho<eta-32}
\\
H(Z_{00},Z_{i0}|Y_{i'},Z_{i'0},Z_{i'i'})
- \hr_0
&\leq
\hr_0 + \tr_1 + \tr_2
- H(Z_{i'0},Z_{ii}|Y_i,Z_{00},Z_{i0})
- H(Z_{00},Z_{i'i'}|Y_{i'},Z_{10},Z_{20})
\label{eq:urho<eta-33}
\\
H(Z_{i0},Z_{ii}|Y_i,Z_{00},Z_{i'0})
- \hr_i
&\leq
\hr_0 + \tr_1 + \tr_2
- H(Z_{i'0},Z_{ii}|Y_i,Z_{00},Z_{i0})
- H(Z_{00},Z_{i'i'}|Y_{i'},Z_{10},Z_{20})
\label{eq:urho<eta-34}
\\
H(Z_{i0},Z_{i'i'}|Y_{i'},Z_{00},Z_{i'0})
- \hr_{i'}
&\leq
\hr_0 + \tr_1 + \tr_2
- H(Z_{i'0},Z_{ii}|Y_i,Z_{00},Z_{i0})
- H(Z_{00},Z_{i'i'}|Y_{i'},Z_{10},Z_{20})
\label{eq:urho<eta-35}
\\
H(Z_{00},Z_{i0},Z_{ii}|Y_i,Z_{i'0})
- \hr_0
- \hr_i
&\leq
\hr_0 + \tr_1 + \tr_2
- H(Z_{i'0},Z_{ii}|Y_i,Z_{00},Z_{i0})
- H(Z_{00},Z_{i'i'}|Y_{i'},Z_{10},Z_{20})
\label{eq:urho<eta-36}
\\
H(Z_{00},Z_{i0},Z_{i'i'}|Y_{i'},Z_{i'0})
- \hr_0
- \hr_{i'}
&\leq
\hr_0 + \tr_1 + \tr_2
- H(Z_{i'0},Z_{ii}|Y_i,Z_{00},Z_{i0})
- H(Z_{00},Z_{i'i'}|Y_{i'},Z_{10},Z_{20})
\label{eq:urho<eta-37}
\\
H(Z_{10},Z_{20},Z_{i'i'}|Y_{i'},Z_{00})
- \tr_{i'}
&\leq
\hr_0 + \tr_1 + \tr_2
- H(Z_{i'0},Z_{ii}|Y_i,Z_{00},Z_{i0})
- H(Z_{00},Z_{i'i'}|Y_{i'},Z_{10},Z_{20})
\label{eq:urho<eta-38}
\\
H(Z_{00},Z_{10},Z_{20},Z_{i'i'}|Y_{i'})
- \hr_0 - \tr_{i'}
&\leq
\hr_0 + \tr_1 + \tr_2
- H(Z_{i'0},Z_{ii}|Y_i,Z_{00},Z_{i0})
- H(Z_{00},Z_{i'i'}|Y_{i'},Z_{10},Z_{20})
\label{eq:urho<eta-39}
\\
H(Z_{i0}|Y_i,Z_{00},Z_{i'0},Z_{ii})
&\leq
2\hr_0 + \tr_1 + \tr_2
- H(Z_{00},Z_{i'0},Z_{ii}|Y_i,Z_{i0})
- H(Z_{00},Z_{i'i'}|Y_{i'},Z_{10},Z_{20})
\label{eq:urho<eta-40}
\\
H(Z_{i0}|Y_{i'},Z_{00},Z_{i'0},Z_{i'i'})
&\leq
2\hr_0 + \tr_1 + \tr_2
- H(Z_{00},Z_{i'0},Z_{ii}|Y_i,Z_{i0})
- H(Z_{00},Z_{i'i'}|Y_{i'},Z_{10},Z_{20})
\label{eq:urho<eta-41}
\\
H(Z_{00},Z_{i0}|Y_i,Z_{i'0},Z_{ii})
- \hr_0
&\leq
2\hr_0 + \tr_1 + \tr_2
- H(Z_{00},Z_{i'0},Z_{ii}|Y_i,Z_{i0})
- H(Z_{00},Z_{i'i'}|Y_{i'},Z_{10},Z_{20})
\label{eq:urho<eta-42}
\\
H(Z_{00},Z_{i0}|Y_{i'},Z_{i'0},Z_{i'i'})
- \hr_0
&\leq
2\hr_0 + \tr_1 + \tr_2
- H(Z_{00},Z_{i'0},Z_{ii}|Y_i,Z_{i0})
- H(Z_{00},Z_{i'i'}|Y_{i'},Z_{10},Z_{20})
\label{eq:urho<eta-43}
\\
H(Z_{i0},Z_{ii}|Y_i,Z_{00},Z_{i'0})
- \hr_i
&\leq
2\hr_0 + \tr_1 + \tr_2
- H(Z_{00},Z_{i'0},Z_{ii}|Y_i,Z_{i0})
- H(Z_{00},Z_{i'i'}|Y_{i'},Z_{10},Z_{20})
\label{eq:urho<eta-44}
\\
H(Z_{i0},Z_{i'i'}|Y_{i'},Z_{00},Z_{i'0})
- \hr_{i'}
&\leq
2\hr_0 + \tr_1 + \tr_2
- H(Z_{00},Z_{i'0},Z_{ii}|Y_i,Z_{i0})
- H(Z_{00},Z_{i'i'}|Y_{i'},Z_{10},Z_{20})
\label{eq:urho<eta-45}
\\
H(Z_{00},Z_{i0},Z_{ii}|Y_i,Z_{i'0})
- \hr_0
- \hr_i
&\leq
2\hr_0 + \tr_1 + \tr_2
- H(Z_{00},Z_{i'0},Z_{ii}|Y_i,Z_{i0})
- H(Z_{00},Z_{i'i'}|Y_{i'},Z_{10},Z_{20})
\label{eq:urho<eta-46}
\\
H(Z_{00},Z_{i0},Z_{i'i'}|Y_{i'},Z_{i'0})
- \hr_0
- \hr_{i'}
&\leq
2\hr_0 + \tr_1 + \tr_2
- H(Z_{00},Z_{i'0},Z_{ii}|Y_i,Z_{i0})
- H(Z_{00},Z_{i'i'}|Y_{i'},Z_{10},Z_{20})
\label{eq:urho<eta-47}
\\
H(Z_{10},Z_{20},Z_{i'i'}|Y_{i'},Z_{00})
- \tr_{i'}
&\leq
2\hr_0 + \tr_1 + \tr_2
- H(Z_{00},Z_{i'0},Z_{ii}|Y_i,Z_{i0})
- H(Z_{00},Z_{i'i'}|Y_{i'},Z_{10},Z_{20})
\label{eq:urho<eta-48}
\\
H(Z_{00},Z_{10},Z_{20},Z_{i'i'}|Y_{i'})
- \hr_0 - \tr_{i'}
&\leq
2\hr_0 + \tr_1 + \tr_2
- H(Z_{00},Z_{i'0},Z_{ii}|Y_i,Z_{i0})
- H(Z_{00},Z_{i'i'}|Y_{i'},Z_{10},Z_{20}),
\label{eq:urho<eta-49}
\end{align}
where
\begin{itemize}
\item
(\ref{eq:urho<eta-00}) is redundant because it comes from
(\ref{eq:hr0+tr2>00102022|2-10|00})$'$ and Lemma \ref{lem:xor} as
\begin{align}
 \hr_0 + \tr_i
 &\geq
 H(Z_{00},Z_{10},Z_{20},Z_{ii}|Y_i) - H(Z_{i'0}|Z_{00})
 \notag
 \\
 &\geq
 H(Z_{00},Z_{i'0},Z_{ii}|Y_i,Z_{i0})
 + H(Z_{i0}|Y_i,Z_{00},Z_{i'0},Z_{ii})
 - H(Z_{i'0}|Z_{00}),
\end{align}
\item
(\ref{eq:urho<eta-01}) is equivalent to
\begin{align}
 \hr_0 + \tr_i
 &\geq
 H(Z_{00},Z_{i'0},Z_{ii}|Y_i,Z_{i0})
 + H(Z_{i0}|Y_{i'},Z_{00},Z_{i'0},Z_{i'i'})
 - H(Z_{i'0}|Z_{00}),
 \label{eq:hr0+tr1>10|2+002011|1-20|00}
\end{align}
\item
(\ref{eq:urho<eta-02}) is redundant because it comes from
(\ref{eq:hr0>00|1}),
(\ref{eq:hr0+tr2>00102022|2-10|00})$'$,
and Lemma \ref{lem:relocation} as
\begin{align}
 \hr_0 + [ \hr_0 + \tr_i ]
 &\geq
 H(Z_{00}|Y_i,Z_{10},Z_{20},Z_{ii})
 + [ H(Z_{00},Z_{10},Z_{20},Z_{ii}|Y_i) - H(Z_{i'0}|Z_{00}) ]
 \notag
 \\
 &\geq
 H(Z_{00},Z_{i'0},Z_{ii}|Y_i,Z_{i0})
 + H(Z_{00},Z_{i0}|Y_i,Z_{i'0},Z_{ii})
 - H(Z_{i'0}|Z_{00}),
\end{align}
\item
(\ref{eq:urho<eta-03}) is equivalent to
\begin{align}
 2\hr_0 + \tr_i
 &\geq
 H(Z_{00},Z_{i'0},Z_{ii}|Y_i,Z_{i0})
 + H(Z_{00},Z_{i0}|Y_{i'},Z_{i'0},Z_{i'i'})
 - H(Z_{i'0}|Z_{00}),
 \label{eq:2hr0+tr1>0010|2+002011|1-20|00}
\end{align}
\item
(\ref{eq:urho<eta-04}) is redundant because it comes from
(\ref{eq:tr1>1011|1}) and
(\ref{eq:hr0+hr2>001022|2-10|00})$'$ as
\begin{align}
 \tr_i + [ \hr_0 + \hr_i ]
 &\geq
 H(Z_{i0},Z_{ii}|Y_i,Z_{00},Z_{i'0})
 + [H(Z_{00},Z_{i'0},Z_{ii}|Y_i,Z_{i0}) - H(Z_{i'0}|Z_{00})],
\end{align}
\item
(\ref{eq:urho<eta-05}) is redundant because it comes from
(\ref{eq:hr1=tr1-10|0011})$'$
and
(\ref{eq:hr0+tr1+tr2>002011|1+1022|2})
as
\begin{align}
 \hr_0
 + \hr_{i'}
 + \tr_{i}
 &=
 [ \hr_0 + \tr_1 + \tr_2 ] - H(Z_{i'0}|Z_{00},Z_{i'i'})
 \notag
 \\
 &\geq
 [
  H(Z_{00},Z_{i'0},Z_{ii}|Y_i,Z_{i0})
  + H(Z_{i0},Z_{i'i'}|Y_{i'},Z_{00},Z_{i'0})
 ]
 - H(Z_{i'0}|Z_{00},Z_{i'i'})
 \notag
 \\
 &\geq
 H(Z_{00},Z_{i'0},Z_{ii}|Y_i,Z_{i0})
 + H(Z_{i0},Z_{i'i'}|Y_{i'},Z_{00},Z_{i'0})
 - H(Z_{i'0}|Z_{00}),
\end{align}
\item
(\ref{eq:urho<eta-06}) is redundant because it comes from
(\ref{eq:hr0+tr1>001011|1}) and (\ref{eq:hr0+hr2>001022|2-10|00})$'$ as
\begin{align}
 [ \hr_0 + \tr_i ] + [ \hr_0 + \hr_i ]
 &\geq
 H(Z_{00},Z_{i0},Z_{ii}|Y_i,Z_{i'0})
 + [ H(Z_{00},Z_{i'0},Z_{ii}|Y_i,Z_{i0}) - H(Z_{i'0}|Z_{00}) ],
\end{align}
\item
(\ref{eq:urho<eta-07}) is redundant because it comes from
(\ref{eq:hr1=tr1-10|0011})$'$
and
(\ref{eq:2hr0+tr1+tr2>002011|1+001022|2})
as
\begin{align}
 2 \hr_0 + \hr_{i'} + \tr_i
 &=
 [ 2 \hr_0 + \tr_1 + \tr_2 ] - H(Z_{i'0}|Z_{00},Z_{i'i'})
 \notag
 \\
 &\geq
 [
  H(Z_{00},Z_{i'0},Z_{ii}|Y_i,Z_{i0})
  + H(Z_{00},Z_{i0},Z_{i'i'}|Y_{i'},Z_{i'0})
 ]
 - H(Z_{i'0}|Z_{00},Z_{i'i'})
 \notag
 \\
 &\geq
 H(Z_{00},Z_{i'0},Z_{ii}|Y_i,Z_{i0})
 + H(Z_{00},Z_{i0},Z_{i'i'}|Y_{i'},Z_{i'0})
 - H(Z_{i'0}|Z_{00}),
\end{align}
\item
(\ref{eq:urho<eta-08}) is redundant because it comes from
(\ref{eq:hr0+tr1+tr2>002011|1+1022|2}) and Lemma \ref{lem:promotion} as
\begin{align}
 \hr_0 + \tr_1 + \tr_2
 &\geq
 H(Z_{00},Z_{i'0},Z_{ii}|Y_i,Z_{i0})
 + H(Z_{i0},Z_{i'i'}|Y_{i'},Z_{00},Z_{i'0})
 \notag
 \\
 &\geq
 H(Z_{00},Z_{i'0},Z_{ii}|Y_i,Z_{i0})
 + H(Z_{10},Z_{20},Z_{i'i'}|Y_{i'},Z_{00})
 - H(Z_{i'0}|Z_{00}),
\end{align}
\item
(\ref{eq:urho<eta-09}) equivalent to
\begin{align}
 2\hr_0 + \tr_1 + \tr_2
 &\geq
 H(Z_{00},Z_{i'0},Z_{ii}|Y_i,Z_{i0})
 + H(Z_{00},Z_{10},Z_{20},Z_{i'i'}|Y_{i'})
 - H(Z_{i'0}|Z_{00}),
 \label{eq:2hr0+tr1+tr2>00102022|2+002011|1-20|00}
\end{align}
\item
(\ref{eq:urho<eta-10}) is redundant because it comes from
(\ref{eq:tr1+tr2>102022|2+11|1})$'$ and Lemma \ref{lem:xor} as
\begin{align}
 \tr_1 + \tr_2
 &\geq
 H(Z_{i'i'}|Y_{i'},Z_{00},Z_{10},Z_{20})
 + H(Z_{10},Z_{20},Z_{ii}|Y_i,Z_{00})
 \notag
 \\
 &\geq
 H(Z_{i'i'}|Y_{i'},Z_{00},Z_{10},Z_{20})
 + H(Z_{i0}|Y_i,Z_{00},Z_{i'0},Z_{ii})
 + H(Z_{i'0},Z_{ii}|Y_i,Z_{00},Z_{i0}),
\end{align}
\item
(\ref{eq:urho<eta-11}) is redundant because it comes from
(\ref{eq:tr1+tr2>2011|1+1022|2}) and Lemma \ref{lem:xor} as
\begin{align}
 \tr_1 + \tr_2
 &\geq
 H(Z_{i0},Z_{i'i'}|Y_{i'},Z_{00},Z_{i'0})
 + H(Z_{i'0},Z_{ii}|Y_i,Z_{00},Z_{i0})
 \notag
 \\
 &\geq
 H(Z_{i0}|Y_{i'},Z_{00},Z_{i'0},Z_{i'i'})
 + H(Z_{i'i'}|Y_{i'},Z_{00},Z_{10},Z_{20})
 + H(Z_{i'0},Z_{ii}|Y_i,Z_{00},Z_{i0}),
\end{align}
\item
(\ref{eq:urho<eta-12}) is redundant because it comes from
(\ref{eq:hr0+tr1+tr2>00102022|2+11|1})$'$ and Lemma \ref{lem:xor} as
\begin{align}
 \hr_0 + \tr_1 + \tr_2
 &\geq
 H(Z_{i'i'}|Y_{i'},Z_{00},Z_{10},Z_{20})
 + H(Z_{00},Z_{10},Z_{20},Z_{ii}|Y_i)
 \notag
 \\
 &\geq
 H(Z_{i'i'}|Y_{i'},Z_{00},Z_{10},Z_{20})
 + H(Z_{00},Z_{i0}|Y_i,Z_{i'0},Z_{ii})
 + H(Z_{i'0},Z_{ii}|Y_i,Z_{00},Z_{i0}),
\end{align}
\item
(\ref{eq:urho<eta-13}) is redundant because it comes from
(\ref{eq:hr0+tr1+tr2>002011|1+1022|2})$'$ and Lemma \ref{lem:xor} as
\begin{align}
 \hr_0 + \tr_1 + \tr_2
 &\geq
 H(Z_{00},Z_{i0},Z_{i'i'}|Y_{i'},Z_{i'0})
 + H(Z_{i'0},Z_{ii}|Y_i,Z_{00},Z_{i0})
 \notag
 \\
 &\geq
 H(Z_{00},Z_{i0}|Y_{i'},Z_{i'0},Z_{i'i'})
 + H(Z_{i'i'}|Y_{i'},Z_{00},Z_{10},Z_{20})
 + H(Z_{i'0},Z_{ii}|Y_i,Z_{00},Z_{i0}),
\end{align}
\item
(\ref{eq:urho<eta-14}) is redundant because it comes from
(\ref{eq:tr1>1011|1}) and
(\ref{eq:hr2+tr1>1022|2+11|1})$'$ as
\begin{align}
 \tr_i + [ \hr_i + \tr_{i'} ]
 &\geq
 H(Z_{i0},Z_{ii}|Y_i,Z_{00},Z_{i'0})
 + [
  H(Z_{i'i'}|Y_{i'},Z_{00},Z_{10},Z_{20})
  + H(Z_{i'0},Z_{ii}|Y_i,Z_{00},Z_{i0})
 ],
\end{align}
\item
(\ref{eq:urho<eta-15}) is redundant because it comes from
(\ref{eq:hr1>11|1})$'$ and
(\ref{eq:tr1+tr2>2011|1+1022|2}) as
\begin{align}
 \hr_{i'} + [ \tr_1 + \tr_2 ]
 &\geq
 H(Z_{i'i'}|Y_{i'},Z_{00},Z_{10},Z_{20})
 + [
  H(Z_{i0},Z_{i'i'}|Y_{i'},Z_{00},Z_{i'0})
  + H(Z_{i'0},Z_{ii}|Y_i,Z_{00},Z_{i0})
 ],
\end{align}
\item
(\ref{eq:urho<eta-16}) is redundant because it comes from
(\ref{eq:hr0+tr1>001011|1}) and
(\ref{eq:hr2+tr1>1022|2+11|1})$'$ as
\begin{align}
 [ \hr_0 + \tr_i ] + [ \hr_i + \tr_{i'} ]
 &\geq
 H(Z_{00},Z_{i0},Z_{ii}|Y_i,Z_{i'0})
 + [
  H(Z_{i'i'}|Y_{i'},Z_{00},Z_{10},Z_{20})
  + H(Z_{i'0},Z_{ii}|Y_i,Z_{00},Z_{i0})
 ],
\end{align}
\item
(\ref{eq:urho<eta-17}) is redundant because it comes from
(\ref{eq:hr1>11|1})$'$ and
(\ref{eq:hr0+tr1+tr2>002011|1+1022|2})$'$ as
\begin{align}
 \hr_{i'} + [ \hr_0 + \tr_1 + \tr_2 ]
 &\geq
 H(Z_{i'i'}|Y_{i'},Z_{00},Z_{10},Z_{20})
 + [ H(Z_{00},Z_{i0},Z_{i'i'}|Y_{i'},Z_{i'0})
  + H(Z_{i'0},Z_{ii}|Y_i,Z_{00},Z_{i0}) ],
\end{align}
\item
(\ref{eq:urho<eta-18}) is equivalent to
\begin{align}
 \tr_i + 2\tr_{i'}
 &\geq
 H(Z_{i'0},Z_{ii}|Y_i,Z_{00},Z_{i0})
 + H(Z_{10},Z_{20},Z_{i'i'}|Y_{i'},Z_{00})
 + H(Z_{i'i'}|Y_{i'},Z_{00},Z_{10},Z_{20}),
 \label{eq:tr1+2tr2>102022|2+2011|1+22|2}
\end{align}
\item
(\ref{eq:urho<eta-19}) is equivalent to
\begin{align}
 \hr_0 + \tr_i + 2\tr_{i'}
 &\geq
 H(Z_{i'0},Z_{ii}|Y_i,Z_{00},Z_{i0})
 + H(Z_{00},Z_{10},Z_{20},Z_{i'i'}|Y_{i'})
 + H(Z_{i'i'}|Y_{i'},Z_{00},Z_{10},Z_{20}),
 \label{eq:hr0+tr1+2tr2>00102022|2+2011|1+22|2}
\end{align}
\item
(\ref{eq:urho<eta-20}) is redundant because it comes from
(\ref{eq:hr0+tr1+tr2>00102022|2+11|1})$'$ and Lemma \ref{lem:xor} as
\begin{align}
 \hr_0 + \tr_1 + \tr_2
 &\geq
 H(Z_{i'i'}|Y_{i'},Z_{00},Z_{10},Z_{20})
 + H(Z_{00},Z_{10},Z_{20},Z_{ii}|Y_i)
 \notag
 \\
 &\geq
 H(Z_{i'i'}|Y_{i'},Z_{00},Z_{10},Z_{10})
 + H(Z_{00},Z_{i'0},Z_{ii}|Y_i,Z_{i0})
 + H(Z_{i0}|Y_i,Z_{00},Z_{i'0},Z_{ii}),
\end{align}
\item
(\ref{eq:urho<eta-21}) is redundant because it comes from
(\ref{eq:hr0+tr1+tr2>102022|2+0011|1})$'$ and Lemma \ref{lem:xor} as
\begin{align}
 \hr_0 + \tr_1 + \tr_2
 &\geq
 H(Z_{i0},Z_{i'i'}|Y_{i'},Z_{00},Z_{i'0})
 + H(Z_{00},Z_{i'0},Z_{ii}|Y_i,Z_{i0})
 \notag
 \\
 &\geq
 H(Z_{i0}|Y_{i'},Z_{00},Z_{i'0},Z_{i'i'})
 + H(Z_{i'i'}|Y_{i'},Z_{00},Z_{10},Z_{20})
 + H(Z_{00},Z_{i'0},Z_{ii}|Y_i,Z_{i0}),
\end{align}
\item
(\ref{eq:urho<eta-22}) is redundant because it comes from
(\ref{eq:hr0>00|1}),
(\ref{eq:hr0+tr1+tr2>00102022|2+11|1})$'$,
and Lemma \ref{lem:relocation} as
\begin{align}
 \hr_0 + [ \hr_0 + \tr_1 + \tr_2 ]
 &\geq
 H(Z_{00}|Y_i,Z_{10},Z_{20},Z_{ii})
 + [
  H(Z_{i'i'}|Y_{i'},Z_{00},Z_{10},Z_{20})
  + H(Z_{00},Z_{10},Z_{20},Z_{ii}|Y_i)
 ]
 \notag
 \\
 &\geq
 H(Z_{00},Z_{i0}|Y_i,Z_{i'0},Z_{ii})
 + H(Z_{i'i'}|Y_{i'},Z_{00},Z_{10},Z_{20})
 + H(Z_{00},Z_{i'0},Z_{ii}|Y_i,Z_{i0}),
\end{align}
\item
(\ref{eq:urho<eta-23}) is redundant because it comes from
(\ref{eq:2hr0+tr1+tr2>002011|1+001022|2}) and Lemma \ref{lem:xor} as
\begin{align}
 2 \hr_0 + \tr_1 + \tr_2
 &\geq
 H(Z_{00},Z_{i0},Z_{i'i'}|Y_{i'},Z_{i'0})
 + H(Z_{00},Z_{i'0},Z_{ii}|Y_i,Z_{i0})
 \notag
 \\
 &\geq
 H(Z_{00},Z_{i0}|Y_{i'},Z_{i'0},Z_{i'i'})
 + H(Z_{i'i'}|Y_{i'},Z_{00},Z_{10},Z_{20})
 + H(Z_{00},Z_{i'0},Z_{ii}|Y_i,Z_{i0}),
\end{align}
\item
(\ref{eq:urho<eta-24}) is redundant because it comes from
(\ref{eq:hr1>11|1}),
(\ref{eq:hr0+tr1+tr2>00102022|2+11|1})$'$,
and Lemma \ref{lem:relocation} as
\begin{align}
 \hr_i + [ \hr_0 + \tr_1 + \tr_2 ]
 &\geq
 H(Z_{ii}|Y_i,Z_{00},Z_{10},Z_{20})
 + [
  H(Z_{i'i'}|Y_{i'},Z_{00},Z_{10},Z_{20})
  + H(Z_{00},Z_{10},Z_{20},Z_{ii}|Y_i)
 ]
 \notag
 \\
 &\geq
 H(Z_{i0},Z_{ii}|Y_i,Z_{00},Z_{i'0})
 + H(Z_{i'i'}|Y_{i'},Z_{00},Z_{10},Z_{20})
 + H(Z_{00},Z_{i'0},Z_{ii}|Y_i,Z_{i0}),
\end{align}
\item
(\ref{eq:urho<eta-25}) is redundant because it comes from
(\ref{eq:hr1>11|1})$'$ and 
(\ref{eq:hr0+tr1+tr2>002011|1+1022|2}) as
\begin{align}
 \hr_{i'} + [ \hr_0 + \tr_1 + \tr_2 ]
 &\geq
 H(Z_{i'i'}|Y_{i'},Z_{00},Z_{10},Z_{20})
 + [ H(Z_{00},Z_{i'0},Z_{ii}|Y_i,Z_{i0}) 
  + H(Z_{i0},Z_{i'i'}|Y_{i'},Z_{00},Z_{i'0}) ],
\end{align}
\item
(\ref{eq:urho<eta-26}) is redundant because it comes from
(\ref{eq:hr0+hr1>0011|1}),
(\ref{eq:hr0+tr1+tr2>00102022|2+11|1})$'$,
and Lemma \ref{lem:relocation} as
\begin{align}
 [ \hr_0 + \hr_i ] + [ \hr_0 + \tr_1 + \tr_2 ]
 &\geq
 H(Z_{00},Z_{ii}|Y_i,Z_{10},Z_{20})
 + [
  H(Z_{i'i'}|Y_{i'},Z_{00},Z_{10},Z_{20})
  + H(Z_{00},Z_{10},Z_{20},Z_{ii}|Y_i)
 ]
 \notag
 \\
 &\geq
 H(Z_{00},Z_{i0},Z_{ii}|Y_i,Z_{i'0})
 + H(Z_{i'i'}|Y_{i'},Z_{00},Z_{10},Z_{20})
 + H(Z_{00},Z_{i'0},Z_{ii}|Y_i,Z_{i0}),
\end{align}
\item
(\ref{eq:urho<eta-27}) is redundant because it comes from
(\ref{eq:hr1>11|1})$'$ and
(\ref{eq:2hr0+tr1+tr2>002011|1+001022|2}) as
\begin{align}
 \hr_{i'} + [ 2 \hr_0 + \tr_1 + \tr_2 ]
 &\geq
 H(Z_{i'i'}|Y_{i'},Z_{00},Z_{10},Z_{20})
 + [ 
  H(Z_{00},Z_{i'0},Z_{ii}|Y_i,Z_{i0})
  + H(Z_{00},Z_{i0},Z_{i'i'}|Y_{i'},Z_{i'0})
 ],
\end{align}
\item
(\ref{eq:urho<eta-28}) is equivalent to
\begin{align}
 \hr_0 + \tr_i + 2\tr_{i'}
 &\geq
 H(Z_{00},Z_{i'0},Z_{ii}|Y_i,Z_{i0})
 + H(Z_{10},Z_{20},Z_{i'i'}|Y_{i'},Z_{00})
 + H(Z_{i'i'}|Y_{i'},Z_{00},Z_{10},Z_{20}),
 \label{eq:hr0+tr1+2tr2>102022|2+002011|1+22|2}
\end{align}
\item
(\ref{eq:urho<eta-29}) is equivalent to
\begin{align}
 2\hr_0 + \tr_i + 2\tr_{i'}
 &\geq
 H(Z_{00},Z_{i'0},Z_{ii}|Y_i,Z_{i0})
 + H(Z_{00},Z_{10},Z_{20},Z_{i'i'}|Y_{i'})
 + H(Z_{i'i'}|Y_{i'},Z_{00},Z_{10},Z_{20}),
 \label{eq:2hr0+tr1+2tr2>00102022|2+002011|1+22|2}
\end{align}
\item
(\ref{eq:urho<eta-30}) is redundant because it comes from
(\ref{eq:hr0+tr1+tr2>102022|2+0011|1})$'$ and Lemma \ref{lem:xor} as
\begin{align}
 \hr_0 + \tr_1 + \tr_2
 &\geq
 H(Z_{00},Z_{i'i'}|Y_{i'},Z_{10},Z_{20})
 + H(Z_{10},Z_{20},Z_{ii}|Y_i,Z_{00})
 \notag
 \\
 &\geq
 H(Z_{00},Z_{i'i'}|Y_{i'},Z_{10},Z_{20})
 + H(Z_{i0}|Y_i,Z_{00},Z_{i'0},Z_{ii})
 + H(Z_{i'0},Z_{ii}|Y_i,Z_{00},Z_{i0}),
\end{align}
\item
(\ref{eq:urho<eta-31}) is redundant because it comes from
(\ref{eq:hr0+tr1+tr2>002011|1+1022|2})$'$
and Lemma \ref{lem:xor} as
\begin{align}
 \hr_0 + \tr_1 + \tr_2
 &\geq
 H(Z_{00},Z_{i0},Z_{i'i'}|Y_{i'},Z_{i'0})
 + H(Z_{i'0},Z_{ii}|Y_i,Z_{00},Z_{i0})
 \notag
 \\
 &\geq
 H(Z_{00},Z_{i'i'}|Y_{i'},Z_{10},Z_{20}) 
 + H(Z_{i0}|Y_{i'},Z_{00},Z_{i'0},Z_{i'i'})
 + H(Z_{i'0},Z_{ii}|Y_i,Z_{00},Z_{i0})
\end{align}
\item
(\ref{eq:urho<eta-32}) is redundant because it comes from
(\ref{eq:2hr0+tr1+tr2>00102022|2+0011|1})$'$ and Lemma \ref{lem:xor} as
\begin{align}
 2 \hr_0 + \tr_1 + \tr_2
 &\geq
 H(Z_{00},Z_{i'i'}|Y_{i'},Z_{10},Z_{20})
 + H(Z_{00},Z_{10},Z_{20},Z_{ii}|Y_i)
 \notag
 \\
 &\geq
 H(Z_{00},Z_{i'i'}|Y_{i'},Z_{10},Z_{20})
 + H(Z_{00},Z_{i0}|Y_i,Z_{i'0},Z_{ii})
 + H(Z_{i'0},Z_{ii}|Y_i,Z_{00},Z_{i0}),
\end{align}
\item
(\ref{eq:urho<eta-33}) is redundant because it comes from
(\ref{eq:hr0>00|1})$'$,
(\ref{eq:hr0+tr1+tr2>002011|1+1022|2})$'$,
and Lemma \ref{lem:relocation} as
\begin{align}
 \hr_0 + [ \hr_0 + \tr_1 + \tr_2 ]
 &\geq
 H(Z_{00}|Y_{i'},Z_{10},Z_{20},Z_{i'i'})
 + [ H(Z_{00},Z_{i0},Z_{i'i'}|Y_{i'},Z_{i'0}) 
  +  H(Z_{i'0},Z_{ii}|Y_i,Z_{00},Z_{i0}) ]
 \notag
 \\
 &\geq
 H(Z_{00},Z_{i0}|Y_{i'},Z_{i'0},Z_{i'i'})
 + H(Z_{00},Z_{i'i'}|Y_{i'},Z_{10},Z_{20})
 + H(Z_{i'0},Z_{ii}|Y_i,Z_{00},Z_{i0}),
\end{align}
\item
(\ref{eq:urho<eta-34}) is redundant because it comes from
(\ref{eq:hr1>11|1}),
(\ref{eq:hr0+tr1+tr2>102022|2+0011|1})$'$,
and Lemma \ref{lem:relocation} as
\begin{align}
 \hr_i + [ \hr_0 + \tr_1 + \tr_2 ]
 &\geq
 H(Z_{ii}|Y_i,Z_{00},Z_{10},Z_{20})
 + [
  H(Z_{00},Z_{i'i'}|Y_{i'},Z_{10},Z_{20}) 
  + H(Z_{10},Z_{20},Z_{ii}|Y_i,Z_{00})
 ]
 \notag
 \\
 &\geq
 H(Z_{i0},Z_{ii}|Y_i,Z_{00},Z_{i'0})
 + H(Z_{00},Z_{i'i'}|Y_{i'},Z_{10},Z_{20})
 + H(Z_{i'0},Z_{ii}|Y_i,Z_{00},Z_{i0}),
\end{align}
\item
(\ref{eq:urho<eta-35}) is redundant because it comes from
(\ref{eq:hr1>11|1})$'$,
(\ref{eq:hr0+tr1+tr2>002011|1+1022|2})$'$,
and Lemma \ref{lem:relocation} as
\begin{align}
 \hr_{i'} + [ \hr_0 + \tr_1 + \tr_2 ]
 &\geq
 H(Z_{i'i'}|Y_{i'},Z_{00},Z_{10},Z_{20})
 + [
  H(Z_{00},Z_{i0},Z_{i'i'}|Y_{i'},Z_{i'0})
  + H(Z_{i'0},Z_{ii}|Y_i,Z_{00},Z_{i0})
 ]
 \notag
 \\
 &\geq
 H(Z_{i0},Z_{i'i'}|Y_{i'},Z_{00},Z_{i'0})
 + H(Z_{00},Z_{i'i'}|Y_{i'},Z_{10},Z_{20})
 + H(Z_{i'0},Z_{ii}|Y_i,Z_{00},Z_{i0}),
\end{align}
\item
(\ref{eq:urho<eta-36}) is redundant because it comes from
(\ref{eq:hr0+tr1>001011|1}) and 
(\ref{eq:hr0+hr2+tr1>0011|1+1022|2})$'$ as
\begin{align}
 [ \hr_0 + \tr_i ] + [ \hr_0 + \hr_1 + \tr_2 ]
 &\geq
 H(Z_{00},Z_{i0},Z_{ii}|Y_i,Z_{i'0})
 + [
  H(Z_{00},Z_{i'i'}|Y_{i'},Z_{10},Z_{20})
  + H(Z_{i'0},Z_{ii}|Y_i,Z_{00},Z_{i0})
 ],
\end{align}
\item
(\ref{eq:urho<eta-37}) is redundant because it comes from
(\ref{eq:hr0+hr1>0011|1})$'$ and 
(\ref{eq:hr0+tr1+tr2>002011|1+1022|2})$'$ as
\begin{align}
 [ \hr_0 + \hr_{i'} ] + [ \hr_0 + \tr_1 + \tr_2 ]
 &\geq
 H(Z_{00},Z_{i'i'}|Y_{i'},Z_{10},Z_{20})
 \notag
 \\*
 &\quad
 + [
  H(Z_{00},Z_{i0},Z_{i'i'}|Y_{i'},Z_{i'0})
  + H(Z_{i'0},Z_{ii}|Y_i,Z_{00},Z_{i0})
 ],
\end{align}
\item
(\ref{eq:urho<eta-38}) is redundant because it comes from
(\ref{eq:hr0+tr1+2tr2>00102022|2+2011|1+22|2})
and Lemma \ref{lem:relocation} as
\begin{align}
 \hr_0 + \tr_i + 2 \tr_{i'}
 &\geq
 H(Z_{i'0},Z_{ii}|Y_i,Z_{00},Z_{i0})
 + H(Z_{00},Z_{10},Z_{20},Z_{i'i'}|Y_{i'})
 + H(Z_{i'i'}|Y_{i'},Z_{00},Z_{10},Z_{20})
 \notag
 \\
 &\geq
 H(Z_{i'0},Z_{ii}|Y_i,Z_{00},Z_{i0})
 + H(Z_{00},Z_{i'i'}|Y_{i'},Z_{10},Z_{20})
 + H(Z_{i0},Z_{i'0},Z_{i'i'}|Y_{i'},Z_{00}),
\end{align}
\item
(\ref{eq:urho<eta-39}) equivalent to
\begin{align}
 2\hr_0 + \tr_i + 2\tr_{i'}
 &\geq
 H(Z_{i'0},Z_{ii}|Y_i,Z_{00},Z_{i0})
 + H(Z_{00},Z_{10},Z_{20},Z_{i'i'}|Y_{i'})
 + H(Z_{00},Z_{i'i'}|Y_{i'},Z_{10},Z_{20}),
 \label{eq:2hr0+tr1+2tr2>00102022|2+2011|1+0022|2}
\end{align}
\item
(\ref{eq:urho<eta-40}) is redundant because it comes from
(\ref{eq:2hr0+tr1+tr2>00102022|2+0011|1})$'$ and Lemma \ref{lem:xor} as
\begin{align}
 2 \hr_0 + \tr_1 + \tr_2
 &\geq
 H(Z_{00},Z_{i'i'}|Y_{i'},Z_{10},Z_{20})
 + H(Z_{00},Z_{10},Z_{20},Z_{ii}|Y_i)
 \notag
 \\
 &\geq
 H(Z_{00},Z_{i'i'}|Y_{i'},Z_{10},Z_{20})
 + H(Z_{00},Z_{i'0},Z_{ii}|Y_i,Z_{i0})
 + H(Z_{i0}|Y_i,Z_{00},Z_{i'0},Z_{ii}),
\end{align}
\item
(\ref{eq:urho<eta-41}) is redundant because it comes from
(\ref{eq:2hr0+tr1+tr2>002011|1+001022|2}) and Lemma \ref{lem:xor} as
\begin{align}
 2 \hr_0 + \tr_1 + \tr_2
 &\geq
 H(Z_{00},Z_{i0},Z_{i'i'}|Y_{i'},Z_{i'0})
 + H(Z_{00},Z_{i'0},Z_{ii}|Y_i,Z_{i0})
 \notag
 \\
 &\geq
 H(Z_{00},Z_{i'i'}|Y_{i'},Z_{10},Z_{20})
 + H(Z_{i0}|Y_{i'},Z_{00},Z_{i'0},Z_{i'i'})
 + H(Z_{00},Z_{i'0},Z_{ii}|Y_i,Z_{i0}),
\end{align}
\item
(\ref{eq:urho<eta-42}) is redundant because it comes from
(\ref{eq:hr0>00|1}),
(\ref{eq:2hr0+tr1+tr2>00102022|2+0011|1})$'$,
and Lemma \ref{lem:relocation} as
\begin{align}
 \hr_0 + [ 2 \hr_0 + \tr_1 + \tr_2 ]
 &\geq
 H(Z_{00}|Y_i,Z_{10},Z_{20},Z_{ii})
 + [
  H(Z_{00},Z_{i'i'}|Y_{i'},Z_{10},Z_{20})
  + H(Z_{00},Z_{10},Z_{20},Z_{ii}|Y_i)
 ]
 \notag
 \\
 &\geq
 H(Z_{00},Z_{i0}|Y_i,Z_{i'0},Z_{ii})
 + H(Z_{00},Z_{i'i'}|Y_{i'},Z_{i0},Z_{i'0})
 + H(Z_{00},Z_{i'0},Z_{ii}|Y_i,Z_{i0}),
\end{align}
\item
(\ref{eq:urho<eta-43}) is redundant because it comes from
(\ref{eq:hr0>00|1})$'$,
(\ref{eq:2hr0+tr1+tr2>002011|1+001022|2}),
and Lemma \ref{lem:relocation} as
\begin{align}
 \hr_0 + [ 2 \hr_0 + \tr_1 + \tr_2 ]
 &\geq
 H(Z_{00}|Y_{i'},Z_{10},Z_{10},Z_{i'i'})
 + [
  H(Z_{00},Z_{i'0},Z_{ii}|Y_i,Z_{i0})
  + H(Z_{00},Z_{i0},Z_{i'i'}|Y_{i'},Z_{i'0})
 ]
 \notag
 \\
 &\geq
 H(Z_{00},Z_{i0}|Y_{i'},Z_{i'0},Z_{i'i'})
 + H(Z_{00},Z_{i'0},Z_{ii}|Y_i,Z_{i0})
 + H(Z_{00},Z_{i'i'}|Y_{i'},Z_{10},Z_{20}),
\end{align}
\item
(\ref{eq:urho<eta-44}) is redundant because it comes from
(\ref{eq:hr1>11|1}),
(\ref{eq:2hr0+tr1+tr2>00102022|2+0011|1})$'$,
and Lemma \ref{lem:relocation} as
\begin{align}
 \hr_i + [ 2 \hr_0 + \tr_1 + \tr_2 ]
 &\geq
 H(Z_{ii}|Y_i,Z_{00},Z_{10},Z_{20})
 + [
  H(Z_{00},Z_{i'i'}|Y_{i'},Z_{10},Z_{20})
  + H(Z_{00},Z_{10},Z_{20},Z_{ii}|Y_i)
 ]
 \notag
 \\
 &\geq
 H(Z_{i0},Z_{ii}|Y_i,Z_{00},Z_{i'0})
 + H(Z_{00},Z_{i'i'}|Y_{i'},Z_{10},Z_{20})
 + H(Z_{00},Z_{i'0},Z_{ii}|Y_i,Z_{i0}),
\end{align}
\item
(\ref{eq:urho<eta-45}) is redundant because it comes from
(\ref{eq:hr1>11|1})$'$,
(\ref{eq:2hr0+tr1+tr2>002011|1+001022|2}),
and Lemma \ref{lem:relocation} as
\begin{align}
 \hr_{i'} + [ 2 \hr_0 + \tr_1 + \tr_2 ]
 &\geq
 H(Z_{i'i'}|Y_{i'},Z_{00},Z_{10},Z_{20})
 + [
  H(Z_{00},Z_{i0},Z_{i'i'}|Y_{i'},Z_{i'0})
  + H(Z_{00},Z_{i'0},Z_{ii}|Y_i,Z_{i0})
 ]
 \notag
 \\
 &\geq
 H(Z_{i0},Z_{i'i'}|Y_{i'},Z_{00},Z_{i'0})
 + H(Z_{00},Z_{i'i'}|Y_{i'},Z_{10},Z_{20})
 + H(Z_{00},Z_{i'0},Z_{ii}|Y_i,Z_{i0}),
\end{align}
\item
(\ref{eq:urho<eta-46}) is redundant because it comes from
(\ref{eq:hr0+hr1>0011|1}),
(\ref{eq:2hr0+tr1+tr2>00102022|2+0011|1})$'$,
and Lemma \ref{lem:relocation} as
\begin{align}
 [ \hr_0 + \hr_i ] + [ 2 \hr_0 + \tr_1 + \tr_2 ]
 &\geq
 H(Z_{00},Z_{ii}|Y_i,Z_{10},Z_{20})
 + [
  H(Z_{00},Z_{i'i'}|Y_{i'},Z_{10},Z_{20})
  + H(Z_{00},Z_{10},Z_{20},Z_{ii}|Y_i)
 ]
 \notag
 \\
 &\geq
 H(Z_{00},Z_{i0},Z_{ii}|Y_i,Z_{i'0})
 + H(Z_{00},Z_{i'i'}|Y_{i'},Z_{10},Z_{20})
 + H(Z_{00},Z_{i'0},Z_{ii}|Y_i,Z_{i0}),
\end{align}
\item
(\ref{eq:urho<eta-47}) is redundant because it comes from
(\ref{eq:hr0+hr1>0011|1})$'$ and 
(\ref{eq:2hr0+tr1+tr2>002011|1+001022|2}) as
\begin{align}
 [ \hr_0 + \hr_{i'} ] + [ 2 \hr_0 + \tr_1 + \tr_2 ]
 &\geq
 H(Z_{00},Z_{i'i'}|Y_{i'},Z_{10},Z_{20})
 \notag
 \\*
 &\quad
 + [ 
  H(Z_{00},Z_{i'0},Z_{ii}|Y_i,Z_{i0})
  + H(Z_{00},Z_{i0},Z_{i'i'}|Y_{i'},Z_{i'0})
 ],
\end{align}
\item
(\ref{eq:urho<eta-48}) is redundant because it comes from
(\ref{eq:2hr0+tr1+2tr2>00102022|2+002011|1+22|2})
and Lemma \ref{lem:relocation} as
\begin{align}
 2 \hr_0 + \tr_i + 2 \tr_{i'}
 &\geq
 H(Z_{00},Z_{10},Z_{20},Z_{i'i'}|Y_{i'})
 + H(Z_{i'i'}|Y_{i'},Z_{00},Z_{10},Z_{20})
 + H(Z_{00},Z_{i'0},Z_{ii}|Y_i,Z_{i0})
 \notag
 \\
 &\geq
 H(Z_{10},Z_{20},Z_{i'i'}|Y_{i'},Z_{00})
 + H(Z_{00},Z_{i'i'}|Y_{i'},Z_{10},Z_{20})
 + H(Z_{00},Z_{i'0},Z_{ii}|Y_i,Z_{i0}),
\end{align}
\item
(\ref{eq:urho<eta-49}) is equivalnt to
\begin{align}
 3\hr_0 + \tr_i + 2\tr_{i'}
 &\geq
 H(Z_{00},Z_{i'0},Z_{ii}|Y_i,Z_{i0})
 + H(Z_{00},Z_{10},Z_{20},Z_{i'i'}|Y_{i'})
 + H(Z_{00},Z_{i'i'}|Y_{i'},Z_{10},Z_{20}).
 \label{eq:3hr0+tr1+2tr2>00102022|2+002011|1+0022|2}
\end{align}
\end{itemize}

From $\sigma_i\leq\orho_1+\orho_2$, we have
\begin{align}
H(Z_{10},Z_{20}|Y_i,Z_{00},Z_{ii})
&\leq
H(Z_{10}|Z_{00}) + H(Z_{20}|Z_{00})
\label{eq:sigma<orho1+orho2-00}
\\
H(Z_{10},Z_{20}|Y_i,Z_{00},Z_{ii})
&\leq
\tr_i + H(Z_{i'0}|Z_{00}) - H(Z_{ii}|Y_i,Z_{00},Z_{10},Z_{20})
\label{eq:sigma<orho1+orho2-01}
\\
H(Z_{10},Z_{20}|Y_i,Z_{00},Z_{ii})
&\leq
\tr_{i'}
+ H(Z_{i0}|Z_{00}) - H(Z_{i'i'}|Y_{i'},Z_{00},Z_{10},Z_{20})
\label{eq:sigma<orho1+orho2-02}
\\
H(Z_{10},Z_{20}|Y_i,Z_{00},Z_{ii})
&\leq
\hr_0 + \tr_i
+ H(Z_{i'0}|Z_{00}) - H(Z_{00},Z_{ii}|Y_i,Z_{10},Z_{20})
\label{eq:sigma<orho1+orho2-03}
\\
H(Z_{10},Z_{20}|Y_i,Z_{00},Z_{ii})
&\leq
\hr_0 + \tr_{i'}
+ H(Z_{i0}|Z_{00}) - H(Z_{00},Z_{i'i'}|Y_{i'},Z_{10},Z_{20})
\label{eq:sigma<orho1+orho2-04}
\\
H(Z_{10},Z_{20}|Y_i,Z_{00},Z_{ii})
&\leq
\tr_1 +\tr_2
- H(Z_{11}|Y_1,Z_{00},Z_{10},Z_{20})
- H(Z_{22}|Y_2,Z_{00},Z_{10},Z_{20})
\label{eq:sigma<orho1+orho2-05}
\\
H(Z_{10},Z_{20}|Y_i,Z_{00},Z_{ii})
&\leq
\hr_0 + \tr_1 + \tr_2
- H(Z_{ii}|Y_i,Z_{00},Z_{10},Z_{20})
- H(Z_{00},Z_{i'i'}|Y_i,Z_{10},Z_{20})
\label{eq:sigma<orho1+orho2-06}
\\
H(Z_{10},Z_{20}|Y_i,Z_{00},Z_{ii})
&\leq
\hr_0 + \tr_1 + \tr_2
- H(Z_{00},Z_{ii}|Y_i,Z_{10},Z_{20})
- H(Z_{i'i'}|Y_{i'},Z_{00},Z_{10},Z_{20})
\label{eq:sigma<orho1+orho2-07}
\\
H(Z_{10},Z_{20}|Y_i,Z_{00},Z_{ii})
&\leq
2\hr_0 + \tr_1 + \tr_2
- H(Z_{00},Z_{11}|Y_1,Z_{10},Z_{20})
- H(Z_{00},Z_{22}|Y_2,Z_{10},Z_{20})
\label{eq:sigma<orho1+orho2-08}
\\
H(Z_{00},Z_{10},Z_{20}|Y_i,Z_{ii}) - \hr_0
&\leq
H(Z_{10}|Z_{00}) + H(Z_{20}|Z_{00})
\label{eq:sigma<orho1+orho2-09}
\\
H(Z_{00},Z_{10},Z_{20}|Y_i,Z_{ii}) - \hr_0
&\leq
\tr_i + H(Z_{i'0}|Z_{00}) - H(Z_{ii}|Y_i,Z_{00},Z_{10},Z_{20})
\label{eq:sigma<orho1+orho2-10}
\\
H(Z_{00},Z_{10},Z_{20}|Y_i,Z_{ii}) - \hr_0
&\leq
\tr_{i'}
+ H(Z_{i0}|Z_{00}) - H(Z_{i'i'}|Y_{i'},Z_{00},Z_{10},Z_{20})
\label{eq:sigma<orho1+orho2-11}
\\
H(Z_{00},Z_{10},Z_{20}|Y_i,Z_{ii}) - \hr_0
&\leq
\hr_0 + \tr_i
+ H(Z_{i'0}|Z_{00})
- H(Z_{00},Z_{ii}|Y_i,Z_{10},Z_{20})
\label{eq:sigma<orho1+orho2-12}
\\
H(Z_{00},Z_{10},Z_{20}|Y_i,Z_{ii}) - \hr_0
&\leq
\hr_0 + \tr_{i'}
+ H(Z_{i0}|Z_{00}) - H(Z_{00},Z_{i'i'}|Y_{i'},Z_{10},Z_{20})
\label{eq:sigma<orho1+orho2-13}
\\
H(Z_{00},Z_{10},Z_{20}|Y_i,Z_{ii}) - \hr_0
&\leq
\tr_1 +\tr_2
- H(Z_{11}|Y_1,Z_{00},Z_{10},Z_{20})
- H(Z_{22}|Y_2,Z_{00},Z_{10},Z_{20})
\label{eq:sigma<orho1+orho2-14}
\\
H(Z_{00},Z_{10},Z_{20}|Y_i,Z_{ii}) - \hr_0
&\leq
\hr_0 + \tr_1 + \tr_2
- H(Z_{ii}|Y_i,Z_{00},Z_{10},Z_{20})
- H(Z_{00},Z_{i'i'}|Y_{i'},Z_{10},Z_{20})
\label{eq:sigma<orho1+orho2-15}
\\
H(Z_{00},Z_{10},Z_{20}|Y_i,Z_{ii}) - \hr_0
&\leq
\hr_0 + \tr_1 + \tr_2
- H(Z_{00},Z_{ii}|Y_i,Z_{10},Z_{20})
- H(Z_{i'i'}|Y_{i'},Z_{00},Z_{10},Z_{20})
\label{eq:sigma<orho1+orho2-16}
\\
H(Z_{00},Z_{10},Z_{20}|Y_i,Z_{ii}) - \hr_0
&\leq
2\hr_0 + \tr_1 + \tr_2
- H(Z_{00},Z_{11}|Y_1,Z_{10},Z_{20})
- H(Z_{00},Z_{22}|Y_2,Z_{10},Z_{20})
\label{eq:sigma<orho1+orho2-17}
\\
H(Z_{10},Z_{20},Z_{ii}|Y_i,Z_{00}) - \hr_i
&\leq
H(Z_{10}|Z_{00}) + H(Z_{20}|Z_{00})
\label{eq:sigma<orho1+orho2-18}
\\
H(Z_{10},Z_{20},Z_{ii}|Y_i,Z_{00}) - \hr_i
&\leq
\tr_i + H(Z_{i'0}|Z_{00}) - H(Z_{ii}|Y_i,Z_{00},Z_{10},Z_{20})
\label{eq:sigma<orho1+orho2-19}
\\
H(Z_{10},Z_{20},Z_{ii}|Y_i,Z_{00}) - \hr_i
&\leq
\tr_{i'}
+ H(Z_{i0}|Z_{00}) - H(Z_{i'i'}|Y_{i'},Z_{00},Z_{10},Z_{20})
\label{eq:sigma<orho1+orho2-20}
\\
H(Z_{10},Z_{20},Z_{ii}|Y_i,Z_{00}) - \hr_i
&\leq
\hr_0 + \tr_i
+ H(Z_{i'0}|Z_{00})
- H(Z_{00},Z_{ii}|Y_i,Z_{10},Z_{20})
\label{eq:sigma<orho1+orho2-21}
\\
H(Z_{10},Z_{20},Z_{ii}|Y_i,Z_{00}) - \hr_i
&\leq
\hr_0 + \tr_{i'}
+ H(Z_{i0}|Z_{00}) - H(Z_{00},Z_{i'i'}|Y_{i'},Z_{10},Z_{20})
\label{eq:sigma<orho1+orho2-22}
\\
H(Z_{10},Z_{20},Z_{ii}|Y_i,Z_{00}) - \hr_i
&\leq
\tr_1 +\tr_2
- H(Z_{11}|Y_1,Z_{00},Z_{10},Z_{20})
- H(Z_{22}|Y_2,Z_{00},Z_{10},Z_{20})
\label{eq:sigma<orho1+orho2-23}
\\
H(Z_{10},Z_{20},Z_{ii}|Y_i,Z_{00}) - \hr_i
&\leq
\hr_0 + \tr_1 + \tr_2
- H(Z_{ii}|Y_i,Z_{00},Z_{10},Z_{20})
- H(Z_{00},Z_{i'i'}|Y_{i'},Z_{10},Z_{20})
\label{eq:sigma<orho1+orho2-24}
\\
H(Z_{10},Z_{20},Z_{ii}|Y_i,Z_{00}) - \hr_i
&\leq
\hr_0 + \tr_1 + \tr_2
- H(Z_{00},Z_{ii}|Y_i,Z_{10},Z_{20})
- H(Z_{i'i'}|Y_{i'},Z_{00},Z_{10},Z_{20})
\label{eq:sigma<orho1+orho2-25}
\\
H(Z_{10},Z_{20},Z_{ii}|Y_i,Z_{00}) - \hr_i
&\leq
2\hr_0 + \tr_1 + \tr_2
- H(Z_{00},Z_{11}|Y_1,Z_{10},Z_{20})
- H(Z_{00},Z_{22}|Y_2,Z_{10},Z_{20})
\label{eq:sigma<orho1+orho2-26}
\\
H(Z_{00},Z_{10},Z_{20},Z_{ii}|Y_i) - \hr_0 - \hr_i
&\leq
H(Z_{10}|Z_{00}) + H(Z_{20}|Z_{00})
\label{eq:sigma<orho1+orho2-27}
\\
H(Z_{00},Z_{10},Z_{20},Z_{ii}|Y_i) - \hr_0 - \hr_i
&\leq
\tr_i + H(Z_{i'0}|Z_{00}) - H(Z_{ii}|Y_i,Z_{00},Z_{10},Z_{20})
\label{eq:sigma<orho1+orho2-28}
\\
H(Z_{00},Z_{10},Z_{20},Z_{ii}|Y_i) - \hr_0 - \hr_i
&\leq
\tr_{i'}
+ H(Z_{i0}|Z_{00}) - H(Z_{i'i'}|Y_{i'},Z_{00},Z_{10},Z_{20})
\label{eq:sigma<orho1+orho2-29}
\\
H(Z_{00},Z_{10},Z_{20},Z_{ii}|Y_i) - \hr_0 - \hr_i
&\leq
\hr_0 + \tr_i
+ H(Z_{i'0}|Z_{00})
- H(Z_{00},Z_{ii}|Y_i,Z_{10},Z_{20})
\label{eq:sigma<orho1+orho2-30}
\\
H(Z_{00},Z_{10},Z_{20},Z_{ii}|Y_i) - \hr_0 - \hr_i
&\leq
\hr_0 + \tr_{i'}
+ H(Z_{i0}|Z_{00}) - H(Z_{00},Z_{i'i'}|Y_{i'},Z_{10},Z_{20})
\label{eq:sigma<orho1+orho2-31}
\\
H(Z_{00},Z_{10},Z_{20},Z_{ii}|Y_i) - \hr_0 - \hr_i
&\leq
\tr_1 +\tr_2
- H(Z_{11}|Y_1,Z_{00},Z_{10},Z_{20})
- H(Z_{22}|Y_2,Z_{00},Z_{10},Z_{20})
\label{eq:sigma<orho1+orho2-32}
\\
H(Z_{00},Z_{10},Z_{20},Z_{ii}|Y_i) - \hr_0 - \hr_i
&\leq
\hr_0 + \tr_1 + \tr_2
- H(Z_{ii}|Y_i,Z_{00},Z_{10},Z_{20})
- H(Z_{00},Z_{i'i'}|Y_{i'},Z_{10},Z_{20})
\label{eq:sigma<orho1+orho2-33}
\\
H(Z_{00},Z_{10},Z_{20},Z_{ii}|Y_i) - \hr_0 - \hr_i
&\leq
\hr_0 + \tr_1 + \tr_2
- H(Z_{00},Z_{ii}|Y_i,Z_{10},Z_{20})
- H(Z_{i'i'}|Y_{i'},Z_{00},Z_{10},Z_{20})
\label{eq:sigma<orho1+orho2-34}
\\
H(Z_{00},Z_{10},Z_{20},Z_{ii}|Y_i) - \hr_0 - \hr_i
&\leq
2\hr_0 + \tr_1 + \tr_2
- H(Z_{00},Z_{11}|Y_1,Z_{10},Z_{20})
- H(Z_{00},Z_{22}|Y_2,Z_{10},Z_{20}),
\label{eq:sigma<orho1+orho2-35}
\end{align}
where
\begin{itemize}
\item
(\ref{eq:sigma<orho1+orho2-00}) is redundant because
\begin{align}
 H(Z_{10}|Z_{00}) + H(Z_{20}|Z_{00})
 &\geq
 H(Z_{10},Z_{20}|Z_{00})
 \notag
 \\
 &\geq
 H(Z_{10},Z_{20}|Y_i,Z_{00},Z_{ii}),
\end{align}
\item
(\ref{eq:sigma<orho1+orho2-01}) is redundant because
it comes from (\ref{eq:tr1>1011|1}) and
Lemmas \ref{lem:xor} and \ref{lem:promotion} as
\begin{align}
 \tr_i
 &\geq
 H(Z_{i0},Z_{ii}|Y_i,Z_{00},Z_{i'0})
 \notag
 \\
 &\geq
 H(Z_{i0}|Y_i,Z_{00},Z_{i'0},Z_{ii}) + H(Z_{ii}|Y_i,Z_{00},Z_{10},Z_{20})
 \notag
 \\
 &\geq
 H(Z_{10},Z_{20}|Y_i,Z_{00},Z_{ii}) + H(Z_{ii}|Y_i,Z_{00},Z_{10},Z_{20})
 - H(Z_{i'0}|Z_{00}),
\end{align}
\item
(\ref{eq:sigma<orho1+orho2-02}) is redundant because
it comes from (\ref{eq:tr1>10|2+11|1})$'$
and Lemma \ref{lem:promotion} as
\begin{align}
 \tr_{i'}
 &\geq
 H(Z_{i'i'}|Y_{i'},Z_{00},Z_{10},Z_{20})
 + H(Z_{i'0}|Y_i,Z_{00},Z_{i0},Z_{ii})
 \notag
 \\
 &\geq
 H(Z_{i'i'}|Y_{i'},Z_{00},Z_{10},Z_{20})
 + H(Z_{10},Z_{20}|Y_i,Z_{00},Z_{ii})
 - H(Z_{i0}|Z_{00}),
\end{align}
\item
(\ref{eq:sigma<orho1+orho2-03}) is redundant because
it comes from (\ref{eq:hr0+tr2>00102022|2-10|00})$'$
and Lemma \ref{lem:xor} as
\begin{align}
 \hr_0 + \tr_i
 &\geq
 H(Z_{00},Z_{10},Z_{20},Z_{ii}|Y_i) - H(Z_{i'0}|Z_{00})
 \notag
 \\
 &\geq
 H(Z_{00},Z_{ii}|Y_i,Z_{10},Z_{10})
 + H(Z_{10},Z_{20}|Y_i,Z_{00},Z_{ii}) 
 - H(Z_{i'0}|Z_{00}),
\end{align}
\item
(\ref{eq:sigma<orho1+orho2-04}) is redundant because it comes from
(\ref{eq:hr0+tr1>10|2+0011|1})$'$ and Lemma \ref{lem:promotion} as
\begin{align}
 \hr_0 + \tr_{i'}
 &\geq
 H(Z_{00},Z_{i'i'}|Y_{i'},Z_{10},Z_{20})
 + H(Z_{i'0}|Y_i,Z_{00},Z_{i0},Z_{ii})
 \notag
 \\
 &\geq
 H(Z_{00},Z_{i'i'}|Y_{i'},Z_{10},Z_{20})
 + H(Z_{10},Z_{20}|Y_i,Z_{00},Z_{ii})
 - H(Z_{i0}|Z_{00}),
\end{align}
\item
(\ref{eq:sigma<orho1+orho2-05}) is redundant because it comes from
(\ref{eq:tr1+tr2>102022|2+11|1})$'$ and Lemma \ref{lem:xor} as
\begin{align}
 \tr_1 + \tr_2
 &\geq
 H(Z_{i'i'}|Y_{i'},Z_{00},Z_{10},Z_{20})
 + H(Z_{10},Z_{20},Z_{ii}|Y_i,Z_{00})
 \notag
 \\
 &\geq
 H(Z_{i'i'}|Y_{i'},Z_{00},Z_{10},Z_{20})
 + H(Z_{10},Z_{20}|Y_i,Z_{00},Z_{ii})
 + H(Z_{ii}|Y_i,Z_{00},Z_{10},Z_{20}),
\end{align}
\item
(\ref{eq:sigma<orho1+orho2-06}) is redundant because it comes from
(\ref{eq:hr0+tr1+tr2>102022|2+0011|1})$'$ and Lemma \ref{lem:xor} as
\begin{align}
 \hr_0 + \tr_1 + \tr_2
 &\geq
 H(Z_{00},Z_{i'i'}|Y_{i'},Z_{10},Z_{20})
 + H(Z_{10},Z_{20},Z_{ii}|Y_i,Z_{00})
 \notag
 \\
 &\geq
 H(Z_{00},Z_{i'i'}|Y_{i'},Z_{10},Z_{20})
 + H(Z_{10},Z_{20}|Y_i,Z_{00},Z_{ii})
 + H(Z_{ii}|Y_i,Z_{00},Z_{10},Z_{20}),
\end{align}
\item
(\ref{eq:sigma<orho1+orho2-07}) is redundant because it comes from
(\ref{eq:hr0+tr1+tr2>00102022|2+11|1})$'$ and Lemma \ref{lem:xor} as
\begin{align}
 \hr_0 + \tr_1 + \tr_2
 &\geq
 H(Z_{i'i'}|Y_{i'},Z_{00},Z_{10},Z_{20})
 + H(Z_{00},Z_{10},Z_{20},Z_{ii}|Y_i)
 \notag
 \\
 &\geq
 H(Z_{i'i'}|Y_{i'},Z_{00},Z_{10},Z_{20})
 + H(Z_{00},Z_{ii}|Y_i,Z_{10},Z_{20})
 + H(Z_{10},Z_{20}|Y_i,Z_{00},Z_{ii}),
\end{align}
\item
(\ref{eq:sigma<orho1+orho2-08}) is redundant because it comes from
(\ref{eq:2hr0+tr1+tr2>00102022|2+0011|1})$'$ and Lemma \ref{lem:xor} as
\begin{align}
 2 \hr_0 + \tr_1 + \tr_2
 &\geq
 H(Z_{00},Z_{i'i'}|Y_{i'},Z_{10},Z_{20})
 + H(Z_{00},Z_{10},Z_{20},Z_{ii}|Y_i)
 \notag
 \\
 &\geq
 H(Z_{00},Z_{i'i'}|Y_{i'},Z_{10},Z_{20})
 + H(Z_{00},Z_{ii}|Y_i,Z_{10},Z_{20})
 + H(Z_{10},Z_{20}|Y_i,Z_{00},Z_{ii}),
\end{align}
\item 
(\ref{eq:sigma<orho1+orho2-09}) is equivalent to
\begin{align}
 \hr_0 
 &\geq
 H(Z_{00},Z_{10},Z_{20}|Y_i,Z_{ii}) - H(Z_{10}|Z_{00}) - H(Z_{20}|Z_{00}),
 \label{eq:hr0>001020|1-10|00-20|00}
\end{align}
\item
(\ref{eq:sigma<orho1+orho2-10}) is redundant because it comes from
(\ref{eq:hr0+tr2>00102022|2-10|00})$'$ and Lemma \ref{lem:xor} as
\begin{align}
 \hr_0 + \tr_i
 &\geq
 H(Z_{00},Z_{10},Z_{20},Z_{ii}|Y_i) - H(Z_{i'0}|Z_{00})
 \notag
 \\
 &\geq
 H(Z_{00},Z_{10},Z_{20}|Y_i,Z_{ii})
 + H(Z_{ii}|Y_i,Z_{00},Z_{10},Z_{20})
 - H(Z_{i'0}|Z_{00}),
\end{align}
\item
(\ref{eq:sigma<orho1+orho2-11}) is equivalent to
\begin{align}
 \hr_0 + \tr_{i'}
 &\geq
 H(Z_{00},Z_{10},Z_{20}|Y_i,Z_{ii})
 +
 H(Z_{i'i'}|Y_{i'},Z_{00},Z_{10},Z_{20})
 - H(Z_{i0}|Z_{00}),
 \label{eq:hr0+tr2>001020|1+22|2-10|00}
\end{align}
\item
(\ref{eq:sigma<orho1+orho2-12}) is redundant because it comes from
(\ref{eq:hr0>00|1}), (\ref{eq:hr0+tr2>00102022|2-10|00})$'$
and Lemma \ref{lem:relocation} as
\begin{align}
 \hr_0 + [ \hr_0 + \tr_i ]
 &\geq
 H(Z_{00}|Y_i,Z_{10},Z_{20},Z_{ii})
 + [
  H(Z_{00},Z_{10},Z_{20},Z_{ii}|Y_i)
  - H(Z_{i'0}|Z_{00})
 ]
 \notag
 \\
 &\geq
 H(Z_{00},Z_{10},Z_{20}|Y_i,Z_{ii})
 + H(Z_{00},Z_{ii}|Y_i,Z_{10},Z_{20})
 - H(Z_{i'0}|Z_{00}),
\end{align}
\item
(\ref{eq:sigma<orho1+orho2-13}) is equivalent to
\begin{align}
 2 \hr_0 + \tr_{i'}
 &\geq
 H(Z_{00},Z_{10},Z_{20}|Y_i,Z_{ii})
 + H(Z_{00},Z_{i'i'}|Y_{i'},Z_{10},Z_{20})
 - H(Z_{i0}|Z_{00}),
 \label{eq:2hr0+tr2>001020|1+0022|2-10|00}
\end{align}
\item
(\ref{eq:sigma<orho1+orho2-14}) is redundant because it comes from
(\ref{eq:hr0+tr1+tr2>00102022|2+11|1})$'$ and Lemma \ref{lem:xor} as
\begin{align}
 \hr_0 + \tr_1 + \tr_2
 &\geq
 H(Z_{i'i'}|Y_{i'},Z_{00},Z_{10},Z_{20})
 + H(Z_{00},Z_{10},Z_{20},Z_{ii}|Y_i)
 \notag
 \\
 &\geq
 H(Z_{i'i'}|Y_{i'},Z_{00},Z_{10},Z_{20})
 + H(Z_{00},Z_{10},Z_{20}|Y_i,Z_{ii})
 + H(Z_{ii}|Y_i,Z_{00},Z_{10},Z_{20}),
\end{align}
\item
(\ref{eq:sigma<orho1+orho2-15}) is redundant because it comes from
(\ref{eq:2hr0+tr1+tr2>00102022|2+0011|1})$'$ and Lemma \ref{lem:xor} as
\begin{align}
 2 \hr_0 + \tr_1 + \tr_2
 &\geq
 H(Z_{00},Z_{i'i'}|Y_{i'},Z_{10},Z_{20})
 + H(Z_{00},Z_{10},Z_{20},Z_{ii}|Y_i)
 \notag
 \\
 &\geq
 H(Z_{00},Z_{i'i'}|Y_{i'},Z_{10},Z_{20})
 + H(Z_{00},Z_{10},Z_{20}|Y_i,Z_{ii})
 + H(Z_{ii}|Y_i,Z_{00},Z_{10},Z_{20}),
\end{align}
\item
(\ref{eq:sigma<orho1+orho2-16}) is redundant because it comes from
(\ref{eq:hr0>00|1}),
(\ref{eq:hr0+tr1+tr2>00102022|2+11|1})$'$ and Lemma \ref{lem:relocation} as
\begin{align}
 \hr_0 + [ \hr_0 + \tr_1 + \tr_2 ]
 &\geq
 H(Z_{00}|Y_i,Z_{10},Z_{20},Z_{ii})
 + [
  H(Z_{i'i'}|Y_{i'},Z_{00},Z_{10},Z_{20})
  + H(Z_{00},Z_{10},Z_{20},Z_{ii}|Y_i)
 ]
 \notag
 \\
 &\geq
 H(Z_{00},Z_{ii}|Y_i,Z_{10},Z_{20})
 + H(Z_{i'i'}|Y_{i'},Z_{00},Z_{10},Z_{20})
 + H(Z_{00},Z_{10},Z_{20}|Y_i,Z_{ii}),
\end{align}
\item
(\ref{eq:sigma<orho1+orho2-17}) is redundant because it comes from
(\ref{eq:hr0>00|1}), (\ref{eq:2hr0+tr1+tr2>00102022|2+0011|1})$'$,
and Lemma \ref{lem:relocation} as
\begin{align}
 \hr_0 + [ 2 \hr_0 + \tr_1 + \tr_2 ]
 &\geq
 H(Z_{00}|Y_i,Z_{10},Z_{20},Z_{ii})
 + [
  H(Z_{00},Z_{i'i'}|Y_{i'},Z_{10},Z_{20})
  + H(Z_{00},Z_{10},Z_{20},Z_{ii}|Y_i)
 ]
 \notag
 \\
 &\geq
 H(Z_{00},Z_{ii}|Y_i,Z_{10},Z_{20})
 + H(Z_{00},Z_{i'i'}|Y_{i'},Z_{10},Z_{20})
 + H(Z_{00},Z_{10},Z_{20}|Y_i,Z_{ii}),
\end{align}
\item
(\ref{eq:sigma<orho1+orho2-18}) is redundant because it comes from
(\ref{eq:hr1>11|1}) and
Lemma \ref{lem:promotion} as
\begin{align}
 \hr_i
 &\geq
 H(Z_{ii}|Y_i,Z_{00},Z_{10},Z_{20})
 \notag
 \\
 &\geq
 H(Z_{10},Z_{20},Z_{ii}|Y_i,Z_{00}) - H(Z_{10},Z_{20}|Z_{00})
 \notag
 \\
 &\geq
 H(Z_{10},Z_{20},Z_{ii}|Y_i,Z_{00}) - H(Z_{10}|Z_{00}) - H(Z_{20}|Z_{00}),
\end{align}
\item
(\ref{eq:sigma<orho1+orho2-19}) is redundant because it comes from
(\ref{eq:hr1>11|1}),
(\ref{eq:tr1>1011|1}),
and Lemma \ref{lem:promotion} as
\begin{align}
 \hr_i + \tr_i
 &\geq
 H(Z_{ii}|Y_i,Z_{00},Z_{10},Z_{20}) + H(Z_{i0},Z_{ii}|Y_i,Z_{00},Z_{i'0})
 \notag
 \\
 &\geq
 H(Z_{ii}|Y_i,Z_{00},Z_{10},Z_{20})
 + H(Z_{10},Z_{20},Z_{ii}|Y_i,Z_{00})
 - H(Z_{i'0}|Z_{00}),
\end{align}
\item
(\ref{eq:sigma<orho1+orho2-20}) is redundant because it comes from
(\ref{eq:hr2+tr1>1022|2+11|1})$'$ and Lemma \ref{lem:promotion} as
\begin{align}
 \hr_i + \tr_{i'}
 &\geq
 H(Z_{i'i'}|Y_{i'},Z_{00},Z_{10},Z_{10})
 + H(Z_{i'0},Z_{ii}|Y_i,Z_{00},Z_{i0})
 \notag
 \\
 &\geq
 H(Z_{i'i'}|Y_{i'},Z_{00},Z_{10},Z_{20})
 + H(Z_{10},Z_{20},Z_{ii}|Y_i,Z_{00}) - H(Z_{i0}|Z_{00}),
\end{align}
\item
(\ref{eq:sigma<orho1+orho2-21}) is redundant because it comes from
(\ref{eq:hr0+hr1>0011|1}), (\ref{eq:tr1>1011|1}),
and Lemma \ref{lem:promotion} as
\begin{align}
 [\hr_0 + \hr_i] + \tr_i
 &\geq
 H(Z_{00},Z_{ii}|Y_i,Z_{10},Z_{20}) + H(Z_{i0},Z_{ii}|Y_i,Z_{00},Z_{i'0})
 \notag
 \\
 &\geq
 H(Z_{00},Z_{ii}|Y_i,Z_{10},Z_{20})
 + H(Z_{10},Z_{20},Z_{ii}|Y_i,Z_{00})
 - H(Z_{i'0}|Z_{00}),
\end{align}
\item
(\ref{eq:sigma<orho1+orho2-22}) is redundant because it comes from
(\ref{eq:hr0+hr2+tr1>0011|1+1022|2})$'$
and Lemma \ref{lem:promotion} as
\begin{align}
 \hr_0 + \hr_i + \tr_{i'}
 &\geq
 H(Z_{00},Z_{i'i'}|Y_{i'},Z_{10},Z_{20})
 + H(Z_{i'0},Z_{ii}|Y_i,Z_{00},Z_{i0})
 \notag
 \\
 &\geq
 H(Z_{00},Z_{i'i'}|Y_{i'},Z_{10},Z_{20})
 + H(Z_{10},Z_{20},Z_{ii}|Y_i,Z_{00}) - H(Z_{i0}|Z_{00}),
\end{align}
\item
(\ref{eq:sigma<orho1+orho2-23}) is redundant because it comes from
(\ref{eq:hr1>11|1}) and (\ref{eq:tr1+tr2>102022|2+11|1})$'$ as
\begin{align}
 \hr_i + [ \tr_1 + \tr_2 ]
 &\geq
 H(Z_{ii}|Y_i,Z_{00},Z_{10},Z_{20})
 + [
  H(Z_{i'i'}|Y_{i'},Z_{00},Z_{10},Z_{20})
  + H(Z_{10},Z_{20},Z_{ii}|Y_i,Z_{00})
 ],
\end{align}
\item
(\ref{eq:sigma<orho1+orho2-24}) is redundant because it comes from
(\ref{eq:hr1>11|1}) and (\ref{eq:hr0+tr1+tr2>102022|2+0011|1})$'$ as
\begin{align}
 \hr_i + [ \hr_0 + \tr_1 + \tr_2 ]
 &\geq
 H(Z_{ii}|Y_i,Z_{00},Z_{10},Z_{20})
 + [
  H(Z_{00},Z_{i'i'}|Y_{i'},Z_{10},Z_{20})
  + H(Z_{10},Z_{20},Z_{ii}|Y_i,Z_{00})
 ],
\end{align}
\item
(\ref{eq:sigma<orho1+orho2-25}) is redundant because it comes from
(\ref{eq:hr0+hr1>0011|1}) and (\ref{eq:tr1+tr2>102022|2+11|1})$'$ as
\begin{align}
 [ \hr_0 + \hr_i ] + [ \tr_i + \tr_{i'} ]
 &\geq
 H(Z_{00},Z_{ii}|Y_i,Z_{10},Z_{20})
 + [
  H(Z_{i'i'}|Y_{i'},Z_{00},Z_{10},Z_{20})
  + H(Z_{10},Z_{20},Z_{ii}|Y_i,Z_{00})
 ],
\end{align}
\item
(\ref{eq:sigma<orho1+orho2-26}) is redundant because it comes from
(\ref{eq:hr0+hr1>0011|1}) and (\ref{eq:hr0+tr1+tr2>102022|2+0011|1})$'$ as
\begin{align}
 [\hr_0 + \hr_i] + [ \hr_0 + \tr_1 + \tr_2 ]
 &\geq
 H(Z_{00},Z_{ii}|Y_i,Z_{10},Z_{20})
 + [
  H(Z_{00},Z_{i'i'}|Y_{i'},Z_{10},Z_{20})
  + H(Z_{10},Z_{20},Z_{ii}|Y_i,Z_{00})
 ],
\end{align}
\item
(\ref{eq:sigma<orho1+orho2-27}) is redundant because it comes from
(\ref{eq:hr1=tr1-10|0011}),
(\ref{eq:hr0+tr2>00102022|2-10|00})$'$
as
\begin{align}
 \hr_0 + \hr_i
 &=
 [ \hr_0 + \tr_i ] - H(Z_{i0}|Z_{00},Z_{ii})
 \notag
 \\
 &\geq
 [ H(Z_{00},Z_{10},Z_{20},Z_{ii}|Y_i) - H(Z_{i'0}|Z_{00}) ]
 - H(Z_{i0}|Z_{00},Z_{ii}),
\end{align}
\item
(\ref{eq:sigma<orho1+orho2-28}) is redundant because it comes from
(\ref{eq:hr1>11|1}) and
(\ref{eq:hr0+tr2>00102022|2-10|00})$'$ as
\begin{align}
 \hr_i + [ \hr_0 + \tr_i ]
 &\geq
 H(Z_{ii}|Y_i,Z_{00},Z_{10},Z_{20})
 + [
  H(Z_{00},Z_{10},Z_{20},Z_{ii}|Y_i)
  - H(Z_{i'0}|Z_{00})
 ],
\end{align}
\item
(\ref{eq:sigma<orho1+orho2-29}) is redundant because it comes from
(\ref{eq:hr1=tr1-10|0011})
and (\ref{eq:hr0+tr1+tr2>00102022|2+11|1})$'$
as
\begin{align}
 \hr_0 + \hr_i + \tr_{i'}
 &=
 [ \hr_0 + \tr_1 + \tr_2 ] - H(Z_{i0}|Z_{00},Z_{ii})
 \notag
 \\
 &\geq
 [
  H(Z_{i'i'}|Y_{i'},Z_{00},Z_{10},Z_{20})
  + H(Z_{00},Z_{10},Z_{20},Z_{ii}|Y_i)
 ]
 - H(Z_{i0}|Z_{00},Z_{ii}) 
 \notag
 \\
 &\geq
 H(Z_{i'i'}|Y_{i'},Z_{00},Z_{10},Z_{20})
 +  H(Z_{00},Z_{10},Z_{20},Z_{ii}|Y_i)
 - H(Z_{i0}|Z_{00}),
\end{align}
\item
(\ref{eq:sigma<orho1+orho2-30}) is redundant because it comes from
(\ref{eq:hr0+hr1>0011|1}) and 
(\ref{eq:hr0+tr2>00102022|2-10|00})$'$ as
\begin{align}
 [ \hr_0 + \hr_i ] + [ \hr_0 + \tr_i ]
 &\geq
 H(Z_{00},Z_{ii}|Y_i,Z_{10},Z_{20})
 + [
  H(Z_{00},Z_{10},Z_{20},Z_{ii}|Y_i) - H(Z_{i'0}|Z_{00})
 ],
\end{align}
\item
(\ref{eq:sigma<orho1+orho2-31}) is redundant because it comes from
(\ref{eq:hr1=tr1-10|0011})
and (\ref{eq:2hr0+tr1+tr2>00102022|2+0011|1})$'$
as
\begin{align}
 2 \hr_0 + \hr_i + \tr_{i'}
 &=
 [2 \hr_0 + \tr_1 + \tr_2] - H(Z_{i0}|Z_{00},Z_{ii})
 \notag
 \\
 &\geq
 [
  H(Z_{00},Z_{i'i'}|Y_{i'},Z_{10},Z_{20})
  + H(Z_{00},Z_{10},Z_{20},Z_{ii}|Y_i)
 ]
 - H(Z_{i0}|Z_{00},Z_{ii})
 \notag
 \\
 &\geq
 H(Z_{00},Z_{i'i'}|Y_{i'},Z_{10},Z_{20})
 + H(Z_{00},Z_{10},Z_{20},Z_{ii}|Y_i)
 - H(Z_{i0}|Z_{00}),
\end{align}
\item
(\ref{eq:sigma<orho1+orho2-32}) is redundant because it comes from
(\ref{eq:hr1>11|1}) and (\ref{eq:hr0+tr1+tr2>00102022|2+11|1})$'$ as
\begin{align}
 \hr_i + [ \hr_0 +  \tr_1 + \tr_2 ]
 &\geq
 H(Z_{ii}|Y_i,Z_{00},Z_{10},Z_{20})
 + [
  H(Z_{i'i'}|Y_{i'},Z_{00},Z_{10},Z_{20})
  + H(Z_{00},Z_{10},Z_{20},Z_{ii}|Y_i)
 ],
\end{align}
\item
(\ref{eq:sigma<orho1+orho2-33}) is redundant because it comes from
(\ref{eq:hr1>11|1}) and (\ref{eq:2hr0+tr1+tr2>00102022|2+0011|1})$'$ as
\begin{align}
 \hr_i + [ 2 \hr_0 + \tr_1 + \tr_2 ]
 &\geq
 H(Z_{ii}|Y_i,Z_{00},Z_{10},Z_{20})
 + [
  H(Z_{00},Z_{i'i'}|Y_{i'},Z_{10},Z_{20})
  + H(Z_{00},Z_{10},Z_{20},Z_{ii}|Y_i)
 ],
\end{align}
\item
(\ref{eq:sigma<orho1+orho2-34}) is redundant because it comes from 
(\ref{eq:hr0+hr1>0011|1}) and (\ref{eq:hr0+tr1+tr2>00102022|2+11|1})$'$ as
\begin{align}
 [ \hr_0 + \hr_i ] + [ \hr_0 + \tr_1 + \tr_2 ]
 &\geq
 H(Z_{00},Z_{ii}|Y_i,Z_{10},Z_{20})
 + [
  H(Z_{i'i'}|Y_{i'},Z_{00},Z_{10},Z_{20})
  + H(Z_{00},Z_{10},Z_{20},Z_{ii}|Y_i)
 ],
\end{align}
\item
(\ref{eq:sigma<orho1+orho2-35}) is redundant because it comes from
(\ref{eq:hr0+hr1>0011|1}) and (\ref{eq:2hr0+tr1+tr2>00102022|2+0011|1})$'$ as
\begin{align}
 [ \hr_0 + \hr_i ] + [ 2 \hr_0 + \tr_1 + \tr_2 ]
 &\geq
 H(Z_{00},Z_{ii}|Y_i,Z_{10},Z_{20})
 \notag
 \\*
 &\quad
 + [ 
  H(Z_{00},Z_{i'i'}|Y_{i'},Z_{10},Z_{20})
  + H(Z_{00},Z_{10},Z_{20},Z_{ii}|Y_i)
 ].
\end{align}
\end{itemize}

From $\sigma_i\leq \zeta_i$, we have
\begin{align}
H(Z_{10},Z_{20}|Y_i,Z_{00},Z_{ii})
&\leq
\hr_0 + \tr_i
+ 2H(Z_{i'0}|Z_{00})
- H(Z_{00},Z_{i'0},Z_{ii}|Y_i,Z_{i0})
\label{eq:sigma<zeta-00}
\\
H(Z_{10},Z_{20}|Y_i,Z_{00},Z_{ii})
&\leq
\hr_0 + \tr_{i'}
+ 2H(Z_{i0}|Z_{00})
- H(Z_{00},Z_{i0},Z_{i'i'}|Y_{i'},Z_{i'0})
\label{eq:sigma<zeta-01}
\\
H(Z_{10},Z_{20}|Y_i,Z_{00},Z_{ii})
&\leq
\tr_i
+ 2\tr_{i'}
- H(Z_{i'0},Z_{ii}|Y_i,Z_{00},Z_{i0}) 
- 2H(Z_{i'i'}|Y_{i'},Z_{00},Z_{10},Z_{20})
\label{eq:sigma<zeta-02}
\\
H(Z_{10},Z_{20}|Y_i,Z_{00},Z_{ii})
&\leq
2\tr_i
+ \tr_{i'}
- 2H(Z_{ii}|Y_i,Z_{00},Z_{10},Z_{20})
- H(Z_{i0},Z_{i'i'}|Y_{i'},Z_{00},Z_{i'0}) 
\label{eq:sigma<zeta-03}
\\
H(Z_{10},Z_{20}|Y_i,Z_{00},Z_{ii})
&\leq
\hr_0 + \tr_i + 2\tr_{i'}
- H(Z_{00},Z_{i'0},Z_{ii}|Y_i,Z_{i0})
- 2H(Z_{i'i'}|Y_{i'},Z_{00},Z_{10},Z_{20})
\label{eq:sigma<zeta-04}
\\
H(Z_{10},Z_{20}|Y_i,Z_{00},Z_{ii})
&\leq
\hr_0 + 2\tr_i + \tr_{i'}
- 2H(Z_{ii}|Y_i,Z_{00},Z_{10},Z_{20})
- H(Z_{00},Z_{i0},Z_{i'i'}|Y_{i'},Z_{i'0})
\label{eq:sigma<zeta-05}
\\
H(Z_{10},Z_{20}|Y_i,Z_{00},Z_{ii})
&\leq
2\hr_0 + \tr_i + 2\tr_{i'}
- H(Z_{i'0},Z_{ii}|Y_i,Z_{00},Z_{i0})
- 2H(Z_{00},Z_{i'i'}|Y_{i'},Z_{10},Z_{20})
\label{eq:sigma<zeta-06}
\\
H(Z_{10},Z_{20}|Y_i,Z_{00},Z_{ii})
&\leq
2\hr_0 + 2\tr_i + \tr_{i'}
- 2H(Z_{00},Z_{ii}|Y_i,Z_{10},Z_{20})
- H(Z_{i0},Z_{i'i'}|Y_{i'},Z_{00},Z_{i'0})
\label{eq:sigma<zeta-07}
\\
H(Z_{10},Z_{20}|Y_i,Z_{00},Z_{ii})
&\leq
3\hr_0 + \tr_i + 2\tr_{i'}
- H(Z_{00},Z_{i'0},Z_{ii}|Y_i,Z_{i0})
- 2H(Z_{00},Z_{i'i'}|Y_{i'},Z_{10},Z_{20})
\label{eq:sigma<zeta-08}
\\
H(Z_{10},Z_{20}|Y_i,Z_{00},Z_{ii})
&\leq
3\hr_0 + 2\tr_i + \tr_{i'}
- 2H(Z_{00},Z_{ii}|Y_i,Z_{10},Z_{20})
- H(Z_{00},Z_{i0},Z_{i'i'}|Y_{i'},Z_{i'0})
\label{eq:sigma<zeta-09}
\\
H(Z_{00},Z_{10},Z_{20}|Y_i,Z_{ii}) - \hr_0
&\leq
\hr_0 + \tr_i
+ 2H(Z_{i'0}|Z_{00})
- H(Z_{00},Z_{i'0},Z_{ii}|Y_i,Z_{i0})
\label{eq:sigma<zeta-10}
\\
H(Z_{00},Z_{10},Z_{20}|Y_i,Z_{ii}) - \hr_0
&\leq
\hr_0 + \tr_{i'}
+ 2H(Z_{i0}|Z_{00})
- H(Z_{00},Z_{i0},Z_{i'i'}|Y_{i'},Z_{i'0})
\label{eq:sigma<zeta-11}
\\
H(Z_{00},Z_{10},Z_{20}|Y_i,Z_{ii}) - \hr_0
&\leq
\tr_i + 2\tr_{i'}
- H(Z_{i'0},Z_{ii}|Y_i,Z_{00},Z_{i0}) 
- 2H(Z_{i'i'}|Y_{i'},Z_{00},Z_{10},Z_{20})
\label{eq:sigma<zeta-12}
\\
H(Z_{00},Z_{10},Z_{20}|Y_i,Z_{ii}) - \hr_0
&\leq
2\tr_i + \tr_{i'}
- 2H(Z_{ii}|Y_i,Z_{00},Z_{10},Z_{20})
- H(Z_{i0},Z_{i'i'}|Y_{i'},Z_{00},Z_{i'0}) 
\label{eq:sigma<zeta-13}
\\
H(Z_{00},Z_{10},Z_{20}|Y_i,Z_{ii}) - \hr_0
&\leq
\hr_0 + \tr_i + 2\tr_{i'}
- H(Z_{00},Z_{i'0},Z_{ii}|Y_i,Z_{i0})
- 2H(Z_{i'i'}|Y_{i'},Z_{00},Z_{10},Z_{20})
\label{eq:sigma<zeta-14}
\\
H(Z_{00},Z_{10},Z_{20}|Y_i,Z_{ii}) - \hr_0
&\leq
\hr_0 + 2\tr_i + \tr_{i'}
- 2H(Z_{ii}|Y_i,Z_{00},Z_{10},Z_{20})
- H(Z_{00},Z_{i0},Z_{i'i'}|Y_{i'},Z_{i'0})
\label{eq:sigma<zeta-15}
\\
H(Z_{00},Z_{10},Z_{20}|Y_i,Z_{ii}) - \hr_0
&\leq
2\hr_0 + \tr_i + 2\tr_{i'}
- H(Z_{i'0},Z_{ii}|Y_i,Z_{00},Z_{i0})
- 2H(Z_{00},Z_{i'i'}|Y_{i'},Z_{10},Z_{20})
\label{eq:sigma<zeta-16}
\\
H(Z_{00},Z_{10},Z_{20}|Y_i,Z_{ii}) - \hr_0
&\leq
2\hr_0 + 2\tr_i + \tr_{i'}
- 2H(Z_{00},Z_{ii}|Y_i,Z_{10},Z_{20})
- H(Z_{i0},Z_{i'i'}|Y_{i'},Z_{00},Z_{i'0})
\label{eq:sigma<zeta-17}
\\
H(Z_{00},Z_{10},Z_{20}|Y_i,Z_{ii}) - \hr_0
&\leq
3\hr_0 + \tr_i + 2\tr_{i'}
- H(Z_{00},Z_{i'0},Z_{ii}|Y_i,Z_{i0})
- 2H(Z_{00},Z_{i'i'}|Y_{i'},Z_{10},Z_{20})
\label{eq:sigma<zeta-18}
\\
H(Z_{00},Z_{10},Z_{20}|Y_i,Z_{ii}) - \hr_0
&\leq
3\hr_0 + 2\tr_i + \tr_{i'}
- 2H(Z_{00},Z_{ii}|Y_i,Z_{10},Z_{20})
- H(Z_{00},Z_{i0},Z_{i'i'}|Y_{i'},Z_{i'0})
\label{eq:sigma<zeta-19}
\\
H(Z_{10},Z_{20},Z_{ii}|Y_i,Z_{00}) - \hr_i
&\leq
\hr_0 + \tr_i
+ 2H(Z_{i'0}|Z_{00})
- H(Z_{00},Z_{i'0},Z_{ii}|Y_i,Z_{i0})
\label{eq:sigma<zeta-20}
\\
H(Z_{10},Z_{20},Z_{ii}|Y_i,Z_{00}) - \hr_i
&\leq
\hr_0 + \tr_{i'}
+ 2H(Z_{i0}|Z_{00})
- H(Z_{00},Z_{i0},Z_{i'i'}|Y_{i'},Z_{i'0})
\label{eq:sigma<zeta-21}
\\
H(Z_{10},Z_{20},Z_{ii}|Y_i,Z_{00}) - \hr_i
&\leq
\tr_i + 2\tr_{i'}
- H(Z_{i'0},Z_{ii}|Y_i,Z_{00},Z_{i0}) 
- 2H(Z_{i'i'}|Y_{i'},Z_{00},Z_{10},Z_{20})
\label{eq:sigma<zeta-22}
\\
H(Z_{10},Z_{20},Z_{ii}|Y_i,Z_{00}) - \hr_i
&\leq
2\tr_i + \tr_{i'}
- 2H(Z_{ii}|Y_i,Z_{00},Z_{10},Z_{20})
- H(Z_{i0},Z_{i'i'}|Y_{i'},Z_{00},Z_{i'0}) 
\label{eq:sigma<zeta-23}
\\
H(Z_{10},Z_{20},Z_{ii}|Y_i,Z_{00}) - \hr_i
&\leq
\hr_0 + \tr_i + 2\tr_{i'}
- H(Z_{00},Z_{i'0},Z_{ii}|Y_i,Z_{i0})
- 2H(Z_{i'i'}|Y_{i'},Z_{00},Z_{10},Z_{20})
\label{eq:sigma<zeta-24}
\\
H(Z_{10},Z_{20},Z_{ii}|Y_i,Z_{00}) - \hr_i
&\leq
\hr_0 + 2\tr_i + \tr_{i'}
- 2H(Z_{ii}|Y_i,Z_{00},Z_{10},Z_{20})
- H(Z_{00},Z_{i0},Z_{i'i'}|Y_{i'},Z_{i'0})
\label{eq:sigma<zeta-25}
\\
H(Z_{10},Z_{20},Z_{ii}|Y_i,Z_{00}) - \hr_i
&\leq
2\hr_0 + \tr_i + 2\tr_{i'}
- H(Z_{i'0},Z_{ii}|Y_i,Z_{00},Z_{i0})
- 2H(Z_{00},Z_{i'i'}|Y_{i'},Z_{10},Z_{20})
\label{eq:sigma<zeta-26}
\\
H(Z_{10},Z_{20},Z_{ii}|Y_i,Z_{00}) - \hr_i
&\leq
2\hr_0 + 2\tr_i + \tr_{i'}
- 2H(Z_{00},Z_{ii}|Y_i,Z_{10},Z_{20})
- H(Z_{i0},Z_{i'i'}|Y_{i'},Z_{00},Z_{i'0})
\label{eq:sigma<zeta-27}
\\
H(Z_{10},Z_{20},Z_{ii}|Y_i,Z_{00}) - \hr_i
&\leq
3\hr_0 + \tr_i + 2\tr_{i'}
- H(Z_{00},Z_{i'0},Z_{ii}|Y_i,Z_{i0})
- 2H(Z_{00},Z_{i'i'}|Y_{i'},Z_{10},Z_{20})
\label{eq:sigma<zeta-28}
\\
H(Z_{10},Z_{20},Z_{ii}|Y_i,Z_{00}) - \hr_i
&\leq
3\hr_0 + 2\tr_i + \tr_{i'}
- 2H(Z_{00},Z_{ii}|Y_i,Z_{10},Z_{20})
- H(Z_{00},Z_{i0},Z_{i'i'}|Y_{i'},Z_{i'0})
\label{eq:sigma<zeta-29}
\\
H(Z_{00},Z_{10},Z_{20},Z_{ii}|Y_i) - \hr_0 - \hr_i
&\leq
\hr_0 + \tr_i
+ 2H(Z_{i'0}|Z_{00})
- H(Z_{00},Z_{i'0},Z_{ii}|Y_i,Z_{i0})
\label{eq:sigma<zeta-30}
\\
H(Z_{00},Z_{10},Z_{20},Z_{ii}|Y_i) - \hr_0 - \hr_i
&\leq
\hr_0 + \tr_{i'}
+ 2H(Z_{i0}|Z_{00})
- H(Z_{00},Z_{i0},Z_{i'i'}|Y_{i'},Z_{i'0})
\label{eq:sigma<zeta-31}
\\
H(Z_{00},Z_{10},Z_{20},Z_{ii}|Y_i) - \hr_0 - \hr_i
&\leq
\tr_i + 2\tr_{i'}
- H(Z_{i'0},Z_{ii}|Y_i,Z_{00},Z_{i0}) 
- 2H(Z_{i'i'}|Y_{i'},Z_{00},Z_{10},Z_{20})
\label{eq:sigma<zeta-32}
\\
H(Z_{00},Z_{10},Z_{20},Z_{ii}|Y_i) - \hr_0 - \hr_i
&\leq
2\tr_i + \tr_{i'}
- 2H(Z_{ii}|Y_i,Z_{00},Z_{10},Z_{20})
- H(Z_{i0},Z_{i'i'}|Y_{i'},Z_{00},Z_{i'0}) 
\label{eq:sigma<zeta-33}
\\
H(Z_{00},Z_{10},Z_{20},Z_{ii}|Y_i) - \hr_0 - \hr_i
&\leq
\hr_0 + \tr_i + 2\tr_{i'}
- H(Z_{00},Z_{i'0},Z_{ii}|Y_i,Z_{i0})
- 2H(Z_{i'i'}|Y_{i'},Z_{00},Z_{10},Z_{20})
\label{eq:sigma<zeta-34}
\\
H(Z_{00},Z_{10},Z_{20},Z_{ii}|Y_i) - \hr_0 - \hr_i
&\leq
\hr_0 + 2\tr_i + \tr_{i'}
- 2H(Z_{ii}|Y_i,Z_{00},Z_{10},Z_{20})
- H(Z_{00},Z_{i0},Z_{i'i'}|Y_{i'},Z_{i'0})
\label{eq:sigma<zeta-35}
\\
H(Z_{00},Z_{10},Z_{20},Z_{ii}|Y_i) - \hr_0 - \hr_i
&\leq
2\hr_0 + \tr_i + 2\tr_{i'}
- H(Z_{i'0},Z_{ii}|Y_i,Z_{00},Z_{i0})
- 2H(Z_{00},Z_{i'i'}|Y_{i'},Z_{10},Z_{20})
\label{eq:sigma<zeta-36}
\\
H(Z_{00},Z_{10},Z_{20},Z_{ii}|Y_i) - \hr_0 - \hr_i
&\leq
2\hr_0 + 2\tr_i + \tr_{i'}
- 2H(Z_{00},Z_{ii}|Y_i,Z_{10},Z_{20})
- H(Z_{i0},Z_{i'i'}|Y_{i'},Z_{00},Z_{i'0})
\label{eq:sigma<zeta-37}
\\
H(Z_{00},Z_{10},Z_{20},Z_{ii}|Y_i) - \hr_0 - \hr_i
&\leq
3\hr_0 + \tr_i + 2\tr_{i'}
- H(Z_{00},Z_{i'0},Z_{ii}|Y_i,Z_{i0})
- 2H(Z_{00},Z_{i'i'}|Y_{i'},Z_{10},Z_{20})
\label{eq:sigma<zeta-38}
\\
H(Z_{00},Z_{10},Z_{20},Z_{ii}|Y_i) - \hr_0 - \hr_i
&\leq
3\hr_0 + 2\tr_i + \tr_{i'}
- 2H(Z_{00},Z_{ii}|Y_i,Z_{10},Z_{20})
- H(Z_{00},Z_{i0},Z_{i'i'}|Y_{i'},Z_{i'0}),
\label{eq:sigma<zeta-39}
\end{align}
where
\begin{itemize}
\item
(\ref{eq:sigma<zeta-00}) is redundant because it comes from
(\ref{eq:hr0+tr2>00102022|2-10|00})$'$
and Lemmas \ref{lem:xor} and \ref{lem:promotion} as
\begin{align}
 \hr_0 + \tr_i
 &\geq
 H(Z_{00},Z_{10},Z_{20},Z_{ii}|Y_i) - H(Z_{i'0}|Z_{00})
 \notag
 \\
 &\geq
 H(Z_{00},Z_{i'0},Z_{ii}|Y_i,Z_{i0})
 + H(Z_{i0}|Y_i,Z_{00},Z_{i'0},Z_{ii})
 - H(Z_{i'0}|Z_{00})
 \notag
 \\
 &\geq
 H(Z_{00},Z_{i'0},Z_{ii}|Y_i,Z_{i0})
 + H(Z_{10},Z_{20}|Y_i,Z_{00},Z_{ii})
 - 2 H(Z_{i'0}|Z_{00}),
\end{align}
\item
(\ref{eq:sigma<zeta-01}) is redundant because it comes from 
(\ref{eq:hr0+tr1>10|2+002011|1-20|00})$'$ and Lemma \ref{lem:promotion} as
\begin{align}
 \hr_0 + \tr_{i'}
 &\geq
 H(Z_{00},Z_{i0},Z_{i'i'}|Y_{i'},Z_{i'0})
 + H(Z_{i'0}|Y_i,Z_{00},Z_{i0},Z_{ii})
 - H(Z_{i0}|Z_{00})
 \notag
 \\
 &\geq
 H(Z_{00},Z_{i0},Z_{i'i'}|Y_{i'},Z_{i'0})
 + H(Z_{10},Z_{20}|Y_i,Z_{00},Z_{ii})
 - 2 H(Z_{i0}|Z_{00}),
\end{align}
\item
(\ref{eq:sigma<zeta-02}) is redundant because
it comes from (\ref{eq:tr1>10|2+11|1})$'$,
(\ref{eq:tr1+tr2>102022|2+11|1})$'$,
and Lemma \ref{lem:relocation} as
\begin{align}
 \tr_{i'} + [ \tr_1 + \tr_2 ]
 &\geq
 [
  H(Z_{i'i'}|Y_{i'},Z_{00},Z_{10},Z_{20})
  + H(Z_{i'0}|Y_i,Z_{00},Z_{i0},Z_{ii})
 ]
 \notag
 \\*
 &\quad
 + [
  H(Z_{i'i'}|Y_{i'},Z_{00},Z_{10},Z_{20}) 
  + H(Z_{10},Z_{20},Z_{ii}|Y_i,Z_{00})
 ]
 \notag
 \\
 &\geq
 2 H(Z_{i'i'}|Y_{i'},Z_{00},Z_{10},Z_{20})
 + H(Z_{i'0},Z_{ii}|Y_i,Z_{00},Z_{i0}) + H(Z_{10},Z_{20}|Y_i,Z_{00},Z_{ii}),
\end{align}
\item
(\ref{eq:sigma<zeta-03}) is redundant because it comes from
(\ref{eq:tr1+2tr2>102022|2+2011|1+22|2})$'$ and Lemma \ref{lem:xor} as
\begin{align}
 2 \tr_i + \tr_{i'}
 &\geq
 H(Z_{i0},Z_{i'i'}|Y_{i'},Z_{00},Z_{i'0})
 + H(Z_{10},Z_{20},Z_{ii}|Y_i,Z_{00})
 + H(Z_{ii}|Y_i,Z_{00},Z_{10},Z_{20})
 \notag
 \\
 &\geq
 H(Z_{i0},Z_{i'i'}|Y_{i'},Z_{00},Z_{i'0})
 + H(Z_{10},Z_{20}|Y_i,Z_{00},Z_{ii})
 + 2 H(Z_{ii}|Y_i,Z_{00},Z_{10},Z_{20}),
\end{align}
\item
(\ref{eq:sigma<zeta-04}) is redundant because it comes from
(\ref{eq:tr1>10|2+11|1})$'$,
(\ref{eq:hr0+tr1+tr2>00102022|2+11|1})$'$,
and Lemma \ref{lem:xor} as
\begin{align}
 \tr_{i'} + [ \hr_0 + \tr_1 + \tr_2 ]
 &\geq
 [
  H(Z_{i'i'}|Y_{i'},Z_{00},Z_{10},Z_{20})
  + H(Z_{i'0}|Y_i,Z_{00},Z_{i0},Z_{ii})
 ]
 \notag
 \\*
 &\quad
 + [
  H(Z_{i'i'}|Y_{i'},Z_{00},Z_{10},Z_{20})
  + H(Z_{00},Z_{10},Z_{20},Z_{ii}|Y_i)
 ]
 \notag
 \\
 &\geq
 2 H(Z_{i'i'}|Y_{i'},Z_{00},Z_{10},Z_{20})
 + H(Z_{00},Z_{i'0},Z_{ii}|Y_i,Z_{i0})
 + H(Z_{10},Z_{20}|Y_i,Z_{00},Z_{ii}),
\end{align}
\item
(\ref{eq:sigma<zeta-05}) is redundant because
it comes from (\ref{eq:hr0+tr1+2tr2>102022|2+002011|1+22|2})$'$
and Lemma \ref{lem:xor} as
\begin{align}
 \hr_0 + 2 \tr_i + \tr_{i'}
 &\geq
 H(Z_{00},Z_{i0},Z_{i'i'}|Y_{i'},Z_{i'0})
 + H(Z_{10},Z_{20},Z_{ii}|Y_i,Z_{00})
 + H(Z_{ii}|Y_i,Z_{00},Z_{10},Z_{20})
 \notag
 \\
 &\geq
 H(Z_{00},Z_{i0},Z_{i'i'}|Y_{i'},Z_{i'0})
 + H(Z_{10},Z_{20}|Y_i,Z_{00},Z_{ii})
 + 2 H(Z_{ii}|Y_i,Z_{00},Z_{10},Z_{20}),
\end{align}
\item
(\ref{eq:sigma<zeta-06}) is redundant because it comes from
(\ref{eq:hr0+tr1>10|2+0011|1})$'$,
(\ref{eq:hr0+tr1+tr2>102022|2+0011|1})$'$,
and Lemma \ref{lem:relocation} as
\begin{align}
 [ \hr_0 + \tr_{i'} ] + [ \hr_0 + \tr_i + \tr_{i'} ]
 &\geq
 [
  H(Z_{00},Z_{i'i'}|Y_{i'},Z_{10},Z_{20})
  + H(Z_{i'0}|Y_i,Z_{00},Z_{i0},Z_{ii})
 ]
 \notag
 \\*
 &\quad
 + [
  H(Z_{00},Z_{i'i'}|Y_{i'},Z_{10},Z_{20})
  + H(Z_{10},Z_{20},Z_{ii}|Y_i,Z_{00})
 ]
 \notag
 \\
 &\geq
 2 H(Z_{00},Z_{i'i'}|Y_{i'},Z_{10},Z_{20})
 + H(Z_{i'0},Z_{ii}|Y_i,Z_{00},Z_{i0})
 \notag
 \\*
 &\quad
 + H(Z_{10},Z_{20}|Y_i,Z_{00},Z_{ii}),
\end{align}
\item
(\ref{eq:sigma<zeta-07}) is redundant because it comes from
(\ref{eq:2hr0+tr1+2tr2>00102022|2+2011|1+0022|2})$'$
and Lemma \ref{lem:xor} as
\begin{align}
 2\hr_0 + 2\tr_i + \tr_{i'}
 &\geq
 H(Z_{i0},Z_{i'i'}|Y_{i'},Z_{00},Z_{i'0})
 + H(Z_{00},Z_{10},Z_{20},Z_{ii}|Y_i)
 + H(Z_{00},Z_{ii}|Y_i,Z_{10},Z_{20})
 \notag
 \\
 &\geq
 H(Z_{i0},Z_{i'i'}|Y_{i'},Z_{00},Z_{i'0})
 + H(Z_{10},Z_{20}|Y_i,Z_{00},Z_{ii})
 + 2 H(Z_{00},Z_{ii}|Y_i,Z_{10},Z_{20}),
\end{align}
\item
(\ref{eq:sigma<zeta-08}) is redundant because it comes from
(\ref{eq:hr0+tr1>10|2+0011|1})$'$,
(\ref{eq:2hr0+tr1+tr2>00102022|2+0011|1})$'$,
and Lemma \ref{lem:relocation} as
\begin{align}
 [\hr_0 + \tr_{i'}]
 + [2\hr_0 + \tr_1 + \tr_2]
 &\geq
 [
  H(Z_{00},Z_{i'i'}|Y_{i'},Z_{10},Z_{20})
  + H(Z_{i'0}|Y_i,Z_{00},Z_{i0},Z_{ii})
 ]
 \notag
 \\*
 &\quad
 + [
  H(Z_{00},Z_{i'i'}|Y_{i'},Z_{10},Z_{20}) 
  + H(Z_{00},Z_{10},Z_{20},Z_{ii}|Y_i)
 ]
 \notag
 \\
 &\geq
 2 H(Z_{00},Z_{i'i'}|Y_{i'},Z_{10},Z_{20})
 + H(Z_{00},Z_{i'0},Z_{ii}|Y_i,Z_{i0})
 \notag
 \\*
 &\quad
 + H(Z_{10},Z_{20}|Y_i,Z_{00},Z_{ii}).
\end{align}
\item
(\ref{eq:sigma<zeta-09}) is redundant because it comes from
(\ref{eq:3hr0+tr1+2tr2>00102022|2+002011|1+0022|2})$'$
and Lemma \ref{lem:xor} as
\begin{align}
 3 \hr_0  + 2 \tr_i + \tr_{i'}
 &\geq
 H(Z_{00},Z_{i0},Z_{i'i'}|Y_{i'},Z_{i'0})
 + H(Z_{00},Z_{10},Z_{20},Z_{ii}|Y_i)
 + H(Z_{00},Z_{ii}|Y_i,Z_{10},Z_{20})
 \notag
 \\
 &\geq
 H(Z_{00},Z_{i0},Z_{i'i'}|Y_{i'},Z_{i'0})
 + H(Z_{10},Z_{20}|Y_i,Z_{00},Z_{ii})
 + 2 H(Z_{00},Z_{ii}|Y_i,Z_{i0},Z_{i'0}),
\end{align}
\item
(\ref{eq:sigma<zeta-10}) is redundant because it comes from
(\ref{eq:hr0>0010|2-10|00})$'$,
(\ref{eq:hr0+tr2>00102022|2-10|00})$'$
and Lemma \ref{lem:relocation} as
\begin{align}
 \hr_0 + [ \hr_0 + \tr_i ]
 &\geq
 [ H(Z_{00},Z_{i'0}|Y_i,Z_{i0},Z_{ii}) - H(Z_{i'0}|Z_{00}) ]
 + [ H(Z_{00},Z_{10},Z_{20},Z_{ii}|Y_i) - H(Z_{i'0}|Z_{00}) ]
 \notag
 \\
 &\geq
 H(Z_{00},Z_{i'0},Z_{ii}|Y_i,Z_{i0})
 + H(Z_{00},Z_{10},Z_{20}|Y_i,Z_{ii})
 - 2 H(Z_{i'0}|Z_{00}),
\end{align}
\item
(\ref{eq:sigma<zeta-11}) is equivalent to
\begin{align}
 2\hr_0 + \tr_{i'}
 &\geq
 H(Z_{00},Z_{10},Z_{20}|Y_i,Z_{ii})
 + H(Z_{00},Z_{i0},Z_{i'i'}|Y_{i'},Z_{i'0})
 -2 H(Z_{i0}|Z_{00}),
 \label{eq:2hr0+tr2>001020|1+001022|2-2*10|00}
\end{align}
\item
(\ref{eq:sigma<zeta-12}) is redundant because it comes from
(\ref{eq:tr1>10|2+11|1})$'$,
(\ref{eq:hr0+tr1+tr2>00102022|2+11|1})$'$,
and Lemma \ref{lem:relocation} as
\begin{align}
 \tr_{i'} + [ \hr_0 + \tr_1 + \tr_2 ]
 &\geq
 [
  H(Z_{i'i'}|Y_{i'},Z_{00},Z_{10},Z_{20})
  + H(Z_{00},Z_{10},Z_{20},Z_{ii}|Y_i)
 ]
 \notag
 \\*
 &\quad
 + [
  H(Z_{i'i'}|Y_{i'},Z_{00},Z_{10},Z_{20})
  + H(Z_{i'0}|Y_i,Z_{00},Z_{i0},Z_{ii})
 ]
 \notag
 \\
 &\geq
 2 H(Z_{i'i'}|Y_{i'},Z_{00},Z_{i0},Z_{i'0})
 + H(Z_{00},Z_{10},Z_{20}|Y_i,Z_{ii})
 + H(Z_{i'0},Z_{ii}|Y_i,Z_{00},Z_{i0}),
\end{align}
\item
(\ref{eq:sigma<zeta-13}) is redundant because it comes from
(\ref{eq:hr0+tr1+2tr2>00102022|2+2011|1+22|2})$'$
and Lemma \ref{lem:xor} as
\begin{align}
 \hr_0 + 2 \tr_i + \tr_{i'}
 &\geq
 H(Z_{i0},Z_{i'i'}|Y_{i'},Z_{00},Z_{i'0})
 + H(Z_{00},Z_{10},Z_{20},Z_{ii}|Y_i)
 + H(Z_{ii}|Y_i,Z_{00},Z_{10},Z_{20})
 \notag
 \\
 &\geq
 H(Z_{i0},Z_{i'i'}|Y_{i'},Z_{00},Z_{i'0}) 
 + H(Z_{00},Z_{10},Z_{20}|Y_i,Z_{ii})
 + 2 H(Z_{ii}|Y_i,Z_{00},Z_{10},Z_{20}),
\end{align}
\item
(\ref{eq:sigma<zeta-14}) is redundant because it comes from
(\ref{eq:2hr0+tr1+2tr2>00102022|2+002011|1+22|2}) and Lemma \ref{lem:xor} as
\begin{align}
 2\hr_0 + \tr_i + 2\tr_{i'}
 &\geq
 H(Z_{00},Z_{i'0},Z_{ii}|Y_i,Z_{i0})
 + H(Z_{00},Z_{10},Z_{20},Z_{i'i'}|Y_{i'})
 + H(Z_{i'i'}|Y_{i'},Z_{00},Z_{10},Z_{20}) 
 \notag
 \\
 &\geq
 H(Z_{00},Z_{i'0},Z_{ii}|Y_i,Z_{i0})
 + H(Z_{00},Z_{10},Z_{20}|Y_i,Z_{ii})
 + 2 H(Z_{i'i'}|Y_{i'},Z_{00},Z_{10},Z_{20}),
\end{align}
\item
(\ref{eq:sigma<zeta-15}) is redundant because it comes from
(\ref{eq:2hr0+tr1+2tr2>00102022|2+002011|1+22|2})$'$
and Lemma \ref{lem:xor} as
\begin{align}
 2 \hr_0 + 2 \tr_i + \tr_{i'}
 &\geq
 H(Z_{00},Z_{i0},Z_{i'i'}|Y_{i'},Z_{i'0})
 + H(Z_{00},Z_{10},Z_{20},Z_{ii}|Y_i)
 + H(Z_{ii}|Y_i,Z_{00},Z_{10},Z_{20})
 \notag
 \\
 &\geq
 H(Z_{00},Z_{i0},Z_{i'i'}|Y_{i'},Z_{i'0})
 + H(Z_{00},Z_{i0},Z_{i'0}|Y_i,Z_{ii})
 + 2 H(Z_{ii}|Y_i,Z_{00},Z_{i0},Z_{i'0}),
\end{align}
\item
(\ref{eq:sigma<zeta-16}) is redundant because it comes from
(\ref{eq:hr0+tr1>10|2+0011|1})$'$,
(\ref{eq:2hr0+tr1+tr2>00102022|2+0011|1})$'$,
and Lemma \ref{lem:relocation} as
\begin{align}
 [ \hr_0 + \tr_{i'} ] + [ 2 \hr_0 + \tr_1 + \tr_2 ]
 &\geq
 [
  H(Z_{00},Z_{i'i'}|Y_{i'},Z_{10},Z_{20})
  + H(Z_{00},Z_{10},Z_{20},Z_{ii}|Y_i)
 ]
 \notag
 \\*
 &\quad
 + [
  H(Z_{00},Z_{i'i'}|Y_{i'},Z_{10},Z_{20})
  + H(Z_{i'0}|Y_i,Z_{00},Z_{i0},Z_{ii})
 ]
 \notag
 \\
 &\geq
 2 H(Z_{00},Z_{i'i'}|Y_{i'},Z_{10},Z_{20})
 + H(Z_{00},Z_{10},Z_{20}|Y_i,Z_{ii})
 \notag
 \\*
 &\quad
 + H(Z_{i'0},Z_{ii}|Y_i,Z_{00},Z_{i0}),
\end{align}
\item
(\ref{eq:sigma<zeta-17}) is redundant because it comes from
(\ref{eq:hr0>00|1}),
(\ref{eq:2hr0+tr1+2tr2>00102022|2+2011|1+0022|2})$'$,
and Lemma \ref{lem:relocation} as
\begin{align}
 \hr_0 + [ 2 \hr_0  + 2 \tr_i + \tr_{i'} ]
 &\geq
 H(Z_{00}|Y_i,Z_{10},Z_{20},Z_{ii})
 + [
  H(Z_{i0},Z_{i'i'}|Y_{i'},Z_{00},Z_{i'0})
  + H(Z_{00},Z_{10},Z_{20},Z_{ii}|Y_i)
  \notag
  \\*
  &\quad
  + H(Z_{00},Z_{ii}|Y_i,Z_{10},Z_{20})
 ]
 \notag
 \\
 &\geq
 2 H(Z_{00},Z_{ii}|Y_i,Z_{i0},Z_{i'0})
 + H(Z_{i0},Z_{i'i'}|Y_{i'},Z_{00},Z_{i'0})
 + H(Z_{00},Z_{10},Z_{20}|Y_i,Z_{ii}),
\end{align}
\item
(\ref{eq:sigma<zeta-18}) is redundant because it comes from
(\ref{eq:2hr0+tr1>0010|2+0011|1})$'$,
(\ref{eq:2hr0+tr1+tr2>00102022|2+0011|1})$'$,
and Lemma \ref{lem:relocation} as
\begin{align}
 [ 2 \hr_0 + \tr_{i'} ] + [ 2 \hr_0 + \tr_1 + \tr_2 ]
 &\geq
 [
  H(Z_{00},Z_{i'i'}|Y_{i'},Z_{10},Z_{20})
  + H(Z_{00},Z_{i'0}|Y_i,Z_{i0},Z_{ii})
 ]
 \notag
 \\*
 &\quad
 + [
  H(Z_{00},Z_{i'i'}|Y_{i'},Z_{10},Z_{20})
  + H(Z_{00},Z_{10},Z_{20},Z_{ii}|Y_i)
 ]
 \notag
 \\
 &\geq
 2 H(Z_{00},Z_{i'i'}|Y_{i'},Z_{10},Z_{20})
 + H(Z_{00},Z_{10},Z_{20}|Y_i,Z_{ii})
 \notag
 \\*
 &\quad
 + H(Z_{00},Z_{i'0},Z_{ii}|Y_i,Z_{i0}),
\end{align}
\item
(\ref{eq:sigma<zeta-19}) is redundant because it comes from
(\ref{eq:hr0>00|1}),
(\ref{eq:3hr0+tr1+2tr2>00102022|2+002011|1+0022|2})$'$,
and Lemma \ref{lem:relocation} as
\begin{align}
 \hr_0 + [ 3 \hr_0  + 2 \tr_i + \tr_{i'} ]
 &\geq
 H(Z_{00}|Y_i,Z_{10},Z_{20},Z_{ii})
 + [
  H(Z_{00},Z_{i0},Z_{i'i'}|Y_{i'},Z_{i'0})
  + H(Z_{00},Z_{10},Z_{20},Z_{ii}|Y_i)
  \notag
  \\*
  &\quad
  + H(Z_{00},Z_{ii}|Y_i,Z_{10},Z_{20})
 ]
 \notag
 \\
 &\geq
 H(Z_{00},Z_{10},Z_{20}|Y_i,Z_{ii})
 + H(Z_{00},Z_{i0},Z_{i'i'}|Y_{i'},Z_{i'0})
 + 2 H(Z_{00},Z_{ii}|Y_i,Z_{10},Z_{20}),
\end{align}
\item
(\ref{eq:sigma<zeta-20}) is redundant because it comes from
(\ref{eq:tr1>1011|1}),
(\ref{eq:hr0+hr2>001022|2-10|00})$'$,
and Lemma \ref{lem:promotion} as
\begin{align}
 \tr_i + [ \hr_0 + \hr_i ]
 &\geq
 H(Z_{i0},Z_{ii}|Y_i,Z_{00},Z_{i'0})
 + [ H(Z_{00},Z_{i'0},Z_{ii}|Y_i,Z_{i0}) - H(Z_{i'0}|Z_{00}) ]
 \notag
 \\
 &\geq
 H(Z_{10},Z_{20},Z_{ii}|Y_i,Z_{00})
 + H(Z_{00},Z_{i'0},Z_{ii}|Y_i,Z_{i0})
 - 2 H(Z_{i'0}|Z_{00}),
\end{align}
\item
(\ref{eq:sigma<zeta-21}) is redundant because it comes from
(\ref{eq:hr1=tr1-10|0011}),
(\ref{eq:hr0+tr1+tr2>002011|1+1022|2})$'$,
and Lemma \ref{lem:promotion} as
\begin{align}
 \hr_0 + \hr_i + \tr_{i'}
 &=
 [ \hr_0 + \tr_1 + \tr_2 ] - H(Z_{i0}|Z_{00},Z_{ii})
 \notag
 \\
 &\geq
 [ H(Z_{00},Z_{i0},Z_{i'i'}|Y_{i'},Z_{i'0})
  + H(Z_{i'0},Z_{ii}|Y_i,Z_{00},Z_{i0}) ] - H(Z_{i0}|Z_{00},Z_{ii}) 
 \notag
 \\
 &\geq
 H(Z_{00},Z_{i0},Z_{i'i'}|Y_{i'},Z_{i'0})
 + H(Z_{10},Z_{20},Z_{ii}|Y_i,Z_{00})
 - 2 H(Z_{i0}|Z_{00}),
\end{align}
\item
(\ref{eq:sigma<zeta-22}) is redundant because it comes from
(\ref{eq:hr2+tr1>1022|2+11|1})$'$ and
(\ref{eq:tr1+tr2>102022|2+11|1})$'$ as
\begin{align}
 [ \hr_i + \tr_{i'} ] + [ \tr_1 + \tr_2 ]
 &\geq
 [ 
  H(Z_{i'i'}|Y_{i'},Z_{00},Z_{10},Z_{20})
  + H(Z_{i'0},Z_{ii}|Y_i,Z_{00},Z_{i0})
 ]
 \notag
 \\*
 &\quad
 + [
  H(Z_{i'i'}|Y_{i'},Z_{00},Z_{10},Z_{20})
  + H(Z_{10},Z_{20},Z_{ii}|Y_i,Z_{00})
 ],
\end{align}
\item
(\ref{eq:sigma<zeta-23}) is redundant because it comes from
(\ref{eq:hr1>11|1}) and (\ref{eq:tr1+2tr2>102022|2+2011|1+22|2})$'$ as
\begin{align}
 \hr_i + [ 2 \tr_i + \tr_{i'} ]
 &\geq
 H(Z_{ii}|Y_i,Z_{00},Z_{10},Z_{20})
 \notag
 \\*
 &\quad
 + [
  H(Z_{i0},Z_{i'i'}|Y_{i'},Z_{00},Z_{i'0})
  + H(Z_{10},Z_{20},Z_{ii}|Y_i,Z_{00})
  + H(Z_{ii}|Y_i,Z_{00},Z_{10},Z_{20})
 ],
\end{align}
\item
(\ref{eq:sigma<zeta-24}) is redundant because it comes from
(\ref{eq:hr0+hr2+tr1>001022|2+11|1})$'$
and (\ref{eq:tr1+tr2>102022|2+11|1})$'$ as
\begin{align}
 [ \hr_0 + \hr_i + \tr_{i'} ] + [ \tr_1 + \tr_2 ]
 &\geq
 [
  H(Z_{i'i'}|Y_{i'},Z_{00},Z_{10},Z_{20})
  + H(Z_{00},Z_{i'0},Z_{ii}|Y_i,Z_{i0})
 ] 
 \notag
 \\*
 &\quad
 + [
  H(Z_{i'i'}|Y_{i'},Z_{00},Z_{10},Z_{20})
  + H(Z_{10},Z_{20},Z_{ii}|Y_i,Z_{00})
 ],
\end{align}
\item
(\ref{eq:sigma<zeta-25}) is redundant because it comes from 
(\ref{eq:hr1>11|1}) and (\ref{eq:hr0+tr1+2tr2>102022|2+002011|1+22|2})$'$ as
\begin{align}
 \hr_i + [ \hr_0 + 2 \tr_i + \tr_{i'} ]
 &\geq
 H(Z_{ii}|Y_i,Z_{00},Z_{10},Z_{20})
 \notag
 \\*
 &\quad
 + [
  H(Z_{00},Z_{i0},Z_{i'i'}|Y_{i'},Z_{i'0}) 
  + H(Z_{10},Z_{20},Z_{ii}|Y_i,Z_{00})
  + H(Z_{ii}|Y_i,Z_{00},Z_{10},Z_{20})
 ],
\end{align}
\item
(\ref{eq:sigma<zeta-26}) is redundant because it comes from
(\ref{eq:hr0+hr2+tr1>0011|1+1022|2})$'$ and
(\ref{eq:hr0+tr1+tr2>102022|2+0011|1})$'$ as
\begin{align}
 [ \hr_0 + \hr_i + \tr_{i'} ] + [ \hr_0 + \tr_1 + \tr_2 ]
 &\geq
 [ 
  H(Z_{00},Z_{i'i'}|Y_{i'},Z_{10},Z_{20})
  + H(Z_{i'0},Z_{ii}|Y_i,Z_{00},Z_{i0})
 ]
 \notag
 \\*
 &\quad
 + [
  H(Z_{00},Z_{i'i'}|Y_{i'},Z_{10},Z_{20})
  + H(Z_{10},Z_{20},Z_{ii}|Y_i,Z_{00})
 ]
 \notag
 \\
 &=
 2 H(Z_{00},Z_{i'i'}|Y_{i'},Z_{10},Z_{20})
 + H(Z_{i'0},Z_{ii}|Y_i,Z_{00},Z_{i0})
 \notag
 \\*
 &\quad
 + H(Z_{10},Z_{20},Z_{ii}|Y_i,Z_{00}),
\end{align}
\item
(\ref{eq:sigma<zeta-27}) is redundant because it comes from
(\ref{eq:hr0+hr1>0011|1}),
(\ref{eq:hr0+tr1+2tr2>00102022|2+2011|1+22|2})$'$,
and Lemma \ref{lem:relocation} as
\begin{align}
 [\hr_0 + \hr_i] + [ \hr_0 + 2 \tr_i + \tr_{i'} ]
 &\geq
 H(Z_{00},Z_{ii}|Y_i,Z_{10},Z_{20})
 + [
  H(Z_{i0},Z_{i'i'}|Y_{i'},Z_{00},Z_{i'0})
  \notag
  \\*
  &\quad
  + H(Z_{00},Z_{10},Z_{20},Z_{ii}|Y_i)
  + H(Z_{ii}|Y_i,Z_{00},Z_{10},Z_{20})
 ]
 \notag
 \\
 &\geq
 2 H(Z_{00},Z_{ii}|Y_i,Z_{10},Z_{20})
 + H(Z_{i0},Z_{i'i'}|Y_{i'},Z_{00},Z_{i'0})
 \notag
 \\*
 &\quad
 + H(Z_{10},Z_{20},Z_{ii}|Y_i,Z_{00}),
\end{align}
\item
(\ref{eq:sigma<zeta-28}) is redundant because it comes from 
(\ref{eq:2hr0+hr2+tr1>001022|2+0011|1})$'$ and
(\ref{eq:hr0+tr1+tr2>102022|2+0011|1})$'$ as
\begin{align}
 [ 2 \hr_0 + \hr_i + \tr_{i'} ] + [ \hr_0 + \tr_1 + \tr_2 ]
 &\geq
 [
  H(Z_{00},Z_{i'i'}|Y_{i'},Z_{10},Z_{20})
  + H(Z_{00},Z_{i'0},Z_{ii}|Y_i,Z_{i0})
 ]
 \notag
 \\*
 &\quad
 + [
  H(Z_{00},Z_{i'i'}|Y_{i'},Z_{10},Z_{20})
  + H(Z_{10},Z_{20},Z_{ii}|Y_i,Z_{00})
 ],
\end{align}
\item
(\ref{eq:sigma<zeta-29}) is redundant because it comes from
(\ref{eq:hr1>11|1}),
(\ref{eq:3hr0+tr1+2tr2>00102022|2+002011|1+0022|2})$'$,
and Lemma \ref{lem:relocation} as
\begin{align}
 \hr_i + [ 3 \hr_0 + 2 \tr_i + \tr_{i'} ]
 &\geq
 H(Z_{ii}|Y_i,Z_{00},Z_{10},Z_{20})
 + [
  H(Z_{00},Z_{i0},Z_{i'i'}|Y_{i'},Z_{i'0})
  \notag
  \\*
  &\quad
  + H(Z_{00},Z_{10},Z_{20},Z_{ii}|Y_i)
  + H(Z_{00},Z_{ii}|Y_i,Z_{10},Z_{20})
 ]
 \notag
 \\
 &\geq
 H(Z_{10},Z_{20},Z_{ii}|Y_i,Z_{00})
 + H(Z_{00},Z_{i0},Z_{i'i'}|Y_{i'},Z_{i'0})
 + 2 H(Z_{00},Z_{ii}|Y_i,Z_{10},Z_{20}),
\end{align}
\item
(\ref{eq:sigma<zeta-30}) is redundant because it comes from
(\ref{eq:hr0+hr2>001022|2-10|00})$'$ and
(\ref{eq:hr0+tr2>00102022|2-10|00})$'$ as
\begin{align}
 [ \hr_0 + \hr_i ] + [ \hr_0 + \tr_i ]
 &\geq
 [ H(Z_{00},Z_{i'0},Z_{ii}|Y_i,Z_{i0}) - H(Z_{i'0}|Z_{00}) ]
 + [ H(Z_{00},Z_{10},Z_{20},Z_{ii}|Y_i) - H(Z_{i'0}|Z_{00}) ],
\end{align}
\item
(\ref{eq:sigma<zeta-31}) is redundant because it comes from
(\ref{eq:hr1=tr1-10|0011})
and 
(\ref{eq:2hr0+tr1+tr2>00102022|2+002011|1-20|00})$'$
as
\begin{align}
 2 \hr_0 + \hr_i + \tr_{i'}
 &=
 [ 2 \hr_0 + \tr_1 + \tr_2 ] - H(Z_{i0}|Z_{00},Z_{ii})
 \notag
 \\
 &\geq
 [
  H(Z_{00},Z_{i0},Z_{i'i'}|Y_{i'},Z_{i'0})
  + H(Z_{00},Z_{10},Z_{20},Z_{ii}|Y_i)
  - H(Z_{i0}|Z_{00})
 ]
 - H(Z_{i0}|Z_{00},Z_{ii})
 \notag
 \\
 &\geq
 H(Z_{00},Z_{i0},Z_{i'i'}|Y_{i'},Z_{i'0}) 
 + H(Z_{00},Z_{10},Z_{20},Z_{ii}|Y_i)
 - 2 H(Z_{i0}|Z_{00}),
\end{align}
\item
(\ref{eq:sigma<zeta-32}) is redundant because it comes from
(\ref{eq:hr2+tr1>1022|2+11|1})$'$ and
(\ref{eq:hr0+tr1+tr2>00102022|2+11|1})$'$ as
\begin{align}
 [ \hr_i + \tr_{i'} ] + [ \hr_0 + \tr_1 + \tr_2 ]
 &\geq
 [
  H(Z_{i'i'}|Y_{i'},Z_{00},Z_{10},Z_{20})
  + H(Z_{00},Z_{10},Z_{20},Z_{ii}|Y_i)
 ]
 \notag
 \\*
 &\quad
 + [
  H(Z_{i'i'}|Y_{i'},Z_{00},Z_{10},Z_{20})
  + H(Z_{i'0},Z_{ii}|Y_i,Z_{00},Z_{i0})
 ],
\end{align}
\item
(\ref{eq:sigma<zeta-33}) is redundant because it comes from
(\ref{eq:hr1>11|1}) and
(\ref{eq:hr0+tr1+2tr2>00102022|2+2011|1+22|2})$'$ as
\begin{align}
 \hr_i + [ \hr_0 + 2 \tr_1 + \tr_2 ]
 &\geq
 H(Z_{ii}|Y_i,Z_{00},Z_{10},Z_{20})
 + [
  H(Z_{i0},Z_{i'i'}|Y_{i'},Z_{00},Z_{i'0})
  + H(Z_{00},Z_{10},Z_{20},Z_{ii}|Y_i)
  \notag
  \\*
  &\quad
  + H(Z_{ii}|Y_i,Z_{00},Z_{10},Z_{20})
 ],
\end{align}
\item
(\ref{eq:sigma<zeta-34}) is redundant because it comes from
(\ref{eq:hr0+hr2+tr1>001022|2+11|1})$'$ and
(\ref{eq:hr0+tr1+tr2>00102022|2+11|1})$'$ as
\begin{align}
 [ \hr_0 + \hr_i  + \tr_{i'} ] + [ \hr_0 + \tr_1 + \tr_2 ]
 &\geq
 [
  H(Z_{i'i'}|Y_{i'},Z_{00},Z_{10},Z_{20})
  +
  H(Z_{00},Z_{i'0},Z_{ii}|Y_i,Z_{i0})
 ]
 \notag
 \\*
 &\quad
 + [
  H(Z_{i'i'}|Y_{i'},Z_{00},Z_{10},Z_{20})
  + H(Z_{00},Z_{10},Z_{20},Z_{ii}|Y_i)
 ]
 \notag
 \\
 &\geq
 2 H(Z_{i'i'}|Y_{i'},Z_{00},Z_{10},Z_{20})
 + H(Z_{00},Z_{i'0},Z_{ii}|Y_i,Z_{i0})
 \notag
 \\*
 &\quad
 + H(Z_{00},Z_{10},Z_{20},Z_{ii}|Y_i),
\end{align}
\item
(\ref{eq:sigma<zeta-35}) is redundant because it comes from
(\ref{eq:hr1>11|1}) and
(\ref{eq:2hr0+tr1+2tr2>00102022|2+002011|1+22|2})$'$ as
\begin{align}
 \hr_i + [ 2 \hr_0 + 2 \tr_i + \tr_{i'} ]
 &\geq
 H(Z_{ii}|Y_i,Z_{00},Z_{10},Z_{20})
 + [
  H(Z_{00},Z_{i0},Z_{i'i'}|Y_{i'},Z_{i'0})
  + H(Z_{00},Z_{10},Z_{20},Z_{ii}|Y_i)
  \notag
  \\*
  &\quad
  + H(Z_{ii}|Y_i,Z_{00},Z_{10},Z_{20})
 ],
\end{align}
\item
(\ref{eq:sigma<zeta-36}) is redundant because it comes from 
(\ref{eq:hr0+hr2+tr1>0011|1+1022|2})$'$ and
(\ref{eq:2hr0+tr1+tr2>00102022|2+0011|1})$'$ as
\begin{align}
 [ \hr_0 + \hr_i  + \tr_{i'} ] + [ 2 \hr_0 + \tr_1 + \tr_2 ]
 &\geq
 [
  H(Z_{00},Z_{i'i'}|Y_{i'},Z_{10},Z_{20})
  + H(Z_{i'0},Z_{ii}|Y_i,Z_{00},Z_{i0})
 ]
 \notag
 \\*
 &\quad
 + [
  H(Z_{00},Z_{i'i'}|Y_{i'},Z_{10},Z_{20})
  + H(Z_{00},Z_{10},Z_{20},Z_{ii}|Y_i)
 ],
\end{align}
\item
(\ref{eq:sigma<zeta-37}) is redundant because it comes from
(\ref{eq:hr0+hr1>0011|1}) and
(\ref{eq:2hr0+tr1+2tr2>00102022|2+2011|1+0022|2})$'$ as
\begin{align}
 [ \hr_0 + \hr_i ] + [ 2 \hr_0 + 2 \tr_i + \tr_{i'} ]
 &\geq
 H(Z_{00},Z_{ii}|Y_i,Z_{10},Z_{20})
 + [
  H(Z_{i0},Z_{i'i'}|Y_{i'},Z_{00},Z_{i'0})
  \notag
  \\*
  &\quad
  + H(Z_{00},Z_{10},Z_{20},Z_{ii}|Y_i)
  + H(Z_{00},Z_{ii}|Y_i,Z_{10},Z_{20})
 ],
\end{align}
\item
(\ref{eq:sigma<zeta-38}) is redundant because it comes from
(\ref{eq:2hr0+hr2+tr1>001022|2+0011|1})$'$ and
(\ref{eq:2hr0+tr1+tr2>00102022|2+0011|1})$'$ as
\begin{align}
 [ 2 \hr_0 + \hr_i  + \tr_{i'} ] + [ 2 \hr_0 + \tr_1 + \tr_2 ]
 &\geq
 [
  H(Z_{00},Z_{i'i'}|Y_{i'},Z_{10},Z_{20})
  +
  H(Z_{00},Z_{i'0},Z_{ii}|Y_i,Z_{i0})
 ]
 \notag
 \\*
 &\quad
 + [
  H(Z_{00},Z_{i'i'}|Y_{i'},Z_{10},Z_{20})
  + H(Z_{00},Z_{10},Z_{20},Z_{ii}|Y_i)
 ],
\end{align}
\item
(\ref{eq:sigma<zeta-39}) is redundant because it comes from
(\ref{eq:hr0+hr1>0011|1}) and
(\ref{eq:3hr0+tr1+2tr2>00102022|2+002011|1+0022|2})$'$ as
\begin{align}
 [ \hr_0 + \hr_i ] + [ 3 \hr_0 + \tr_{i'} + 2 \tr_i ]
 &\geq
 H(Z_{00},Z_{ii}|Y_i,Z_{10},Z_{20})
 + [
  H(Z_{00},Z_{i0},Z_{i'i'}|Y_{i'},Z_{i'0})
  \notag
  \\*
  &\quad
  + H(Z_{00},Z_{10},Z_{20},Z_{ii}|Y_i)
  + H(Z_{00},Z_{ii}|Y_i,Z_{10},Z_{20})
 ]
 \notag
 \\
 &\geq
 2 H(Z_{00},Z_{ii}|Y_i,Z_{10},Z_{20})
 + H(Z_{00},Z_{i0},Z_{i'i'}|Y_{i'},Z_{i'0})
 \notag
 \\*
 &\quad
 + H(Z_{00},Z_{10},Z_{20},Z_{ii}|Y_i).
\end{align}
\end{itemize}

Finally, from (\ref{eq:def-hr0})--(\ref{eq:def-tri}),
and Lemmas \ref{lem:c-independence-H(A,B|C)}
and \ref{lem:c-independence-I(A;C|D)}
with the fact that $I(Z_{10},Z_{11};Z_{20},Z_{22}|Z_{00})=0$,
we have
\begin{itemize}
\item 
(\ref{eq:hr0>00|1}) is equivalent to
\begin{align}
 R_0
 &\leq  
 H(Z_{00}) - H(Z_{00}|Y_i,Z_{10},Z_{20},Z_{ii})
 \notag
 \\
 &= 
 I(Z_{00};Y_i,Z_{10},Z_{20},Z_{ii}),
\end{align}
\item 
(\ref{eq:hr0>0010|1-10|00}) is equivalent to
\begin{align}
 R_0
 &\leq
 H(Z_{00})
 + H(Z_{i0}|Z_{00})
 - H(Z_{00},Z_{i0}|Y_i,Z_{i'0},Z_{ii})
 \notag
 \\
 &= 
 I(Z_{00},Z_{i0};Y_i,Z_{i'0},Z_{ii}),
\end{align}
\item
(\ref{eq:hr0>0010|2-10|00})$'$ is equivalent to
\begin{align}
 R_0
 &\leq
 H(Z_{00})
 + H(Z_{i'0}|Z_{00})
 - H(Z_{00},Z_{i'0}|Y_i,Z_{i0},Z_{ii})
 \notag
 \\
 &= 
 I(Z_{00},Z_{i'0};Y_i,Z_{i0},Z_{ii}),
\end{align}
\item
(\ref{eq:hr0>001020|1-10|00-20|00}) is equivalent to
\begin{align}
 R_0
 &\leq
 H(Z_{00}) + H(Z_{10}|Z_{00}) + H(Z_{20}|Z_{00})
 - H(Z_{00},Z_{10},Z_{20}|Y_i,Z_{ii})
 \notag
 \\
 &= 
 H(Z_{00}) + H(Z_{10},Z_{20}|Z_{00})
 - H(Z_{00},Z_{10},Z_{20}|Y_i,Z_{ii})
 \notag
 \\
 &= 
 I(Z_{00},Z_{10},Z_{20};Y_i,Z_{ii}),
\end{align}
\item
(\ref{eq:hr1>11|1}) is equivalent to
\begin{align}
 R_i
 &\leq
 H(Z_{ii}|Z_{00}) - H(Z_{ii}|Y_i,Z_{00},Z_{10},Z_{20})
 \notag
 \\
 &=
 I(Z_{ii};Y_i,Z_{10},Z_{20}|Z_{00}),
\end{align}
\item
(\ref{eq:tr1>1011|1}) is equivalent to
\begin{align}
 R_i
 &\leq
 H(Z_{i0},Z_{ii}|Z_{00}) - H(Z_{i0},Z_{ii}|Y_i,Z_{00},Z_{i'0})
 \notag
 \\
 &=
 I(Z_{i0},Z_{ii};Y_i,Z_{i'0}|Z_{00}),
\end{align}
\item
(\ref{eq:tr1>10|2+11|1}) is equivalent to
\begin{align}
 R_i
 &\leq
 H(Z_{i0},Z_{ii}|Z_{00})
 - H(Z_{ii}|Y_i,Z_{00},Z_{10},Z_{20})
 - H(Z_{i0}|Y_{i'},Z_{00},Z_{i'0},Z_{i'i'})
 \notag
 \\
 &=
 H(Z_{i0}|Z_{00}) + H(Z_{ii}|Z_{00}) - I(Z_{i0};Z_{ii}|Z_{00})
 - H(Z_{ii}|Y_i,Z_{00},Z_{10},Z_{20})
 - H(Z_{i0}|Y_{i'},Z_{00},Z_{i'0},Z_{i'i'})
 \notag
 \\
 &=
 I(Z_{ii};Y_i,Z_{10},Z_{20}|Z_{00})
 + I(Z_{i0};Y_{i'},Z_{i'0},Z_{i'i'}|Z_{00})
 - I(Z_{i0};Z_{ii}|Z_{00}),
\end{align}
\item
(\ref{eq:hr0+hr1>0011|1}) is equivalent to
\begin{align}
 R_0 + R_i
 &\leq
 H(Z_{00}) + H(Z_{ii}|Z_{00}) - H(Z_{00},Z_{ii}|Y_i,Z_{10},Z_{20})
 \notag
 \\
 &=
 I(Z_{00},Z_{ii};Y_i,Z_{10},Z_{20}),
\end{align}
\item
(\ref{eq:hr0+tr1>001011|1}) is equivalent to
\begin{align}
 R_0 + R_i
 &\leq
 H(Z_{00}) + H(Z_{i0},Z_{ii}|Z_{00}) - H(Z_{00},Z_{i0},Z_{ii}|Y_i,Z_{i'0})
 \notag
 \\
 &=
 I(Z_{00},Z_{i0},Z_{ii};Y_i,Z_{i'0}),
\end{align}
\item
(\ref{eq:hr0+hr2>001022|2-10|00})$'$ is equivalent to
\begin{align}
 R_0 + R_i
 &\leq
 H(Z_{00}) + H(Z_{i'0}|Z_{00}) + H(Z_{ii}|Z_{00}) 
 - H(Z_{00},Z_{i'0},Z_{ii}|Y_i,Z_{i0})
 \notag
 \\
 &=
 H(Z_{00}) + H(Z_{i'0},Z_{ii}|Z_{00}) 
 - H(Z_{00},Z_{i'0},Z_{ii}|Y_i,Z_{i0})
 \notag
 \\
 &=
 I(Z_{00},Z_{i'0},Z_{ii};Y_i,Z_{i0}),
\end{align}
\item
(\ref{eq:hr0+tr2>00102022|2-10|00})$'$ is equivalent to
\begin{align}
 R_0 + R_i
 &\leq
 H(Z_{00})
 + H(Z_{i0},Z_{ii}|Z_{00})
 + H(Z_{i'0}|Z_{00})
 - H(Z_{00},Z_{10},Z_{20},Z_{ii}|Y_i)
 \notag
 \\
 &=
 H(Z_{00})
 + H(Z_{i0},Z_{i'0},Z_{ii}|Z_{00})
 - H(Z_{00},Z_{10},Z_{20},Z_{ii}|Y_i)
 \notag
 \\
 &=
 I(Z_{00},Z_{10},Z_{20},Z_{ii};Y_i),
\end{align}
\item
(\ref{eq:hr0+tr1>0010|2+11|1}) is equivalent to
\begin{align}
 R_0 + R_i
 &\leq
 H(Z_{00}) + H(Z_{i0},Z_{ii}|Z_{00})
 - H(Z_{ii}|Y_i,Z_{00},Z_{10},Z_{20})
 - H(Z_{00},Z_{i0}|Y_{i'},Z_{i'0},Z_{i'i'})
 \notag
 \\
 &=
 H(Z_{00})
 + H(Z_{i0}|Z_{00})
 + H(Z_{ii}|Z_{00})
 - I(Z_{i0};Z_{ii}|Z_{00})
 \notag
 \\*
 &\quad
 - H(Z_{ii}|Y_i,Z_{00},Z_{10},Z_{20})
 - H(Z_{00},Z_{i0}|Y_{i'},Z_{i'0},Z_{i'i'})
 \notag
 \\
 &=
 I(Z_{ii};Y_i,Z_{10},Z_{20}|Z_{00})
 + I(Z_{00},Z_{i0};Y_{i'},Z_{i'0},Z_{i'i'})
 - I(Z_{i0};Z_{ii}|Z_{00}),
\end{align}
\item
(\ref{eq:hr0+tr2>001020|1+22|2-10|00})$'$ is equivalent to
\begin{align}
 R_0 + R_i
 &\leq
 H(Z_{00})
 + H(Z_{i0},Z_{ii}|Z_{00})
 + H(Z_{i'0}|Z_{00})
 - H(Z_{ii}|Y_i,Z_{00},Z_{10},Z_{20})
 - H(Z_{00},Z_{10},Z_{20}|Y_{i'},Z_{i'i'})
 \notag
 \\
 &=
 H(Z_{00})
 + H(Z_{i0}|Z_{00})
 + H(Z_{ii}|Z_{00})
 - I(Z_{i0};Z_{ii}|Z_{00})
 + H(Z_{i'0}|Z_{00})
 \notag
 \\*
 &\quad
 - H(Z_{ii}|Y_i,Z_{00},Z_{10},Z_{20})
 - H(Z_{00},Z_{10},Z_{20}|Y_{i'},Z_{i'i'})
 \notag
 \\
 &=
 H(Z_{00})
 + H(Z_{i0},Z_{i'0}|Z_{00})
 + H(Z_{ii}|Z_{00})
 - I(Z_{i0};Z_{ii}|Z_{00})
 \notag
 \\*
 &\quad
 - H(Z_{ii}|Y_i,Z_{00},Z_{10},Z_{20})
 - H(Z_{00},Z_{10},Z_{20}|Y_{i'},Z_{i'i'})
 \notag
 \\
 &=
 I(Z_{ii};Y_i,Z_{10},Z_{20}|Z_{00})
 + I(Z_{00},Z_{10},Z_{20};Y_{i'},Z_{i'i'})
 - I(Z_{i0};Z_{ii}|Z_{00}),
\end{align}
\item
(\ref{eq:hr0+tr1>10|2+0011|1}) is equivalent to
\begin{align}
 R_0 + R_i
 &\leq
 H(Z_{00}) + H(Z_{i0},Z_{ii}|Z_{00})
 - H(Z_{00},Z_{ii}|Y_i,Z_{10},Z_{20})
 - H(Z_{i0}|Y_{i'},Z_{00},Z_{i'0},Z_{i'i'})
 \notag
 \\
 &=
 H(Z_{00})
 + H(Z_{i0}|Z_{00})
 + H(Z_{ii}|Z_{00})
 - I(Z_{i0};Z_{ii}|Z_{00})
 \notag
 \\*
 &\quad
 - H(Z_{00},Z_{ii}|Y_i,Z_{10},Z_{20})
 - H(Z_{i0}|Y_{i'},Z_{00},Z_{i'0},Z_{i'i'})
 \notag
 \\
 &=
 I(Z_{00},Z_{ii};Y_i,Z_{10},Z_{20})
 + I(Z_{i0};Y_{i'},Z_{i'0},Z_{i'i'}|Z_{00})
 - I(Z_{i0};Z_{ii}|Z_{00}),
\end{align}
\item
(\ref{eq:hr0+tr1>10|2+002011|1-20|00}) is equivalent to
\begin{align}
 R_0 + R_i
 &\leq
 H(Z_{00})
 + H(Z_{i0},Z_{ii}|Z_{00})
 + H(Z_{i'0}|Z_{00})
 - H(Z_{00},Z_{i'0},Z_{ii}|Y_i,Z_{i0})
 - H(Z_{i0}|Y_{i'},Z_{00},Z_{i'0},Z_{i'i'})
 \notag
 \\
 &=
 H(Z_{00})
 + H(Z_{i0}|Z_{00})
 + H(Z_{ii}|Z_{00})
 - I(Z_{i0};Z_{ii}|Z_{00})
 + H(Z_{i'0}|Z_{00})
 \notag
 \\*
 &\quad
 - H(Z_{00},Z_{i'0},Z_{ii}|Y_i,Z_{i0})
 - H(Z_{i0}|Y_{i'},Z_{00},Z_{i'0},Z_{i'i'})
 \notag
 \\
 &=
 H(Z_{00})
 + H(Z_{i0}|Z_{00})
 + H(Z_{i'0},Z_{ii}|Z_{00})
 - I(Z_{i0};Z_{ii}|Z_{00})
 \notag
 \\*
 &\quad
 - H(Z_{00},Z_{i'0},Z_{ii}|Y_i,Z_{i0})
 - H(Z_{i0}|Y_{i'},Z_{00},Z_{i'0},Z_{i'i'})
 \notag
 \\
 &=
 I(Z_{00},Z_{i'0},Z_{ii};Y_i,Z_{i0})
 + I(Z_{i0};Y_{i'},Z_{i'0},Z_{i'i'}|Z_{00})
 - I(Z_{i0};Z_{ii}|Z_{00}),
\end{align}
\item
(\ref{eq:hr2+tr1>1022|2+11|1}) is equivalent to
\begin{align}
 R_1 + R_2
 &\leq
 H(Z_{i0},Z_{ii}|Z_{00})
 + H(Z_{i'i'}|Z_{00})
 - H(Z_{ii}|Y_i,Z_{00},Z_{10},Z_{20})
 - H(Z_{i0},Z_{i'i'}|Y_{i'},Z_{00},Z_{i'0})
 \notag
 \\
 &=
 H(Z_{i0}|Z_{00})
 + H(Z_{ii}|Z_{00})
 - I(Z_{i0};Z_{ii}|Z_{00})
 + H(Z_{i'i'}|Z_{00})
 \notag
 \\*
 &\quad
 - H(Z_{ii}|Y_i,Z_{00},Z_{10},Z_{20})
 - H(Z_{i0},Z_{i'i'}|Y_{i'},Z_{00},Z_{i'0})
 \notag
 \\
 &=
 H(Z_{ii}|Z_{00})
 + H(Z_{i0},Z_{i'i'}|Z_{00})
 - I(Z_{i0};Z_{ii}|Z_{00})
 \notag
 \\*
 &\quad
 - H(Z_{ii}|Y_i,Z_{00},Z_{10},Z_{20})
 - H(Z_{i0},Z_{i'i'}|Y_{i'},Z_{00},Z_{i'0})
 \notag
 \\
 &=
 I(Z_{ii};Y_i,Z_{10},Z_{20}|Z_{00})
 + I(Z_{i0},Z_{i'i'};Y_{i'},Z_{i'0}|Z_{00})
 - I(Z_{i0};Z_{ii}|Z_{00}),
\end{align}
\item
(\ref{eq:tr1+tr2>102022|2+11|1}) is equivalent to
\begin{align}
 R_1 + R_2
 &\leq
 H(Z_{10},Z_{11}|Z_{00})
 + H(Z_{20},Z_{22}|Z_{00})
 - H(Z_{ii}|Y_i,Z_{00},Z_{10},Z_{20})
 - H(Z_{10},Z_{20},Z_{i'i'}|Y_{i'},Z_{00})
 \notag
 \\
 &=
 H(Z_{i0}|Z_{00})
 + H(Z_{ii}|Z_{00})
 - I(Z_{i0};Z_{ii}|Z_{00})
 + H(Z_{i'0},Z_{i'i'}|Z_{00})
 \notag
 \\*
 &\quad
 - H(Z_{ii}|Y_i,Z_{00},Z_{10},Z_{20})
 - H(Z_{10},Z_{20},Z_{i'i'}|Y_{i'},Z_{00})
 \notag
 \\
 &=
 H(Z_{ii}|Z_{00})
 + H(Z_{i0},Z_{i'0},Z_{i'i'}|Z_{00})
 - I(Z_{i0};Z_{ii}|Z_{00})
 \notag
 \\*
 &\quad
 - H(Z_{ii}|Y_i,Z_{00},Z_{10},Z_{20})
 - H(Z_{10},Z_{20},Z_{i'i'}|Y_{i'},Z_{00})
 \notag
 \\
 &=
 I(Z_{ii};Y_i,Z_{10},Z_{20}|Z_{00})
 + I(Z_{10},Z_{20},Z_{i'i'};Y_{i'}|Z_{00})
 - I(Z_{i0};Z_{ii}|Z_{00}),
\end{align}
\item
(\ref{eq:tr1+tr2>2011|1+1022|2}) is equivalent to
\begin{align}
 R_1 + R_2
 &\leq
 H(Z_{10},Z_{11}|Z_{00}) + H(Z_{20},Z_{22}|Z_{00})
 - H(Z_{20},Z_{11}|Y_1,Z_{00},Z_{10})
 - H(Z_{10},Z_{22}|Y_2,Z_{00},Z_{20})
 \notag
 \\
 &=
 H(Z_{10}|Z_{00})
 + H(Z_{11}|Z_{00})
 - I(Z_{10};Z_{11}|Z_{00})
 + H(Z_{20}|Z_{00})
 + H(Z_{22}|Z_{00})
 - I(Z_{20};Z_{22}|Z_{00})
 \notag
 \\*
 &\quad
 - H(Z_{20},Z_{11}|Y_1,Z_{00},Z_{10})
 - H(Z_{10},Z_{22}|Y_2,Z_{00},Z_{20})
 \notag
 \\
 &=
 H(Z_{20},Z_{11}|Z_{00})
 + H(Z_{10},Z_{22}|Z_{00})
 - I(Z_{10};Z_{11}|Z_{00})
 - I(Z_{20};Z_{22}|Z_{00})
 \notag
 \\*
 &\quad
 - H(Z_{20},Z_{11}|Y_1,Z_{00},Z_{10})
 - H(Z_{10},Z_{22}|Y_2,Z_{00},Z_{20})
 \notag
 \\
 &=
 I(Z_{20},Z_{11};Y_1,Z_{10}|Z_{00})
 + I(Z_{10},Z_{22};Y_2,Z_{20}|Z_{00})
 - I(Z_{10};Z_{11}|Z_{00}) - I(Z_{20};Z_{22}|Z_{00}),
\end{align}
\item
(\ref{eq:2hr0+tr1>0010|2+0011|1}) is equivalent to
\begin{align}
 2R_0 + R_i
 &\leq
 2H(Z_{00})
 + H(Z_{i0},Z_{ii}|Z_{00})
 - H(Z_{00},Z_{ii}|Y_i,Z_{10},Z_{20})
 - H(Z_{00},Z_{i0}|Y_{i'},Z_{i'0},Z_{i'i'})
 \notag
 \\
 &=
 2H(Z_{00})
 + H(Z_{i0}|Z_{00})
 + H(Z_{ii}|Z_{00})
 - I(Z_{i0};Z_{ii}|Z_{00})
 \notag
 \\*
 &\quad
 - H(Z_{00},Z_{ii}|Y_i,Z_{10},Z_{20})
 - H(Z_{00},Z_{i0}|Y_{i'},Z_{i'0},Z_{i'i'})
 \notag
 \\
 &=
 I(Z_{00},Z_{ii};Y_i,Z_{10},Z_{20})
 + I(Z_{00},Z_{i0};Y_{i'},Z_{i'0},Z_{i'i'})
 -I(Z_{i0};Z_{ii}|Z_{00}),
\end{align}
\item
(\ref{eq:2hr0+tr2>001020|1+0022|2-10|00})$'$ is equivalent to
\begin{align}
 2R_0 + R_i
 &\leq
 2H(Z_{00})
 + H(Z_{i0},Z_{ii}|Z_{00})
 + H(Z_{i'0}|Z_{00})
 - H(Z_{00},Z_{ii}|Y_i,Z_{10},Z_{20})
 - H(Z_{00},Z_{10},Z_{20}|Y_{i'},Z_{i'i'})
 \notag
 \\
 &=
 2H(Z_{00})
 + H(Z_{i0}|Z_{00})
 + H(Z_{ii}|Z_{00})
 - I(Z_{i0};Z_{ii}|Z_{00})
 + H(Z_{i'0}|Z_{00})
 \notag
 \\*
 &\quad
 - H(Z_{00},Z_{ii}|Y_i,Z_{10},Z_{20})
 - H(Z_{00},Z_{10},Z_{20}|Y_{i'},Z_{i'i'})
 \notag
 \\
 &=
 2H(Z_{00})
 + H(Z_{ii}|Z_{00})
 + H(Z_{10},Z_{20}|Z_{00})
 - I(Z_{i0};Z_{ii}|Z_{00})
 \notag
 \\*
 &\quad
 - H(Z_{00},Z_{ii}|Y_i,Z_{10},Z_{20})
 - H(Z_{00},Z_{10},Z_{20}|Y_{i'},Z_{i'i'})
 \notag
 \\
 &=
 I(Z_{00},Z_{ii};Y_i,Z_{10},Z_{20})
 + I(Z_{00},Z_{10},Z_{20};Y_{i'},Z_{i'i'})
 - I(Z_{i0};Z_{ii}|Z_{00}),
\end{align}
\item
(\ref{eq:2hr0+tr1>0010|2+002011|1-20|00}) is equivalent to
\begin{align}
 2R_0 + R_i
 &\leq
 2H(Z_{00})
 + H(Z_{i0},Z_{ii}|Z_{00})
 + H(Z_{i'0}|Z_{00})
 - H(Z_{00},Z_{i'0},Z_{ii}|Y_i,Z_{i0})
 - H(Z_{00},Z_{i0}|Y_{i'},Z_{i'0},Z_{i'i'})
 \notag
 \\
 &=
 2H(Z_{00})
 + H(Z_{i0}|Z_{00})
 + H(Z_{ii}|Z_{00})
 - I(Z_{i0};Z_{ii}|Z_{00})
 + H(Z_{i'0}|Z_{00})
 \notag
 \\*
 &\quad
 - H(Z_{00},Z_{i'0},Z_{ii}|Y_i,Z_{i0})
 - H(Z_{00},Z_{i0}|Y_{i'},Z_{i'0},Z_{i'i'})
 \notag
 \\
 &=
 2H(Z_{00})
 + H(Z_{i'0},Z_{ii}|Z_{00})
 + H(Z_{i0}|Z_{00})
 - I(Z_{i0};Z_{ii}|Z_{00})
 \notag
 \\*
 &\quad
 - H(Z_{00},Z_{i'0},Z_{ii}|Y_i,Z_{i0})
 - H(Z_{00},Z_{i0}|Y_{i'},Z_{i'0},Z_{i'i'})
 \notag
 \\
 &=
 I(Z_{00},Z_{i'0},Z_{ii};Y_i,Z_{i0})
 + I(Z_{00},Z_{i0};Y_{i'},Z_{i'0},Z_{i'i'})
 - I(Z_{i0};Z_{ii}|Z_{00}),
\end{align}
\item
(\ref{eq:2hr0+tr2>001020|1+001022|2-2*10|00})$'$ is equivalent to
\begin{align}
 2R_0 + R_i
 &\leq
 2H(Z_{00})
 + H(Z_{i0},Z_{ii}|Z_{00})
 + 2 H(Z_{i'0}|Z_{00})
 - H(Z_{00},Z_{i'0},Z_{ii}|Y_i,Z_{i0})
 - H(Z_{00},Z_{10},Z_{20}|Y_{i'},Z_{i'i'})
 \notag
 \\
 &=
 2H(Z_{00})
 + H(Z_{i0}|Z_{00})
 + H(Z_{ii}|Z_{00})
 - I(Z_{i0};Z_{ii}|Z_{00})
 + 2 H(Z_{i'0}|Z_{00})
 \notag
 \\*
 &\quad
 - H(Z_{00},Z_{i'0},Z_{ii}|Y_i,Z_{i0})
 - H(Z_{00},Z_{10},Z_{20}|Y_{i'},Z_{i'i'})
 \notag
 \\
 &=
 2H(Z_{00})
 + H(Z_{i'0},Z_{ii}|Z_{00})
 + H(Z_{i0},Z_{i'0}|Z_{00})
 - I(Z_{i0};Z_{ii}|Z_{00})
 \notag
 \\*
 &\quad
 - H(Z_{00},Z_{i'0},Z_{ii}|Y_i,Z_{i0})
 - H(Z_{00},Z_{10},Z_{20}|Y_{i'},Z_{i'i'})
 \notag
 \\
 &=
 I(Z_{00},Z_{i'0},Z_{ii};Y_i,Z_{i0})
 + I(Z_{00},Z_{10},Z_{20};Y_{i'},Z_{i'i'})
 - I(Z_{i0};Z_{ii}|Z_{00}),
\end{align}
\item
(\ref{eq:tr1+2tr2>102022|2+2011|1+22|2})$'$ is equivalent to
\begin{align}
 2 R_i + R_{i'}
 &\leq
 2H(Z_{i0},Z_{ii}|Z_{00})
 + H(Z_{i'0},Z_{i'i'}|Z_{00})
 \notag
 \\*
 &\quad
 - H(Z_{10},Z_{20},Z_{ii}|Y_i,Z_{00})
 - H(Z_{ii}|Y_i,Z_{00},Z_{10},Z_{20})
 - H(Z_{i0},Z_{i'i'}|Y_{i'},Z_{00},Z_{i'0}) 
 \notag
 \\
 &=
 H(Z_{i0},Z_{ii}|Z_{00})
 + H(Z_{i0}|Z_{00})
 + H(Z_{ii}|Z_{00})
 - I(Z_{i0};Z_{ii}|Z_{00})
 \notag
 \\*
 &\quad
 + H(Z_{i'0}|Z_{00})
 + H(Z_{i'i'}|Z_{00})
 - I(Z_{i'0};Z_{i'i'}|Z_{00})
 \notag
 \\*
 &\quad
 \notag
 - H(Z_{10},Z_{20},Z_{ii}|Y_i,Z_{00})
 - H(Z_{ii}|Y_i,Z_{00},Z_{10},Z_{20})
 - H(Z_{i0},Z_{i'i'}|Y_{i'},Z_{00},Z_{i'0}) 
 \\
 &=
 H(Z_{i0},Z_{i'0},Z_{ii}|Z_{00})
 + H(Z_{ii}|Z_{00})
 + H(Z_{i0},Z_{i'i'}|Z_{00})
 - I(Z_{i0};Z_{ii}|Z_{00})
 - I(Z_{i'0};Z_{i'i'}|Z_{00})
 \notag
 \\*
 &\quad
 \notag
 - H(Z_{10},Z_{20},Z_{ii}|Y_i,Z_{00})
 - H(Z_{ii}|Y_i,Z_{00},Z_{10},Z_{20})
 - H(Z_{i0},Z_{i'i'}|Y_{i'},Z_{00},Z_{i'0}) 
 \\
 &=
 I(Z_{10},Z_{20},Z_{ii};Y_i|Z_{00})
 + I(Z_{ii};Y_i,Z_{10},Z_{20}|Z_{00})
 + I(Z_{i0},Z_{i'i'};Y_{i'},Z_{i'0}|Z_{00}) 
 \notag
 \\*
 &\quad
 - I(Z_{10};Z_{11}|Z_{00})
 - I(Z_{20};Z_{22}|Z_{00}),
\end{align}
\item
(\ref{eq:hr0+hr2+tr1>001022|2+11|1}) is equivalent to
\begin{align}
 R_0 + R_1 + R_2
 &\leq
 H(Z_{00})
 + H(Z_{i0},Z_{ii}|Z_{00})
 + H(Z_{i'i'}|Z_{00})
 - H(Z_{ii}|Y_i,Z_{00},Z_{10},Z_{20})
 - H(Z_{00},Z_{i0},Z_{i'i'}|Y_{i'},Z_{i'0})
 \notag
 \\
 &=
 H(Z_{00})
 + H(Z_{i0}|Z_{00})
 + H(Z_{ii}|Z_{00})
 - I(Z_{i0};Z_{ii}|Z_{00})
 + H(Z_{i'i'}|Z_{00})
 \notag
 \\*
 &\quad
 - H(Z_{ii}|Y_i,Z_{00},Z_{10},Z_{20})
 - H(Z_{00},Z_{i0},Z_{i'i'}|Y_{i'},Z_{i'0})
 \notag
 \\
 &=
 H(Z_{00})
 + H(Z_{ii}|Z_{00})
 + H(Z_{i0},Z_{i'i'}|Z_{00})
 - I(Z_{i0};Z_{ii}|Z_{00})
 \notag
 \\*
 &\quad
 - H(Z_{ii}|Y_i,Z_{00},Z_{10},Z_{20})
 - H(Z_{00},Z_{i0},Z_{i'i'}|Y_{i'},Z_{i'0})
 \notag
 \\
 &=
 I(Z_{ii};Y_i,Z_{10},Z_{20}|Z_{00})
 + I(Z_{00},Z_{i0},Z_{i'i'};Y_{i'},Z_{i'0})
 - I(Z_{i0};Z_{ii}|Z_{00}),
\end{align}
\item
(\ref{eq:hr0+tr1+tr2>00102022|2+11|1}) is equivalent to
\begin{align}
 R_0 + R_1 + R_2
 &\leq
 H(Z_{00})
 + H(Z_{10},Z_{11}|Z_{00})
 + H(Z_{20},Z_{22}|Z_{00})
 \notag
 \\*
 &\quad
 - H(Z_{ii}|Y_i,Z_{00},Z_{10},Z_{20})
 - H(Z_{00},Z_{10},Z_{20},Z_{i'i'}|Y_{i'})
 \notag
 \\
 &=
 H(Z_{00})
 + H(Z_{i0}|Z_{00})
 + H(Z_{ii}|Z_{00})
 - I(Z_{i0};Z_{ii}|Z_{00})
 + H(Z_{i'0},Z_{i'i'}|Z_{00})
 \notag
 \\*
 &\quad
 - H(Z_{ii}|Y_i,Z_{00},Z_{10},Z_{20})
 - H(Z_{00},Z_{10},Z_{20},Z_{i'i'}|Y_{i'})
 \notag
 \\
 &=
 H(Z_{00})
 + H(Z_{ii}|Z_{00})
 + H(Z_{i0},Z_{i'0},Z_{i'i'}|Z_{00})
 - I(Z_{i0};Z_{ii}|Z_{00})
 \notag
 \\*
 &\quad
 - H(Z_{ii}|Y_i,Z_{00},Z_{10},Z_{20})
 - H(Z_{00},Z_{10},Z_{20},Z_{i'i'}|Y_{i'})
 \notag
 \\
 &=
 I(Z_{ii};Y_i,Z_{10},Z_{20}|Z_{00})
 + I(Z_{00},Z_{10},Z_{20},Z_{i'i'};Y_{i'})
 - I(Z_{i0};Z_{ii}|Z_{00}),
\end{align}
\item
(\ref{eq:hr0+hr2+tr1>0011|1+1022|2}) is equivalent to
\begin{align}
 R_0 + R_1 + R_2
 &\leq
 H(Z_{00})
 + H(Z_{i0},Z_{ii}|Z_{00})
 + H(Z_{i'i'}|Z_{00})
 - H(Z_{00},Z_{ii}|Y_i,Z_{10},Z_{20})
 - H(Z_{i0},Z_{i'i'}|Y_{i'},Z_{00},Z_{i'0})
 \notag
 \\
 &=
 H(Z_{00})
 + H(Z_{i0}|Z_{00})
 + H(Z_{ii}|Z_{00})
 - I(Z_{i0};Z_{ii}|Z_{00})
 + H(Z_{i'i'}|Z_{00})
 \notag
 \\*
 &\quad
 - H(Z_{00},Z_{ii}|Y_i,Z_{10},Z_{20})
 - H(Z_{i0},Z_{i'i'}|Y_{i'},Z_{00},Z_{i'0})
 \notag
 \\
 &=
 H(Z_{00})
 + H(Z_{ii}|Z_{00})
 + H(Z_{i0},Z_{i'i'}|Z_{00})
 - I(Z_{i0};Z_{ii}|Z_{00})
 \notag
 \\*
 &\quad
 - H(Z_{00},Z_{ii}|Y_i,Z_{10},Z_{20})
 - H(Z_{i0},Z_{i'i'}|Y_{i'},Z_{00},Z_{i'0})
 \notag
 \\
 &=
 I(Z_{00},Z_{ii};Y_i,Z_{10},Z_{20})
 + I(Z_{i0},Z_{i'i'};Y_{i'},Z_{i'0}|Z_{00})
 - I(Z_{i0};Z_{ii}|Z_{00}),
\end{align}
\item
(\ref{eq:hr0+tr1+tr2>102022|2+0011|1}) is equivalent to
\begin{align}
 R_0 + R_1 + R_2
 &\leq
 H(Z_{00})
 + H(Z_{10},Z_{11}|Z_{00})
 + H(Z_{20},Z_{22}|Z_{00})
 \notag
 \\*
 &\quad
 - H(Z_{00},Z_{ii}|Y_i,Z_{10},Z_{20})
 - H(Z_{10},Z_{20},Z_{i'i'}|Y_{i'},Z_{00})
 \notag
 \\
 &=
 H(Z_{00})
 + H(Z_{i0}|Z_{00})
 + H(Z_{ii}|Z_{00})
 - I(Z_{i0},Z_{ii}|Z_{00})
 + H(Z_{i'0},Z_{i'i'}|Z_{00})
 \notag
 \\*
 &\quad
 - H(Z_{00},Z_{ii}|Y_i,Z_{10},Z_{20})
 - H(Z_{10},Z_{20},Z_{i'i'}|Y_{i'},Z_{00})
 \notag
 \\
 &=
 H(Z_{00})
 + H(Z_{ii}|Z_{00})
 + H(Z_{i0},Z_{i'0},Z_{i'i'}|Z_{00})
 - I(Z_{i0},Z_{ii}|Z_{00})
 \notag
 \\*
 &\quad
 - H(Z_{00},Z_{ii}|Y_i,Z_{10},Z_{20})
 - H(Z_{10},Z_{20},Z_{i'i'}|Y_{i'},Z_{00})
 \notag
 \\
 &=
 I(Z_{00},Z_{ii};Y_i,Z_{10},Z_{20})
 + I(Z_{10},Z_{20},Z_{i'i'};Y_{i'}|Z_{00})
 - I(Z_{i0};Z_{ii}|Z_{00}),
\end{align}
\item
(\ref{eq:hr0+tr1+tr2>002011|1+1022|2}) is equivalent to
\begin{align}
 R_0 + R_1 + R_2
 &\leq
 H(Z_{00}) + H(Z_{10},Z_{11}|Z_{00}) + H(Z_{20},Z_{22}|Z_{00})
 \notag
 \\*
 &\quad
 - H(Z_{00},Z_{i'0},Z_{ii}|Y_i,Z_{i0})
 - H(Z_{i0},Z_{i'i'}|Y_{i'},Z_{00},Z_{i'0})
 \notag
 \\
 &=
 H(Z_{00})
 + H(Z_{i0}|Z_{00})
 + H(Z_{ii}|Z_{00})
 - I(Z_{i0};Z_{ii}|Z_{00})
 \notag
 \\*
 &\quad
 + H(Z_{i'0}|Z_{00})
 + H(Z_{i'i'}|Z_{00})
 - I(Z_{i'0},Z_{i'i'}|Z_{00})
 \notag
 \\*
 &\quad
 - H(Z_{00},Z_{i'0},Z_{ii}|Y_i,Z_{i0})
 - H(Z_{i0},Z_{i'i'}|Y_{i'},Z_{00},Z_{i'0})
 \notag
 \\
 &=
 H(Z_{00})
 + H(Z_{i'0},Z_{ii}|Z_{00})
 + H(Z_{i0},Z_{i'i'}|Z_{00})
 - I(Z_{i0};Z_{ii}|Z_{00})
 - I(Z_{i'0},Z_{i'i'}|Z_{00})
 \notag
 \\*
 &\quad
 - H(Z_{00},Z_{i'0},Z_{ii}|Y_i,Z_{i0})
 - H(Z_{i0},Z_{i'i'}|Y_{i'},Z_{00},Z_{i'0})
 \notag
 \\
 &=
 I(Z_{00},Z_{i'0},Z_{ii};Y_i,Z_{i0})
 + I(Z_{i0},Z_{i'i'};Y_{i'},Z_{i'0}|Z_{00})
 - I(Z_{i0};Z_{ii}|Z_{00})
 - I(Z_{i'0};Z_{i'i'}|Z_{00}),
\end{align}
\item
(\ref{eq:2hr0+hr2+tr1>001022|2+0011|1}) is equivalent to
\begin{align}
 2R_0 + R_1 + R_2
 &\leq
 2H(Z_{00})
 + H(Z_{i0},Z_{ii}|Z_{00})
 + H(Z_{i'i'}|Z_{00})
 - H(Z_{00},Z_{ii}|Y_i,Z_{10},Z_{20})
 - H(Z_{00},Z_{i0},Z_{i'i'}|Y_{i'},Z_{i'0})
 \notag
 \\
 &=
 2H(Z_{00})
 + H(Z_{i0}|Z_{00})
 + H(Z_{ii}|Z_{00})
 - I(Z_{i0};Z_{ii}|Z_{00})
 + H(Z_{i'i'}|Z_{00})
 \notag
 \\*
 &\quad
 - H(Z_{00},Z_{ii}|Y_i,Z_{10},Z_{20})
 - H(Z_{00},Z_{i0},Z_{i'i'}|Y_{i'},Z_{i'0})
 \notag
 \\
 &=
 2H(Z_{00})
 + H(Z_{ii}|Z_{00})
 + H(Z_{i0},Z_{i'i'}|Z_{00})
 - I(Z_{i0};Z_{ii}|Z_{00})
 \notag
 \\*
 &\quad
 - H(Z_{00},Z_{ii}|Y_i,Z_{10},Z_{20})
 - H(Z_{00},Z_{i0},Z_{i'i'}|Y_{i'},Z_{i'0})
 \notag
 \\
 &=
 I(Z_{00},Z_{ii};Y_i,Z_{10},Z_{20})
 + I(Z_{00},Z_{i0},Z_{i'i'};Y_{i'},Z_{i'0})
 - I(Z_{i0};Z_{ii}|Z_{00}),
\end{align}
\item
(\ref{eq:2hr0+tr1+tr2>00102022|2+0011|1}) is equivalent to
\begin{align}
 2R_0 + R_1 + R_2
 &\leq
 2H(Z_{00}) + H(Z_{10},Z_{11}|Z_{00}) + H(Z_{20},Z_{22}|Z_{00})
 \notag
 \\*
 &\quad
 - H(Z_{00},Z_{ii}|Y_i,Z_{10},Z_{20})
 - H(Z_{00},Z_{10},Z_{20},Z_{i'i'}|Y_{i'})
 \notag
 \\
 &=
 2H(Z_{00})
 + H(Z_{i0}|Z_{00})
 + H(Z_{ii}|Z_{00})
 - I(Z_{i0};Z_{ii}|Z_{00})
 + H(Z_{i'0},Z_{i'i'}|Z_{00})
 \notag
 \\*
 &\quad
 - H(Z_{00},Z_{ii}|Y_i,Z_{10},Z_{20})
 - H(Z_{00},Z_{10},Z_{20},Z_{i'i'}|Y_{i'})
 \notag
 \\
 &=
 2H(Z_{00})
 + H(Z_{ii}|Z_{00})
 + H(Z_{10},Z_{20},Z_{i'i'}|Z_{00})
 - I(Z_{i0};Z_{ii}|Z_{00})
 \notag
 \\*
 &\quad
 - H(Z_{00},Z_{ii}|Y_i,Z_{10},Z_{20})
 - H(Z_{00},Z_{10},Z_{20},Z_{i'i'}|Y_{i'})
 \notag
 \\
 &=
 I(Z_{00},Z_{ii};Y_i,Z_{10},Z_{20})
 + I(Z_{00},Z_{10},Z_{20},Z_{i'i'};Y_{i'})
 - I(Z_{i0};Z_{ii}|Z_{00}),
\end{align}
\item
(\ref{eq:2hr0+tr1+tr2>00102022|2+002011|1-20|00}) is equivalent to
\begin{align}
 2R_0 + R_1 + R_2
 &\leq
 2H(Z_{00}) + H(Z_{10},Z_{11}|Z_{00}) + H(Z_{20},Z_{22}|Z_{00})
 + H(Z_{i'0}|Z_{00})
 \notag
 \\*
 &\quad
 - H(Z_{00},Z_{i'0},Z_{ii}|Y_i,Z_{i0})
 - H(Z_{00},Z_{10},Z_{20},Z_{i'i'}|Y_{i'})
 \notag
 \\
 &=
 2H(Z_{00})
 + H(Z_{i0}|Z_{00})
 + H(Z_{ii}|Z_{00})
 - I(Z_{i0};Z_{ii}|Z_{00})
 + H(Z_{i'0},Z_{i'i'}|Z_{00})
 + H(Z_{i'0}|Z_{00})
 \notag
 \\*
 &\quad
 - H(Z_{00},Z_{i'0},Z_{ii}|Y_i,Z_{i0})
 - H(Z_{00},Z_{10},Z_{20},Z_{i'i'}|Y_{i'})
 \notag
 \\
 &=
 2H(Z_{00})
 + H(Z_{i'0},Z_{ii}|Z_{00})
 + H(Z_{10},Z_{20},Z_{i'i'}|Z_{00})
 - I(Z_{i0};Z_{ii}|Z_{00})
 \notag
 \\*
 &\quad
 - H(Z_{00},Z_{i'0},Z_{ii}|Y_i,Z_{i0})
 - H(Z_{00},Z_{10},Z_{20},Z_{i'i'}|Y_{i'})
 \notag
 \\
 &=
 I(Z_{00},Z_{i'0},Z_{ii};Y_i,Z_{i0})
 + I(Z_{00},Z_{10},Z_{20},Z_{i'i'};Y_{i'})
 - I(Z_{i0};Z_{ii}|Z_{00}),
\end{align}
\item
(\ref{eq:2hr0+tr1+tr2>002011|1+001022|2}) is equivalent to
\begin{align}
 2R_0 + R_1 + R_2
 &\leq
 2H(Z_{00}) + H(Z_{10},Z_{11}|Z_{00}) + H(Z_{20},Z_{22}|Z_{00})
 \notag
 \\*
 &\quad
 - H(Z_{00},Z_{20},Z_{11}|Y_1,Z_{10})
 - H(Z_{00},Z_{10},Z_{22}|Y_2,Z_{20})
 \notag
 \\
 &=
 2H(Z_{00})
 + H(Z_{10}|Z_{00})
 + H(Z_{11}|Z_{00})
 - I(Z_{10};Z_{11}|Z_{00})
 \notag
 \\*
 &\quad
 + H(Z_{20}|Z_{00})
 + H(Z_{22}|Z_{00})
 - I(Z_{20};Z_{22}|Z_{00})
 \notag
 \\*
 &\quad
 - H(Z_{00},Z_{20},Z_{11}|Y_1,Z_{10})
 - H(Z_{00},Z_{10},Z_{22}|Y_2,Z_{20})
 \notag
 \\
 &=
 2H(Z_{00})
 + H(Z_{20},Z_{11}|Z_{00})
 + H(Z_{10},Z_{22}|Z_{00})
 - I(Z_{10};Z_{11}|Z_{00})
 - I(Z_{20};Z_{22}|Z_{00})
 \notag
 \\*
 &\quad
 - H(Z_{00},Z_{20},Z_{11}|Y_1,Z_{10})
 - H(Z_{00},Z_{10},Z_{22}|Y_2,Z_{20})
 \notag
 \\
 &=
 I(Z_{00},Z_{20},Z_{11};Y_1,Z_{10})
 + I(Z_{00},Z_{10},Z_{22};Y_2,Z_{20})
 - I(Z_{10};Z_{11}|Z_{00}) - I(Z_{20};Z_{22}|Z_{00}),
\end{align}
\item
(\ref{eq:hr0+tr1+2tr2>102022|2+002011|1+22|2})$'$ is equivalent to
\begin{align}
 R_0 + 2R_i + R_{i'}
 &\leq
 H(Z_{00})
 + 2H(Z_{i0},Z_{ii}|Z_{00})
 + H(Z_{i'0},Z_{i'i'}|Z_{00})
 \notag
 \\*
 &\quad
 - H(Z_{10},Z_{20},Z_{ii}|Y_i,Z_{00})
 - H(Z_{ii}|Y_i,Z_{00},Z_{10},Z_{20})
 - H(Z_{00},Z_{i0},Z_{i'i'}|Y_{i'},Z_{i'0})
 \notag
 \\
 &=
 H(Z_{00})
 + H(Z_{i0},Z_{ii}|Z_{00})
 + H(Z_{i0}|Z_{00})
 + H(Z_{ii}|Z_{00})
 - I(Z_{i0};Z_{ii}|Z_{00})
 \notag
 \\*
 &\quad
 + H(Z_{i'0}|Z_{00})
 + H(Z_{i'i'}|Z_{00})
 - I(Z_{i'0};Z_{i'i'}|Z_{00})
 \notag
 \\*
 &\quad
 - H(Z_{10},Z_{20},Z_{ii}|Y_i,Z_{00})
 - H(Z_{ii}|Y_i,Z_{00},Z_{10},Z_{20})
 - H(Z_{00},Z_{i0},Z_{i'i'}|Y_{i'},Z_{i'0})
 \notag
 \\
 &=
 H(Z_{00})
 + H(Z_{10},Z_{20},Z_{ii}|Z_{00})
 + H(Z_{ii}|Z_{00})
 + H(Z_{i0},Z_{i'i'}|Z_{00})
 \notag
 \\*
 &\quad
 - I(Z_{i0};Z_{ii}|Z_{00})
 - I(Z_{i'0};Z_{i'i'}|Z_{00})
 \notag
 \\*
 &\quad
 - H(Z_{10},Z_{20},Z_{ii}|Y_i,Z_{00})
 - H(Z_{ii}|Y_i,Z_{00},Z_{10},Z_{20})
 - H(Z_{00},Z_{i0},Z_{i'i'}|Y_{i'},Z_{i'0})
 \notag
 \\
 &=
 I(Z_{10},Z_{20},Z_{ii};Y_i|Z_{00})
 + I(Z_{ii};Y_i,Z_{10},Z_{20}|Z_{00})
 + I(Z_{00},Z_{i0},Z_{i'i'};Y_{i'},Z_{i'0})
 \notag
 \\*
 &\quad
 - I(Z_{10};Z_{11}|Z_{00})
 - I(Z_{20};Z_{22}|Z_{00}),
\end{align}
\item
(\ref{eq:hr0+tr1+2tr2>00102022|2+2011|1+22|2})$'$ is equivalent to
\begin{align}
 R_0 + 2R_i + R_{i'}
 &\leq
 H(Z_{00})
 + 2H(Z_{i0},Z_{ii}|Z_{00})
 + H(Z_{i'0},Z_{i'i'}|Z_{00})
 \notag
 \\*
 &\quad
 - H(Z_{00},Z_{10},Z_{20},Z_{ii}|Y_i)
 - H(Z_{ii}|Y_i,Z_{00},Z_{10},Z_{20})
 - H(Z_{i0},Z_{i'i'}|Y_{i'},Z_{00},Z_{i'0}) 
 \notag
 \\
 &=
 H(Z_{00})
 + H(Z_{i0},Z_{ii}|Z_{00})
 + H(Z_{i0}|Z_{00})
 + H(Z_{ii}|Z_{00})
 - I(Z_{i0};Z_{ii}|Z_{00})
 \notag
 \\*
 &\quad
 + H(Z_{i'0}|Z_{00})
 + H(Z_{i'i'}|Z_{00})
 - I(Z_{i'0};Z_{i'i'}|Z_{00})
 \notag
 \\*
 &\quad
 - H(Z_{00},Z_{10},Z_{20},Z_{ii}|Y_i)
 - H(Z_{ii}|Y_i,Z_{00},Z_{10},Z_{20})
 - H(Z_{i0},Z_{i'i'}|Y_{i'},Z_{00},Z_{i'0}) 
 \notag
 \\
 &=
 H(Z_{00})
 + H(Z_{10},Z_{20},Z_{ii}|Z_{00})
 + H(Z_{ii}|Z_{00})
 + H(Z_{i0},Z_{i'i'}|Z_{00})
 \notag
 \\*
 &\quad
 - I(Z_{i0};Z_{ii}|Z_{00})
 - I(Z_{i'0};Z_{i'i'}|Z_{00})
 \notag
 \\*
 &\quad
 - H(Z_{00},Z_{10},Z_{20},Z_{ii}|Y_i)
 - H(Z_{ii}|Y_i,Z_{00},Z_{10},Z_{20})
 - H(Z_{i0},Z_{i'i'}|Y_{i'},Z_{00},Z_{i'0}) 
 \notag
 \\
 &=
 I(Z_{00},Z_{10},Z_{20},Z_{ii};Y_i)
 + I(Z_{ii};Y_i,Z_{10},Z_{20}|Z_{00})
 + I(Z_{i0},Z_{i'i'};Y_{i'},Z_{i'0}|Z_{00}) 
 \notag
 \\*
 &\quad
 - I(Z_{10};Z_{11}|Z_{00})
 - I(Z_{20};Z_{22}|Z_{00}),
\end{align}
\item
(\ref{eq:2hr0+tr1+2tr2>00102022|2+002011|1+22|2})$'$ is equivalent to
\begin{align}
 2R_0 + 2R_i + R_{i'}
 &\leq
 2H(Z_{00})
 + 2H(Z_{i0},Z_{ii}|Z_{00})
 + H(Z_{i'0},Z_{i'i'}|Z_{00})
 \notag
 \\*
 &\quad
 - H(Z_{00},Z_{10},Z_{20},Z_{ii}|Y_i)
 - H(Z_{ii}|Y_i,Z_{00},Z_{10},Z_{20})
 - H(Z_{00},Z_{i0},Z_{i'i'}|Y_{i'},Z_{i'0})
 \notag
 \\
 &=
 2H(Z_{00})
 + H(Z_{i0},Z_{ii}|Z_{00})
 + H(Z_{i0}|Z_{00})
 + H(Z_{ii}|Z_{00})
 - I(Z_{i0};Z_{ii}|Z_{00})
 \notag
 \\*
 &\quad
 + H(Z_{i'0}|Z_{00})
 + H(Z_{i'i'}|Z_{00})
 - I(Z_{i'0};Z_{i'i'}|Z_{00})
 \notag
 \\*
 &\quad
 - H(Z_{00},Z_{10},Z_{20},Z_{ii}|Y_i)
 - H(Z_{ii}|Y_i,Z_{00},Z_{10},Z_{20})
 - H(Z_{00},Z_{i0},Z_{i'i'}|Y_{i'},Z_{i'0})
 \notag
 \\
 &=
 2H(Z_{00})
 + H(Z_{i0},Z_{i'0},Z_{ii}|Z_{00})
 + H(Z_{ii}|Z_{00})
 + H(Z_{i0},Z_{i'i'}|Z_{00})
 \notag
 \\*
 &\quad
 - I(Z_{i0};Z_{ii}|Z_{00})
 - I(Z_{i'0};Z_{i'i'}|Z_{00})
 \notag
 \\*
 &\quad
 - H(Z_{00},Z_{10},Z_{20},Z_{ii}|Y_i)
 - H(Z_{ii}|Y_i,Z_{00},Z_{10},Z_{20})
 - H(Z_{00},Z_{i0},Z_{i'i'}|Y_{i'},Z_{i'0})
 \notag
 \\
 &=
 I(Z_{00},Z_{10},Z_{20},Z_{ii};Y_i)
 + I(Z_{ii};Y_i,Z_{10},Z_{20}|Z_{00})
 + I(Z_{00},Z_{i0},Z_{i'i'};Y_{i'},Z_{i'0})
 \notag
 \\*
 &\quad
 - I(Z_{10};Z_{11}|Z_{00})
 - I(Z_{20};Z_{22}|Z_{00}),
\end{align}
\item
(\ref{eq:2hr0+tr1+2tr2>00102022|2+2011|1+0022|2})$'$ is equivalent to
\begin{align}
 2R_0 + 2R_i + R_{i'}
 &\leq
 2H(Z_{00})
 + 2H(Z_{i0},Z_{ii}|Z_{00})
 + H(Z_{i'0},Z_{i'i'}|Z_{00})
 \notag
 \\*
 &\quad
 - H(Z_{00},Z_{10},Z_{20},Z_{ii}|Y_i)
 - H(Z_{00},Z_{ii}|Y_i,Z_{10},Z_{20})
 - H(Z_{i0},Z_{i'i'}|Y_{i'},Z_{00},Z_{i'0})
 \notag
 \\
 &=
 2H(Z_{00})
 + H(Z_{i0},Z_{ii}|Z_{00})
 + H(Z_{i0}|Z_{00})
 + H(Z_{ii}|Z_{00})
 - I(Z_{i0};Z_{ii}|Z_{00})
 \notag
 \\*
 &\quad
 + H(Z_{i'0}|Z_{00})
 + H(Z_{i'i'}|Z_{00})
 - I(Z_{i'0};Z_{i'i'}|Z_{00})
 \notag
 \\*
 &\quad
 - H(Z_{00},Z_{10},Z_{20},Z_{ii}|Y_i)
 - H(Z_{00},Z_{ii}|Y_i,Z_{10},Z_{20})
 - H(Z_{i0},Z_{i'i'}|Y_{i'},Z_{00},Z_{i'0})
 \notag
 \\
 &=
 2H(Z_{00})
 + H(Z_{i0},Z_{i'0},Z_{ii}|Z_{00})
 + H(Z_{ii}|Z_{00})
 + H(Z_{i0},Z_{i'i'}|Z_{00})
 \notag
 \\*
 &\quad
 - I(Z_{i0};Z_{ii}|Z_{00})
 - I(Z_{i'0};Z_{i'i'}|Z_{00})
 \notag
 \\*
 &\quad
 - H(Z_{00},Z_{10},Z_{20},Z_{ii}|Y_i)
 - H(Z_{00},Z_{ii}|Y_i,Z_{10},Z_{20})
 - H(Z_{i0},Z_{i'i'}|Y_{i'},Z_{00},Z_{i'0})
 \notag
 \\
 &=
 I(Z_{00},Z_{10},Z_{20},Z_{ii};Y_i)
 + I(Z_{00},Z_{ii};Y_i,Z_{10},Z_{20})
 + I(Z_{i0},Z_{i'i'};Y_{i'},Z_{i'0}|Z_{00})
 \notag
 \\*
 &\quad
 - I(Z_{10};Z_{11}|Z_{00})
 - I(Z_{20};Z_{22}|Z_{00}),
\end{align}
\item
(\ref{eq:3hr0+tr1+2tr2>00102022|2+002011|1+0022|2})$'$ is equivalent to
\begin{align}
 3R_0 + 2R_i + R_{i'}
 &\leq
 3H(Z_{00})
 + 2H(Z_{i0},Z_{ii}|Z_{00})
 + H(Z_{i'0},Z_{i'i'}|Z_{00})
 \notag
 \\*
 &\quad
 - H(Z_{00},Z_{10},Z_{20},Z_{ii}|Y_i)
 - H(Z_{00},Z_{ii}|Y_i,Z_{10},Z_{20})
 - H(Z_{00},Z_{i0},Z_{i'i'}|Y_{i'},Z_{i'0})
 \notag
 \\
 &=
 3H(Z_{00})
 + H(Z_{i0},Z_{ii}|Z_{00})
 + H(Z_{i0}|Z_{00})
 + H(Z_{ii}|Z_{00})
 - I(Z_{i0};Z_{ii}|Z_{00})
 \notag
 \\*
 &\quad
 + H(Z_{i'0}|Z_{00})
 + H(Z_{i'i'}|Z_{00})
 - I(Z_{i'0};Z_{i'i'}|Z_{00})
 \notag
 \\*
 &\quad
 - H(Z_{00},Z_{10},Z_{20},Z_{ii}|Y_i)
 - H(Z_{00},Z_{ii}|Y_i,Z_{10},Z_{20})
 - H(Z_{00},Z_{i0},Z_{i'i'}|Y_{i'},Z_{i'0})
 \notag
 \\
 &=
 3H(Z_{00})
 + H(Z_{i0},Z_{i'0},Z_{ii}|Z_{00})
 + H(Z_{ii}|Z_{00})
 + H(Z_{i0},Z_{i'i'}|Z_{00})
 \notag
 \\*
 &\quad
 - I(Z_{i0};Z_{ii}|Z_{00})
 - I(Z_{i'0};Z_{i'i'}|Z_{00})
 \notag
 \\*
 &\quad
 - H(Z_{00},Z_{10},Z_{20},Z_{ii}|Y_i)
 - H(Z_{00},Z_{ii}|Y_i,Z_{10},Z_{20})
 - H(Z_{00},Z_{i0},Z_{i'i'}|Y_{i'},Z_{i'0})
 \notag
 \\
 &=
 I(Z_{00},Z_{10},Z_{20},Z_{ii};Y_i)
 + I(Z_{00},Z_{ii};Y_i,Z_{10},Z_{20})
 + I(Z_{00},Z_{i0},Z_{i'i'};Y_{i'},Z_{i'0})
 \notag
 \\*
 &\quad
 - I(Z_{10};Z_{11}|Z_{00})
 - I(Z_{20};Z_{22}|Z_{00}).
\end{align}
\end{itemize}

\subsection{Proof of (\ref{eq:crng-fme0})--(\ref{eq:crng-fme3})}
\label{sec:proof-crng-crng-0-3}

We have
\begin{align}
R_0
&\leq
H(Q)
\notag
\\
&=
I(Q;Q)
\notag
\\
&\leq
I(Q,V_{00};Q,V_{i0},V_{ii})
\notag
\\
&=
I(Z_{00};Z_{ii})
\notag
\\
&\leq
\min\lrb{
 \begin{aligned}
  &I(Z_{00};Y_i,Z_{10},Z_{20},Z_{ii}),
  \\
  &I(Z_{00},Z_{i0};Y_i,Z_{i'0},Z_{ii}),
  \\
  &I(Z_{00},Z_{i'0};Y_i,Z_{i0},Z_{ii}),
  \\
  &I(Z_{00},Z_{10},Z_{20};Y_i,Z_{ii})
 \end{aligned}
},
\end{align}
where the first inequality comes from (\ref{eq:crng-H(Q)}),
and the second equality comes from
(\ref{eq:def-Z00}) and (\ref{eq:def-Zii}).
\hfill\IEEEQED

\subsection{Details of Proof for Case 0}
\label{sec:proof-crng-crng-case0-appendix}

In the following, we use (\ref{eq:Qindependence})
(\ref{eq:def-Zi0}), (\ref{eq:def-Zii}),
(\ref{eq:def-Zi0-case0}),
and Lemma \ref{lem:Q} without notice.

We have (\ref{eq:crng-case0-I(V11;Y1|V00,V10,V20)}) as
\begin{align}
I(V_{ii};Y_i|V_{00},V_{10},V_{20})
&=
I(V_{i0},V_{ii};Y_i|V_{00},V_{10},V_{20})
\notag
\\
&=
I(Q,V_{i0},V_{ii};Y_i|Q,V_{00},V_{10},V_{20})
\notag
\\
&=
I(Z_{ii};Y_i|Z_{00},Z_{10},Z_{20})
\notag
\\
&=
I(Z_{ii};Y_i,Z_{10},Z_{20}|Z_{00})
- I(Z_{ii};Z_{10},Z_{20}|Z_{00})
\notag
\\
&=
I(Z_{ii};Y_i,Z_{10},Z_{20}|Z_{00})
- I(Z_{i0};Z_{ii}|Z_{00}),
\end{align}
where the last equality comes from
(\ref{eq:I(ZX1;ZX2|Z0)=0}) 
and Lemmas \ref{lem:c-independence-I(A;C|D)}
and \ref{lem:c-independence-I(A;B,C|D)}.

We have (\ref{eq:crng-case0-H(Q)+I(V11;Y1|V00,V10,V20)}) as
\begin{align}
H(Q) + I(V_{ii};Y_i|V_{00},V_{10},V_{20})
&=
H(Q) + I(V_{i0},V_{ii};Y_i|V_{00},V_{10},V_{20})
\notag
\\
&=
H(Q) + I(V_{i0},V_{ii};Y_i,V_{i0}|V_{00},V_{i'0})
- I(V_{i0},V_{ii};V_{i0}|V_{00},V_{i'0})
\notag
\\
&=
H(Q) + I(V_{i0},V_{ii};Y_i,V_{i0}|V_{00},V_{i'0})
- I(V_{i0};V_{i0},V_{ii}|V_{00})
\notag
\\
&\leq
H(Q)
+ \min\lrb{
 \begin{aligned}
  &I(V_{00},V_{i0},V_{ii};Y_i,V_{10},V_{20}),
  \\
  &I(V_{00},V_{10},V_{20},V_{ii};Y_i,V_{i0})
 \end{aligned}
}
- I(V_{i0};V_{i0},V_{ii}|V_{00})
\notag
\\
&=
\min\lrb{
 \begin{aligned}
  &I(Q,V_{00},V_{i0},V_{ii};Y_i,Q,V_{10},V_{20}),
  \\
  &I(Q,V_{00},V_{10},V_{20},V_{ii};Y_i,Q,V_{i0})
 \end{aligned}
}
- I(Q,V_{i0};Q,V_{i0},V_{ii}|Q,V_{00})
\notag
\\
&=
\min\lrb{
 \begin{aligned}
  &I(Z_{00},Z_{ii};Y_i,Z_{10},Z_{20}),
  \\
  &I(Z_{00},Z_{i'0},Z_{ii};Y_i,Z_{i0})
 \end{aligned}
}
- I(Z_{i0};Z_{ii}|Z_{00}),
\end{align}
where the third equality comes
from (\ref{eq:I(VX1;VX2|V0)=0})
and Lemmas \ref{lem:c-independence-I(A;C|D)}
and \ref{lem:c-independence-I(A;B|C,D)}.

We have (\ref{eq:crng-case0-I(V10,V11;Y1|V00,V20)}) as
\begin{align}
I(V_{i0},V_{ii};Y_i|V_{00},V_{i'0})
&=
I(Q,V_{i0},V_{ii};Y_i|Q,V_{00},V_{i'0})
\notag
\\
&=
I(Z_{ii};Y_i|Z_{00},Z_{i'0})
\notag
\\
&\leq
\min\lrb{
 \begin{aligned}
  &I(Z_{ii};Y_i,Z_{10},Z_{20}|Z_{00}),
  \\
  &I(Z_{i0},Z_{ii};Y_i,Z_{i'0}|Z_{00})
 \end{aligned}
}.
\end{align}

We have (\ref{eq:crng-case0-H(Q)+I(V10,V11;Y1|V00,V20)}) as
\begin{align}
H(Q) + I(V_{i0},V_{ii};Y_i|V_{00},V_{i'0})
&\leq
H(Q)
+ \min\lrb{
 \begin{aligned}
  &I(V_{00},V_{i0},V_{ii};Y_i,V_{i'0}),
  \\
  &I(V_{00},V_{10},V_{20},V_{ii};Y_i,V_{i0})
 \end{aligned}
}
\notag
\\
&=
\min\lrb{
 \begin{aligned}
  &I(Q,V_{00},V_{i0},V_{ii};Y_i,Q,V_{i'0}),
  \\
  &I(Q,V_{00},V_{10},V_{20},V_{ii};Y_i,Q,V_{i0})
 \end{aligned}
}
\notag
\\
&=
\min\lrb{
 \begin{aligned}
  &I(Z_{00},Z_{ii};Y_i,Z_{i'0}),
  \\
  &I(Z_{00},Z_{i'0},Z_{ii};Y_i,Z_{i0})
 \end{aligned}
}
\notag
\\
&\leq
\min\lrb{
 \begin{aligned}
  &I(Z_{00},Z_{ii};Y_i,Z_{10},Z_{20}),
  \\
  &I(Z_{00},Z_{i0},Z_{ii};Y_i,Z_{i'0}),
  \\
  &I(Z_{00},Z_{i'0},Z_{ii};Y_i,Z_{i0})
 \end{aligned}
}.
\end{align}

We have (\ref{eq:crng-case0-I(V20,V11;Y1|V00,V10)}) as
\begin{align}
I(V_{i'0},V_{ii};Y_i|V_{00},V_{i0})
&=
I(V_{10},V_{20},V_{ii};Y_i|V_{00},V_{i0})
\notag
\\
&=
I(Q,V_{10},V_{20},V_{ii};Y_i|Q,V_{00},V_{i0})
\notag
\\
&=
I(Z_{i'0},Z_{ii};Y_i|Z_{00},Z_{i0})
\notag
\\
&=
I(Z_{i'0},Z_{ii};Y_i,Z_{i0}|Z_{00})
- I(Z_{i'0},Z_{ii};Z_{i0}|Z_{00})
\notag
\\
&=
I(Z_{i'0},Z_{ii};Y_i,Z_{i0}|Z_{00})
- I(Z_{i0};Z_{ii}|Z_{00}),
\end{align}
where the last quality comes from
(\ref{eq:I(ZX1;ZX2|Z0)=0})
and Lemmas \ref{lem:c-independence-I(A;C|D)}
and \ref{lem:c-independence-I(A;B,C|D)}.

We have (\ref{eq:crng-case0-H(Q)+I(V20,V11;Y1|V00,V10)}) as
\begin{align}
H(Q) + I(V_{i'0},V_{ii};Y_i|V_{00},V_{i0})
&=
H(Q) + I(V_{10},V_{20},V_{ii};Y_i|V_{00},V_{i0})
\notag
\\
&=
H(Q)
+ I(V_{10},V_{20},V_{ii};Y_i,V_{i0}|V_{00})
- I(V_{10},V_{20},V_{ii};V_{i0}|V_{00})
\notag
\\
&=
H(Q)
+ I(V_{10},V_{20},V_{ii};Y_i,V_{i0}|V_{00})
- I(V_{i0};V_{i0},V_{ii}|V_{00})
\notag
\\
&\leq
H(Q)
+ I(V_{00},V_{10},V_{20},V_{ii};Y_i,V_{i0})
- I(V_{i0};V_{i0},V_{ii}|V_{00})
\notag
\\
&=
I(Q,V_{00},V_{10},V_{20},V_{ii};Y_i,Q,V_{i0})
- I(Q,V_{i0};Q,V_{i0},V_{ii}|Q,V_{00})
\notag
\\
&=
I(Z_{00},Z_{i'0},Z_{ii};Y_i,Z_{i0})
- I(Z_{i0};Z_{ii}|Z_{00}),
\end{align}
where the third equality comes
from
(\ref{eq:I(VX1;VX2|V0)=0})
and Lemmas \ref{lem:c-independence-I(A;C|D)}
and \ref{lem:c-independence-I(A;B,C|D)}.

We have (\ref{eq:crng-case0-I(V10,V20,V11;Y1|V00)}) as
\begin{align}
I(V_{10},V_{20},V_{ii};Y_i|V_{00})
&=
I(Q,V_{10},V_{20},V_{ii};Y_i|Q,V_{00})
\notag
\\
&=
I(Z_{i'0},Z_{ii};Y_i|Z_{00})
\notag
\\
&\leq
\min\lrb{
 \begin{aligned}
  &I(Z_{i'0},Z_{ii};Y_i,Z_{i0}|Z_{00}),
  \\
  &I(Z_{10},Z_{20},Z_{ii};Y_i|Z_{00})
 \end{aligned}
}.
\end{align}

We have (\ref{eq:crng-case0-I(V00,V10,V20,V11;Y1)}) as
\begin{align}
I(V_{00},V_{10},V_{20},V_{ii};Y_i)
&=
I(Q,V_{00},V_{10},V_{20},V_{ii};Y_i)
\notag
\\
&=
I(Z_{00},Z_{i'0},Z_{ii};Y_i)
\notag
\\
&\leq
\min\lrb{
 \begin{aligned}
  &I(Z_{00},Z_{i'0},Z_{ii};Y_i,Z_{i0}),
  \\
  &I(Z_{00},Z_{10},Z_{20},Z_{ii};Y_i)
 \end{aligned}
}.
\end{align}

We have (\ref{eq:crng-case0-I(V10;Y2|V00,V20,V22)}) as
\begin{align}
I(V_{i0};Y_{i'}|V_{00},V_{i'0},V_{i'i'})
&=
I(Q,V_{i0};Y_{i'}|Q,V_{00},V_{i'0},V_{i'i'})
\notag
\\
&=
I(Z_{i0};Y_{i'}|Z_{00},Z_{i'i'})
\notag
\\
&\leq
I(Z_{i0};Y_{i'},Z_{i'0},Z_{i'i'}|Z_{00}).
\end{align}

We have (\ref{eq:crng-case0-H(Q)+I(V10;Y2|V00,V20,V22)}) as
\begin{align}
H(Q) + I(V_{i0};Y_{i'}|V_{00},V_{i'0},V_{i'i'})
&\leq
H(Q) + I(V_{00},V_{i0};Y_{i'},V_{i'0},V_{i'i'})
\notag
\\
&=
I(Q,V_{00},V_{i0};Y_{i'},Q,V_{i'0},V_{i'i'})
\notag
\\
&=
I(Z_{00},Z_{i0};Y_{i'},Z_{i'i'})
\notag
\\
&\leq
\min\lrb{
 \begin{aligned}
  &I(Z_{00},Z_{i0};Y_{i'},Z_{i'0},Z_{i'i'}),
  \\
  &I(Z_{00},Z_{10},Z_{20};Y_{i'},Z_{i'i'})
 \end{aligned}
}.
\end{align}

We have (\ref{eq:crng-fme4}) and (\ref{eq:crng-fme5}) as
\begin{align}
R_i
&\leq
I(V_{i0},V_{ii};Y_i|V_{00},V_{i'0})
\notag
\\
&\leq
\min\lrb{
 \begin{aligned}
  &I(Z_{ii};Y_i,Z_{10},Z_{20}|Z_{00}),
  \\
  &I(Z_{i0},Z_{ii};Y_i,Z_{i'0}|Z_{00})
 \end{aligned}
},
\end{align}
where the first inequality comes from (\ref{eq:crng-Ri})
and the second inequality comes from 
(\ref{eq:crng-case0-I(V10,V11;Y1|V00,V20)}).

We have (\ref{eq:crng-fme6}) as
\begin{align}
R_i
&\leq
I(V_{ii};Y_i|V_{00},V_{10},V_{20})
+ I(V_{i0};Y_{i'}|V_{00},V_{i'0},V_{i'i'})
\notag
\\
&\leq
I(Z_{ii};Y_i,Z_{10},Z_{20}|Z_{00})
+ I(Z_{i0};Y_{i'},Z_{i'0},Z_{i'i'}|Z_{00})
- I(Z_{i0};Z_{ii}|Z_{00}),
\end{align}
where the first inequality comes from
(\ref{eq:crng-case0}),
and the second inequality comes from 
(\ref{eq:crng-case0-I(V11;Y1|V00,V10,V20)})
and (\ref{eq:crng-case0-I(V10;Y2|V00,V20,V22)}).

We have
(\ref{eq:crng-fme7})--(\ref{eq:crng-fme9}) as
\begin{align}
R_0 + R_i
&\leq
H(Q) + I(V_{i0},V_{ii};Y_i|V_{00},V_{i'0})
\notag
\\
&\leq
\min\lrb{
 \begin{aligned}
  &I(Z_{00},Z_{ii};Y_i,Z_{10},Z_{20}),
  \\
  &I(Z_{00},Z_{i0},Z_{ii};Y_i,Z_{i'0}),
  \\
  &I(Z_{00},Z_{i'0},Z_{ii};Y_i,Z_{i0})
 \end{aligned}
},
\end{align}
where the first inequality comes from (\ref{eq:crng-Ri})
and (\ref{eq:crng-H(Q)})
and the second inequality comes from
(\ref{eq:crng-case0-H(Q)+I(V10,V11;Y1|V00,V20)}).

We have (\ref{eq:crng-fme10}) as
\begin{align}
R_0 + R_i
&\leq
I(V_{00},V_{10},V_{20},V_{ii};Y_i)
\notag
\\
&=
I(Z_{00},Z_{10},Z_{20},Z_{ii};Y_i),
\end{align}
where the first inequality comes from (\ref{eq:crng-R0+Ri})
and the second inequality comes from
(\ref{eq:crng-case0-I(V00,V10,V20,V11;Y1)}).

We have (\ref{eq:crng-fme11}) and (\ref{eq:crng-fme12}) as
\begin{align}
R_0 + R_i
&\leq
H(Q) + I(V_{ii};Y_i|V_{00},V_{10},V_{20})
+ I(V_{i0};Y_{i'}|V_{00},V_{i'0},V_{i'i'})
\notag
\\
&\leq
I(Z_{ii};Y_i,Z_{10},Z_{20}|Z_{00})
+ \min\lrb{
 \begin{aligned}
  &I(Z_{00},Z_{i0};Y_{i'},Z_{i'0},Z_{i'i'}),
  \\
  &I(Z_{00},Z_{10},Z_{20};Y_{i'},Z_{i'i'})
 \end{aligned}
}
-I(Z_{i0};Z_{ii}|Z_{00}),
\end{align}
where the first inequality comes from
(\ref{eq:crng-H(Q)})
and (\ref{eq:crng-case0}),
and the second inequality comes from
(\ref{eq:crng-case0-I(V11;Y1|V00,V10,V20)})
and (\ref{eq:crng-case0-H(Q)+I(V10;Y2|V00,V20,V22)}).

We have (\ref{eq:crng-fme13}) and (\ref{eq:crng-fme14}) as
\begin{align}
R_0 + R_i
&\leq
H(Q) + I(V_{ii};Y_i|V_{00},V_{10},V_{20})
+ I(V_{i0};Y_{i'}|V_{00},V_{i'0},V_{i'i'})
\notag
\\
&\leq
\min\lrb{
 \begin{aligned}
  &I(Z_{00},Z_{ii};Y_i,Z_{10},Z_{20})
  \\
  &I(Z_{00},Z_{i'0},Z_{ii};Y_i,Z_{i0})
 \end{aligned}
}
+ I(Z_{i0};Y_{i'},Z_{i'0},Z_{i'i'}|Z_{00})
-I(Z_{i0};Z_{ii}|Z_{00}),
\end{align}
where the first inequality comes from
(\ref{eq:crng-H(Q)})
and (\ref{eq:crng-case0}),
and the second inequality comes from
(\ref{eq:crng-case0-H(Q)+I(V11;Y1|V00,V10,V20)})
and (\ref{eq:crng-case0-I(V10;Y2|V00,V20,V22)}).

We have (\ref{eq:crng-fme15}) and (\ref{eq:crng-fme16}) as
\begin{align}
R_1 + R_2
&\leq
I(V_{ii};Y_i|V_{00},V_{10},V_{20})
+ I(V_{10},V_{20},V_{i'i'};Y_{i'}|V_{00})
\notag
\\
&\leq
I(Z_{ii};Y_i,Z_{10},Z_{20}|Z_{00})
+ \min\lrb{
 \begin{aligned}
  &I(Z_{i0},Z_{i'i'};Y_{i'},Z_{i'0}|Z_{00}),
  \\
  &I(Z_{10},Z_{20},Z_{i'i'};Y_{i'}|Z_{00})
 \end{aligned}
}
- I(Z_{i0};Z_{ii}|Z_{00}),
\end{align}
where the first inequality comes from (\ref{eq:crng-R1+R2-2})$'$
and the second inequality comes from
(\ref{eq:crng-case0-I(V11;Y1|V00,V10,V20)})
and (\ref{eq:crng-case0-I(V10,V20,V11;Y1|V00)})$'$.

We have (\ref{eq:crng-fme17}) from the fact that
\begin{align}
R_1 + R_2
&\leq
I(V_{20},V_{11};Y_1|V_{00},V_{10})
+ I(V_{10},V_{22};Y_2|V_{00},V_{20})
\notag
\\
&=
I(Z_{20},Z_{11};Y_1,Z_{10}|Z_{00})
+ I(Z_{10},Z_{22};Y_2,Z_{20}|Z_{00})
- I(Z_{10};Z_{11}|Z_{00})
- I(Z_{20};Z_{22}|Z_{00}),
\end{align}
where the first inequality comes from (\ref{eq:crng-R1+R2-1})
and the equality comes from 
(\ref{eq:crng-case0-I(V20,V11;Y1|V00,V10)}).

We have (\ref{eq:crng-fme18})--(\ref{eq:crng-fme21}) from the fact that
\begin{align}
2 R_0 + R_i
&\leq
2 H(Q) + I(V_{ii};Y_i|V_{00},V_{10},V_{20})
+ I(V_{i0};Y_{i'}|V_{00},V_{i'0},V_{i'i'})
\notag
\\
&\leq
\min\lrb{
 \begin{aligned}
  &I(Z_{00},Z_{ii};Y_i,Z_{10},Z_{20}),
  \\
  &I(Z_{00},Z_{i'0},Z_{ii};Y_i,Z_{i0})
 \end{aligned}
}
+ \min\lrb{
 \begin{aligned}
  &I(Z_{00},Z_{i0};Y_{i'},Z_{i'0},Z_{i'i'}),
  \\
  &I(Z_{00},Z_{10},Z_{20};Y_{i'},Z_{i'i'})
 \end{aligned}
}
- I(Z_{i0};Z_{ii}|Z_{00}),
\end{align}
where the first inequality comes from
(\ref{eq:crng-H(Q)})
and (\ref{eq:crng-case0}),
and the second inequality comes from
(\ref{eq:crng-case0-H(Q)+I(V11;Y1|V00,V10,V20)})
and (\ref{eq:crng-case0-H(Q)+I(V10;Y2|V00,V20,V22)}).

We have (\ref{eq:crng-fme22}) as
\begin{align}
2 R_i + R_{i'}
&\leq
I(V_{10},V_{20},V_{ii};Y_i|V_{00})
+ I(V_{ii};Y_i|V_{00},V_{10},V_{20})
+ I(V_{i0},V_{i'i'};Y_{i'}|V_{00},V_{i'0})
\notag
\\
&\leq
I(Z_{10},Z_{20},Z_{ii};Y_i|Z_{00})
+ I(Z_{ii};Y_i,Z_{10},Z_{20}|Z_{00}) 
+ I(Z_{i0},Z_{i'i'};Y_{i'},Z_{i'0}|Z_{00})
\notag
\\*
&\quad
- I(Z_{i0};Z_{ii}|Z_{00})
- I(Z_{i'0};Z_{i'i'}|Z_{00}),
\end{align}
where the first inequality comes from (\ref{eq:crng-2Ri+Ri'})
and the second inequality comes from
(\ref{eq:crng-case0-I(V11;Y1|V00,V10,V20)}),
(\ref{eq:crng-case0-I(V20,V11;Y1|V00,V10)})$'$,
and (\ref{eq:crng-case0-I(V10,V20,V11;Y1|V00)}).

We have (\ref{eq:crng-fme23}) and (\ref{eq:crng-fme24}) as
\begin{align}
R_0 + R_1 + R_2
&\leq
I(V_{ii};Y_i|V_{00},V_{10},V_{20})
+ I(V_{00},V_{10},V_{20},V_{i'i'};Y_{i'})
\notag
\\
&\leq
I(Z_{ii};Y_i,Z_{10},Z_{20}|Z_{00})
+ \min\lrb{
 \begin{aligned}
  &I(Z_{00},Z_{i0},Z_{i'i'};Y_{i'},Z_{i'0}),
  \\
  &I(Z_{00},Z_{10},Z_{20},Z_{i'i'};Y_{i'})
 \end{aligned}
}
- I(Z_{i0};Z_{ii}|Z_{00}),
\end{align}
where the first inequality comes from (\ref{eq:crng-R0+R1+R2})$'$,
and the second inequality comes from 
(\ref{eq:crng-case0-I(V11;Y1|V00,V10,V20)})
and (\ref{eq:crng-case0-I(V00,V10,V20,V11;Y1)})$'$.

We have (\ref{eq:crng-fme25}) and (\ref{eq:crng-fme26}) from the fact that
\begin{align}
R_0 + [R_1 + R_2]
&\leq
H(Q)
+ I(V_{ii};Y_i|V_{00},V_{10},V_{20})
+ I(V_{10},V_{20},V_{i'i'};Y_{i'}|V_{00})
\notag
\\
&\leq
I(Z_{00},Z_{ii};Y_i,Z_{10},Z_{20})
+\min\lrb{
 \begin{aligned}
  &I(Z_{i0},Z_{i'i'};Y_{i'},Z_{i'0}|Z_{00}),
  \\
  &I(Z_{10},Z_{20},Z_{i'i'};Y_{i'}|Z_{00})
 \end{aligned}
}
- I(Z_{i0};Z_{ii}|Z_{00}),
\end{align}
where the first inequality comes from (\ref{eq:crng-R1+R2-2})$'$
and (\ref{eq:crng-H(Q)}),
and the second inequality comes from
(\ref{eq:crng-case0-I(V10,V20,V11;Y1|V00)})$'$.

We have (\ref{eq:crng-fme27}) as
\begin{align}
R_0 + [R_1 + R_2]
&\leq
H(Q)
+ I(V_{i'0},V_{ii};Y_i|V_{00},V_{i0})
+ I(V_{i0},V_{i'i'};Y_{i'}|V_{00},V_{i'0})
\notag
\\
&\leq
I(Z_{00},Z_{i'0},Z_{ii};Y_i,Z_{i0})
+ I(Z_{i0},Z_{i'i'};Y_{i'},Z_{i'0}|Z_{00})
- I(Z_{i0};Z_{ii}|Z_{00})
- I(Z_{i'0};Z_{i'i'}|Z_{00}),
\end{align}
where the first inequality comes from (\ref{eq:crng-R1+R2-1})
and (\ref{eq:crng-H(Q)}),
and the second inequality comes from
(\ref{eq:crng-case0-I(V20,V11;Y1|V00,V10)})$'$
and (\ref{eq:crng-case0-H(Q)+I(V20,V11;Y1|V00,V10)}).

We have (\ref{eq:crng-fme28}) as
\begin{align}
2R_0 + R_1 + R_2
&\leq
R_0 + [R_0 + R_1 + R_2]
\notag
\\
&\leq
H(Q) + I(V_{ii};Y_i|V_{00},V_{10},V_{20})
+ I(V_{00},V_{10},V_{20},V_{i'i'};Y_{i'})
\notag
\\
&\leq
I(Z_{00},Z_{ii};Y_i,Z_{10},Z_{20})
+ I(Z_{00},Z_{i0},Z_{i'i'};Y_{i'},Z_{i'0})
- I(Z_{i0};Z_{ii}|Z_{00}),
\end{align}
where the first inequality comes from (\ref{eq:crng-R0+R1+R2})$'$
and (\ref{eq:crng-H(Q)}),
and the second inequality comes from
(\ref{eq:crng-case0-H(Q)+I(V11;Y1|V00,V10,V20)})
and (\ref{eq:crng-case0-I(V00,V10,V20,V11;Y1)})$'$.

We have (\ref{eq:crng-fme29}) and (\ref{eq:crng-fme30}) as
\begin{align}
2R_0 + R_1 + R_2
&=
R_0 + [R_0 + R_1 + R_2]
\notag
\\
&\leq
H(Q)
+ I(V_{ii};Y_i|V_{00},V_{10},V_{20})
+ I(V_{00},V_{10},V_{20},V_{i'i'};Y_{i'})
\notag
\\
&\leq
\min\lrb{
 \begin{aligned}
  &I(Z_{00},Z_{ii};Y_i,Z_{10},Z_{20})
  \\
  &I(Z_{00},Z_{i'0},Z_{ii};Y_i,Z_{i0})
 \end{aligned}
}
+ I(Z_{00},Z_{10},Z_{20},Z_{i'i'};Y_{i'})
- I(Z_{i0};Z_{ii}|Z_{00}),
\end{align}
where the first inequality comes from
(\ref{eq:crng-R0+R1+R2})$'$
and (\ref{eq:crng-H(Q)}),
and the second inequality comes from
(\ref{eq:crng-case0-H(Q)+I(V11;Y1|V00,V10,V20)})
and (\ref{eq:crng-case0-I(V00,V10,V20,V11;Y1)})$'$.

We have (\ref{eq:crng-fme31}) as
\begin{align}
2R_0 + [R_1 + R_2]
&\leq
2H(Q)
+ I(V_{20},V_{11};Y_1|V_{00},V_{10})
+ I(V_{10},V_{22};Y_2|V_{00},V_{20})
\notag
\\
&\leq
I(Z_{00},Z_{20},Z_{11};Y_1,Z_{10})
+ I(Z_{00},Z_{10},Z_{22};Y_2,Z_{20})
- I(Z_{10};Z_{11}|Z_{00}) - I(Z_{20};Z_{22}|Z_{00}),
\end{align}
where the first inequality comes from (\ref{eq:crng-R1+R2-1})
and (\ref{eq:crng-H(Q)}),
and the second inequality comes from
(\ref{eq:crng-case0-H(Q)+I(V20,V11;Y1|V00,V10)}).

We have (\ref{eq:crng-fme32}) as
\begin{align}
R_0 + [2 R_i + R_{i'}]
&\leq
H(Q)
+ I(V_{10},V_{20},V_{ii};Y_i|V_{00})
+ I(V_{ii};Y_i|V_{00},V_{10},V_{20})
+ I(V_{i0},V_{i'i'};Y_{i'}|V_{00},V_{i'0})
\notag
\\
&\leq
I(Z_{10},Z_{20},Z_{ii};Y_i|Z_{00})
+ I(Z_{ii};Y_i,Z_{10},Z_{20}|Z_{00})
+ I(Z_{00},Z_{i0},Z_{i'i'};Y_{i'},Z_{i'0})
\notag
\\*
&\quad
- I(Z_{i0};Z_{ii}|Z_{00})
- I(Z_{i'0};Z_{i'i'}|Z_{00}),
\end{align}
where the first inequality comes from (\ref{eq:crng-2Ri+Ri'}) 
and (\ref{eq:crng-H(Q)}),
and the second inequality comes from
(\ref{eq:crng-case0-I(V11;Y1|V00,V10,V20)}),
(\ref{eq:crng-case0-H(Q)+I(V20,V11;Y1|V00,V10)})$'$,
and (\ref{eq:crng-case0-I(V10,V20,V11;Y1|V00)}).

We have (\ref{eq:crng-fme33}) as
\begin{align}
R_0 + 2R_i + R_{i'}
&\leq
I(V_{00},V_{10},V_{20},V_{ii};Y_i)
+ I(V_{ii};Y_i|V_{00},V_{10},V_{20})
+ I(V_{i0},V_{i'i'};Y_{i'}|V_{00},V_{i'0})
\notag
\\
&=
I(Z_{00},Z_{10},Z_{20},Z_{ii};Y_i)
+ I(Z_{ii};Y_i,Z_{10},Z_{20}|Z_{00})
+ I(Z_{i0},Z_{i'i'};Y_{i'},Z_{i'0}|Z_{00}) 
\notag
\\*
&\quad
- I(Z_{i0};Z_{ii}|Z_{00})
- I(Z_{i'0};Z_{i'i'}|Z_{00}),
\end{align}
where the first inequality comes from (\ref{eq:crng-R0+2Ri+Ri'}),
and the second inequality comes from
(\ref{eq:crng-case0-I(V11;Y1|V00,V10,V20)}),
(\ref{eq:crng-case0-I(V20,V11;Y1|V00,V10)})$'$,
and (\ref{eq:crng-case0-I(V00,V10,V20,V11;Y1)}).

We have (\ref{eq:crng-fme34}) as
\begin{align}
2 R_0 + 2R_i + R_{i'}
&=
R_0 + [R_0 + 2R_i + R_{i'}]
\notag
\\
&\leq
H(Q)
+ I(V_{00},V_{10},V_{20},V_{ii};Y_i)
+ I(V_{ii};Y_i|V_{00},V_{10},V_{20})
+ I(V_{i0},V_{i'i'};Y_{i'}|V_{00},V_{i'0})
\notag
\\
&\leq
I(Z_{00},Z_{10},Z_{20},Z_{ii};Y_i)
+ I(Z_{ii};Y_i,Z_{10},Z_{20}|Z_{00})
+ I(Z_{00},Z_{i0},Z_{i'i'};Y_{i'},Z_{i'0})
\notag
\\*
&\quad
- I(Z_{i0};Z_{ii}|Z_{00})
- I(Z_{i'0};Z_{i'i'}|Z_{00}),
\end{align}
where the first inequality comes from (\ref{eq:crng-R0+2Ri+Ri'}) 
and (\ref{eq:crng-H(Q)}),
and the second inequality comes from
(\ref{eq:crng-case0-I(V11;Y1|V00,V10,V20)}),
(\ref{eq:crng-case0-H(Q)+I(V20,V11;Y1|V00,V10)})$'$,
and (\ref{eq:crng-case0-I(V00,V10,V20,V11;Y1)}).

We have (\ref{eq:crng-fme35}) as
\begin{align}
2 R_0 + 2R_i + R_{i'}
&=
R_0 + [R_0 + 2R_i + R_{i'}]
\notag
\\
&\leq
H(Q)
+ I(V_{00},V_{10},V_{20},V_{ii};Y_i)
+ I(V_{ii};Y_i|V_{00},V_{10},V_{20})
+ I(V_{i0},V_{i'i'};Y_{i'}|V_{00},V_{i'0})
\notag
\\
&\leq
I(Z_{00},Z_{10},Z_{20},Z_{ii};Y_i)
+ I(Z_{00},Z_{ii};Y_i,Z_{10},Z_{20})
+ I(Z_{i0},Z_{i'i'};Y_{i'},Z_{i'0}|Z_{00})
\notag
\\*
&\quad
- I(Z_{i0};Z_{ii}|Z_{00})
- I(Z_{i'0};Z_{i'i'}|Z_{00}),
\end{align}
where the first inequality comes from (\ref{eq:crng-R0+2Ri+Ri'}) 
and (\ref{eq:crng-H(Q)}),
and the second inequality comes from
(\ref{eq:crng-case0-H(Q)+I(V11;Y1|V00,V10,V20)}),
(\ref{eq:crng-case0-I(V20,V11;Y1|V00,V10)})$'$,
and (\ref{eq:crng-case0-I(V00,V10,V20,V11;Y1)}).

We have (\ref{eq:crng-fme36}) as
\begin{align}
3 R_0 + 2 R_i + R_{i'}
&=
2R_0+ [R_0 + 2 R_i + R_{i'}]
\notag
\\
&\leq
2H(Q)
+ I(V_{00},V_{10},V_{20},V_{ii};Y_i)
+ I(V_{ii};Y_i|V_{00},V_{10},V_{20})
+ I(V_{i0},V_{i'i'};Y_{i'}|V_{00},V_{i'0})
\notag
\\
&\leq
I(Z_{00},Z_{10},Z_{20},Z_{ii};Y_i)
+ I(Z_{00},Z_{ii};Y_i,Z_{10},Z_{20})
+ I(Z_{00},Z_{i0},Z_{i'i'};Y_{i'},Z_{i'0})
\notag
\\*
&\quad
- I(Z_{i0};Z_{ii}|Z_{00})
- I(Z_{i'0};Z_{i'i'}|Z_{00}),
\end{align}
where the first inequality comes from (\ref{eq:crng-R0+2Ri+Ri'}) 
and (\ref{eq:crng-H(Q)}),
and the second inequality comes from
(\ref{eq:crng-case0-H(Q)+I(V11;Y1|V00,V10,V20)}),
(\ref{eq:crng-case0-H(Q)+I(V20,V11;Y1|V00,V10)})$'$,
and (\ref{eq:crng-case0-I(V00,V10,V20,V11;Y1)}).

\subsection{Details of Proof for Case 1}
\label{sec:proof-crng-crng-case1-appendix}

In the following, we use (\ref{eq:Qindependence}),
(\ref{eq:def-Z00}), (\ref{eq:def-Zii}),
(\ref{eq:def-Z10-case1}), (\ref{eq:def-Z20-case1}),
and Lemma \ref{lem:Q} without notice.

We have (\ref{eq:crng-case1-I(Z20;Z22|Z00)}) as
\begin{align}
I(Z_{20};Z_{22}|Z_{00})
&=
I(Q;Q,V_{20},V_{22}|Q,V_{00})
\notag
\\
&=
0.
\end{align}

We have (\ref{eq:crng-case1-R1-1})
from (\ref{eq:crng-R1+R2-1}) and (\ref{eq:crng-case1-R2lower})
as
\begin{align}
R_1
&=
R_1 + R_2 - R_2
\notag
\\
&\leq
I(V_{20},V_{11};Y_1|V_{00},V_{10})
+ I(V_{10},V_{22};Y_2|V_{00},V_{20})
- I(V_{22};Y_2|V_{00},V_{10},V_{20})
- I(V_{20};Y_1|V_{00},V_{10},V_{11})
\notag
\\
&=
I(V_{11};Y_1|V_{00},V_{10})
+ I(V_{10};Y_2|V_{00},V_{20}).
\end{align}

We have
(\ref{eq:crng-case1-R1-2})
from (\ref{eq:crng-R1+R2-2}) and (\ref{eq:crng-case1-R2lower})
as
\begin{align}
R_1
&=
R_1 + R_2 - R_2
\notag
\\
&\leq
I(V_{10},V_{20},V_{11};Y_1|V_{00})
+ I(V_{22};Y_2|V_{00},V_{10},V_{20})
- I(V_{22};Y_2|V_{00},V_{10},V_{20})
- I(V_{20};Y_1|V_{00},V_{10},V_{11})
\notag
\\
&=
I(V_{10},V_{11};Y_1|V_{00}).
\end{align}

We have
(\ref{eq:crng-case1-R0+R1})
from (\ref{eq:crng-R0+R1+R2}) and (\ref{eq:crng-case1-R2lower})
as
\begin{align}
R_0 + R_1
&=
R_0 + R_1 + R_2 - R_2
\notag
\\
&\leq
I(V_{00},V_{10},V_{20},V_{11};Y_1)
+ I(V_{22};Y_2|V_{00},V_{10},V_{20})
- I(V_{22};Y_2|V_{00},V_{10},V_{20})
- I(V_{20};Y_1|V_{00},V_{10},V_{11})
\notag
\\
&=
I(V_{00},V_{10},V_{11};Y_1).
\end{align}

We have
(\ref{eq:crng-case1-R1+R2})
from (\ref{eq:crng-2Ri+Ri'}) and (\ref{eq:crng-case1-R2lower})
as
\begin{align}
R_1 + R_2
&=
R_1 + 2 R_2 - R_2
\notag
\\
&\leq
I(V_{10},V_{20},V_{22};Y_2|V_{00})
+ I(V_{22};Y_2|V_{00},V_{10},V_{20})
+ I(V_{20},V_{11};Y_1|V_{00},V_{10})
\notag
\\*
&\quad
- I(V_{22};Y_2|V_{00},V_{10},V_{20})
- I(V_{20};Y_1|V_{00},V_{10},V_{11})
\notag
\\
&=
I(V_{11};Y_1|V_{00},V_{10})
+ I(V_{10},V_{20},V_{22};Y_2|V_{00}).
\end{align}

We have (\ref{eq:crng-case1-R0+R1+R2})
from (\ref{eq:crng-R0+2Ri+Ri'}) and (\ref{eq:crng-case1-R2lower})
as
\begin{align}
R_0 + R_1 + R_2
&=
R_0 + R_1 + 2 R_2 - R_2
\notag
\\
&\leq
I(V_{00},V_{10},V_{20},V_{22};Y_2)
+ I(V_{22};Y_2|V_{00},V_{10},V_{20})
+ I(V_{20},V_{11};Y_1|V_{00},V_{10})
\notag
\\*
&\quad
- I(V_{22};Y_2|V_{00},V_{10},V_{20})
- I(V_{20};Y_1|V_{00},V_{10},V_{11})
\notag
\\
&=
I(V_{11};Y_1|V_{00},V_{10})
+ I(V_{00},V_{10},V_{20},V_{22};Y_2).
\end{align}

We have (\ref{eq:crng-case1-H(Q)}) as
\begin{align}
H(Q)
&=
I(Q,Q)
\notag
\\
&\leq
\min\lrb{
 \begin{aligned}
  &I(Q,V_{00};Y_1,Q,V_{10},V_{11}),
  \\
  &I(Q,V_{00},V_{10};Y_1,Q,V_{10},V_{11})
 \end{aligned}
}
\notag
\\
&=
\min\lrb{
 \begin{aligned}
  &I(Z_{00},Z_{20};Y_1,Z_{10},Z_{11}),
  \\
  &I(Z_{00},Z_{10},Z_{20};Y_1,Z_{11})
 \end{aligned}
}.
\end{align}

We have (\ref{eq:crng-case1-I(V11;Y1|V00,V10)}) as
\begin{align}
I(V_{11};Y_1|V_{00},V_{10})
&=
I(Q,V_{10},V_{11};Y_1|Q,V_{00},V_{10})
\notag
\\
&=
I(Z_{11};Y_1|Z_{00},Z_{10})
\notag
\\
&=
I(Z_{11};Y_1,Z_{10}|Z_{00})
- I(Z_{10};Z_{11}|Z_{00})
\notag
\\
&\leq
\min\lrb{
 \begin{aligned}
  &I(Z_{11};Y_1,Z_{10},Z_{20}|Z_{00}),
  \\
  &I(Z_{20},Z_{11};Y_1,Z_{10}|Z_{00})
 \end{aligned}
}
- I(Z_{10};Z_{11}|Z_{00}).
\end{align}

We have (\ref{eq:crng-case1-H(Q)+I(V11;Y1|V00,V10)}) as
\begin{align}
H(Q) + I(V_{11};Y_1|V_{00},V_{10})
&=
H(Q) + I(V_{10},V_{11};Y_1|V_{00},V_{10})
\notag
\\
&=
H(Q)
+ I(V_{10},V_{11};Y_1,V_{10}|V_{00})
- I(V_{10},V_{11};V_{10}|V_{00})
\notag
\\
&\leq
H(Q)
+ I(V_{00},V_{10},V_{11};Y_1,V_{10})
- I(V_{10},V_{11};V_{10}|V_{00})
\notag
\\
&=
I(Q,V_{00},V_{10},V_{11};Y_1,Q,V_{10})
- I(Q,V_{10};Q,V_{10},V_{11}|Q,V_{00})
\notag
\\
&=
I(Z_{00},Z_{11};Y_1,Z_{10})
- I(Z_{10};Z_{11}|Z_{00})
\notag
\\
&\leq
\min\lrb{
 \begin{aligned}
  &I(Z_{00},Z_{11};Y_1,Z_{10},Z_{20}),
  \\
  &I(Z_{00},Z_{20},Z_{11};Y_1,Z_{10})
 \end{aligned}
}
- I(Z_{10};Z_{11}|Z_{00}).
\end{align}

We have (\ref{eq:crng-case1-I(V10,V11;Y1|V00)}) as
\begin{align}
I(V_{10},V_{11};Y_1|V_{00})
&=
I(Q,V_{10},V_{11};Y_1|Q,V_{00})
\notag
\\
&=
I(Z_{11};Y_1|Z_{00})
\notag
\\
&\leq
\min\lrb{
 \begin{aligned}
  &I(Z_{11};Y_1,Z_{10},Z_{20}|Z_{00}),
  \\
  &I(Z_{10},Z_{11};Y_1,Z_{20}|Z_{00}),
  \\
  &I(Z_{20},Z_{11};Y_1,Z_{10}|Z_{00}),
  \\
  &I(Z_{10},Z_{20},Z_{11};Y_1|Z_{00})
 \end{aligned}
}.
\end{align}

We have (\ref{eq:crng-case1-H(Q)+I(V10,V11;Y1|V00)}) as
\begin{align}
H(Q) + I(V_{10},V_{11};Y_1|V_{00})
&\leq
H(Q) + I(V_{00},V_{10},V_{11};Y_1,V_{10})
\notag
\\
&=
I(Q,V_{00},V_{10},V_{11};Y_1,Q,V_{10})
\notag
\\
&=
I(Z_{00},Z_{20},Z_{11};Y_1,Z_{10}).
\end{align}

We have (\ref{eq:crng-case1-I(V00,V10,V11;Y1)}) as
\begin{align}
I(V_{00},V_{10},V_{11};Y_1)
&=
I(Q,V_{00},V_{10},V_{11};Y_1)
\notag
\\
&=
I(Z_{00},Z_{11};Y_1)
\notag
\\
&\leq
\min\lrb{
 \begin{aligned}
  &I(Z_{00},Z_{11};Y_1,Z_{10},Z_{20}),
  \\
  &I(Z_{00},Z_{10},Z_{11};Y_1,Z_{20}),
  \\
  &I(Z_{00},Z_{20},Z_{11};Y_1,Z_{10}),
  \\
  &I(Z_{00},Z_{10},Z_{20},Z_{11};Y_1)
 \end{aligned}
}.
\end{align}

We have (\ref{eq:crng-case1-I(V00,V10,V20,V11;Y1)}) as
\begin{align}
I(V_{00},V_{10},V_{20},V_{11};Y_1)
&=
I(Q,V_{00},V_{10},V_{20},V_{11};Y_1)
\notag
\\
&=
I(Z_{00},Z_{10},Z_{20},Z_{11};Y_1).
\end{align}

We have (\ref{eq:crng-case1-I(V10;Y2|V00,V20)}) as
\begin{align}
I(V_{10};Y_2|V_{00},V_{20})
&\leq
I(V_{10};Y_2,V_{20},V_{22}|V_{00})
\notag
\\
&=
I(Q,V_{10};Y_2,Q,V_{20},V_{22}|Q,V_{00})
\notag
\\
&=
I(Z_{10};Y_2,Z_{20},Z_{22}|Z_{00}).
\end{align}

We have (\ref{eq:crng-case1-H(Q)+I(V10;Y2|V00,V20)}) as
\begin{align}
H(Q)
+ I(V_{10};Y_2|V_{00},V_{20})
&\leq
H(Q)
+ I(V_{00},V_{10};Y_2,V_{20},V_{22})
\notag
\\
&=
I(Q,V_{00},V_{10};Y_2,Q,V_{20},V_{22})
\notag
\\
&=
\min\lrb{
 \begin{aligned}
  &I(Z_{00},Z_{10};Y_2,Z_{20},Z_{22}),
  \\
  &I(Z_{00},Z_{10},Z_{20},;Y_2,Z_{22})
 \end{aligned}
}.
\end{align}

We have (\ref{eq:crng-case1-I(V20,V22;Y2|V00,V10)}) as
\begin{align}
I(V_{20},V_{22};Y_2|V_{00},V_{10})
&=
I(Q,V_{20},V_{22};Y_2|Q,V_{00},V_{10})
\notag
\\
&=
I(Z_{22};Y_2|Z_{00},Z_{10})
\notag
\\
&\leq
\min\lrb{
 \begin{aligned}
  &I(Z_{22};Y_2,Z_{10},Z_{20}|Z_{00}),
  \\
  &I(Z_{20},Z_{22};Y_2,Z_{10}|Z_{00})
 \end{aligned}
}.
\end{align}

We have (\ref{eq:crng-case1-H(Q)+I(V20,V22;Y2|V00,V10)}) as
\begin{align}
H(Q) + I(V_{20},V_{22};Y_2|V_{00},V_{10})
&\leq
H(Q)
+ \min\lrb{
 \begin{aligned}
  &I(V_{00},V_{20},V_{22};Y_2,V_{10}),
  \\
  &I(V_{00},V_{10},V_{20},V_{22};Y_2)
 \end{aligned}
}
\notag
\\
&\leq
\min\lrb{
 \begin{aligned}
  &I(Q,V_{00},V_{20},V_{22};Y_2,Q,V_{10}),
  \\
  &I(Q,V_{00},V_{10},V_{20},V_{22};Y_2,Q)
 \end{aligned}
}
\notag
\\
&=
\min\lrb{
 \begin{aligned}
  &I(Z_{00},Z_{22};Y_2,Z_{10},Z_{20}),
  \\
  &I(Z_{00},Z_{20},Z_{22};Y_2,Z_{10}),
  \\
  &I(Z_{00},Z_{10},Z_{22};Y_2,Z_{20})
 \end{aligned}
}.
\end{align}

We have (\ref{eq:crng-case1-I(V10,V20,V22;Y2|V00)}) as
\begin{align}
I(V_{10},V_{20},V_{22};Y_2|V_{00})
&=
I(Q,V_{10},V_{20},V_{22};Y_2|Q,V_{00})
\notag
\\
&=
I(Z_{10},Z_{22};Y_2|Z_{00})
\notag
\\
&\leq
\min\lrb{
 \begin{aligned}
  &I(Z_{10},Z_{22};Y_2,Z_{20}|Z_{00}),
  \\
  &I(Z_{10},Z_{20},Z_{22};Y_2|Z_{00})
 \end{aligned}
}.
\end{align}

We have (\ref{eq:crng-case1-H(Q)+I(V10,V20,V22;Y2|V00)}) as
\begin{align}
H(Q)+ I(V_{10},V_{20},V_{22};Y_2|V_{00})
&\leq
H(Q) + I(V_{00},V_{10},V_{20},V_{22};Y_2)
\notag
\\
&\leq
I(Q,V_{00},V_{10},V_{20},V_{22};Y_2,Q)
\notag
\\
&=
I(Z_{00},Z_{10},Z_{22};Y_2,Z_{20}).
\end{align}

We have (\ref{eq:crng-case1-I(V00,V10,V20,V22;Y2)}) as
\begin{align}
I(V_{00},V_{10},V_{20},V_{22};Y_2)
&=
I(Q,V_{00},V_{10},V_{20},V_{22};Y_2)
\notag
\\
&=
I(Z_{00},Z_{10},Z_{22};Y_2)
\notag
\\
&\leq
\min\lrb{
 \begin{aligned}
  &I(Z_{00},Z_{10},Z_{22};Y_2,Z_{20})
  \\
  &I(Z_{00},Z_{10},Z_{20},Z_{22};Y_2)
 \end{aligned}
}.
\end{align}

We have (\ref{eq:crng-fme4}) and (\ref{eq:crng-fme5}) as
\begin{align}
R_1
&\leq
I(V_{10},V_{11};Y_1|V_{00})
\notag
\\
&\leq
\min\lrb{
 \begin{aligned}
  &I(Z_{11};Y_1,Z_{10},Z_{20}|Z_{00})
  \\
  &I(Z_{10},Z_{11};Y_1,Z_{20}|Z_{00})
 \end{aligned}
}
\label{eq:crng-case1-4-5-1}
\\
R_2
&\leq
I(V_{20},V_{22};Y_2|V_{00},V_{10})
\notag
\\
&\leq
\min\lrb{
 \begin{aligned}
  &I(Z_{22};Y_2,Z_{10},Z_{20}|Z_{00})
  \\
  &I(Z_{20},Z_{22};Y_2,Z_{10}|Z_{00})
 \end{aligned}
},
\label{eq:crng-case1-4-5-2}
\end{align}
where
the first inequality of (\ref{eq:crng-case1-4-5-1})
comes from (\ref{eq:crng-case1-R1-2}),
the second inequality of (\ref{eq:crng-case1-4-5-1})
comes from (\ref{eq:crng-case1-I(V10,V11;Y1|V00)}),
the first inequality of (\ref{eq:crng-case1-4-5-2})
comes from (\ref{eq:crng-Ri})$'$,
and the second inequality of (\ref{eq:crng-case1-4-5-2})
comes from (\ref{eq:crng-case1-I(V20,V22;Y2|V00,V10)}),

We have (\ref{eq:crng-fme6}) as
\begin{align}
R_1
&\leq
I(V_{11};Y_1|V_{00},V_{10})
+ I(V_{10};Y_2|V_{00},V_{20})
\notag
\\
&\leq
I(Z_{11};Y_1,Z_{10},Z_{20}|Z_{00})
+ I(Z_{10};Y_2,Z_{20},Z_{22}|Z_{00})
- I(Z_{10};Z_{11}|Z_{00})
\label{eq:crng-case1-6-1}
\\
R_2
&\leq
I(V_{20},V_{22};Y_2|V_{00},V_{10})
\notag
\\
&\leq
I(Z_{22};Y_2,Z_{10},Z_{20}|Z_{00})
+ I(Z_{20};Y_1,Z_{10},Z_{11}|Z_{00})
- I(Z_{20};Z_{22}|Z_{00}),
\label{eq:crng-case1-6-2}
\end{align}
where
the first inequality of (\ref{eq:crng-case1-6-1})
comes from (\ref{eq:crng-case1-R1-1}),
the second inequality of (\ref{eq:crng-case1-6-1})
comes from (\ref{eq:crng-case1-I(V11;Y1|V00,V10)})
and (\ref{eq:crng-case1-I(V10;Y2|V00,V20)}),
the first inequality of (\ref{eq:crng-case1-6-2})
comes from (\ref{eq:crng-Ri})$'$,
and the second inequality of (\ref{eq:crng-case1-6-2})
comes from (\ref{eq:crng-case1-I(Z20;Z22|Z00)})
and (\ref{eq:crng-case1-I(V20,V22;Y2|V00,V10)}).

We have (\ref{eq:crng-fme7})--(\ref{eq:crng-fme9}) as
\begin{align}
R_0 + R_1
&\leq
I(V_{00},V_{10},V_{11};Y_1)
\notag
\\
&\leq
\min\lrb{
 \begin{aligned}
  &I(Z_{00},Z_{11};Y_1,Z_{10},Z_{20}),
  \\
  &I(Z_{00},Z_{10},Z_{11};Y_1,Z_{20}),
  \\
  &I(Z_{00},Z_{20},Z_{11};Y_1,Z_{10})
 \end{aligned}
}
\label{eq:crng-case1-7-9-1}
\\
R_0 + R_2
&\leq
H(Q)
+ I(V_{20},V_{22};Y_2|V_{00},V_{10})
\notag
\\
&\leq
\min\lrb{
 \begin{aligned}
  &I(Z_{00},Z_{22};Y_2,Z_{10},Z_{20}),
  \\
  &I(Z_{00},Z_{20},Z_{22};Y_2,Z_{10}),
  \\
  &I(Z_{00},Z_{10},Z_{22};Y_2,Z_{20})
 \end{aligned}
},
\label{eq:crng-case1-7-9-2}
\end{align}
where
the first inequality of (\ref{eq:crng-case1-7-9-1})
comes from (\ref{eq:crng-case1-R0+R1}),
the second inequality of (\ref{eq:crng-case1-7-9-1})
comes from (\ref{eq:crng-case1-I(V00,V10,V11;Y1)}),
the first inequality of (\ref{eq:crng-case1-7-9-2})
comes from (\ref{eq:crng-Ri})$'$ and (\ref{eq:crng-H(Q)}),
and the second inequality of (\ref{eq:crng-case1-7-9-2})
comes from (\ref{eq:crng-case1-H(Q)+I(V20,V22;Y2|V00,V10)}).

We have (\ref{eq:crng-fme10}) as
\begin{align}
R_0 + R_i
&\leq
I(V_{00},V_{10},V_{20},V_{ii};Y_i)
\notag
\\
&\leq
I(Z_{00},Z_{10},Z_{20},Z_{ii};Y_i),
\end{align}
where
the first inequality comes from (\ref{eq:crng-R0+Ri}),
and the second inequality comes from
(\ref{eq:crng-case1-I(V00,V10,V20,V11;Y1)})
and (\ref{eq:crng-case1-I(V00,V10,V20,V22;Y2)}).

We have (\ref{eq:crng-fme11}) and (\ref{eq:crng-fme12}) as
\begin{align}
R_0 + R_1
&\leq
H(Q) + I(V_{11};Y_1|V_{00},V_{10}) + I(V_{10};Y_2|V_{00},V_{20})
\notag
\\
&\leq
I(Z_{11};Y_1,Z_{10},Z_{20}|Z_{00})
+ \min\lrb{
 \begin{aligned}
  &I(Z_{00},Z_{10};Y_2,Z_{20},Z_{22}),
  \\
  &I(Z_{00},Z_{10},Z_{20};Y_2,Z_{22})
 \end{aligned}
}
-I(Z_{10};Z_{11}|Z_{00})
\label{eq:crng-case1-11-12-1}
\\
R_0 + R_2
&\leq
H(Q) + I(V_{20},V_{22};Y_2|V_{00},V_{10})
\notag
\\
&\leq
I(Z_{22};Y_2,Z_{10},Z_{20}|Z_{00})
+ \min\lrb{
 \begin{aligned}
  &I(Z_{00},Z_{20};Y_1,Z_{10},Z_{11}),
  \\
  &I(Z_{00},Z_{10},Z_{20};Y_1,Z_{11})
 \end{aligned}
}
-I(Z_{20};Z_{22}|Z_{00}),
\label{eq:crng-case1-11-12-2}
\end{align}
where
the first inequality of (\ref{eq:crng-case1-11-12-1})
comes from 
(\ref{eq:crng-H(Q)}) and (\ref{eq:crng-case1-R1-1}),
the second inequality of (\ref{eq:crng-case1-11-12-1})
comes from (\ref{eq:crng-case1-I(V11;Y1|V00,V10)})
and (\ref{eq:crng-case1-H(Q)+I(V10;Y2|V00,V20)}),
the first inequality of (\ref{eq:crng-case1-11-12-2})
comes from (\ref{eq:crng-Ri})$'$ and (\ref{eq:crng-H(Q)}),
and the second inequality of (\ref{eq:crng-case1-11-12-2})
comes from (\ref{eq:crng-case1-I(Z20;Z22|Z00)}),
(\ref{eq:crng-case1-H(Q)}),
and (\ref{eq:crng-case1-I(V20,V22;Y2|V00,V10)}).

We have (\ref{eq:crng-fme13}) and (\ref{eq:crng-fme14}) as
\begin{align}
R_0 + R_1
&\leq
H(Q) + I(V_{11};Y_1|V_{00},V_{10}) + I(V_{10};Y_2|V_{00},V_{20})
\notag
\\
&\leq
\min\lrb{
 \begin{aligned}
  &I(Z_{00},Z_{11};Y_1,Z_{10},Z_{20}),
  \\
  &I(Z_{00},Z_{20},Z_{11};Y_1,Z_{10})
 \end{aligned}
}
+ I(Z_{10};Y_2,Z_{20},Z_{22}|Z_{00})
-I(Z_{10};Z_{11}|Z_{00})
\label{eq:crng-case1-13-14-1}
\\
R_0 + R_2
&\leq
H(Q) + I(V_{20},V_{22};Y_2|V_{00},V_{10})
\notag
\\
&\leq
\min\lrb{
 \begin{aligned}
  &I(Z_{00},Z_{22};Y_2,Z_{10},Z_{20})
  \\
  &I(Z_{00},Z_{10},Z_{22};Y_2,Z_{20})
 \end{aligned}
}
+ I(Z_{20};Y_1,Z_{10},Z_{11}|Z_{00})
- I(Z_{20};Z_{22}|Z_{00}),
\label{eq:crng-case1-13-14-2}
\end{align}
where
the first inequality of (\ref{eq:crng-case1-13-14-1})
comes from (\ref{eq:crng-H(Q)})
and (\ref{eq:crng-case1-R1-1}),
the second inequality of (\ref{eq:crng-case1-13-14-1})
comes from (\ref{eq:crng-case1-H(Q)+I(V11;Y1|V00,V10)})
and (\ref{eq:crng-case1-I(V10;Y2|V00,V20)}),
the first inequality of (\ref{eq:crng-case1-13-14-2})
comes from (\ref{eq:crng-Ri})$'$ and (\ref{eq:crng-H(Q)}),
and the second inequality of (\ref{eq:crng-case1-13-14-2})
comes from (\ref{eq:crng-case1-I(Z20;Z22|Z00)}),
and (\ref{eq:crng-case1-H(Q)+I(V20,V22;Y2|V00,V10)}).

We have (\ref{eq:crng-fme15}) and (\ref{eq:crng-fme16}) as
\begin{align}
R_1 + R_2
&\leq
I(V_{11};Y_1|V_{00},V_{10})
+ I(V_{10},V_{20},V_{22};Y_2|V_{00})
\notag
\\
&\leq
I(Z_{11};Y_1,Z_{10},Z_{20}|Z_{00})
+ \min\lrb{
 \begin{aligned}
  &I(Z_{10},Z_{22};Y_2,Z_{20}|Z_{00}),
  \\
  &I(Z_{10},Z_{20},Z_{22};Y_2|Z_{00})
 \end{aligned}
}
- I(Z_{10};Z_{11}|Z_{00})
\label{eq:crng-case1-15-16-1}
\\
R_1 + R_2
&\leq
I(V_{10},V_{11};Y_1|V_{00})
+ I(V_{20},V_{22};Y_2|V_{00},V_{10})
\notag
\\
&\leq
I(Z_{22};Y_2,Z_{10},Z_{20}|Z_{00})
+ \min\lrb{
 \begin{aligned}
  &I(Z_{20},Z_{11};Y_1,Z_{10}|Z_{00}),
  \\
  &I(Z_{10},Z_{20},Z_{11};Y_1|Z_{00})
 \end{aligned}
}
- I(Z_{20};Z_{22}|Z_{00}),
\label{eq:crng-case1-15-16-2}
\end{align}
where
the first inequality of (\ref{eq:crng-case1-15-16-1})
comes from (\ref{eq:crng-case1-R1+R2}),
the second inequality of (\ref{eq:crng-case1-15-16-1})
comes from (\ref{eq:crng-case1-I(V11;Y1|V00,V10)})
and (\ref{eq:crng-case1-I(V10,V20,V22;Y2|V00)}),
the first inequality of (\ref{eq:crng-case1-15-16-2})
comes from (\ref{eq:crng-Ri})$'$ and (\ref{eq:crng-case1-R1-2}),
and the second inequality of (\ref{eq:crng-case1-15-16-2})
comes from (\ref{eq:crng-case1-I(Z20;Z22|Z00)}),
(\ref{eq:crng-case1-I(V10,V11;Y1|V00)}),
and (\ref{eq:crng-case1-I(V20,V22;Y2|V00,V10)}).

We have (\ref{eq:crng-fme17}) as
\begin{align}
R_1 + R_2
&\leq
I(V_{11};Y_1|V_{00},V_{10})
+ I(V_{10},V_{20},V_{22};Y_2|V_{00})
\notag
\\
&\leq
I(Z_{20},Z_{11};Y_1,Z_{10}|Z_{00})
+ I(Z_{10},Z_{22};Y_2,Z_{20}|Z_{00})
- I(Z_{10};Z_{11}|Z_{00})
- I(Z_{20};Z_{22}|Z_{00}),
\end{align}
where the first equality comes from (\ref{eq:crng-case1-R1+R2})
and the second inequality comes from
(\ref{eq:crng-case1-I(Z20;Z22|Z00)}),
(\ref{eq:crng-case1-I(V11;Y1|V00,V10)}),
and (\ref{eq:crng-case1-I(V10,V20,V22;Y2|V00)}).

We have (\ref{eq:crng-fme18})--(\ref{eq:crng-fme21}) from the fact that
\begin{align}
2 R_0 + R_1
&\leq
2 H(Q) + I(V_{11};Y_1|V_{00},V_{10}) + I(V_{10};Y_2|V_{00},V_{20})
\notag
\\
&\leq
\min\lrb{
 \begin{aligned}
  &I(Z_{00},Z_{11};Y_1,Z_{10},Z_{20}),
  \\
  &I(Z_{00},Z_{20},Z_{11};Y_1,Z_{10})
 \end{aligned}
}
+ \min\lrb{
 \begin{aligned}
  &I(Z_{00},Z_{10};Y_2,Z_{20},Z_{22}),
  \\
  &I(Z_{00},Z_{10},Z_{20};Y_2,Z_{22})
 \end{aligned}
}
- I(Z_{10};Z_{11}|Z_{00})
\label{eq:crng-case1-18-21-1}
\\
2 R_0 + R_2
&\leq
2 H(Q) + I(V_{20},V_{22};Y_2|V_{00},V_{10})
\notag
\\
&\leq
\min\lrb{
 \begin{aligned}
  &I(Z_{00},Z_{22};Y_2,Z_{10},Z_{20})
  \\
  &I(Z_{00},Z_{10},Z_{22};Y_2,Z_{20})
 \end{aligned}
}
+
\min\lrb{
 \begin{aligned}
  &I(Z_{00},Z_{20};Y_1,Z_{10},Z_{11}),
  \\
  &I(Z_{00},Z_{10},Z_{20};Y_1,Z_{11})
 \end{aligned}
}
- I(Z_{20};Z_{22}|Z_{00}),
\label{eq:crng-case1-18-21-2}
\end{align}
where
the first inequality of (\ref{eq:crng-case1-18-21-1})
comes from (\ref{eq:crng-H(Q)})
and (\ref{eq:crng-case1-R0+R1}),
the second inequality of (\ref{eq:crng-case1-18-21-1})
comes from (\ref{eq:crng-case1-H(Q)+I(V11;Y1|V00,V10)})
and (\ref{eq:crng-case1-H(Q)+I(V10;Y2|V00,V20)}),
the first inequality of (\ref{eq:crng-case1-18-21-2})
comes from (\ref{eq:crng-Ri})$'$ and (\ref{eq:crng-H(Q)}),
and the second inequality of (\ref{eq:crng-case1-18-21-2})
comes from (\ref{eq:crng-case1-I(Z20;Z22|Z00)}),
(\ref{eq:crng-case1-H(Q)}),
and (\ref{eq:crng-case1-H(Q)+I(V20,V22;Y2|V00,V10)}).

We have (\ref{eq:crng-fme22}) as
\begin{align}
2 R_1 + R_2
&=
R_1 + [R_1 + R_2]
\notag
\\
&\leq
I(V_{10},V_{11};Y_1|V_{00})
+ I(V_{11};Y_1|V_{00},V_{10})
+ I(V_{10},V_{20},V_{22};Y_2|V_{00})
\notag
\\
&\leq
I(Z_{10},Z_{20},Z_{11};Y_1|Z_{00})
+ I(Z_{11};Y_1,Z_{10},Z_{20}|Z_{00}) 
+ I(Z_{10},Z_{22};Y_2,Z_{20}|Z_{00})
\notag
\\*
&\quad
- I(Z_{10};Z_{11}|Z_{00})
- I(Z_{20};Z_{22}|Z_{00})
\label{eq:crng-case1-22-1}
\\
R_1 + 2 R_2
&=
[R_1 + R_2] + R_2
\notag
\\
&\leq 
I(V_{11};Y_1|V_{00},V_{10})
+ I(V_{10},V_{20},V_{22};Y_2|V_{00})
+ I(V_{20},V_{22};Y_2|V_{00},V_{10})
\notag
\\
&\leq
I(Z_{10},Z_{20},Z_{22};Y_2|Z_{00})
+ I(Z_{22};Y_2,Z_{10},Z_{20}|Z_{00})
+ I(Z_{20},Z_{11};Y_1,Z_{10}|Z_{00})
\notag
\\*
&\quad
- I(Z_{10};Z_{11}|Z_{00}) - I(Z_{20};Z_{22}|Z_{00}),
\label{eq:crng-case1-22-2}
\end{align}
where
the first inequality of (\ref{eq:crng-case1-22-1})
comes from (\ref{eq:crng-case1-R1-2})
and (\ref{eq:crng-case1-R1+R2}),
the second inequality of (\ref{eq:crng-case1-22-1})
comes from (\ref{eq:crng-case1-I(Z20;Z22|Z00)}),
(\ref{eq:crng-case1-I(V11;Y1|V00,V10)}),
(\ref{eq:crng-case1-I(V10,V11;Y1|V00)}),
and (\ref{eq:crng-case1-I(V10,V20,V22;Y2|V00)}),
the first inequality of (\ref{eq:crng-case1-22-2})
comes from (\ref{eq:crng-Ri})$'$ and (\ref{eq:crng-case1-R1+R2}),
and the second inequality of (\ref{eq:crng-case1-22-2})
comes from (\ref{eq:crng-case1-I(Z20;Z22|Z00)}),
(\ref{eq:crng-case1-I(V11;Y1|V00,V10)}),
(\ref{eq:crng-case1-I(V20,V22;Y2|V00,V10)}),
and (\ref{eq:crng-case1-I(V10,V20,V22;Y2|V00)}).

We have (\ref{eq:crng-fme23}) and (\ref{eq:crng-fme24}) as
\begin{align}
R_0 + R_1 + R_2
&\leq
I(V_{11};Y_1|V_{00},V_{10})
+ I(V_{00},V_{10},V_{20},V_{22};Y_2)
\notag
\\
&\leq
I(Z_{11};Y_1,Z_{10},Z_{20}|Z_{00})
+\min\lrb{
 \begin{aligned}
  &I(Z_{00},Z_{10},Z_{22};Y_2,Z_{20})
  \\
  &I(Z_{00},Z_{10},Z_{20},Z_{22};Y_2)
 \end{aligned}
}
- I(Z_{10};Z_{11}|Z_{00})
\label{eq:crng-case1-23-24-1}
\\
[R_0 + R_1] + R_2
&\leq
I(V_{00},V_{10},V_{11};Y_1)
+ I(V_{20},V_{22};Y_2|V_{00},V_{10})
\notag
\\
&\leq
I(Z_{22};Y_2,Z_{10},Z_{20}|Z_{00})
+ \min\lrb{
 \begin{aligned}
  &I(Z_{00},Z_{20},Z_{11};Y_1,Z_{10}),
  \\
  &I(Z_{00},Z_{10},Z_{20},Z_{11};Y_1)
 \end{aligned}
}
- I(Z_{20};Z_{22}|Z_{00}),
\label{eq:crng-case1-23-24-2}
\end{align}
where
the first inequality of (\ref{eq:crng-case1-23-24-1})
comes from (\ref{eq:crng-case1-R0+R1+R2}),
and the second inequality of (\ref{eq:crng-case1-23-24-1})
comes from (\ref{eq:crng-case1-I(V11;Y1|V00,V10)})
and (\ref{eq:crng-case1-I(V00,V10,V20,V22;Y2)}),
the first inequality of (\ref{eq:crng-case1-23-24-2})
comes from (\ref{eq:crng-Ri})$'$ and
(\ref{eq:crng-case1-R0+R1}),
and the second inequality of (\ref{eq:crng-case1-23-24-2})
comes from (\ref{eq:crng-case1-I(Z20;Z22|Z00)}),
(\ref{eq:crng-case1-I(V00,V10,V11;Y1)}),
and (\ref{eq:crng-case1-I(V20,V22;Y2|V00,V10)}).

We have (\ref{eq:crng-fme25}) and (\ref{eq:crng-fme26}) as
\begin{align}
R_0 + [R_1 + R_2]
&\leq
H(Q)
+ I(V_{11};Y_1|V_{00},V_{10})
+ I(V_{10},V_{20},V_{22};Y_2|V_{00})
\notag
\\
&\leq
I(Z_{00},Z_{11};Y_1,Z_{10},Z_{20})
+ \min\lrb{
 \begin{aligned}
  &I(Z_{10},Z_{22};Y_2,Z_{20}|Z_{00}),
  \\
  &I(Z_{10},Z_{20},Z_{22};Y_2|Z_{00})
 \end{aligned}
}
- I(Z_{10};Z_{11}|Z_{00})
\label{eq:crng-case1-25-26-1}
\\
R_0 + R_1 + R_2
&\leq
H(Q)
+ I(V_{10},V_{11};Y_1|V_{00})
+ I(V_{20},V_{22};Y_2|V_{00},V_{10})
\notag
\\
&\leq
I(Z_{00},Z_{22};Y_2,Z_{10},Z_{20})
+ \min\lrb{
 \begin{aligned}
  &I(Z_{20},Z_{11};Y_1,Z_{10}|Z_{00}),
  \\
  &I(Z_{10},Z_{20},Z_{11};Y_1|Z_{00})
 \end{aligned}
}
- I(Z_{20};Z_{22}|Z_{00}),
\label{eq:crng-case1-25-26-2}
\end{align}
where
the first inequality of (\ref{eq:crng-case1-25-26-1})
comes from (\ref{eq:crng-H(Q)})
and (\ref{eq:crng-case1-R1+R2}),
the second inequality of (\ref{eq:crng-case1-25-26-1})
comes from (\ref{eq:crng-case1-H(Q)+I(V11;Y1|V00,V10)})
and (\ref{eq:crng-case1-I(V10,V20,V22;Y2|V00)}),
the first inequality of (\ref{eq:crng-case1-25-26-2})
comes from (\ref{eq:crng-Ri})$'$, (\ref{eq:crng-H(Q)}),
(\ref{eq:crng-case1-R1-2}),
and the second inequality of (\ref{eq:crng-case1-25-26-2})
comes from (\ref{eq:crng-case1-I(Z20;Z22|Z00)}),
(\ref{eq:crng-case1-I(V10,V11;Y1|V00)}),
and (\ref{eq:crng-case1-H(Q)+I(V20,V22;Y2|V00,V10)}).

We have (\ref{eq:crng-fme27}) as
\begin{align}
R_0 + R_1 + R_2
&\leq
H(Q)
+ I(V_{11};Y_1|V_{00},V_{10})
+ I(V_{10},V_{20},V_{22};Y_2|V_{00})
\notag
\\
&\leq
I(Z_{00},Z_{20},Z_{11};Y_1,Z_{10})
+ I(Z_{10},Z_{22};Y_2,Z_{20}|Z_{00})
- I(Z_{10};Z_{11}|Z_{00})
- I(Z_{20};Z_{22}|Z_{00})
\label{eq:crng-case1-27-1}
\\
R_0 + R_1 + R_2
&\leq
H(Q)
+ I(V_{11};Y_1|V_{00},V_{10})
+ I(V_{10},V_{20},V_{22};Y_2|V_{00})
\notag
\\
&\leq
I(Z_{00},Z_{10},Z_{22};Y_2,Z_{20})
+ I(Z_{20},Z_{11};Y_1,Z_{10}|Z_{00})
- I(Z_{10};Z_{11}|Z_{00})
- I(Z_{20};Z_{22}|Z_{00}),
\label{eq:crng-case1-27-2}
\end{align}
where the first inequality of (\ref{eq:crng-case1-27-1})
and (\ref{eq:crng-case1-27-2})
come from (\ref{eq:crng-H(Q)})
and (\ref{eq:crng-case1-R1+R2}),
the second inequality of (\ref{eq:crng-case1-27-1})
comes from (\ref{eq:crng-case1-I(Z20;Z22|Z00)}),
(\ref{eq:crng-case1-H(Q)+I(V11;Y1|V00,V10)}),
and (\ref{eq:crng-case1-I(V10,V20,V22;Y2|V00)}),
and the second inequality of (\ref{eq:crng-case1-27-2})
comes from (\ref{eq:crng-case1-I(Z20;Z22|Z00)}),
(\ref{eq:crng-case1-I(V11;Y1|V00,V10)}),
and (\ref{eq:crng-case1-H(Q)+I(V10,V20,V22;Y2|V00)}).

We have (\ref{eq:crng-fme28}) as
\begin{align}
2 R_0 + [R_1 + R_2]
&\leq
2H(Q) + I(V_{11};Y_1|V_{00},V_{10}) + I(V_{10},V_{20},V_{22};Y_2|V_{00})
\notag
\\
&\leq
I(Z_{00},Z_{11};Y_1,Z_{10},Z_{20})
+ I(Z_{00},Z_{10},Z_{22};Y_2,Z_{20})
- I(Z_{10};Z_{11}|Z_{00})
\label{eq:crng-case1-28-1}
\\
2 R_0 + R_1 + R_2
&\leq
2 H(Q)
+ I(V_{10},V_{11};Y_1|V_{00})
+ I(V_{20},V_{22};Y_2|V_{00},V_{10})
\notag
\\
&\leq
I(Z_{00},Z_{22};Y_2,Z_{10},Z_{20})
+ I(Z_{00},Z_{20},Z_{11};Y_1,Z_{10})
- I(Z_{20};Z_{22}|Z_{00}),
\label{eq:crng-case1-28-2}
\end{align}
where
the first inequality of (\ref{eq:crng-case1-28-1})
comes from (\ref{eq:crng-H(Q)})
and (\ref{eq:crng-case1-R1+R2}),
the second inequality of (\ref{eq:crng-case1-28-1})
comes from (\ref{eq:crng-case1-H(Q)+I(V11;Y1|V00,V10)}),
and (\ref{eq:crng-case1-H(Q)+I(V10,V20,V22;Y2|V00)}),
the first inequality of (\ref{eq:crng-case1-28-2})
comes from (\ref{eq:crng-Ri})$'$, (\ref{eq:crng-H(Q)}),
and (\ref{eq:crng-case1-R1-2}),
and the second inequality of (\ref{eq:crng-case1-28-2})
comes from (\ref{eq:crng-case1-I(Z20;Z22|Z00)}),
(\ref{eq:crng-case1-H(Q)+I(V10,V11;Y1|V00)}),
and (\ref{eq:crng-case1-H(Q)+I(V20,V22;Y2|V00,V10)}).

We have (\ref{eq:crng-fme29}) and (\ref{eq:crng-fme30}) as
\begin{align}
2 R_0 + R_1 + R_2
&=
R_0 + [R_0 + R_1 + R_2]
\notag
\\
&\leq
H(Q)
+ I(V_{11};Y_1|V_{00},V_{10})
+ I(V_{00},V_{10},V_{20},V_{22};Y_2)
\notag
\\
&\leq
\min\lrb{
 \begin{aligned}
  &I(Z_{00},Z_{11};Y_1,Z_{10},Z_{20}),
  \\
  &I(Z_{00},Z_{20},Z_{11};Y_1,Z_{10})
 \end{aligned}
}
+ I(Z_{00},Z_{10},Z_{20},Z_{22};Y_2)
- I(Z_{10};Z_{11}|Z_{00})
\label{eq:crng-case1-29-1}
\\
2 R_0 + R_1 + R_2
&=
R_0 + [R_0 + R_1] + R_2
\notag
\\
&\leq
H(Q)
+ I(V_{00},V_{10},V_{11};Y_1)
+ I(V_{20},V_{22};Y_2|V_{00},V_{10})
\notag
\\
&\leq
\min\lrb{
 \begin{aligned}
  &I(Z_{00},Z_{22};Y_2,Z_{10},Z_{20}),
  \\
  &I(Z_{00},Z_{10},Z_{22};Y_2,Z_{20})
 \end{aligned}
}
+ I(Z_{00},Z_{10},Z_{20},Z_{11};Y_1)
- I(Z_{20};Z_{22}|Z_{00}),
\label{eq:crng-case1-29-2}
\end{align}
where
the first inequality of (\ref{eq:crng-case1-29-1})
comes from (\ref{eq:crng-H(Q)})
and (\ref{eq:crng-case1-R0+R1+R2}),
the second inequality of (\ref{eq:crng-case1-29-1})
comes from (\ref{eq:crng-case1-H(Q)+I(V11;Y1|V00,V10)}),
and (\ref{eq:crng-case1-I(V00,V10,V20,V22;Y2)}),
the first inequality of (\ref{eq:crng-case1-29-2})
comes from (\ref{eq:crng-Ri})$'$, (\ref{eq:crng-H(Q)}),
and (\ref{eq:crng-case1-R0+R1}),
and the second inequality of (\ref{eq:crng-case1-29-2})
comes from (\ref{eq:crng-case1-I(Z20;Z22|Z00)}),
(\ref{eq:crng-case1-I(V00,V10,V11;Y1)}),
and (\ref{eq:crng-case1-H(Q)+I(V20,V22;Y2|V00,V10)}).

We have (\ref{eq:crng-fme31}) as
\begin{align}
2 R_0 + R_1 + R_2
&=
R_0 + [R_0 + R_1 + R_2]
\notag
\\
&\leq
2H(Q)
+ I(V_{11};Y_1|V_{00},V_{10})
+ I(V_{10},V_{20},V_{22};Y_2|V_{00})
\notag
\\
&\leq
I(Z_{00},Z_{20},Z_{11};Y_1,Z_{10})
+ I(Z_{00},Z_{10},Z_{22};Y_2,Z_{20})
- I(Z_{10};Z_{11}|Z_{00})
- I(Z_{20};Z_{22}|Z_{00}),
\end{align}
where
the first inequality
comes from (\ref{eq:crng-H(Q)})
and (\ref{eq:crng-case1-R0+R1+R2}),
and the second inequality
comes from (\ref{eq:crng-case1-I(Z20;Z22|Z00)}),
(\ref{eq:crng-case1-H(Q)+I(V11;Y1|V00,V10)}),
and (\ref{eq:crng-case1-H(Q)+I(V10,V20,V22;Y2|V00)}).

We have (\ref{eq:crng-fme32}) as
\begin{align}
R_0 + 2R_1 + R_2
&=
R_0 + R_1 + [R_1 + R_2]
\notag
\\
&\leq
H(Q)
+ I(V_{10},V_{11};Y_1|V_{00})
+ I(V_{11};Y_1|V_{00},V_{10})
+ I(V_{10},V_{20},V_{22};Y_2|V_{00})
\notag
\\
&\leq
I(Z_{10},Z_{20},Z_{11};Y_1|Z_{00})
+ I(Z_{11};Y_1,Z_{10},Z_{20}|Z_{00})
+ I(Z_{00},Z_{10},Z_{22};Y_2,Z_{20})
\notag
\\*
&\quad
- I(Z_{10};Z_{11}|Z_{00})
- I(Z_{20};Z_{22}|Z_{00})
\label{eq:crng-case1-32-1}
\\
R_0 + 2R_2 + R_1
&=
R_0 + [R_1+ R_2] + R_2
\notag
\\
&\leq
H(Q)
+ I(V_{11};Y_1|V_{00},V_{10})
+ I(V_{10},V_{20},V_{22};Y_2|V_{00})
+ I(V_{20},V_{22};Y_2|V_{00},V_{10})
\notag
\\
&\leq
I(Z_{10},Z_{20},Z_{22};Y_2|Z_{00})
+ I(Z_{22};Y_2,Z_{10},Z_{20}|Z_{00})
+ I(Z_{00},Z_{20},Z_{11};Y_1,Z_{10})
\notag
\\*
&\quad
- I(Z_{10};Z_{11}|Z_{00})
- I(Z_{20};Z_{22}|Z_{00}),
\label{eq:crng-case1-32-2}
\end{align}
where
the first inequality of (\ref{eq:crng-case1-32-1})
comes from (\ref{eq:crng-H(Q)}),
(\ref{eq:crng-case1-R1-2}), 
and (\ref{eq:crng-case1-R1+R2}),
the second inequality of (\ref{eq:crng-case1-32-1})
comes from (\ref{eq:crng-case1-I(Z20;Z22|Z00)}),
(\ref{eq:crng-case1-I(V11;Y1|V00,V10)}),
(\ref{eq:crng-case1-I(V10,V11;Y1|V00)}),
and (\ref{eq:crng-case1-H(Q)+I(V10,V20,V22;Y2|V00)}),
the first inequality of (\ref{eq:crng-case1-32-2})
comes from (\ref{eq:crng-Ri})$'$, (\ref{eq:crng-H(Q)}),
and (\ref{eq:crng-case1-R1+R2}),
and the second inequality of (\ref{eq:crng-case1-32-2})
comes from (\ref{eq:crng-case1-I(Z20;Z22|Z00)}),
(\ref{eq:crng-case1-H(Q)+I(V11;Y1|V00,V10)}),
(\ref{eq:crng-case1-I(V20,V22;Y2|V00,V10)}),
and (\ref{eq:crng-case1-I(V10,V20,V22;Y2|V00)}).

We have (\ref{eq:crng-fme33}) as
\begin{align}
R_0 + 2R_1 + R_2
&=
[R_0 + R_1] + [R_1 + R_2]
\notag
\\
&\leq
I(V_{00},V_{10},V_{20},V_{11};Y_1)
+ I(V_{11};Y_1|V_{00},V_{10})
+ I(V_{10},V_{20},V_{22};Y_2|V_{00})
\notag
\\
&\leq
I(Z_{00},Z_{10},Z_{20},Z_{11};Y_1)
+ I(Z_{11};Y_1,Z_{10},Z_{20}|Z_{00})
+ I(Z_{10},Z_{22};Y_2,Z_{20}|Z_{00}) 
\notag
\\*
&\quad
- I(Z_{10};Z_{11}|Z_{00})
- I(Z_{20};Z_{22}|Z_{00})
\label{eq:crng-case1-33-1}
\\
R_0 + 2 R_2 + R_1
&=
[R_0 + R_1 + R_2] + R_2
\notag
\\
&\leq
I(V_{11};Y_1|V_{00},V_{10})
+ I(V_{00},V_{10},V_{20},V_{22};Y_2)
+ I(V_{20},V_{22};Y_2|V_{00},V_{10})
\notag
\\
&\leq
I(Z_{00},Z_{10},Z_{20},Z_{22};Y_2)
+ I(Z_{22};Y_2,Z_{10},Z_{20}|Z_{00})
+ I(Z_{20},Z_{11};Y_1,Z_{10}|Z_{00}) 
\notag
\\*
&\quad
- I(Z_{10};Z_{11}|Z_{00})
- I(Z_{20};Z_{22}|Z_{00}),
\label{eq:crng-case1-33-2}
\end{align}
where
the first inequality of (\ref{eq:crng-case1-33-1})
comes from (\ref{eq:crng-R0+Ri}) and (\ref{eq:crng-case1-R1+R2}),
the second inequality of (\ref{eq:crng-case1-33-1})
comes from (\ref{eq:crng-case1-I(Z20;Z22|Z00)}),
(\ref{eq:crng-case1-I(V11;Y1|V00,V10)}),
(\ref{eq:crng-case1-I(V00,V10,V20,V11;Y1)}),
and (\ref{eq:crng-case1-I(V10,V20,V22;Y2|V00)}),
the first inequality of (\ref{eq:crng-case1-33-2})
comes from (\ref{eq:crng-Ri})$'$
and (\ref{eq:crng-case1-R0+R1+R2}),
and the second inequality of (\ref{eq:crng-case1-33-2})
comes from (\ref{eq:crng-case1-I(Z20;Z22|Z00)}),
(\ref{eq:crng-case1-I(V11;Y1|V00,V10)}),
(\ref{eq:crng-case1-I(V20,V22;Y2|V00,V10)}),
and (\ref{eq:crng-case1-I(V00,V10,V20,V22;Y2)}).

We have (\ref{eq:crng-fme34}) as
\begin{align}
2R_0 + 2R_1 + R_2
&=
R_0 + [R_0 + R_1] + [R_1 + R_2]
\notag
\\
&\leq
H(Q)
+ I(V_{00},V_{10},V_{11};Y_1)
+ I(V_{11};Y_1|V_{00},V_{10})
+ I(V_{10},V_{20},V_{22};Y_2|V_{00})
\notag
\\
&\leq
I(Z_{00},Z_{10},Z_{20},Z_{11};Y_1)
+ I(Z_{11};Y_1,Z_{10},Z_{20}|Z_{00})
+ I(Z_{00},Z_{10},Z_{22};Y_2,Z_{20})
\notag
\\*
&\quad
- I(Z_{10};Z_{11}|Z_{00})
- I(Z_{20};Z_{22}|Z_{00})
\label{eq:crng-case1-34-1}
\\
2R_0 + 2R_2 + R_1
&=
R_0 + [R_0 + R_1 + R_2] + R_2
\notag
\\
&\leq
H(Q)
+ I(V_{11};Y_1|V_{00},V_{10})
+ I(V_{00},V_{10},V_{20},V_{22};Y_2)
+ I(V_{20},V_{22};Y_2|V_{00},V_{10})
\notag
\\
&\leq
I(Z_{00},Z_{10},Z_{20},Z_{22};Y_2)
+ I(Z_{22};Y_2,Z_{10},Z_{20}|Z_{00})
+ I(Z_{00},Z_{20},Z_{11};Y_1,Z_{10})
\notag
\\*
&\quad
- I(Z_{10};Z_{11}|Z_{00})
- I(Z_{20};Z_{22}|Z_{00}),
\label{eq:crng-case1-34-2}
\end{align}
where
the first inequality of (\ref{eq:crng-case1-34-1})
comes from (\ref{eq:crng-H(Q)}),
(\ref{eq:crng-case1-R0+R1}),
and (\ref{eq:crng-case1-R1+R2}),
the second inequality of (\ref{eq:crng-case1-34-1})
comes from (\ref{eq:crng-case1-I(Z20;Z22|Z00)}),
(\ref{eq:crng-case1-I(V11;Y1|V00,V10)}),
(\ref{eq:crng-case1-I(V00,V10,V11;Y1)}),
and (\ref{eq:crng-case1-H(Q)+I(V10,V20,V22;Y2|V00)}),
the first inequality of (\ref{eq:crng-case1-34-2})
comes from (\ref{eq:crng-Ri})$'$, (\ref{eq:crng-H(Q)}),
and (\ref{eq:crng-case1-R0+R1+R2}),
and the second inequality of (\ref{eq:crng-case1-34-2})
comes from (\ref{eq:crng-case1-I(Z20;Z22|Z00)}),
(\ref{eq:crng-case1-H(Q)+I(V11;Y1|V00,V10)}),
(\ref{eq:crng-case1-I(V20,V22;Y2|V00,V10)}),
and (\ref{eq:crng-case1-I(V00,V10,V20,V22;Y2)}).

We have (\ref{eq:crng-fme35}) as
\begin{align}
2R_0 + 2R_1 + R_2
&=
R_0 + [R_0 + R_1] + [R_1 + R_2]
\notag
\\
&\leq
H(Q)
+ I(V_{00},V_{10},V_{11};Y_1)
+ I(V_{11};Y_1|V_{00},V_{10})
+ I(V_{10},V_{20},V_{22};Y_2|V_{00})
\notag
\\
&\leq
I(Z_{00},Z_{10},Z_{20},Z_{11};Y_1)
+ I(Z_{00},Z_{11};Y_1,Z_{10},Z_{20})
+ I(Z_{10},Z_{22};Y_2,Z_{20}|Z_{00})
\notag
\\*
&\quad
- I(Z_{10};Z_{11}|Z_{00})
- I(Z_{20};Z_{22}|Z_{00})
\label{eq:crng-case1-35-1}
\\
2R_0 + 2R_2 + R_1
&=
R_0 + [R_0 + R_1 + R_2] + R_2
\notag
\\
&\leq
H(Q)
+ I(V_{11};Y_1|V_{00},V_{10})
+ I(V_{00},V_{10},V_{20},V_{22};Y_2)
+ I(V_{20},V_{22};Y_2|V_{00},V_{10})
\notag
\\
&\leq
I(Z_{00},Z_{10},Z_{20},Z_{22};Y_2)
+ I(Z_{00},Z_{22};Y_2,Z_{10},Z_{20})
+ I(Z_{20},Z_{11};Y_1,Z_{10}|Z_{00})
\notag
\\*
&\quad
- I(Z_{10};Z_{11}|Z_{00})
- I(Z_{20};Z_{22}|Z_{00}),
\label{eq:crng-case1-35-2}
\end{align}
where
the first inequality of (\ref{eq:crng-case1-35-1})
comes from (\ref{eq:crng-H(Q)}),
(\ref{eq:crng-case1-R0+R1}),
and (\ref{eq:crng-case1-R1+R2})
the second inequality of (\ref{eq:crng-case1-35-1})
comes from (\ref{eq:crng-case1-I(Z20;Z22|Z00)}),
(\ref{eq:crng-case1-H(Q)+I(V11;Y1|V00,V10)}),
(\ref{eq:crng-case1-I(V00,V10,V11;Y1)}),
and (\ref{eq:crng-case1-I(V10,V20,V22;Y2|V00)}),
the first inequality of (\ref{eq:crng-case1-35-2})
comes from (\ref{eq:crng-Ri})$'$, (\ref{eq:crng-H(Q)}),
and (\ref{eq:crng-case1-R0+R1+R2}),
and the second inequality of (\ref{eq:crng-case1-35-2})
comes from (\ref{eq:crng-case1-I(Z20;Z22|Z00)}),
(\ref{eq:crng-case1-I(V11;Y1|V00,V10)}),
(\ref{eq:crng-case1-H(Q)+I(V20,V22;Y2|V00,V10)}),
and (\ref{eq:crng-case1-I(V00,V10,V20,V22;Y2)}).

We have (\ref{eq:crng-fme36}) as
\begin{align}
3 R_0 + 2 R_1 + R_2
&=
2 R_0 + [R_0 + R_1] + [R_1 + R_2]
\notag
\\
&\leq
2H(Q)
+ I(V_{00},V_{10},V_{11};Y_1)
+ I(V_{11};Y_1|V_{00},V_{10})
+ I(V_{10},V_{20},V_{22};Y_2|V_{00})
\notag
\\
&\leq
I(Z_{00},Z_{10},Z_{20},Z_{11};Y_1)
+ I(Z_{00},Z_{11};Y_1,Z_{10},Z_{20})
+ I(Z_{00},Z_{10},Z_{22};Y_{2},Z_{20})
\notag
\\*
&\quad
- I(Z_{10};Z_{11}|Z_{00})
- I(Z_{20};Z_{22}|Z_{00}),
\label{eq:crng36-1}
\\
3 R_0 + 2 R_2 + R_1
&=
2 R_0 + [R_0 + R_1 + R_2] + R_2
\notag
\\
&\leq
2H(Q)
+ I(V_{11};Y_1|V_{00},V_{10})
+ I(V_{00},V_{10},V_{20},V_{22};Y_2)
+ I(V_{20},V_{22};Y_2|V_{00},V_{10})
\notag
\\
&\leq
I(Z_{00},Z_{10},Z_{20},Z_{22};Y_2)
+ I(Z_{00},Z_{22};Y_2,Z_{10},Z_{20})
+ I(Z_{00},Z_{20},Z_{11};Y_1,Z_{10})
\notag
\\*
&\quad
- I(Z_{10};Z_{11}|Z_{00})
- I(Z_{20};Z_{22}|Z_{00}),
\label{eq:crng36-2}
\end{align}
where
the first inequality of (\ref{eq:crng36-1})
comes from (\ref{eq:crng-H(Q)}),
(\ref{eq:crng-case1-R0+R1}),
and (\ref{eq:crng-case1-R1+R2}),
the second inequality of (\ref{eq:crng36-1})
comes from (\ref{eq:crng-case1-I(Z20;Z22|Z00)}),
(\ref{eq:crng-case1-H(Q)+I(V11;Y1|V00,V10)}),
(\ref{eq:crng-case1-I(V00,V10,V11;Y1)}),
(\ref{eq:crng-case1-H(Q)+I(V10,V20,V22;Y2|V00)}),
the first inequality of (\ref{eq:crng36-2})
comes from (\ref{eq:crng-Ri})$'$, (\ref{eq:crng-H(Q)}),
and (\ref{eq:crng-case1-R0+R1+R2}),
and the second inequality of (\ref{eq:crng36-2})
comes from (\ref{eq:crng-case1-I(Z20;Z22|Z00)}),
(\ref{eq:crng-case1-H(Q)+I(V11;Y1|V00,V10)}),
(\ref{eq:crng-case1-H(Q)+I(V20,V22;Y2|V00,V10)}),
and (\ref{eq:crng-case1-I(V00,V10,V20,V22;Y2)}).


\begin{thebibliography}{99}
\addcontentsline{toc}{chapter}{Bibliography}
\bibitem{A74}
R.\ Ahlswede,
``The capacity region of a channel with two senders and two receivers,''
{\it Annals of Prob.},
vol.\ 2, no.\ 5, pp.\ 805--814, 1974.
\bibitem{AC98}
R.\ Ahlswede and I.\ Csisz\'{a}r,
``Common randomness in information theory and cryptography
--- Part II: CR capacity,''
{\it IEEE Trans.\ Inform.\ Theory},
vol.\ IT-44, no.1,
pp.\ 225--240, Jan.\ 1998.
\bibitem{C78}
A.\ B.\ Carleial,
``Interference channels,''
{\it IEEE Trans.\ Inform.\ Theory},
vol.\ IT-24, no.\ 1, pp.\ 60--70, 1978.
\bibitem{CW}
J.\ L.\ Carter and M.\ N.\ Wegman,
``Universal classes of hash functions,''
{\it J.\ Comput.\ Syst.\ Sci.},
vol.\ 18, pp.\ 143--154, 1979.
\bibitem{CMG06}
H.\ F. Chong, M. Motani, and H. K. Garg,
``A comparison of two
achievable rate regions for the interference channel,''
{\it Proc.\ ITA Workshop}, San Diego, USA, Feb. 2006.
\bibitem{CMGE08}
H.\ F.\ Chong, M.\ Motani, H.\ K.\ Garg, and H.\ El Gamal,
``On the Han-Kobayashi region for the interference channel,''
{\it IEEE Trans.\ Inform.\ Theory},
vol.\ IT-54, no.\ 7, pp.\ 3188--3195, 2008.
\bibitem{C75}
T.\ M.\ Cover,
``A proof of the data compression theorem of Slepian and Wolf
for ergodic sources,''
{\it IEEE Trans. Inform Theory},
vol.~IT-21, no.~2, pp.~226--228, Mar.\ 1975.
\bibitem{CSI82}
I.\ Csisz\'{a}r,
``Linear codes for sources and source networks:
Error exponents, universal coding,''
{\it IEEE Trans.\ Inform.\ Theory},
vol.\ IT-28, no.\ 4, pp.\ 585--592, Jul.\ 1982.
\bibitem{CK11}
I.\ Csisz\'{a}r and J.\ K\"{o}rner,
{\it Information Theory: Coding Theorems for Discrete Memoryless Systems
2nd Ed.},
Cambridge University Press, 2011.
\bibitem{EK11}
A.\ El Gamal and Y.H. Kim,
{\it Network Information Theory},
Cambridge University Press, 2011.
\bibitem{GGP15}
I.\ B.\ Gattegno, Z.\ Goldfeld, H. H. Permuter,
``Fourier-Motzkin elimination software for information theoretic inequalities,''
{\it IEEE Inform.\ Theory Society Newsletter},
pp.~25--28, Sep. 2015.
\bibitem{GP17}
I. B. Gattegno and H. H. Permuter,
``Extracting analytic proofs from numerically solved
Shannon-type inequalities,''
available at {\tt arXiv:1707.01656 [cs.IT]}, 2017.
\bibitem{HLTY20}
S.W. Ho, L. Ling, C. W. Tan, and R. W. Yeung,
``Proving and disproving information inequalities:
theory and scalable algorithms,''
{\it IEEE Trans.\ Inform.\ Theory},
vol.~IT-66, no.~9, pp.~5522--5536, 2020.
\bibitem{HAN}
T.S.\ Han,
{\it Information-Spectrum Methods in Information Theory},
Springer, 2003.
\bibitem{HK81}
T.S.\ Han, and K.\ Kobayashi;
``A new achievable rate region for the interference channel,''
{\it IEEE Trans. Info. Theory},
vol.\ IT-27, no.\ 1, pp.\ 49 - 60, Jan.\ 1981.
\bibitem{IZ89}
R.\ Impagliazzo and D.\ Zuckerman,
``How to recycle random bits,''
{\it 30th IEEE Symp. Fund. Computer Sci.},
Oct.~30--Nov.~1, 1989, pp.~248--253.
\bibitem{JXG08}
J.\ Jiang, Y.\ Xin, and H.\ K.\ Garg,
``Interference channels with common information,''
{\it IEEE Trans.\ Inform.\ Theory},
vol.~IT-54, no.~1, pp.~171--187, Jan.\ 2008.
\bibitem{KH07}
K.\ Kobayashi and T.S. Han,
``A further consideration on the HK and the CMG
regions for the interference channel,''
{\it Proc. ITA Workshop}, San Diego, USA, Jan.\ 2007.
\bibitem{L21}
C.\ T.\ Li,
``An automated theorem proving framework for information-theoretic results,''
{\it Proc.\ 2021 IEEE Int.\ Symp\. Inform.\ Theory},
Melbourne, Australia, 12-20 Jul.\ 12--20, 2021, pp. 2750--2755.
\bibitem{CRNG}
J.\ Muramatsu,
``Channel coding and lossy source coding using a generator
of constrained random numbers,''
{\it IEEE Trans.\ Inform.\ Theory},
vol.~IT-60, no.~5, pp.~2667--2686, May 2014.
\bibitem{HASH}
J.\ Muramatsu and S.\ Miyake,
``Hash property and coding theorems for sparse matrices and
maximal-likelihood coding,''
{\it IEEE Trans.\ Inform.\ Theory},
vol.\ IT-56, no. 5, pp.\ 2143--2167, May 2010.
Corrections: vol.\ IT-56, no.\ 9, p.\ 4762, Sep.\ 2010,
vol.\ IT-59, no.\ 10, pp.\ 6952--6953, Oct.\ 2013.
\bibitem{ISIT2010}
J.\ Muramatsu and S.\ Miyake,
``Construction of broadcast channel code
based on hash property,''
{\it Proc.\ 2010 IEEE Int.\ Symp.\ Inform.\ Theory},
Austin, U.S.A., June 13--18, pp.\ 575--579, 2010.
Extended version is available at {\tt arXiv:1006.5271[cs.IT]}, 2010.
\bibitem{SDECODING}
J.\ Muramatsu and S.\ Miyake,
``On the error probability of stochastic decision and stochastic
decoding,''
{\it Proc.\ 2017 IEEE Int.\ Symp.\ Inform.\ Theory},
Aachen, Germany, Jun.\ 25--30, 2017, pp.\ 1643--1647.
Extended version is available at
{\tt arXiv:1701.04950[cs.IT]}.
\bibitem{CRNG-MULTI}
J.\ Muramatsu and S.\ Miyake,
``Multi-terminal codes using constrained-random-number generator,''
{\it Proceedings of the 2018 International Symposium on Information
Theory and its Applications},
Singapore, Oct.~28--31, 2018, pp.\ 612--616.
Extended version is available at {\tt arXiv:1801.02875v2[cs.IT]}.
\bibitem{SW2CC}
J.\ Muramatsu and S.\ Miyake,
``Channel code using constrained-random-number generator revisited,''
{\it IEEE Trans.\ Inform.\ Theory},
vol.\ IT-65, no.\ 1, pp.\ 500--508, Jan.\ 2019.
\bibitem{NXY15}
C.\ Nair, L.\ Xia, and M.\ Yazdanpanah,
``Sub-optimality of Han-Kobayashi achievable region for interference
channels,''
{\it Proc.\ 2015 IEEE Int.\ Symp.\ Inform.\ Theory},
Hong Kong, China, Jun.\ 14--19, 2015, pp. 2416--2420.
\bibitem{PPD}
R. Pulikkoonattu, E. Perron, and S. Diggavi,
``X Information Theoretic Inequalities Prover,''
{\tt http://xitip.epfl.ch/}
\bibitem{SV06}
A.\ Somekh-Baruch and S.\ Verd\'u,
``General relayless networks: representation of the capacity region,''
{\it Proc.\ 2006 IEEE Int.\ Symp.\ Inform.\ Theory},
Seattle, USA, 6--12 July, 2006, pp.\ 2408--2412.
\bibitem{VH94}
S. Verd\'u and T.S.\ Han,
``A general formula for channel capacity,''
{\it IEEE Trans. Inform. Theory},
vol.\ IT-40, no.\ 4, pp.\ 1147--1157, Jul. 1994.
\bibitem{Y08}
R.\ W.\ Yeung,
{\it Information Theory and Network Coding},
Springer, 2008.
\end{thebibliography}
\end{document}